\documentclass[twoside]{thesis}
\usepackage{amsmath}  
\usepackage{epsfig}   
\usepackage{eqnlabel} 
\usepackage{cite}     
\usepackage{bibquote} 

\newcommand{\bbox}{\boldsymbol}
\def\NEG#1{{\rlap/#1}}
\def\doquote#1#2{\begin{quotation} ``#1'' \\ ---\textit{#2}\end{quotation}}

\begin{document}

\prelimpages
\Title{Light Front Field Theory Calculation of Deuteron Properties}
\Author{Jason Randolph Cooke}
\Year{2001}
\Program{Physics}
\titlepage

\Chair{Gerald A. Miller}{Professor}{Department of Physics}
\Signature{Gerald A. Miller}
\Signature{Daniel R. Phillips}
\Signature{Stephen R. Sharpe}
\signaturepage

\doctoralquoteslip

\setcounter{page}{-1}
\abstract{
Light front dynamics is a promising approach for solving bound state
problems in nuclear physics. It is also ideal for calculating the
deuteron form factors at high momentum transfers. However, in
light-front dynamics rotational invariance is not manifest, which
results in a splitting in the binding energies of states with different
magnetic quantum numbers and a breaking of the angular condition for the
matrix elements of the deuteron current operator. The objective of this
work is to investigate the symptoms of broken rotational invariance for
deuterons calculated using one-meson-exchange (OME) and
two-meson-exchange (TME) potentials for various models.

We first consider the Wick-Cutkosky model. The binding energies of
states with different $m$ values are split when calculated with the OME
potential, but this splitting is reduced for all states when the TME
potential is included. In addition, we find that appropriate OME+TME
potentials give almost identical results to the ladder and crossed
ladder Bethe-Salpeter equation. 

Next, we derive light-front nucleon-nucleon OME and TME potentials from
an effective nuclear Lagrangian. We consider first the potentials
generated by the exchange of pions only. There is a large splitting in
the binding energies between the $m=0$ and $m=1$ deuteron states when
the one-pion-exchange (OPE) potential is used. Including the chiral
two-pion-exchange (TPE) potential in the calculation reduces this
splitting. We then proceed to use the potentials generated by all the
mesons, and find that the deuteron mass splittings are smaller for both
the OME and OME+TME potentials than in the pion-only model. 

The deuteron wave functions are used to calculate the electromagnetic
and axial current matrix elements and form factors. The matrix elements
of the electromagnetic current operator have better transformation
properties under rotations when we use the OME potential instead of the
OME+TME potential to calculate the wave functions. The axial current
matrix elements have about the same transformation properties regardless
of whether the OME or OME+TME potential is used. Finally, at momentum
transfers greater than about 2~GeV$^2$, the breaking of rotational
invariance causes less uncertainty in the $A$ and $B$ form factors than
do the uncertainties in the nucleon form factors. 
}

\tableofcontents
\listoffigures
\listoftables

\chapter*{Glossary}      
\addcontentsline{toc}{chapter}{Glossary}
\thispagestyle{plain}
\begin{glossary}
\item[BSE] The Bethe-Salpeter equation.
\item[deuteron] The only stable bound state of a proton and a
neutron. Also a life giving material (for thesises at least).
\item[equal-time dynamics] A Hamiltonian theory based on quantization of
fields on a hyperplane defined at one instant in time. Such a plane
contains all points in space at the same time.
\item[FSR] The Feynman-Schwinger representation of the two-particle
Green's function.
\item[LFTOPT] Light-front time-ordered perturbation theory.
\item[light-front dynamics] A Hamiltonian theory based on quantization
of fields on a hyperplane defined at one instant in light-front
time. Such a plane is traced out by the front of a plane wave of light.
\item[light-front time] The light-front time $x^+$ of a space-time
location is defined by $x^+=x^0+x^3$, where $x^0$ and $x^3$ are the
time and $z$ direction ordinates of the event.
\item[MGF] Modified Green's Function (potential).
\item[ncTME] Non-chiral two-meson-exchange (potential).
\item[ncTPE] Non-chiral two-pion-exchange (potential).
\item[OBE] One-boson-exchange (potential).
\item[OME] One-meson-exchange (potential).
\item[OPE] One-pion-exchange (potential).
\item[TBE] Two-boson-exchange (potential).
\item[TME] Two-meson-exchange (potential).
\item[TPE] Two-pion-exchange (potential).
\end{glossary}

\acknowledgments{
\vskip2pc
{\narrower\noindent
Most of the work described here was done in collaboration with 
Gerald A.\ Miller (my advisor) and with Daniel R.\ Phillips. I want to
express my deepest gratitude to Jerry for teaching me about nuclear
theory and light-front dynamics, and for guiding me through this work.
I am grateful to Daniel for providing useful advice and productive
discussions on many occasions.

This work is supported in part by the U.S.\ Dept.\ of Energy under Grant
No.\ DE-FG03-97ER4014. 

I have also benefited from numerous conversations about our respective
research topics with Daniel Arndt, Jiunn-Wei Chen, Patrick Fox, Gautam
Rupak, and Brian Tiburzi. I also owe a lot to my friends H.\ Pieter Mumm
and Paul Bedrosian, who ``experienced'' the first-year with me, and to
my undergraduate research advisor and friend, Madappa Prakash.

Most of all, I thank my wife for her patience and love during graduate
school, and my son for supporting me by saying ``You can do it, Daddy!''
\par}}

\dedication{\begin{center}For Sam\end{center}}

\textpages
\chapter{Introduction} \label{ch:int}

\doquote
{If you always do what interests you, at least one person is pleased.}
{Katharine Hepburn} 

Recent experiments at Thomas Jefferson National Accelerator Facility
have measured the $A(Q^2)$ structure function of the deuteron for
momentum transfers up to 6 (GeV/c)$^2$ \cite{Alexa:1999fe}, and
measurements for $B(Q^2)$ are planned.  At such large momentum
transfers, a relativistic description of the deuteron is required.  Even
at lower momentum transfers, a relativistic description is important to
understand the details of the form factors. In addition, incorporating
relativity is important for the deuteron wave function to transform
correctly under boosts to large momentum, which is important for
calculating form factors.

One approach that gives a relativistic description of the deuteron is
light-front dynamics. The subject of this work is to investigate the
consequences of combining light front dynamics with various nuclear
models to calculate bound state wave functions, and using them to
calculate the deuteron form factors. 

\section{Theoretical Descriptions of the Deuteron}

In principle, the two-particle deuteron wave function is obtained by
solving the full quantum field theory that describes the interacting
nucleons for the lowest energy bound state. In practice, many
approximations must be made before even a numeric solution is possible.

There are many approximations that can be used to obtain relativistic
two-particle wave functions. One approach is to use the
Feynman-Schwinger representation (FSR) of the two-particle Green's
function \cite{Simonov:1993kp}, which effectively includes {\em all}
two-particle to two-particle ladder and crossed ladder diagrams.
However, it requires a path integral to be done numerically, and
intensive computation is required to obtain an accurate answers. While
the FSR has been applied to scalar theories, such as the Wick-Cutkosky
model, it has not been used with more realistic models of the deuteron
involving spin.

Another approach comes from using the Bethe-Salpeter equation (BSE)
\cite{Nambu:1950vt,Schwinger:1951ex,Schwinger:1951hq,Gell-Mann:1951rw,%
Salpeter:1951sz}.
Solving the BSE is equivalent to solving the FSR with a truncated
kernel. This simplification allows the BSE to be solved much more
quickly than the FSR, however, the truncation of the BSE kernel causes
the BSE results to differ from the FSR results.  It is well known that
any finite truncation of the kernel yields bound-state wave functions
with problems, such as the incorrect one-body limit
\cite{Gross:1982nz}. In addition, other approximations can be made to
the propagators used in the BSE to obtain a quasi-potential equation
\cite{Brown:1976,Blankenbecler:1966gx,Gross:1969rv,Gross:1982nz} which
addresses some of the problems with the BSE.

Constructing potentials from a relativistic Hamiltonian theory provides
yet another approach. It can be shown that the potentials derived in
this manner are physically equivalent to the Bethe-Salpeter
equation. There are several different forms of Hamiltonian dynamics
\cite{Dirac:1949cp}, each associated with a hypersurface on which the
commutation relations for the fields are defined. Equal-time dynamics is
the most commonly known form of Hamiltonian dynamics, but we will
consider light-front dynamics in this work, where the orientation of the
light-front is fixed.

There is another light-front dynamics approach that is in use,
explicitly covariant light-front dynamics
\cite{Carbonell:1998rj,Schoonderwoerd:1998pk}. In that approach,
manifest covariance is kept by using a null-plane whose orientation is
variable. Although that approach has some benefits, we are interested in
numerical calculations, for which the usual light-front dynamics is best
suited.

\section{Benefits of Light-front Dynamics}

The utility of the light-front dynamics was first discussed by Dirac
\cite{Dirac:1949cp}. Light-front dynamics makes use of the light-front
coordinate system, where a four-vector $x^\mu$ is expressed as
$x^\mu=(x^+,x^-,x^1,x^2)$, with $x^\pm=x^0\pm x^3$.  This is simply a
change of variables, but an especially convenient one.  Using this
coordinate system and defining the commutation relations at equal
light-front time ($x^+=t_{\text{LF}}$), we obtain a light-front
Hamiltonian \cite{Brodsky:1998de,Harindranath:1996hq,Heinzl:1998kz}.
We use Hamiltonian in the light-front Schr\"odinger equation to solve
for bound states. There are many desirable features of the light-front
dynamics and the use of light-front coordinates.

First of all, high-energy experiments are naturally described using
light-front coordinates.  The wave front of a beam of high-energy
particles traveling in the (negative) three-direction is defined by a
surface where $x^+$ is (approximately) constant.  Such a beam can probe
the wave function of a target described in terms of light-front
variables \cite{Brodsky:1998de,Miller:1997cr}: the Bjorken $x$ variable
used to describe high-energy experiments is simply the ratio of the plus
momentum of the struck constituent particle to the total plus momentum 
($p^+$) of the bound state.

Secondly, the vacuum for a theory with massive particles can be very
simple on the light front.  This is because all massive particles and
anti-particles have positive plus momentum, and the total plus momentum
is a conserved quantity.  Thus, the na\"{\i}ve vacuum (with $p^+=0$) is
empty, and diagrams that couple to this vacuum are zero.  This greatly
reduces the number of non-trivial light-front time-ordered diagrams.

Thirdly, the generators of boosts in the one, two, and plus directions
are kinematic, meaning they are  independent of the interaction.
Thus, even when the Hamiltonian is truncated, the wave functions will
transform correctly under boosts.  Thus, light-front dynamics is useful
for describing form factors at high momentum transfers.

Finally, it is easy to perform relativistic calculations using
light-front dynamics. This is partly due to the simplicity of the
vacuum, and partly due the the fact that, with light-front variables,
center-of-momentum variables can be cleanly separated from the relative
momentum variables. This allows us to write relativistic equations which
have the simple form of a non-relativistic Schr\"odinger equation.

\section{Breaking of Rotational Invariance}

One serious drawback of light-front dynamics is that rotational
invariance is not manifest in any light-front Hamiltonian\footnote{This
problem is not unique to light-front dynamics. In any form of Hamiltonian
dynamics, at least three of the Poincar\'{e} group generators are
dynamic, and thus complicated.}.
This is a result of selecting a particular direction in space for the
orientation of the light-front.

An untruncated light-front Hamiltonian will commute with the total
relative angular momentum operator, since the total momentum commutes
with the relative momentum.  Thus, eigenstates of the full Hamiltonian
will also be eigenstates of the angular momentum.  However, as mentioned
earlier, a Fock-space truncation of the light-front  Hamiltonian results
in the momentum operator four-vector losing covariance under rotations.
Hence $J^2$ and the truncated Hamiltonian do not commute and this
implies that the eigenstates of the truncated Hamiltonian will not be
eigenstates of the angular momentum.

How will this violation of rotational invariance affect physical
observables? One way to observe this violation is to note that on the
light front, rotational invariance about the $z$-axis is maintained.
This allows us to classify states as eigenstates of $J_3$ with
eigenvalues $m$. We compare the energies of states with the same
angular momentum quantum number $j$ but different $m$ values. If the
Hamiltonian were rotationally invariant, the energies should be the
same; the breaking of rotational invariance causes the energies to be
different \cite{Trittmann:1997ga}.

We expect that the higher Fock-space components of the full Hamiltonian
will be small if the coupling constant is small enough.  Thus,
truncation at successively higher orders in the Fock-space expansion
should reduce the violation of
rotational invariance of the truncated Hamiltonian.  In particular, by
retaining enough terms in the perturbation expansion of the Hamiltonian,
the violation of rotational invariance can be reduced to an arbitrarily
small amount, provided only that a perturbation expansion is valid.
Thus, the only real question is: How many terms are required?

\section{Outline}

In chapter~\ref{wc:ch:wcmodel}, the massive Wick-Cutkosky model is
used to investigate the degree to which rotational invariance is broken
for deeply bound states. We use light-front dynamics to obtain
one-meson-exchange (OME) and two-meson-exchange (TME) potentials, and
then we use those potentials to calculate the bound states. We develop
several methods for quantifying the extent to which rotational
invariance is broken by the bound states. The binding energies we
calculate are compared with those calculated using other methods in the
literature, such as the Bethe-Salpeter equation
\cite{Phillips:1996ed,dandata} and the Feynman-Schwinger representation
of the Greens function \cite{Simonov:1993kp,Nieuwenhuis:1996mc}.

A brief discussion about the approximation we use for the Wick-Cutkosky
potentials is in order. It is well known that the vacuum of the full
Wick-Cutkosky model is unstable \cite{Baym} due to the cubic coupling
which provides the interaction. However, when the bound-state
calculation is restricted to the two-particle sector, the quenched
approximation is used, and the self-energy and vertex-renormalization
diagrams are neglected, then the theory has a well defined ground state.
In this chapter, we compare the results of our light-front Hamiltonian
calculation to the Bethe-Salpeter and Feynman-Schwinger representation
calculations, both of which
use the same approximations.  The use of these simplifying
approximations allows us to highlight the differences between the
various approaches. The inclusion of the self-energy diagrams and
counterterms for the light-front Hamiltonian \cite{Wivoda:1993qr} and
for the Feynman-Schwinger representation
\cite{Savkli:1999rw,Rosenfelder:1996bd,Rosenfelder:1996ra}
will not be discussed here. 

In chapter~\ref{ch:pionly}, we introduce a model Lagrangian for nuclear
physics which includes chiral symmetry \cite{Miller:1997cr}. The methods
introduced in chapter~\ref{wc:ch:wcmodel} are generalized for use with
this nuclear model. The Hamiltonian is derived and used to calculate the
OME and TME potentials for a new light-front nucleon-nucleon
potential. We also formulate a new light-front pion-only model. The
potentials are used to calculate the binding energy of the deuteron. We
have some freedom in how to choose the TME potentials, and we consider
several different choices.

We are interested mainly in the rotational properties of the potentials
and wave functions, not in achieving detailed agreement with experimental
data. This is because detailed agreement between calculations in
light-front dynamics and experimental data cannot be achieved until the
problems of rotational invariance have been addressed. From this
standpoint, we may simplify our calculations by omitting effects that
are formally rotationally invariant, such at crossed TME potentials, but
are required for precise calculations.

The wave functions obtained in chapter~\ref{ch:pionly} are used in
chapter~\ref{ch:ffdeut} to calculate the electromagnetic and axial form
factors of the deuteron. Although rotational invariance demands that
there be only three independent matrix elements of the deuteron current,
the light-front calculation of the deuteron current results in four
independent matrix elements. This is a result of the lack of manifest
rotational invariance on the light front. There are several
prescriptions for choosing which deuteron current matrix element should
be eliminated, and in principle this choice will affect the form
factors. We attempt to find currents that transform correctly (or well
enough) under rotations so that the choice of ``bad'' component does not
matter too much. 
 
\chapter{Wick-Cutkosky Model} \label{wc:ch:wcmodel}

The Wick-Cutkosky model is an ideal starting point for any investigation
of the deuteron. The model, introduced with an analytic solution in 1954
by Wick \cite{Wick:1954eu} and Cutkosky \cite{Cutkosky:1954ru}, is
concerned with the interaction of two heavy scalar particles mediated by
a massless scalar particle. The model can be extended by allowing the
exchange particle to have a mass. The massive Wick-Cutkosky model cannot
be solved analytically, but it is amenable to numeric calculations
\cite{Karmanov:2000hr,Mangin-Brinet:2001tc,Sales:2001gk,Schoonderwoerd:1998pk,%
LevineWright1,LevineWright2,LevineWright3,Nieuwenhuis:1996mc}.

A large amount of the material presented in this chapter is based on
previously published work by the author in
Refs.~\cite{Cooke:1999yi,Cooke:2000ef}.

\section{Formalism} \label{wc:ch:wcformalism}

We consider an isospin doublet of two uncharged scalars
$\phi=(\phi_1,\phi_2)$ with mass $M$ (which we will refer to as {\em
nucleons}), that couple to a third, uncharged scalar $\chi$ with mass
$\mu$ (which we will refer to as a {\em meson}) by a $\phi^2 \chi$
interaction. This is denoted as a $\phi^2 \chi$ theory, which is an
extension of the Wick-Cutkosky model \cite{Wick:1954eu,Cutkosky:1954ru}
to include massive mesons. This model has been used on the light front
to study scattering states \cite{Schoonderwoerd:1998pk} as well as bound
states \cite{Sales:1999ec}.  The Lagrangian is 
\begin{eqnarray}
{\mathcal L} &=& 
\frac{1}{2}\left(\partial_\mu\phi\partial^\mu\phi-  M^2\phi^2\right)+
\frac{1}{2}\left(\partial_\mu\chi\partial^\mu\chi-\mu^2\chi^2\right)+
g\frac{M}{2} \phi^2 \chi, \label{wc:thelagrangian}
\end{eqnarray}
where $g$ is a dimensionless coupling constant and
$\phi^2=\phi_1^2+\phi_2^2$.
This Lagrangian will be used in two formalisms, the Bethe-Salpeter
equation and the Hamiltonian approach.

\subsection{Light-front Hamiltonian}

To obtain the light-front Hamiltonian from the Lagrangian in
Eq.~(\ref{wc:thelagrangian}), we follow the approach of by Miller
\cite{Miller:1997cr} and many others (see the review in
Ref.~\cite{Brodsky:1998de}) to write the light-front Hamiltonian ($P^-$)
as the sum of a free, non-interacting part and a term containing the
interactions.  We use the conventions given in
Appendix~\ref{ch:notation}. The operators can be expressed in
terms of Fock space operators since for this theory in light-front
dynamics, the physical vacuum is the Fock space vacuum, and thus the
Hilbert space is simply the Fock space.  The Hamiltonian is obtained by
using the energy-momentum tensor $T^{\mu\nu}$ in
\begin{eqnarray}
P^\mu=\frac{1}{2}\int dx^-d^2x_\perp\;
T^{+\mu}(x^+=0,x^-,\bbox{x}_\perp).
\end{eqnarray}
The usual relations determine  $T^{+\mu}$, with
\begin{eqnarray}
T^{\mu\nu}=-g^{\mu\nu}{\cal L} +\sum_r
\frac{\partial{\cal L}}{\partial (\partial_\mu\phi_r)}\partial^\nu\phi_r,
\label{wc:tmunu}
\end{eqnarray}
in which the degrees of freedom (the fields $\phi$ and $\chi$) are
labeled by $\phi_r$.

It is worthwhile to consider the limit in which the interactions between the 
fields are removed.  This will allow us to define the free Hamiltonian $P^-_0$
and to display  the necessary commutation relations.  The energy-momentum
tensor of the non-interacting fields is defined as $T_0^{\mu\nu}$.  Use of
Eq.~(\ref{wc:tmunu})  leads to the result
\begin{eqnarray}
T^{\mu\nu}_0 &=&
\partial^\mu \phi \partial^\nu \phi - \frac{g^{\mu\nu}}{2}
\left[\partial_\sigma \phi \partial^\sigma \phi-M^2\phi^2\right] 
+ \partial^\mu \chi \partial^\nu \chi - \frac{g^{\mu\nu}}{2}
\left[\partial_\sigma \chi \partial^\sigma \chi-\mu^2\chi^2\right],
\end{eqnarray}
with
\begin{eqnarray}
T^{+-}_0 =
\bbox{\nabla}_\perp\phi\cdot\bbox{\nabla}_\perp\phi + M^2 \phi^2 +
\bbox{\nabla}_\perp\chi\cdot\bbox{\nabla}_\perp\chi + \mu^2 \chi^2.
\end{eqnarray}

The scalar nucleon fields can be expressed in terms of creation and
destruction operators:
\begin{eqnarray}
\phi_i(x)&=&
\int \frac{d^2 k_\perp dk^+ \, \theta(k^+)}{(2\pi)^{3/2}\sqrt{2k^+}} \left[
a_i(\bbox{k})e^{-ik\cdot x} +a_i^\dagger(\bbox{k})e^{ik\cdot x}\right],
\end{eqnarray}
where $i=1,2$ is a particle index,
$k\cdot x= \frac{1}{2}(k^-x^++k^+x^-)-\bbox{k}_\perp\cdot\bbox{x}_\perp$
with $k^-= \frac{M^2 + \bbox{k}_\perp^2}{k^+}$, and
$\bbox{k}\equiv(k^+,\bbox{k}_\perp)$. 
Note that $k^-$ is such that the particles are on the mass shell, which
is a consequence of using a Hamiltonian theory.
The $\theta$ function restricts $k^+$ to positive values.
Likewise, the scalar meson field is given by
\begin{eqnarray}
\chi(x)&=&
\int \frac{d^2 k_\perp dk^+ \, \theta(k^+)}{(2\pi)^{3/2}\sqrt{2k^+}} \left[
a_\chi(\bbox{k})e^{-ik\cdot x} +a_\chi^\dagger(\bbox{k})e^{ik\cdot x}\right],
\end{eqnarray}
where $k^-= \frac{\mu^2 + \bbox{k}_\perp^2}{k^+}$,
so that the mesons are also on the
mass shell.  The non-vanishing commutation relations  are
\begin{eqnarray}
\left[a_\alpha(\bbox{k}),a_\alpha^\dagger(\bbox{k}')\right] &=&
\delta(\bbox{k}_\perp-\bbox{k}'_\perp)
\delta(k^+-{k'}^+), \label{wc:comm}
\end{eqnarray}
where $\alpha = 1,2,\chi$ is a particle index.
The commutation relations are defined at equal light-front
time, $x^+=0$.  It is useful to define
\begin{eqnarray}
\delta^{(2,+)}(\bbox{k}-\bbox{k}') &\equiv&
\delta(\bbox{k}_\perp-\bbox{k}'_\perp)
\delta(k^+-{k'}^+), \label{wc:eq:deltwoplus}
\end{eqnarray}
which will be used throughout this work.

We write a ket in the two-distinguishable-particle
sector of the Fock space as 
\begin{eqnarray}
|k_1,k_2 \rangle &=& a_1^\dagger(k_1) a_2^\dagger(k_2) |0\rangle.
\end{eqnarray}
This implies that the identity operator in this Fock space sector can be
written as
\begin{eqnarray}
I_{2} &=& \int d^2k_{1,\perp}dk_1^+ \int d^2k_{2,\perp}dk_2^+
|k_1,k_2 \rangle \langle k_1,k_2 | \label{wc:twopartident}.
\end{eqnarray}

The derivatives appearing in the quantity $T^{+-}_0$ are evaluated and
then one sets $x^+$ to 0 to obtain the result
\begin{eqnarray}
P^-_0 &=& \int_k \, \left[
\frac{M^2 + \bbox{k}_\perp^2}{k^+}
\left( a_1^\dagger(k) a_1(k) + a_2^\dagger(k) a_2(k) \right) 
+ \frac{\mu^2 + \bbox{k}_\perp^2}{k^+}
a_\chi^\dagger(k) a_\chi(k)
\right], \label{wc:p0minusop}
\end{eqnarray}
with $\int_k = \int d^2 k_\perp dk^+ \, \theta(k^+)$.
Eq.~(\ref{wc:p0minusop}) has the interpretation of an operator that
counts the light-front energy $k^-$ (which is $\frac{M^2 +
\bbox{k}_\perp^2}{k^+}$ for the nucleons and $\frac{\mu^2 +
\bbox{k}_\perp^2}{k^+}$ for the mesons) of all of the particles.

We now consider the interacting part of the Lagrangian, ${\mathcal L}_I$.
An analysis similar to that for the non-interacting parts
yields the interacting part of the light-front
Hamiltonian $P_I^-$;
\begin{eqnarray}
P_I^- &=& \sum_{i=1,2} \frac{M}{2}\int_k \int_{k'}
\frac{1}{(2\pi)^{3/2} \sqrt{2 k^+ {k'}^+(k^++{k'}^+)} } \nonumber\\ 
&& \phantom{\sum_{i=1,2} \frac{gM}{2} \int_k \int_{k'}}
\times \left\{ \left[
2 a_i^\dagger(k+k') a_\chi(k') a_i(k   )
+ a_\chi^\dagger(k+k') a_i(k') a_i(k   )
\right] \right. \nonumber\\
&& \phantom{\sum_{i=1,2} \frac{gM}{2} \int_k \int_{k'}
\times \left\{ \right. }
+ \mbox{Hermitian conjugate} \Big\}. 
\phantom{ \left.\right\} } \label{wc:intham}
\end{eqnarray}
The interaction Hamiltonian is self-adjoint since the Hilbert space is
the Fock space. The total light-front Hamiltonian is given by
$P^-=P^-_0+gP^-_I$.

\subsection{Hamiltonian Bound-state Equations}
\label{wc:uxsec}

We will be studying the bound states of two distinguishable nucleons.
The technology of time-ordered (old-fashioned) perturbation theory is
used to construct the light-front time-ordered perturbation theory (LFTOPT)
for our Hamiltonian.  We start with the light-front Schr\"odinger
equation in the full Fock space, 
\begin{eqnarray}
\left( P_0^- + g P_I^- \right)
| \psi_F \rangle &=&
| \psi_F \rangle P^-, \label{wc:fullfock}
\end{eqnarray}
where $P_0^- + g P_I^-$ is the Hamiltonian in the full Fock-space basis,
$|\psi_F\rangle$ is the wave function in the
full Fock space, and $P^-$ is the light-front energy of that
state.  Recall that $P^-_0$, the non-interacting part of the
Hamiltonian, is diagonal in the momentum basis, while $P^-_I$, which
contains the interaction, has only off-diagonal elements.

A serious drawback of this equation is that the wave function
$|\psi_F\rangle$ has support from infinitely many sectors of
the Fock space, since $P^-_I$ changes the total number of particles.
However, the components of the wave function with many particles will be
small compared to the two-particle component if the coupling constant is
not too large and the exchange particle $\chi$ is massive.  We will
construct the two-particle light-front
Schr\"odinger equation which the two-particle component of the
wave function satisfies.  From this construction, we will obtain the
rules for the LFTOPT.

We start by introducing the projection operators ${\mathcal P}$ and
${\mathcal Q}$.  The operator ${\mathcal P}$ projects out the sector of
Fock space with two distinguishable nucleons and no mesons, while
${\mathcal Q}=I-{\mathcal P}$ projects out all the other sectors.  We
define
\begin{eqnarray}
{\mathcal P} | \psi_F \rangle & \equiv & | \psi   \rangle, \\
{\mathcal Q} | \psi_F \rangle & \equiv & | \psi_Q \rangle, 
\end{eqnarray}
so that
$|\psi_F\rangle=|\psi\rangle+|\psi_Q\rangle$.
Since the free Hamiltonian does not change the number of particles, 
$[{\mathcal P},P_0^-]=[{\mathcal Q},P_0^-]=0$.  The interaction
Hamiltonian changes the particle number, so it cannot
connect the two-particle sector to itself, thus
${\mathcal P} P_I^- {\mathcal P} = 0$. 

Using these projection operators, Eq.~(\ref{wc:fullfock}) can be broken up
into two parts,
\begin{eqnarray}
P_0^- | \psi \rangle 
+ g {\mathcal P} P_I^- {\mathcal Q} | \psi_Q \rangle &=& |
\psi \rangle P^-, \\ 
\left( P_0^- + g {\mathcal Q} P_I^- {\mathcal Q} \right) |
\psi_Q \rangle 
+ g {\mathcal Q} P_I^- {\mathcal P} | \psi \rangle &=& |
\psi_Q \rangle P^-. 
\end{eqnarray}
Eliminating the $|\psi_Q \rangle$ and using the expression
of the identity given in Eq.~(\ref{wc:twopartident}) we obtain the
two-particle effective light-front Schr\"odinger equation
\begin{eqnarray}
& &\int d^2p_{1,\perp} dp_1^+ \int d^2p_{2,\perp} dp_2^+
\langle \bbox{k}_1, \bbox{k}_2 |
\left[ P_0^- + V(g,P^-) \right] 
| \bbox{p}_1, \bbox{p}_2 \rangle
\langle \bbox{p}_1, \bbox{p}_2 | \psi \rangle \nonumber \\
& & \quad =
\langle \bbox{k}_1, \bbox{k}_2 | \psi \rangle P^-,
\label{wc:projfock}
\end{eqnarray}
where $P_0^-$ and the potential $V$ act in the two-nucleon basis.  The
two-particle potential is given by
\begin{eqnarray}
V(g,P^-) &=&  g^2 {\mathcal P} P_I^- \frac{{\mathcal Q}}{P^- - P_0^-
- g {\mathcal Q} P_I^- {\mathcal Q}} P_I^- {\mathcal P}.
\end{eqnarray}
Note that Eq.~(\ref{wc:projfock}) is similar to Eq.~(\ref{wc:fullfock}),
except for two main differences.  Here we have a two-nucleon
wave function, which makes it simpler.  However, the potential depends
on the light-front energy $P^-$, thus making it more complicated. 

The denominator in the definition of the potential is non-diagonal
in the full Fock space, so the matrix inversion that it represents is
highly non-trivial.  This problem is avoided by expanding the inversion
in powers of the coupling constant $g$ to get
\begin{eqnarray}
V(g,P^-) &=& {\mathcal P} P_I^- \left[
\frac{g^2 {\mathcal Q}}{P^- - P_0^-} \sum_{n=0}^\infty 
\left( P_I^- \frac{g {\mathcal Q}}{P^- - P_0^-} \right)^n \right]
P_I^- {\mathcal P}.
\end{eqnarray}
This can be simplified further by noting that in the two-nucleon sector
of our theory, every meson
emitted must be absorbed, so there must be an even number of
interactions.  Thus, the full potential can be written as the sum of $n$
meson exchange potentials,
\begin{eqnarray}
V(P^-,g) &=& \sum_{n=1}^\infty g^{2n} V_{(2n)}(P^-), \label{wc:fullpot}
\end{eqnarray}
where $V_{(2n)}$ is the potential due to the exchange of $n$ mesons,
given by
\begin{eqnarray}
V_{(2n)}(P^-) &=& {\mathcal P}
\left( P_I^- \frac{{\mathcal Q}}{P^- - P_0^-} \right)^{2n-1} P_I^-
{\mathcal P}.  \label{wc:LFTOPpotdet}
\end{eqnarray}

To see how to write a sum of diagrams for the potential, we express what
what Eq.~(\ref{wc:LFTOPpotdet}) represents in words. We start off with
two particles, then the interaction occurs.  There are two possibilities
of what can happen; nucleon 1 or 2 can emit a meson.  Each possibility
has a separate diagram.  After the interaction, there is propagation
with the light-front Green's function,
\begin{eqnarray}
G_{\text{LF}}(P^-) &=& \frac{1}{P^--P_0^-},
\end{eqnarray}
until another interaction occurs, and so on.  We simply sum up all of
the possible orderings of the interaction to get the full potential.
The $n^{\text{th}}$ order potential is simply the sum of all
possible diagrams with $n$-meson exchanges.

Each intermediate state in Eq.~(\ref{wc:LFTOPpotdet}) has more than two
particles, so the diagrams are two-particle irreducible with respect to
the two-particle Green's function $G_{2\text{LF}}={\mathcal
P}G_{\text{LF}}{\mathcal P}$.  We can represent $G_{2\text{LF}}$ by its
diagonal matrix elements,
\begin{eqnarray}
G_{2\text{LF}}(\bbox{k}_1,\bbox{k}_2;P^-) = \frac{1}{P^- - k_1^- - k_2^-}.
\end{eqnarray}

In the diagrams we draw, the nucleons will be represented by solid lines
and the mesons by dashed lines.  Although the states we will be
considering consist of two distinguishable nucleons, we will not
label the nucleon lines.  Energy denominator terms are represented by
vertical, thin, dotted lines. We will be using the quenched
approximation (so there are no nucleon loops) and neglect the mass and
vertex renormalization diagrams (so the physical masses and coupling
constant are used, and each meson emitted from one nucleon must
be absorbed by the other nucleon).  
It is not expected that these restrictions will lead to qualitatively
different results than the true full solution when the states are not
too deeply bound.  The quenched approximation is reasonable when the
masses of the nucleon fields are large compared to the binding energy.
Use of the physical masses and coupling constant are reasonable as well
when the momenta are not too large.

A truncation must be made for the potential in Eq.~(\ref{wc:fullpot}),
since in a Hamiltonian theory the infinite sum of graphs cannot be
calculated. However, all the graphs are required to obtain manifest
Lorentz invariance. We stress again that we will compare various
truncations of the light-front Hamiltonian to other calculations which
do not include renormalization diagrams.  This is because we want to
determine the effect of truncation on the light-front Hamiltonian.  The
differences between our calculation and those which include the
self-energy graphs \cite{Wivoda:1993qr,Sales:2001gk}, which may be
large for deeply-bound states, are not considered here.

The rules for drawing the $n$-meson-exchange graphs that
correspond to this approximation are:
\begin{enumerate}
\item{} Draw all topologically distinct time-ordered diagrams with $n$
mesons.  Use solid lines for the nucleons and dashed lines for the
mesons.
\item{} Delete all graphs which couple particles to the vacuum.  In the
massive theory we consider here, these diagrams always vanish since the
vacuum has zero plus momentum and massive particles always have
positive plus momentum.
\item{} Our quenched approximation and use of the physical masses and
coupling constant requires us to delete all graphs that have nucleon
loops or have mesons that are emitted and absorbed from the same
nucleon.
\end{enumerate}

Once the diagrams are drawn, we use the following rules to convert the
sum of diagrams into the potential
$\langle\bbox{k}_{1f},\bbox{k}_{2f}|V_{(2n)}(P^-)|\bbox{k}_{1i},\bbox{k}_{2i}\rangle$:
\begin{enumerate}
\item{} Overall factor of
$\frac{\delta^{(2,+)}(\bbox{k}_{1f}+\bbox{k}_{2f}-\bbox{k}_{1i}-\bbox{k}_{2i})}{2(2\pi)^3\sqrt{k_{1f}^+k_{2f}^+k_{1i}^+k_{2i}^+}}$.
This delta function says that the total light-front three-momentum is
conserved.  We define the light-front three-momentum
$\bbox{P}\equiv\bbox{k}_{1f}+\bbox{k}_{2f}$.
\item{} To each internal line, assign a light-front three-momentum
$\bbox{q}_i$ where $1\leq i \leq N$ and $N$ is the number of internal
lines. The light-front energy for particle $i$ with mass $m_i$ is  
$q_i=\frac{m_1^2+\bbox{q}^2_{i,\perp}}{q_i^+}$.  It is useful to define
$z_i=q_i^+/P^+$.
\item{} A factor of $\frac{\theta(z_i^+)}{z_i^+}$ for each internal
line.
\item{} An extra factor of $\frac{M^2}{P^+P^-}$ for each internal meson
line.
\item{} A factor of $\frac{P^-}{\left( P^- - \sum_i q_i^-\right)}$
between consecutive vertices, where the sum is over only the particles
that exist in the intermediate time between those vertices.
\item{} Use light-front three-momentum conservation to eliminate all the
independent momenta.
\item{} Integrate with 
$\int \frac{d^2q_{i,\perp} dz_i}{2 P^+P^- (2\pi)^3}$ over all remaining
free internal momenta.
\item{} Symmetry factor of $\frac{1}{2}$ when two nucleons are created
or destroyed at the same time.
\end{enumerate} \label{wc:rules}
With these rules, one can calculate the effective potential for any
order.

\subsection{Further Development of the Light-front Schr\"odinger
Equation} \label{wc:sec:furtherdev}

Once the potential is calculated, we can plug it into
Eq.~(\ref{wc:projfock}), which we write as
\begin{eqnarray}
& &\int d^2k_{1i,\perp} dk_{1i}^+ \int d^2k_{2i,\perp} dk_{2i}^+
\langle \bbox{k}_{1f}, \bbox{k}_{2f} |
\left[ P_0^- + V(g(P^-),P^-) \right] 
| \bbox{k}_{1i}, \bbox{k}_{2i} \rangle
\langle \bbox{k}_{1i}, \bbox{k}_{2i} | \psi \rangle \nonumber \\
& & \quad =
\langle \bbox{k}_{1f}, \bbox{k}_{2f} | \psi \rangle P^-,
\label{wc:fullse}
\end{eqnarray}
where $P^-$ is an arbitrary light-front energy and $g(P^-)$ is the
coupling constant which yields the bound-state wave function with $P^-$
as the bound-state energy.  We call this $g(P^-)$ the spectrum of the
light-front Schr\"odinger equation for the corresponding wave function.

The total momentum $\bbox{P}=\bbox{k}_{1f}+\bbox{k}_{2f}$ is conserved by the
potential given in Eq.~(\ref{wc:fullpot}), so the wave function in
Eq.~(\ref{wc:fullse}) can be parameterized by the total momentum.  To make
the calculations easier later, we choose to be in the center-of-momentum
frame, where the components of the total momentum can be written as
$\bbox{P}_\perp=0$ and $P^+=P^-=E$.  The energy, $E$, is
the same as the mass of the bound state.   In terms of the binding
energy $B$, $E=2M-B$.  In the center-of-momentum frame, the wave
function is parameterized by $E$, so we can define 
\begin{eqnarray}
\langle \bbox{k}_{1f},\bbox{k}_{2f} |\psi_M\rangle
&=&
\delta^{(2,+)}(\bbox{k}_{1f}+\bbox{k}_{2f}-\bbox{P})
\psi(\bbox{k}_{1f}), \label{wc:twopartredpsi} \\
\langle \bbox{k}_{1f},\bbox{k}_{2f}| V_{(2n)}(P^-) 
| \bbox{k}_{1i},\bbox{k}_{2i} \rangle
&=& \delta^{(2,+)}(\bbox{k}_{1f}+\bbox{k}_{2f}-\bbox{k}_{1i}-\bbox{k}_{2i})
V_{(2n)}(E;\bbox{k}_{1f};\bbox{k}_{1i}) \label{wc:twopartredpot}.
\end{eqnarray}
With these, Eq.~(\ref{wc:fullse}) effectively becomes a one-particle
equation, where particle 2's momentum is determined by
$\bbox{k}_{2f}=\bbox{P}-\bbox{k}_{1f}$.  The minus component (the
light-front energy) of particle 2 is defined by the requirement that the
particle 2 is on mass shell, so
$k^-_{2f}=(M^2+\bbox{k}_{2f,\perp}^2)/k_{2f}^+$. We also define
\begin{eqnarray}
x &\equiv& k_{1f}^+/P^+ = x_{Bj},
\end{eqnarray}
where $x_{Bj}$ is the Bjorken $x$
variable, so that $k_{2f}^+/P^+=1-x$.  Likewise, we write the Bjorken
variables that correspond to the momenta $\bbox{k}_{1i}$ and $\bbox{q}_1$
as $y\equiv k_{1i}^+/P^+$ and $z\equiv q_1^+/P^+$.

Using Eqs.~(\ref{wc:twopartredpsi}), (\ref{wc:twopartredpot}), and
the fact that the plus momentum of both nucleons is positive, we can
write the light-front Schr\"odinger equation Eq.~(\ref{wc:fullse}) as
\begin{eqnarray}
\int d^2k_{1i,\perp} \int_0^E dk^+_{1i} V(g(E),E;\bbox{k}_{1f};\bbox{k}_{1i})
\psi(\bbox{k}_{1i})
&=& 
\psi(\bbox{k}_{1f}) (E-k_{1f}^--k_{2f}^-). \label{wc:fullse3}
\end{eqnarray}

It is useful to convert from light-front coordinates
$\bbox{k}_1=(k_1^+,\bbox{k}_\perp)$ to ``equal-time'' coordinates
$\bbox{k}_{\text{ET}}=(\bbox{k}_\perp,k^3)$, using an implicit
definition of $k^3$ 
\cite{Terentev:1976jk}
\begin{eqnarray}
k_1^+ &=& \frac{E}{2k^0(\bbox{k}_{\text{ET}})}
\left[ k^0(\bbox{k}_{\text{ET}}) + k^3 \right], \label{wc:eteq} 
\\
k^0(\bbox{k}_{\text{ET}}) &=& \sqrt{ M^2 + \bbox{k}_{\text{ET}}^2 },
\end{eqnarray}
or an explicit transformation \cite{Tiburzi:2000je}
\begin{eqnarray}
k^0(k_1^+,\bbox{k}_\perp) &=&
\frac{E}{2} \sqrt{\frac{M^2+\bbox{k}_\perp^2}{k_1^+k_2^+}} =
\sqrt{\frac{M^2+\bbox{k}_\perp^2}{4x(1-x)}}, \\
k^3 &=& (k_1^+-k_2^+) \frac{k^0}{E} = (2x-1) k^0.
\end{eqnarray}

Often the explicit dependence of $k^0$ on $\bbox{k}_{\text{ET}}$ will
not be shown.  It is worth emphasizing that this is just a convenient
change of variables; $\psi(\bbox{k}_{\text{ET}})$ is not the
usual equal-time wave function.  With this transformation,
we can express $k_1^-$ and $k_2^-$ as 
\begin{eqnarray}
k_1^- &=& \left(1-\frac{k^3}{k^0}\right)\frac{2(k^0)^2}{E},
\label{wc:pminet1} \\
k_2^- &=& \left(1+\frac{k^3}{k^0}\right)\frac{2(k^0)^2}{E}. 
\end{eqnarray}
We may also write
\begin{eqnarray}
k_1^+ &=& \left(1+\frac{k^3}{k^0}\right)\frac{E}{2}, \\
k_2^+ &=& \left(1-\frac{k^3}{k^0}\right)\frac{E}{2}. \label{wc:pminet2}
\end{eqnarray}
Using Eqs.~(\ref{wc:pminet1}-\ref{wc:pminet2}), we can rewrite
Eq.~(\ref{wc:fullse3}) as
\begin{eqnarray}
\int d^3k_{i,\text{ET}}
\frac{2 k_{1i}^+ k_{2i}^+}{k_i^0}
V(g(E),E,\bbox{k}_{f,\text{ET}};\bbox{k}_{i,\text{ET}})
\psi(\bbox{k}_{i,\text{ET}})
&=& 
\psi(\bbox{k}_{f,\text{ET}})
\left[E^2 - (2k_f^0)^2 \right].
\label{wc:fullse4}
\end{eqnarray}

Now consider the exchange of the particle labels 1 and 2.  This causes
\begin{eqnarray}
\bbox{k}_{1,\perp} &\rightarrow& \bbox{k}_{2,\perp} = -\bbox{k}_{1,\perp}, \\
k_1^+ &\rightarrow& k_2^+ = E-k_1^+,
\end{eqnarray}
which means that $k^3$ as defined in Eq.~(\ref{wc:eteq}) transforms as
$k^3 \rightarrow -k^3$, so $\bbox{k}_{\text{ET}} \rightarrow
-\bbox{k}_{\text{ET}}$.  Consequently, exchange of particle labels 1 and
2 is the same as parity for $\bbox{k}_{\text{ET}}$.

For scattering states, the OBE potential computed for on-shell nucleons and
used in the Weinberg integral equation \cite{Weinberg:1966jm} (which is
essentially the scattering analogue of Eq.~(\ref{wc:fullse})) leads to
manifestly rotationally
invariant results when written in terms of the relative momenta
\cite{Miller:1997cr}.  The similarity between that rotationally invariant 
result and the usual equal-time result implies that this equal-time momentum
$\bbox{k}$ can be interpreted as the relative momentum of the two particles.
For bound states, this exact simplification does not occur, since for
bound states the potential is, of necessity, evaluated off the energy shell
(but on the mass shell).  However, we expect that the OBE potential, written in
terms of the relative momentum, is {\em approximately} spherically symmetric
for lightly-bound states.  Thus, the wave functions are approximate
eigenfunctions of the ``relative angular momentum'', which we define
by the operator $\bbox{L}=\bbox{x}\times\bbox{k}$.
Our ``relative angular momentum operator'' is not the same as the
true orbital angular momentum operator which is obtained from the
Lagrangian via the energy-momentum tensor in a way similar to the
Hamiltonian.

Since the two nucleons are identical except for the particle label, the
effective potential commutes with parity to all orders in $g^2$. Furthermore,
the light-front Hamiltonian is explicitly invariant under rotations about the
three-axis. These considerations allow us to classify the wave functions as
having eigenvalues $p$ of parity (${\mathcal P}$) and $m$ of the
three-component of the angular momentum operator ($J_3$).  We label the
wave function as $\langle k_1 |\psi_{m,p}\rangle$, where 
\begin{eqnarray}
\langle k_1 |J_3|\psi^n_{m,p}\rangle &=&
\langle k_1 |    \psi^n_{m,p}\rangle m, \\
\langle k_1 | {\mathcal P} |\psi^n_{m,p}\rangle &=&
\langle k_2   |\psi^n_{m,p}\rangle, \\
&=&
\langle k_1 |    \psi^n_{m,p}\rangle p.
\end{eqnarray}

With this, we may write the OBE truncation of full uncrossed Hamiltonian as
\begin{eqnarray}
P^-_{\text{OBE}}(g,E) &=& P_0^- + g^2 V_{\text{OBE}}(E),
\end{eqnarray}
which gives OBE light-front Schr\"odinger equation
\begin{eqnarray}
P^-_{\text{OBE}} \Big( g^{n,m,p}_{\text{OBE}}(E),E \Big)
|\psi^{n,\text{OBE}}_{m,p} \rangle &=&
|\psi^{n,\text{OBE}}_{m,p} \rangle E, \label{wc:obeham}
\end{eqnarray}
where $E$ is an arbitrary energy, $|\psi^{n,\text{OBE}}_{m,p} \rangle$
is the $n^{\text{th}}$ wave function with parity $p$ and $J_3$ quantum
number $m$, and $g^{n,m,p}_{\text{OBE}}(E)$ is the coupling constant
which yields that bound-state wave function with $E$ as the bound-state
energy. 

For the OBE+TBE truncation, we have
\begin{eqnarray}
P^-_{\text{TBE}}(g,E) &=& P_0^- + g^2 V_{\text{OBE}}(E)
+ g^4 V_{\text{TBE}}(E),
\end{eqnarray}
which gives TBE light-front Schr\"odinger equation
\begin{eqnarray}
P^-_{\text{TBE}} \Big( g^{n,m,p}_{\text{TBE}}(E),E \Big)
|\psi^{n,\text{TBE}}_{m,p} \rangle &=&
|\psi^{n,\text{TBE}}_{m,p} \rangle E, \label{wc:tbeham}
\end{eqnarray}
where the quantities here are defined in an analogous way to
Eq.~(\ref{wc:obeham}).  By comparing the spectra
$g^{n,m,p}_{\text{OBE}}(E)$ and $g^{n,m,p}_{\text{TBE}}(E)$, we can see
what effect adding the TBE potential to the OBE potential has on the
coupling constant for a given bound-state energy.

The quantum numbers $m$ and $p$ of the wave function can be used to
rewrite Eq.~(\ref{wc:fullse4}) as 
\begin{eqnarray}
& & \int_0^\infty dk_{i,\text{ET}} \int_0^{\pi/2} d\theta_i
\frac{2 k_{1i}^+ k_{2i}^+  k^2_{i,\text{ET}} \sin\theta_i}{k^0_i}
V_{p,m}(k_{f,\text{ET}},\theta_f;k_{i,\text{ET}},\theta_f)
\psi_{p,m}(k_{i,\text{ET}},\theta_i)
\nonumber \\ & & \qquad = 
\psi_{p,m}(k_{f,\text{ET}},\theta_f)
\left[E^2 - 4 (k_f^0)^2 \right],
\label{wc:fullse6}
\end{eqnarray}
where
\begin{eqnarray}
&&V_{p,m}(k_{f,\text{ET}},\theta_f;k_{i,\text{ET}},\theta_i) \nonumber\\
&&\qquad = \frac{1}{2} \Big[
V_m(k_{f,\text{ET}},\theta_f;k_{i,\text{ET}},\theta_i) + p
V_m(k_{f,\text{ET}},\theta_f;k_{i,\text{ET}},\pi-\theta_i) \Big]\\
&&V_m(k_{f,\text{ET}},\theta_f;k_{i,\text{ET}},\theta_i) \nonumber\\
&&\qquad = \int \! \! \!\int_0^{2\pi} \frac{d\phi_f\,d\phi_i}{2\pi}
e^{im(\phi_f-\phi_i)}
V(\bbox{k}_{f,\text{ET}};\bbox{k}_{i,\text{ET}}). \label{wc:angintet}
\end{eqnarray}
We call $V(k_{f,\text{ET}},\theta_f;k_{i,\text{ET}},\theta_i)$ the
azimuthal-angle-integrated potential. The light-front vectors can also
be expressed in terms of the azimuthal angle $\phi$, the angle between
$\bbox{k}_\perp$ and the $x$-axis.  This allows the
azimuthal-angle-integrated potential to be written in light-front
coordinates 
\begin{eqnarray}
V_m(k^+_{1f},k_{1f,\perp};k_{1i}^+,k_{1i,\perp}) &=& \int\!\!\!\int_0^{2\pi}
\frac{d\phi_f\,d\phi_i}{2\pi} e^{im(\phi_f-\phi_i)}
V(\bbox{k}_{1f};\bbox{k}_{1i}). \label{wc:angintlf} 
\end{eqnarray}

All of the simplifications of Eq.~(\ref{wc:fullse6}) based on physical
considerations have been addressed.  However, further rearrangements need
to be done before Eq.~(\ref{wc:fullse6}) is fit to be solved on the
computer.  Since these involve only numerical techniques, they are
relegated to Appendix~\ref{ch:numericaleq}.

\subsection{Bethe-Salpeter Equation} \label{wc:bseequiv}

\begin{figure}
\begin{center}
\epsfig{angle=0,width=5.0in,height=1.1in,file=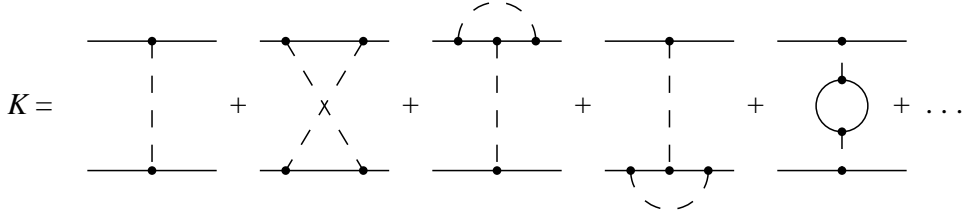}
\caption{The full kernel for the Bethe-Salpeter equation of the
Wick-Cutkosky model.
\label{fig:wc.fullbseker}}
\end{center}
\end{figure}

The Bethe-Salpeter equation \cite{Nambu:1950vt,Schwinger:1951ex,%
Schwinger:1951hq,Gell-Mann:1951rw,Salpeter:1951sz} provides a way of
describing bound states based on Feynman propagators and kernels
constructed from covariant quantum field theory, and thus is manifestly
covariant.  The equation for the bound state of nucleons 1 and 2 can be
written as
\begin{eqnarray}
G K \psi &=& \psi, \label{wc:bse}
\end{eqnarray}
$G$ is the free two-particle propagator, which is the product of two
one-particle propagators, $G=S_1 S_2$, $\psi$ is the four-dimensional
Bethe-Salpeter amplitude, and $K$ is the sum of all two-particle
irreducible two-to-two Feynman graphs, shown in
Fig.~\ref{fig:wc.fullbseker}. There are no nucleon exchange graphs since
we treat only the case where the two nucleons are distinguishable.

\begin{figure}
\begin{center}
\epsfig{angle=0,width=2.0in,height=0.8in,file=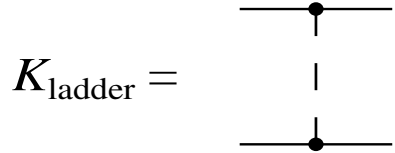}
\caption{The ladder (one-boson-exchange) kernel for the Bethe-Salpeter
equation.
\label{fig:wc.ladderbseker}}
\end{center}
\end{figure}

We consider the ladder approximation to the Bethe-Salpeter equation,
\begin{eqnarray}
g^2 G \widetilde{K}_{\text{ladder}} \psi &=& \psi, \label{wc:lbseforget}
\end{eqnarray}
which is obtained by replacing $K$ in Eq.~(\ref{wc:bse}) with
$g^2\widetilde{K}_{\text{ladder}}$. We define 
$\widetilde{K}_{\text{ladder}}=\frac{1}{g^2}K_{\text{ladder}}$, where 
$K_{\text{ladder}}$ is the graph due to one-boson-exchange, shown in
Fig~\ref{fig:wc.ladderbseker}.
This definition of $\widetilde{K}_{\text{ladder}}$ makes it independent of the coupling
constant. Making this approximation leaves the Bethe-Salpeter equation
covariant.  This implies that the Bethe-Salpeter amplitudes $\psi$ have
definite angular momentum $l$, and hence the energies of the states are
degenerate for different $m$ projections of the same angular momentum.

This equation can be simplified in the center-of-momentum frame.  In
that frame, once the total energy is defined as $P^0$, the four-momentum
of the second particle is given by $k_2^\mu = P^\mu - k_1^\mu$, and thus
the Bethe-Salpeter equation is effectively a one-particle equation.  The
remainder of the discussion in this section will be done using the
center-of-momentum frame.  We then can rewrite
Eq.~(\ref{wc:lbseforget}), taking into account the explicit symmetries
of the equation, as
\begin{eqnarray}
\left[ g_{\text{LBSE}}^{n,l,m}(E) \right]^2 G(E)
\widetilde{K}_{\text{ladder}}(E) \psi_{n,l,m}
&=& \psi_{n,l,m}, \label{wc:lbse}
\end{eqnarray}
where $E$ is an arbitrary energy, $\psi_{n,l,m}$ is the $n^{\text{th}}$
Bethe-Salpeter amplitude with angular momentum $l$ and three-projection
$m$, and $g_{\text{LBSE}}^{n,l,m}(E)$ is the coupling constant which
yields that bound-state Bethe-Salpeter amplitude with $E$ as the
bound-state energy. We call this $g(E)$ the spectrum of the ladder
Bethe-Salpeter equation for the corresponding Bethe-Salpeter amplitude.
The Greens function $G(E)$ and the ladder kernel
$\widetilde{K}_{\text{ladder}}(E)$ are
functions of the energy in the c.m.\ frame and are implicitly effective
one-particle operators.

\begin{figure}
\begin{center}
\epsfig{angle=0,width=2.0in,height=0.8in,file=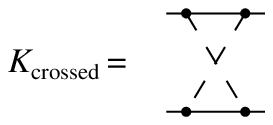}
\caption{The crossed (two-boson-exchange) kernel for the Bethe-Salpeter
equation.
\label{fig:wc.crossedbseker}}
\end{center}
\end{figure}

We may also take the kernel to be
$\widetilde{K}=\widetilde{K}_{\text{ladder}}+\widetilde{K}_{\text{crossed}}$,
where 
$\widetilde{K}_{\text{crossed}}=\frac{1}{g^2}K_{\text{crossed}}$, and 
$K_{\text{crossed}}$ is the crossed diagram is
shown in Fig.~\ref{fig:wc.crossedbseker}. Since this kernel is also
covariant, the discussion given above for the ladder kernel also applies
for the ladder plus crossed kernel. In particular, the bound states
calculated by using kernel with the Bethe-Salpeter equation will be
angular momentum eigenstates.

The comparison we draw is only between the ladder (or ladder plus
crossed) Bethe-Salpeter
equation and the light-front Hamiltonian that corresponds to that
approximation. The exact nature of the correspondence is discussed in
section~\ref{wc:uxsec}.  Since we are not looking at the solutions to
the full theory, for our purposes it does not matter that there are
sizeable differences between the solutions to the full Bethe-Salpeter
equation and the ladder approximation when the coupling constant is
large
\cite{LevineWright1,LevineWright2,LevineWright3,Nieuwenhuis:1996mc}.

\subsection{Comparison Between the Light-front Schr\"odinger Equation
and the Bethe-Salpeter Equation} \label{wc:bseequivcomp}

The solution to the untruncated light-front Schr\"odinger equation in
the uncrossed approximation is equivalent to the solution of the ladder
Bethe-Salpeter equation.  When the full uncrossed Hamiltonian is
truncated, differences will be introduced.  Thus, here we think of the
ladder Bethe-Salpeter equation as the exact theory which the truncated
light-front Schr\"odinger equations approximate.  As more graphs are
included in the truncated light-front Hamiltonian potential, the
agreement with the Bethe-Salpeter equation will obviously be better.
The question we wish to answer is how well the spectra $g(E)$ for the
two different truncations [Eqs.~(\ref{wc:obeham}) and (\ref{wc:tbeham})]
approximate the spectra for the ``exact'' theory, the BSE results.

The lack of manifest rotational invariance of the truncated Hamiltonian
theory causes a breaking of the degeneracy of the spectra of the
truncated light-front Schr\"odinger equations for different $m$ states
--- unlike the case for the Bethe-Salpeter equation.  The wave functions
from the Hamiltonian approach are classified by their dominant angular
momentum contribution $l$, so that we can compare the spectra for
different $m$ projections of the same total angular momentum $l$.  By
doing this, we can compare how the degeneracy of the spectra is broken
for the  OBE and the OBE+TBE truncations, and also compare to the
spectra obtained from the ladder Bethe-Salpeter equation. 

We want to approximate our potential $V$ so that Eq.~(\ref{wc:fullse})
is physically equivalent to the ladder Bethe-Salpeter equation.  This
approximation of $V$ will be called the {\em uncrossed
approximation}. By physically equivalent, we mean that the spectra of
the potential $V$ should reproduce the spectrum for the states of the
Bethe-Salpeter equation, excluding the so-called ``abnormal'' states
\cite{Wick:1954eu,Cutkosky:1954ru}. 
It is well known how to reduce the
Bethe-Salpeter equation to a physically equivalent Hamiltonian
(Schr\"odinger-type) equation.  For an extensive discussion of this
issue of defining the potential equivalent to a sum of Feynman graphs in
the equal-time case see, for instance, Klein \cite{Klein:1953}, Phillips
and Wallace \cite{Phillips:1996eb}, Lahiff and Afnan
\cite{Lahiff:1997bj}, and, for examples on the light front, Ligterink and
Bakker \cite{Ligterink:1995tm} and Schoonderwoerd and Bakker
\cite{Schoonderwoerd:1996kj}. The general procedure to get the
effective potential due to $n$ boson exchange takes two steps.  First,
write the sum of all Feynman graphs obtained from iteration of the
Bethe-Salpeter kernel with $n$ boson exchanges. Then, write that sum in
terms of LFTOPT graphs, and discard all graphs which are not
two-particle-irreducible with respect to the light-front two-body
propagator,
\begin{eqnarray}
G_{\text{LF}}\left(P^-\right) &=& \frac{1}{P^- - P^-_0}.
\end{eqnarray}
The graphs which remain after this procedure constitute the
$V_{\text{nBE}}$.

As an example, we construct the TBE potential.  When the ladder
Bethe-Salpeter equation is used, only one Feynman graph contributes, the
box diagram arising from the iteration of the Feynman OBE kernel.  This
gives six non-vanishing LFTOPT diagrams.  (The other diagrams vanish
because the vacuum is simple on the light front.)

\begin{figure}
\begin{center}
\epsfig{angle=0,width=5.5in,height=1.5in,file=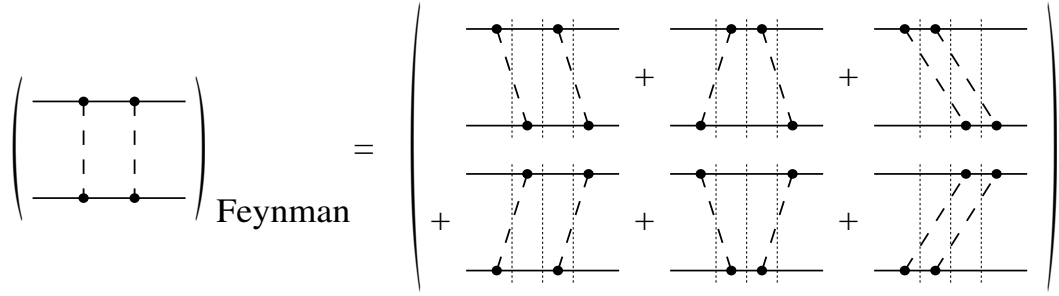}
\caption{The relation between the Feynman box diagram and non-vanishing
light-front time-ordered perturbation theory diagrams.
\label{wc:diag:feynbox}}
\end{center}
\end{figure}

The first four diagrams are iterations of the OBE potential and are
reducible with respect to $G_{\text{LF}}$.  The last two are
two-particle-irreducible and thus constitute the TBE potential,
$V_{\text{TBE}}$.

\section{Wick-Cutkosky Perturbative Potentials} \label{wc:ch:wcpertpot}

A truncation must be made of the expansion of the potential given in
Eq.~(\ref{wc:fullpot}), since it is not feasible to calculate the
infinite sum of graphs for the potential in this Hamiltonian theory. For
this investigation  we consider three truncations of the potential
derived from the field theory.  First, the OBE potential and the TBE
potentials are 
calculated.  Then, we note that a subset of the TBE diagrams, the
stretched-box diagrams, correspond to the truncated potential derived
from the ladder Bethe-Salpeter equation.  Thus, three truncated
potentials are obtained that have a physical interpretation.

The matrix elements of these potentials are written in the two-particle
momentum basis, denoting the momentum of the incoming particles by
$\bbox{k}_{1i}$ and $\bbox{k}_{2i}$, and the outgoing particles
$\bbox{k}_{1f}$ and $\bbox{k}_{2f}$.  For simplicity, we choose to work
in the center-of-momentum frame.  By inspecting the rules for converting
a light-front time-ordered diagram into a potential given in
section~\ref{wc:rules}, and looking at Eq.~(\ref{wc:twopartredpot}), we
find that each piece of the effective-one-particle potential
$V(E;\bbox{k}_{1f},\bbox{k}_{1i})$ is proportional to
\begin{eqnarray}
\frac{E}{2(2\pi)^3
\sqrt{k_{1f}^+k_{2f}^+k_{1i}^+k_{2i}^+}}. \label{wc:supressfactor} 
\end{eqnarray}
(The factor of $E$ in the numerator is included to simplify later
equations.) This term will by suppressed in all of the potentials
written in this chapter.

\subsection{OBE Potential}

\begin{figure}
\begin{center}
\epsfig{angle=0,width=5.0in,height=1.173in,file=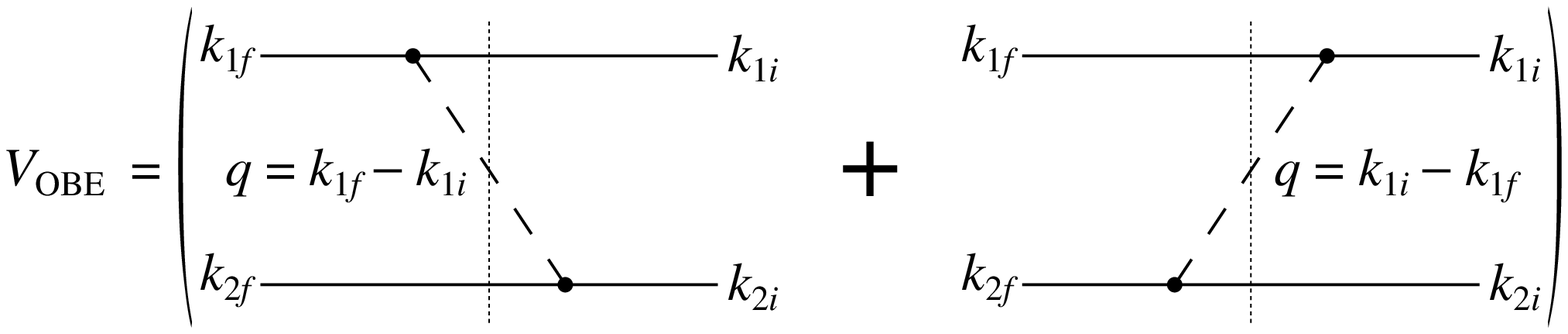}
\caption{The diagrams for the OBE potential.
\label{wc:obediagrams}}
\end{center}
\end{figure}

We start by drawing all the allowed and non-vanishing time-ordered
diagrams with one meson exchange.  These diagrams are shown in 
Fig.~\ref{wc:obediagrams}.
The light-front time-ordered perturbation theory rules given
in section~\ref{wc:rules} are used to calculate the potential due to OBE
potential,
\begin{eqnarray}
V_{\text{OBE}}(E;\bbox{k}_{1f};\bbox{k}_{1i})
&=& \left(\frac{M}{E}\right)^2 \left[
\frac{ \theta(x-y)/|x-y|}{E - k^-_{1i} - k^-_{2f} -
\omega^-(\bbox{k}_{1f}-\bbox{k}_{1i})} 
\right. \nonumber \\
&& 
\phantom{\left(\frac{M}{E}\right)^2 \left[ \right.} \left. 
+
\frac{ \theta(y-x)/|y-x|}{E - k^-_{1f} - k^-_{2i} -
\omega^-(\bbox{k}_{1i}-\bbox{k}_{1f})} 
\right] \label{wc:OBEpot}.
\end{eqnarray}
We have introduced the notation that meson with light-front
three-momentum $\bbox{q}$ has a light-front energy given by
\begin{eqnarray}
\omega^-(\bbox{q}) &=& \frac{\mu^2 + \bbox{q}_\perp^2}{q^+}.
\end{eqnarray}
The azimuthal-angle integration of $V_{\text{OBE}}$ is discussed in
Appendix~\ref{a:ints:angint}.

The potential given in Eq.~(\ref{wc:OBEpot}) can also be used for
scattering states.  In that case,
$E=k_{1f}^-+k_{2f}^-=k_{1i}^-+k_{2i}^-$, which allows the potential to
be written as  
\begin{eqnarray}
V_{\text{OBE}}(E_{\text{scat}};\bbox{k}_1;\bbox{p}_1)
&=& \frac{M^2/E_{\text{scat}}}{(k_{1f}-k_{1i})^2 - \mu^2 }.
\label{wc:OBEpotscat}
\end{eqnarray}
The scattering potential is the same as the usual equal-time OBE
potential.  This must be the case, since the scattering potential is
also given by covariant Feynman diagrams, which have the same form
independent of the form of dynamics.

Returning to the bound-state regime, we note that the OBE potential can
be written in terms of the equal-time coordinates.  A
reorganization of Eq.~(\ref{wc:OBEpot}) yields
\begin{eqnarray}
&&V_{\text{OBE}}(E;\bbox{k}_{f,\text{ET}};\bbox{k}_{i,\text{ET}}) \nonumber\\
&&\qquad= \left(\frac{M}{E}\right)^2 E \left[
\frac{\theta(x-y)}
{(k_{1f}^+-k_{1i}^+)(E - k^-_{1i} - k^-_{2f}) - \mu^2 -
(\bbox{k}_{i\perp}-\bbox{k}_{f\perp})^2} 
\right. \nonumber \\
&& \qquad
\phantom{=\left(\frac{M}{E}\right)^2 E \left[ \right.} \left. 
+
\frac{\theta(y-x)}
{(k_{1i}^+-k_{1f}^+)(E - k^-_{1f} - k^-_{2i}) - \mu^2 -
(\bbox{k}_{i\perp}-\bbox{k}_{f\perp})^2} 
\right]. \label{wc:OBEpot2}
\end{eqnarray}
Using the relations in Eqs.~(\ref{wc:pminet1}-\ref{wc:pminet2}), we find
\begin{eqnarray}
(k_{1f}^+-k_{1i}^+)(E - k^-_{1i} - k^-_{2f})
&=&
\left(\frac{k_f^3}{k_f^0}-\frac{k_i^3}{k_i^0}\right)
\left(\frac{E^2-4M^2}{2} -k_{i,\text{ET}}^2 -k_{f,\text{ET}}^2 \right)
\nonumber \\ & & 
+ \frac{k_f^3}{k_f^0} \frac{k_i^3}{k_i^0} \left(k_f^0-k_i^0\right)^2
- (k_f^3-k_i^3)^2. \label{wc:etred}
\end{eqnarray}
Under the exchange $k_f\leftrightarrow k_i$, the only thing that changes
in Eq.~(\ref{wc:etred}) is that the first term picks up a minus sign.
This observation allows Eq.~(\ref{wc:OBEpot2}) to be rewritten as
\begin{eqnarray}
V_{\text{OBE}}(E;\bbox{k}_{f,\text{ET}};\bbox{k}_{i,\text{ET}})
&=& \left(\frac{M}{E}\right)^2
\frac{E}
{\left|\frac{k_f^3}{k_f^0}-\frac{k_i^3}{k_i^0}\right|
\Delta
+ \frac{k_f^3}{k_f^0} \frac{k_i^3}{k_i^0} (q^0_{\text{ET}})^2 
- \bbox{q}_{\text{ET}}^2
- \mu^2
},
\label{wc:OBEpotet}
\end{eqnarray}
where
\begin{eqnarray}
\Delta &=& \frac{E^2-4M^2}{2} -k_{i,\text{ET}}^2 -k_{f,\text{ET}}^2, \\
q_{\text{ET}}^\mu &=& k^\mu_{f,\text{ET}} - k^\mu_{i,\text{ET}}.
\end{eqnarray}
Note that $q_{\text{ET}}^-$ is not the light-front energy of the meson,
since in a Hamiltonian theory only the light-front three-momenta are
conserved; the four-momenta are not conserved.  Equation~(\ref{wc:OBEpotet}) 
will be useful in the context of approximations based on the physical
arguments that we will discuss in section~\ref{wc:appphys}.

\subsection{TBE Potentials}
\label{wc:deftbepots}

\begin{figure}
\begin{center}
\epsfig{angle=0,width=6.0in,height=7.25in,file=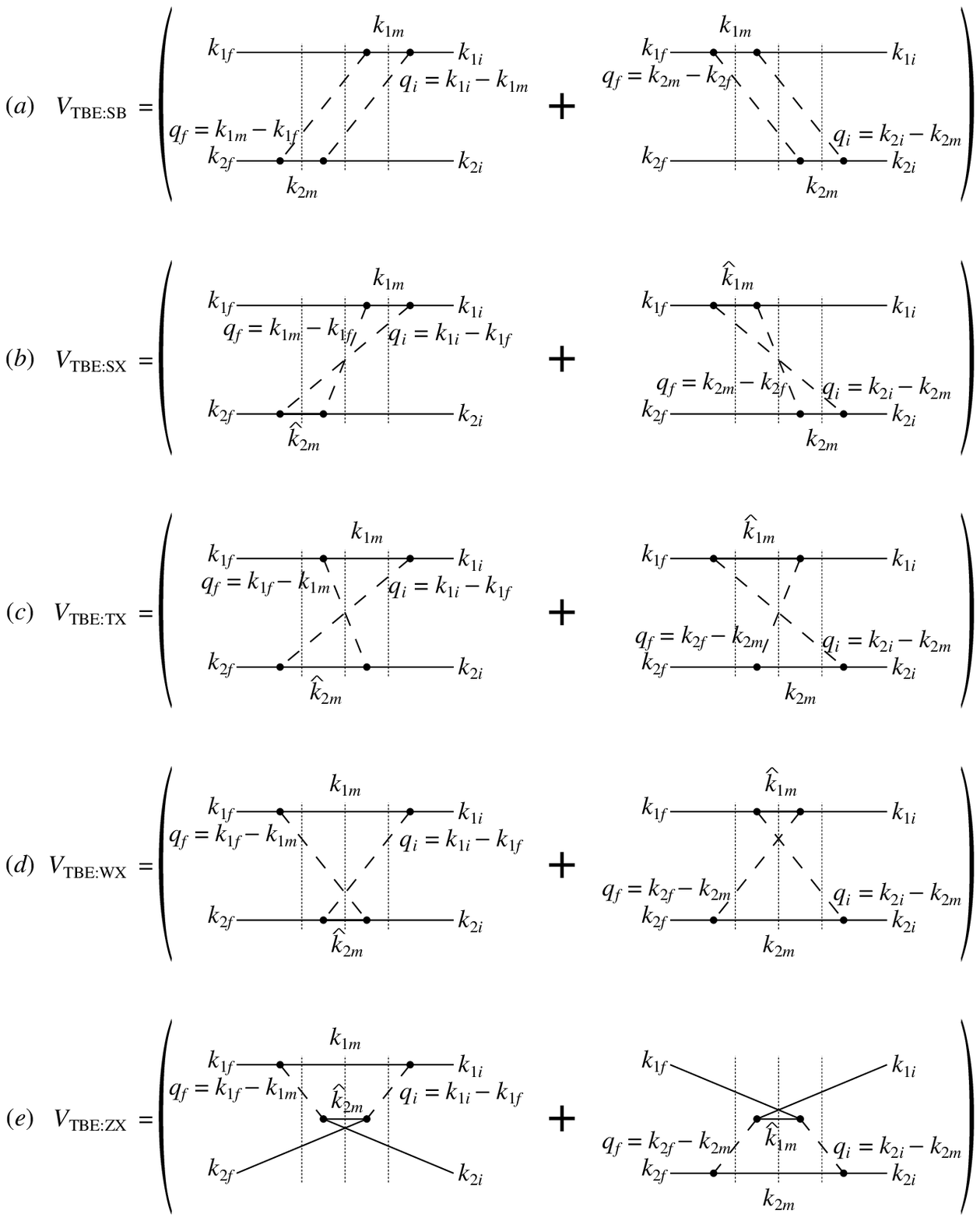}
\caption[The components of the TBE potential, the stretched-box,
stretched-crossed, T-crossed, wide-crossed, and Z-crossed diagrams.]
{The components of the TBE potential, (a) the stretched-box,
(b) stretched-crossed, (c) T-crossed, (d) wide-crossed, and (e)
Z-crossed diagrams.  Here, $\widehat{k}_{1m}=k_{1f}+k_{1i}-k_{1m}$
and $\widehat{k}_{2m}=k_{2f}+k_{2i}-k_{2m}$.
\label{wc:tbediagrams}}
\end{center}
\end{figure}

As in the previous section, we start by drawing all the allowed,
non-vanishing time-ordered diagrams with two meson exchanges shown in
Fig.~\ref{wc:tbediagrams}.  The diagrams are classified according to the
behavior of the intermediate particles.  The total TBE potential is
given by the sum of all the diagrams, so
\begin{eqnarray}
V_{\text{TBE}} &=& 
V_{\text{TBE:SB}} +
V_{\text{TBE:SX}} +
V_{\text{TBE:TX}} +
V_{\text{TBE:WX}} +
V_{\text{TBE:ZX}}.
\end{eqnarray}

In the diagrams for the TBE potential in Fig.~\ref{wc:tbediagrams}, the
intermediate loop momenta can be parameterized by $\bbox{k}_{1m}$ or
$\bbox{k}_{2m}$.  The dependent variable is defined by the relation
$\bbox{P}=\bbox{k}_{1i}+\bbox{k}_{2i}$.  The Bjorken $x$ variable that
corresponds to $\bbox{k}_{1m}$ ($\bbox{k}_{2m}$) is labeled with $z$
($1-z$). We use the Feynman rules to calculate all of these potentials,
starting with
\begin{eqnarray}
&&V_{\text{TBE:SB}}(E;\bbox{k}_{1f};\bbox{k}_{1i}) \nonumber \\
&&=
\left(\frac{M}{E}\right)^4
\int \frac{d^2 k_{m\perp}}{2(2\pi)^3} \left[
\int_0^1 dz \frac{\theta(z-y)\theta(x-z)}{z(1-z)(z-y)(x-z)}
\right. \nonumber \\ && \phantom{\left(\frac{M}{E}\right)^4 \int
\frac{d^2 k_{m\perp}}{2(2\pi)^3} \left[\right.} \left. \times
\frac{1}{E -k_{1m}^- - k_{2f}^- - \omega^-(\bbox{k}_{1f}-\bbox{k}_{1m})}
\right. \nonumber \\ && \phantom{\left(\frac{M}{E}\right)^4 \int
\frac{d^2 k_{m\perp}}{2(2\pi)^3} \left[\right.} \left. \times
\frac{1}{E -k_{1i}^- - k_{2f}^- - \omega^-(\bbox{k}_{1f}-\bbox{k}_{1m}) -
\omega^-(\bbox{k}_{1m}-\bbox{k}_{1i})}
\right. \nonumber \\ && \phantom{\left(\frac{M}{E}\right)^4 \int
\frac{d^2 k_{m\perp}}{2(2\pi)^3} \left[\right.} \left. \times
\frac{1}{E -k_{1i}^- - k_{2m}^- - \omega^-(\bbox{k}_{1m}-\bbox{k}_{1i})}
\right]
\nonumber \\ && \phantom{\left(\frac{M}{E}\right)^4 \int
\frac{d^2 q_\perp}{2(2\pi)^3}} 
+ \{ 1\leftrightarrow 2\} \label{wc:TBESBpot}.
\end{eqnarray}
The symbol $\{ 1\leftrightarrow 2\}$ means that all labels 1 are replaced with
2 and vice versa, as well as replacing the Bjorken variables $x$, $y$,
and $z$ with $1-x$, $1-y$, and $1-z$.  This is a way of explicitly
stating the symmetry of the potential under exchange of particles 1 and
2.  A detailed discussion of the evaluation of the loop integral in
Eq.~(\ref{wc:TBESBpot}) is given in Appendix~\ref{a:ints:loopints}. 

It is straightforward to calculate the other parts of the TBE potential,
\begin{eqnarray}
&&V_{\text{TBE:SX}}(E;\bbox{k}_{1f};\bbox{k}_{1i}) \nonumber \\
&&=
\left(\frac{M}{E}\right)^4
\int \frac{d^2 k_{m\perp}}{2(2\pi)^3} \left[
\int_0^1 dz \frac{\theta(x-z)\theta(z-y)}{z(x-z)(1+z-x-y)(z-y)}
\right. \nonumber \\ && \phantom{\left(\frac{M}{E}\right)^4 \int
\frac{d^2 q_\perp}{2(2\pi)^3} \left[\right.} \left. \times
\frac{1}{E -k_{1m}^- - k_{2f}^- - 
\omega^-(\bbox{k}_{1f}-\bbox{k}_{1m})}
\right. \nonumber \\ && \phantom{\left(\frac{M}{E}\right)^4 \int
\frac{d^2 k_{m\perp}}{2(2\pi)^3} \left[\right.} \left. \times
\frac{1}{E -k_{1i}^- - k_{2f}^- - 
\omega^-(\bbox{k}_{1f}-\bbox{k}_{1m}) -
\omega^-(\bbox{k}_{1m}-\bbox{k}_{1i})}
\right. \nonumber \\ && \phantom{\left(\frac{M}{E}\right)^4 \int
\frac{d^2 k_{m\perp}}{2(2\pi)^3} \left[\right.} \left. \times
\frac{1}{E -k_{1i}^- - \widehat{k}_{2m}^- -
\omega^-(\bbox{k}_{1f}-\bbox{k}_{1m})}
\right]
\nonumber \\ && \phantom{\left(\frac{M}{E}\right)^4 \int
\frac{d^2 k_{m\perp}}{2(2\pi)^3}} 
+ \{ 1\leftrightarrow 2\} \label{wc:TBESXpot},
\end{eqnarray}
where we have denoted the light-front energy of particle 2 by
$\widehat{k}_{2m}^-$, given by
\begin{eqnarray}
\widehat{k}_{2m}^- &=&
\epsilon^-(\bbox{P}+\bbox{k}_{1m}-\bbox{k}_{1i}-\bbox{k}_{1f}), \\
\epsilon^-(\bbox{k}) &=& \frac{M^2 + \bbox{k}_\perp^2}{k^+}.
\end{eqnarray}
The rest of the TBE potential is given by
\begin{eqnarray}
&&V_{\text{TBE:TX}}(E;\bbox{k}_{1f};\bbox{k}_{1i}) \nonumber \\
&&=
\left(\frac{M}{E}\right)^4
\int \frac{d^2 k_{m\perp}}{2(2\pi)^3} \left[
\int_0^1 dz \frac{
\theta(x-z)\theta(1+z-x-y)\theta(y-z)}{z(x-z)(1+z-x-y)(y-z)}
\right. \nonumber \\ && \phantom{\left(\frac{M}{E}\right)^4 \int
\frac{d^2 k_{m\perp}}{2(2\pi)^3} \left[\right.} \left. \times
\frac{1}{E -k_{1m}^- - k_{2f}^- - 
\omega^-(\bbox{k}_{1f}-\bbox{k}_{1m})}
\right. \nonumber \\ && \phantom{\left(\frac{M}{E}\right)^4 \int
\frac{d^2 k_{m\perp}}{2(2\pi)^3} \left[\right.} \left. \times
\frac{1}{E -k_{1m}^- - \widehat{k}_{2m}^- -
\omega^-(\bbox{k}_{1f}-\bbox{k}_{1m}) -
\omega^-(\bbox{k}_{1i}-\bbox{k}_{1m})}
\right. \nonumber \\ && \phantom{\left(\frac{M}{E}\right)^4 \int
\frac{d^2 k_{m\perp}}{2(2\pi)^3} \left[\right.} \left. \times
\frac{1}{E -k_{1i}^- - \widehat{k}_{2m}^- -
\omega^-(\bbox{k}_{1f}-\bbox{k}_{1m})}
\right]
\nonumber \\ && \phantom{\left(\frac{M}{E}\right)^4 \int
\frac{d^2 k_{m\perp}}{2(2\pi)^3}} 
+ \{ 1\leftrightarrow 2\}, \label{wc:TBETXpot}
\end{eqnarray}
\begin{eqnarray}
&&V_{\text{TBE:WX}}(E;\bbox{k}_{1f};\bbox{k}_{1i}) \nonumber \\
&&=
\left(\frac{M}{E}\right)^4
\int \frac{d^2 k_{m\perp}}{2(2\pi)^3} \left[
\int_0^1 dz \frac{\theta(x-z)\theta(1+z-x-y)\theta(y-z)}{z(x-z)(1+z-x-y)(y-z)}
\right. \nonumber \\ && \phantom{\left(\frac{M}{E}\right)^4 \int
\frac{d^2 k_{m\perp}}{2(2\pi)^3} \left[\right.} \left. \times
\frac{1}{E -k_{1m}^- - k_{2f}^- -
\omega^-(\bbox{k}_{1f}-\bbox{k}_{1m})}
\right. \nonumber \\ && \phantom{\left(\frac{M}{E}\right)^4 \int
\frac{d^2 k_{m\perp}}{2(2\pi)^3} \left[\right.} \left. \times
\frac{1}{E -k_{1m}^- - \widehat{k}_{2m}^- -
\omega^-(\bbox{k}_{1f}-\bbox{k}_{1m}) -
\omega^-(\bbox{k}_{1i}-\bbox{k}_{1m})}
\right. \nonumber \\ && \phantom{\left(\frac{M}{E}\right)^4 \int
\frac{d^2 k_{m\perp}}{2(2\pi)^3} \left[\right.} \left. \times
\frac{1}{E - k_{1m}^- - k_{2i}^- -
\omega^-(\bbox{k}_{1i}-\bbox{k}_{1m})}
\right]
\nonumber \\ && \phantom{\left(\frac{M}{E}\right)^4 \int
\frac{d^2 k_{m\perp}}{2(2\pi)^3}} 
+ \{ 1\leftrightarrow 2\}, \label{wc:TBEWXpot}
\end{eqnarray}
\begin{eqnarray}
&&V_{\text{TBE:ZX}}(E;\bbox{k}_{1f};\bbox{k}_{1i}) \nonumber \\
&&=
\left(\frac{M}{E}\right)^4
\int \frac{d^2 k_{m\perp}}{2(2\pi)^3} \left[
\int_0^1 dz \frac{\theta(x+y-1-z)}{z(x-z)(x+y-1-z)(y-z)}
\right. \nonumber \\ && \phantom{\left(\frac{M}{E}\right)^4 \int
\frac{d^2 k_{m\perp}}{2(2\pi)^3} \left[\right.} \left. \times
\frac{1}{E -k_{1m}^- - k_{2f}^- -
\omega^-(\bbox{k}_{1f}-\bbox{k}_{1m})}
\right. \nonumber \\ && \phantom{\left(\frac{M}{E}\right)^4 \int
\frac{d^2 k_{m\perp}}{2(2\pi)^3} \left[\right.} \left. \times
\frac{1}{E - k_{1m}^- - k_{2f}^- - k_{2i}^- -
\epsilon^-(\bbox{k}_{1i}+\bbox{k}_{1i}-\bbox{P}-\bbox{k}_{1m})}
\right. \nonumber \\ && \phantom{\left(\frac{M}{E}\right)^4 \int
\frac{d^2 k_{m\perp}}{2(2\pi)^3} \left[\right.} \left. \times
\frac{1}{E - k_{1m}^- - k_{2i}^- -
\omega^-(\bbox{k}_{1i}-\bbox{k}_{1m})}
\right]
\nonumber \\ && \phantom{\left(\frac{M}{E}\right)^4 \int
\frac{d^2 k_{m\perp}}{2(2\pi)^3}} 
+ \{ 1\leftrightarrow 2\} \label{wc:TBEZXpot}.
\end{eqnarray}
The loop integrals in the expressions for the TBE potentials and the
azimuthal-angle integrations are discussed in Appendix~\ref{a:ints:loopints}.

\subsection{TBE:SB Potential: Connection to the Ladder Bethe-Salpeter
Equation}
\label{wc:tbe:sbSec} 

As discussed in Section~\ref{wc:bseequivcomp}, a truncated kernel of the
Bethe-Salpeter equation is physically equivalent to a Hamiltonian
potential which does not include all the graphs that the full theory
allows.
In particular, consider the Bethe-Salpeter equation when the
ladder kernel is used.  The physically equivalent light-front potential
will not 
include any graphs where the meson lines cross, so to order $g^4$, the
potential is given by $g^2 V_{\text{OBE}}+ g^4 V_{\text{TBE:SB}}$.
Therefore, by 
considering the TBE:SB truncation, we can test how well the light-front
Hamiltonian approach approximates the full ladder Bethe-Salpeter
equation.  This concept is discussed more throughly in
Refs.~\cite{Cooke:1999yi,Sales:1999ec,Ligterink:1995tm}.

\section{Wick-Cutkosky Non-perturbative Potentials} \label{wc:appphys}

The potentials discussed in this section are derived from the OBE field
theory potential, but additional approximations are made to simplify
the expressions.

\subsection{Symmetrized-mass Approximation}

Krautg\"artner, Pauli and W\"olz \cite{Krautgartner:1992xz}, and
Trittmann and Pauli \cite{Trittmann:1997ga} studied positronium with a
large coupling constant in light-front dynamics.  The
one-photon-exchange potential they obtain has a colinear singularity due
to the sum of the instantaneous photon exchange graph and a
gauge-dependent factor from the spin sum.  They argue that the
singularity is not physical, and therefore must be canceled by
higher-order terms in the potential.  The effect of those terms can be
simulated by choosing the bound-state energy so that the coefficient of
the singular term vanishes.  They find that this condition is met when
the light-front energy $P^-$ in the one-photon-exchange potential is
replaced with the operator $\omega$, expressed here in the two-particle
basis,
\begin{eqnarray}
P^- \Rightarrow
\omega\left(\bbox{k}_{1f},\bbox{k}_{2f};\bbox{k}_{1i},\bbox{k}_{2i}\right)
&\equiv&
\frac{1}{2} \left(k_{1i}^- + k_{2i}^- + k_{1f}^- +k_{2f}^- \right).
\end{eqnarray}
This approximation is called the {\it symmetrized mass}
\cite{Krautgartner:1992xz}, 
the average of the total $P^-$ in the initial and final states.  It is
important to note that this approximation affects not only the singular
term, but also the energy denominator in the rest of the OBE potential.
The modified denominators simulate the effects of the non-perturbative
higher-order terms that are not included explicitly in the OBE
potential.  Potentials 
obtained with this approximation are similar to those given by the
unitary transformation method \cite{Eden:1996ey,Okubo:1954}, where the
potentials depend explicitly on the initial- and final-state energies.

In our model, there are no singularities associated with the OBE graphs
because we deal only with scalar fields.  However, we may use their
approximation to obtain a new light-front OBE potential that should
incorporate some non-perturbative effects.  Recalling that the only
place where $P^-$ occurred in Eq.~(\ref{wc:OBEpot}) was in the denominator,
the $E$ in the denominator of the OBE potential is replaced with
$\omega$ to get
\begin{eqnarray}
&&V_\omega(\bbox{k}_{1f};\bbox{k}_{1i}) \nonumber \\
&&\qquad = \left(\frac{M}{E}\right)^2 \left[
\frac{ \theta(x-y)/|x-y|}{
\frac{1}{2}\left(- k_{1i}^- + k_{2i}^- + k_{1f}^- - k_{2f}^- \right)
- \omega^-(\bbox{k}_{1f}-\bbox{k}_{1i})}
\right. \nonumber \\
&& \qquad
\phantom{=\left(\frac{M}{E}\right)^2 \left[ \right.} \left. 
+
\frac{ \theta(y-x)/|y-x|}{
\frac{1}{2}\left(+ k_{1i}^- - k_{2i}^- - k_{1f}^- + k_{2f}^- \right)
- \omega^-(\bbox{k}_{1i}-\bbox{k}_{1f})}
\right] \\
&& \qquad =\left(\frac{M}{E}\right)^2
\frac{E}{\frac{1}{2} (k_{1f}^+-k_{1i}^+)\left(- k_{1i}^- + k_{2i}^- +
k_{1f}^- - k_{2f}^- \right) 
- \mu^2 - (\bbox{k}_{f\perp}-\bbox{k}_{i\perp})^2}.
\label{wc:presymmmass}
\end{eqnarray}
Writing the light-front variables in the denominator in terms of the
equal-time variables, as prescribed in
Eqs.~(\ref{wc:pminet1}-\ref{wc:pminet2}), we find 
\begin{eqnarray}
\frac{1}{2}(k_{1f}^+-k_{1i}^+)\left(- k_{1i}^- + k_{2i}^- + k_{1f}^- -
k_{2f}^- \right) 
&=& -(k_f^3 - k_i^3)^2 + \frac{k_f^3}{k_f^0} \frac{k_i^3}{k_i^0}
(k_f^0-k_i^0)^2. 
\end{eqnarray}
Thus, Eq.~(\ref{wc:presymmmass}) can be rewritten as
\begin{eqnarray}
V_\omega(\bbox{k}_{f,\text{ET}};\bbox{k}_{i,\text{ET}})
&=& \left(\frac{M}{E}\right)^2
\frac{E}{
\frac{k_f^3}{k_f^0} \frac{k_i^3}{k_i^0} (k_f^0-k_i^0)^2
 - (\bbox{k}_{f,\text{ET}}-\bbox{k}_{i,\text{ET}})^2 - \mu^2}
\label{wc:symmmass}.
\end{eqnarray}

This result can also be obtained more directly by considering the first
term in Eq.~(\ref{wc:etred}).  Recall that the $E^2$ that appears in the
denominator is written as  $P^+P^-$ in an arbitrary frame, so in the
symmetrized-mass approximation, the $E^2$ term is replaced with
$E\omega$.  This causes the $\Delta$ term in the denominator of 
Eq.~(\ref{wc:OBEpotet}) to vanish, so that the equation reduces to
Eq.~(\ref{wc:symmmass}).  Also, note that by writing this new potential,
we attempt to incorporate physics from higher-order graphs than just the
OBE graphs.

The singularity structure of the symmetrized-mass potential is easily
analyzed. When 
scattering states are used, in the center-of-momentum frame the total
energy of the state is $E=2k_f^0=2k_i^0$, so the relations in
Eqs.~(\ref{wc:pminet1}-\ref{wc:pminet2}) become
\begin{eqnarray}
k_1^\pm &=& k^0 \pm k^3, \\
k_2^\pm &=& k^0 \mp k^3.
\end{eqnarray}
Using these relations, the symmetrized mass is $\omega = E$.  Thus, for
scattering states, this potential is same as the OBE scattering
potential and the singularity structure is the same.

\subsection{Instantaneous and Retarded Approximations}

For our bound states, $k_{\text{ET}}^2 \ll M^2$, so that the
$\frac{k_f^3 k_i^3}{k_f^0 k_i^0}(k_f^0-k_i^0)^2$ will be much smaller
than the other 
terms in the denominator of  Eq.~(\ref{wc:symmmass}).  Therefore, we may
approximate the symmetrized-mass potential $V_\omega$ by the
instantaneous potential,
\begin{eqnarray}
V_{\text{Inst}}(\bbox{k}_{f,\text{ET}};\bbox{k}_{i,\text{ET}})
&=& \left(\frac{M}{E}\right)^2
\frac{-E}{(\bbox{k}_{f,\text{ET}}-\bbox{k}_{i,\text{ET}})^2 + \mu^2}
\label{wc:instpot}.
\end{eqnarray}
Alternatively, we may also argue that since the energy difference term
is small, we can also approximate $V_\omega$ by the retarded potential,
\begin{eqnarray}
V_{\text{Ret}}(\bbox{k}_{f,\text{ET}};\bbox{k}_{i,\text{ET}})
&=& \left(\frac{M}{E}\right)^2
\frac{E}{(k_f^0-k_i^0)^2 -
(\bbox{k}_{f,\text{ET}}-\bbox{k}_{i,\text{ET}})^2 - \mu^2}\\ 
&=& \left(\frac{M}{E}\right)^2
\frac{E}{(k_{f,\text{ET}}-k_{i,\text{ET}})^2 - \mu^2},
\label{wc:retpot}
\end{eqnarray}
where $k_{f,\text{ET}}$ and  $k_{i,\text{ET}}$ represent four-vectors,
defined by the equal-time three-vectors and the condition that
$k_{f,\text{ET}}^2=k_{i,\text{ET}}^2=M^2$.  These potentials resemble the
three-dimensional Blankenbecler-Sugar \cite{Blankenbecler:1966gx} or
Gross \cite{Gross:1969rv} quasi-potentials.

Both of these approximations are reasonable if the energy difference
between the initial and final states is small, which is valid for
lightly-bound states.  The instantaneous
potential is a better approximation of the symmetrized-mass potential,
since, if we expand the symmetrized-mass potential to second-order in
perturbation theory about $k_f^0=k_i^0$, we get $V_\omega=V_{\text{Inst}}$.
Also, note that these potentials are explicitly rotationally invariant
in terms of our equal-time parameterization, which provides significant
computational advantages.

\subsection{Three-dimensional Reduction of the Bethe-Salpeter Equation}
\label{wc:sec:3dredbasic}

We now consider a non-perturbative approximation used by Wallace and
Mandelzweig \cite{Wallace:1989,Wallace:1989nm}.  The basic idea is to first
make an approximation of the Bethe-Salpeter equation, then reduce that
modified Bethe-Salpeter equation to the physically 
equivalent Hamiltonian equation.
This approach was used by Phillips and Wallace \cite{Phillips:1996eb}
for the model we use, however, they obtained an equal-time Hamiltonian,
while we seek a light-front Hamiltonian.  Before we do this,  we first
review the basic mechanics of the three-dimensional reduction, as
presented in Sales {\it et al.} \cite{Sales:1999ec} and specialized to
our particular case.  We postpone the discussion of the approximation
until section~\ref{wc:sect:3dred}.

The Bethe-Salpeter equation can be written in matrix form as
$\Gamma=KG_0\Gamma$, or explicitly in function form in the momentum
basis as
\begin{eqnarray}
\Gamma(k_{1f};P) &=& \int \frac{d^4 k_{1i}}{(2\pi)^4} K(k_{1f},k_{1i};P)
G_0(k_{1i};P) \Gamma(k_{1i};P).
\end{eqnarray}
Here, $\Gamma$ is the four-dimensional vertex function,
$K$ is the four-dimensional kernel, and $G_0$ is the two-particle
four-dimensional Green's function.  The momenta are $P$, the
total four-momentum, and $k_1$, the four-momentum of particle
1.  Particle 2's momentum is implicitly $P-k_1$. The four-dimensional 
Green's function is given by
\begin{eqnarray}
G_0(k_1;P) &=& i d(k_1) d(P-k_1),
\end{eqnarray}
where $d$ is the one-particle Green's function.  On the light front,
$d$ can be written as
\begin{eqnarray}
d(p) &=& \left(\frac{1}{p^+}\right)
\frac{1}{p^- - \mbox{Sign}(p^+)\epsilon^-(\bbox{p})},
\label{wc:onepartprop}
\end{eqnarray}
where the light-front energy $\epsilon^-$ is given by 
\begin{eqnarray}
\epsilon^-(\bbox{p}) &=& \frac{M^2 + \bbox{p}_\perp^2}{|p^+|} - i \eta.
\end{eqnarray}
The real part of $\epsilon^-$ is a positive definite quantity, and
$\eta$ is positive infinitesimal. 

The Bethe-Salpeter equation can be rewritten \cite{Woloshyn:1973} as
\begin{eqnarray}
\Gamma &=& W \widehat{G}_0 \Gamma, \label{wc:modbse}
\end{eqnarray}
where $\widehat{G}_0$ is an auxiliary Green's function, and $W$ is
defined by
\begin{eqnarray}
W &=& K + K(G_0-\widehat{G}_0)W.
\end{eqnarray}
The advantage of this rearrangement is that we are free to choose the
form of the auxiliary Green's function, $\widehat{G}_0$.  The choice of
$\widehat{G}_0$ advocated in Ref.~\cite{Sales:1999ec} is
\begin{eqnarray}
\widehat{G}_0(k_{1f},k_{1i};P)
&=& 
G_0(k_{1f};P)
\frac{\delta^{(2,+)}(\bbox{k}_{1f}-\bbox{k}_{1i})}{g_0(\bbox{k}_{1f};P)}
G_0(k_{1i};P), \label{wc:defghat}
\end{eqnarray}
where
\begin{eqnarray}
g_0(\bbox{k}_1,P) &=& \int \frac{dk_1^-}{2(2\pi)} G_0(k_1;P).
\label{wc:defg0} 
\end{eqnarray}
There is an extra factor of 2 in the denominator of Eq.~(\ref{wc:defg0})
when compared to the equal-time formalism.  This is due to the Jacobian
of the light-front coordinates.

Using the definition of $\widehat{G}_0$ given in Eq.~(\ref{wc:defghat}),
we can integrate the modified Bethe-Salpeter equation, in
Eq.~(\ref{wc:modbse}), over the light-front energy to get
\begin{eqnarray}
\gamma(\bbox{k}_{1f};P) &=& \int \frac{d^2k_{1i,\perp} \,
dk^+_{1i}}{(2\pi)^3} 
w(\bbox{k}_{1f},\bbox{k}_{1i};P) g_0(\bbox{k}_{1i};P)
\gamma(\bbox{k}_{1i};P),
\label{wc:eqn:3ered}
\end{eqnarray}
where
\begin{eqnarray}
w(\bbox{k}_{1f},\bbox{k}_{1i};P) &\equiv& 
\frac{1}{g_0(\bbox{k}_{1f};P)}
\langle G_0 W G_0 \rangle (\bbox{k}_{1f},\bbox{k}_{1i};P)
\frac{1}{g_0(\bbox{k}_{1i};P)}, \\
\gamma(\bbox{k}_1;P) &\equiv & \frac{1}{g_0(\bbox{k}_1;P)}
\int \frac{dk^-_1}{2(2\pi)} G_0(k_1;P) \Gamma(k_1;P).
\end{eqnarray}
The functional $\langle f \rangle$ is defined by its action on an
arbitrary function $f(k_{1f},k_{1i})$, where $k_{1f}$ and $k_{1i}$ are
four-vectors, as
\begin{eqnarray}
\langle f \rangle (\bbox{k}_{1f},\bbox{k}_{1i}) &=& \int
\frac{dk^-_{1f}}{2(2\pi)} \, \frac{dk^-_{1i}}{2(2\pi)} f(k_{1f},k_{1i}).
\end{eqnarray}

We proceed by calculating the specific form of $g_0$,
\begin{eqnarray}
g_0(\bbox{k}_1;P)
&=& \frac{\theta(k_1^+)\theta(k_2^+)}{2 k_1^+ k_2^+}
\frac{1}{P^- -k_1^- -k_2^-},
\end{eqnarray}
where $k_i^-=\epsilon^-(\bbox{k}_i)$.  As $g_0$ is a three-dimensional
quantity, it is clear that $k_i^-$ is not the independent minus
component of a momentum four-vector.  With this expression for $g_0$, we
can specialize Eq.~(\ref{wc:eqn:3ered}) to the center-of-momentum frame and
obtain 
\begin{eqnarray}
&&\left( E -k_{1f}^- -k_{2f}^- \right) \psi(\bbox{k}_{1f};E) \nonumber\\
&&\qquad= \int d^2k_{1i,\perp} \int_0^{E} dk_{1i}^+ \,
\frac{w(\bbox{k}_{1f},\bbox{k}_{1i};E)}
{2(2\pi)^3 \sqrt{k_{1f}^+ k_{2f}^+ k_{1i}^+ k_{2i}^+}}
\psi(\bbox{k}_{1i};E),
\end{eqnarray}
where
\begin{eqnarray}
\psi(\bbox{k}_1;E) &=&
\frac{g_0(\bbox{k}_1;E)}{\sqrt{k_1^+k_2^+}} 
\gamma(\bbox{k}_1;E).
\end{eqnarray}
By comparing this equation to Eq.~(\ref{wc:fullse3}), we find that the 
full light-front two-nucleon effective potential that corresponds to
the kernel $K$, after suppressing the coefficient given in
Eq.~(\ref{wc:supressfactor}), is 
\begin{eqnarray}
V(\bbox{k}_{1f},\bbox{k}_{1f};E) &=& \frac{1}{E}
w(\bbox{k}_{1f},\bbox{k}_{1f};E). \label{wc:convredtopot}
\end{eqnarray}
Thus, we can calculate light-front potentials directly from the
Bethe-Salpeter equation using this method.

The potential $V$ can be expanded in powers of the coupling constant, as
done in LFTOPT.  We find that when the auxiliary Green's function given
in Eq.~(\ref{wc:defghat}) and the OBE kernel are used, the lowest order
parts of the potential (as calculated in Ref.~\cite{Sales:1999ec}) are
the same as our OBE and TBE:SB potentials. Thus, we conclude that this
method produces the physically 
equivalent Hamiltonian theory to the Bethe-Salpeter equation
being used.  We will use this in the next section to derive a
Hamiltonian potential for a situation where LFTOPT cannot be applied.

\subsection{The Modified-Green's-function Approach}
\label{wc:sect:3dred}

Now that the technology for the three-dimensional reduction has been
reviewed, we derive an approximate kernel for the Bethe-Salpeter
equation.  We will follow the approach of Phillips and Wallace
\cite{Phillips:1996eb} and works cited therein.  The idea is to start
with the Bethe-Salpeter equation where the kernel is truncated to only
include ladder (one-boson-exchange, see Fig.~\ref{fig:wc.ladderbseker})
and crossed (two-boson-exchange, see Fig.~\ref{fig:wc.crossedbseker})
parts,
\begin{eqnarray}
\Gamma &=& (K_{\text{ladder}}+K_{\text{cross}}) G_0 \Gamma.
\label{wc:laddercrossBSE}
\end{eqnarray}
An uncrossed approximation is used where the crossed part of the kernel
is approximated by
$K_{\text{cross}}\approx K_{\text{ladder}}G_CK_{\text{ladder}}$.
Our job is to find a valid modified Green's function, $G_C$.  Using this
uncrossed approximation,
\begin{eqnarray}
\Gamma &\approx& (K_{\text{ladder}}+
K_{\text{ladder}} G_C K_{\text{ladder}}) G_0 \Gamma.
\label{wc:ladderuncrossBSEstart}
\end{eqnarray}

One can attempt to rewrite Eq.~(\ref{wc:laddercrossBSE}) as an equation
linear in $K_{\text{ladder}}$, to obtain the modified-Green's-function
Bethe-Salpeter equation,
\begin{eqnarray}
\Gamma_{\text{MGF}} &=& K_{\text{ladder}} \Big( G_0 + G_C \Big) 
\Gamma_{\text{MGF}}.
\label{wc:ladderuncrossBSE}
\end{eqnarray}
By iterating this integral equation for $\Gamma_C$, we obtain
\begin{eqnarray}
\Gamma_{\text{MGF}} &=&
\left[
K_{\text{ladder}} + \sum_{n=1}^\infty
K_{\text{ladder}} \left(G_C K_{\text{ladder}}\right)^n \right] 
G_0 \Gamma_{\text{MGF}}.
\end{eqnarray}
The part of Eq.~(\ref{wc:ladderuncrossBSE}) that plays the role of the
kernel includes the uncrossed approximation of the original kernel
$K_{\text{ladder}}+K_{\text{ladder}}G_C K_{\text{ladder}}$ as well as
many more terms.  We note that the higher-order terms approximate some
of the higher-order terms that should be included in the full kernel,
such as three-boson-exchange diagrams where several meson lines cross.
However, this approach undercounts the higher-order
terms which it approximates, and also leaves out some terms completely.
Therefore, this new Bethe-Salpeter equation will give results that are
closer to the full solution than Eq.~(\ref{wc:laddercrossBSE}), but will
not give the exact solution.  The articles by Wallace and
Mandelzweig \cite{Wallace:1989,Wallace:1989nm} demonstrate that this
approach, by effectively summing an infinite set of interactions, gives
the correct one-body limit, which is something that the usual
Bethe-Salpeter equation with a truncated kernel cannot do.

The modified Bethe-Salpeter equation in Eq.~(\ref{wc:ladderuncrossBSE}) is
reduced to a Hamiltonian equation via the technique discussed in the
previous section.  
The equal-time Hamiltonian has been derived by Phillips and Wallace
\cite{Phillips:1996eb}.   They found that this modified-Green's-function
approach gave a spectra that lies closer to the full ground-state
spectra than the other approximations they considered.  We will use the
light-front reduction to obtain the light-front potential for the
Hamiltonian equation physically equivalent to Eq.~(\ref{wc:ladderuncrossBSE}).

\begin{figure}
\begin{center}
\epsfig{angle=0,width=5.0in,height=2.0in,file=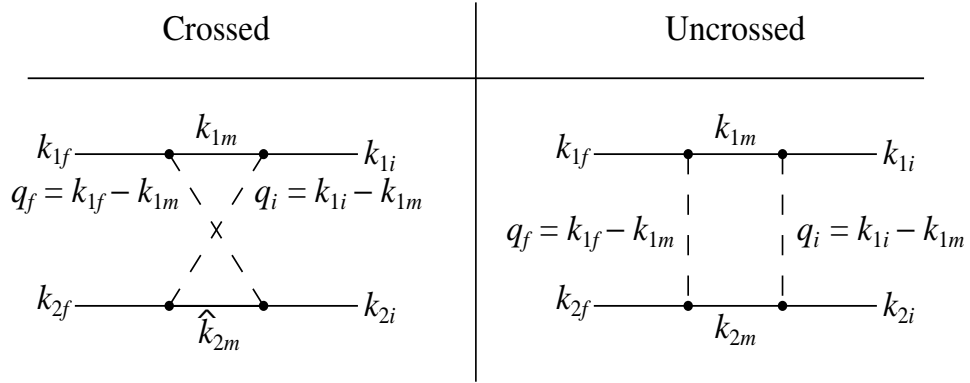}
\caption
[The two-boson crossed and uncrossed Feynman graphs.]
{The two-boson crossed and uncrossed Feynman graphs. Here,
$\widehat{k}_{2m}=-k_{2m}+k_{2i}+k_{2f}$.
\label{wc:crosseduncrosseddiagrams}} 
\end{center}
\end{figure}

To clearly see what role $G_C$ plays, we compare the crossed and
uncrossed Feynman graphs in Fig.~\ref{wc:crosseduncrosseddiagrams}.
Using the Feynman rules,
\begin{eqnarray}
K_{\text{crossed}} &\propto& \int \frac{d^4k_{1m}}{(2\pi)^4}
\frac{1}{(k_{1f}-k_{1m})^2-\mu^2}
d(k_{1m}) d(\widehat{k}_{2m}) 
\frac{1}{(k_{1i}-k_{1m})^2-\mu^2}, \\
K_{\text{uncrossed}} &\propto& \int \frac{d^4k_{1m}}{(2\pi)^4}
\frac{1}{(k_{1f}-k_{1m})^2-\mu^2}
d(k_{1m}) d(k_{2m})
\frac{1}{(k_{1i}-k_{1m})^2-\mu^2},
\end{eqnarray}
where $d$ is the one-particle propagator given in
Eq.~(\ref{wc:onepartprop}), and
$\widehat{k}_{2m}=-k_{2m}+k_{2i}+k_{2f}$.  The only difference between
these two graphs is that the crossed one has $d(\widehat{k}_{2m})$ while
the uncrossed one has $d(k_{2m})$.

We want an approximate one-particle Green's function
$d_C$ that only depends on $q_1$ and $P$, so that
\begin{eqnarray}
d_C(k_1;P) &\approx& d(\widehat{k}_2).
\end{eqnarray}
Substitution of $d_C(k_1;P)$ for $d(\widehat{k}_2)$ in the crossed
graph causes the graph to become uncrossed.  The penalty for this
simplification is that a modified Green's function propagates in the
intermediate state, namely $i d(k_1) d_C(k_1;P)$.  It is important
that this approximation be invariant under relabeling particle labels,
so we explicitly symmetrize by defining 
\begin{eqnarray}
G_C &=& \frac{i}{2}\left[ d(k_1) d_C(k_1;P) + d(k_2) d_C(k_2;P) \right] .
\end{eqnarray}

How should we approximate $d_C$?  Since we are interested in obtaining a
bound state, a low-energy approximation is chosen.  Specializing to the
center-of-momentum frame in this limit, the external momenta are half
the total momentum, so $k_{2i}=k_{2f}=P/2$ and
$\widehat{k}_{2m}=P-k_{2m}=k_{1m}$.  This approximation is similar to
the one used by Phillips and Wallace.  Thus, we define
$d_C(k_1;P)\equiv d(k_1)$ so
\begin{eqnarray}
G_C(k_1;P) &=& G_1(k_1;P) + G_2(k_1;P),
\end{eqnarray}
where we define $G_1$ and $G_2$ by 
\begin{eqnarray}
G_1(k_1;P) &=& \frac{i}{2} d(k_1)^2, \\
G_2(k_1;P) &=& \frac{i}{2} d(P-k_1)^2.
\end{eqnarray}
This approximation for $G_C$ is valid for this model for the
energy range we study, as discussed in Appendix~\ref{a:uxcheck}.  

We can write the modified-Green's-function Bethe-Salpeter equation as
\begin{eqnarray}
\Gamma_{\text{MGF}} = K_{\text{ladder}} \widetilde{G}_0 \Gamma_{\text{MGF}}, \label{wc:eqn:bsemod}
\end{eqnarray}
where
\begin{eqnarray}
\widetilde{G}_0 &=& G_0 + G_1 + G_2.
\end{eqnarray}
This is used as the starting point of three-dimensional reduction
discussed in Section~\ref{wc:sec:3dredbasic}, where $\widetilde{G}_0$ is
considered as the Green's function.  Before doing the reduction,
note that the two poles of $G_1$ and $G_2$ lie in the same half plane
for each function, so
\begin{eqnarray}
\widetilde{g}_0(\bbox{k}_1,P)
&\equiv& \int \frac{dk_1^-}{2(2\pi)} \widetilde{G}_0(k_1,;P) \\
&=& g_0(\bbox{k}_1,P).
\end{eqnarray}

Proceeding with the three-dimensional reduction of
Eq.~(\ref{wc:eqn:bsemod}) in the center-of-momentum frame, we obtain
\begin{eqnarray}
&&\left( E -k_{1f}^- -k_{2f}^- \right) \widetilde{\psi}(\bbox{k}_{1f};E)
\nonumber\\&& \qquad = \int d^2k_{1i,\perp} \int_0^{E} dk_{1i}^+ \,
\frac{\widetilde{w}(\bbox{k}_{1f},\bbox{k}_{1i};E)}
{2(2\pi)^3 \sqrt{k_{1f}^+ k_{2f}^+ k_{1i}^+ k_{2i}^+}}
\widetilde{\psi}(\bbox{k}_{1i};E),
\label{wc:eqn:3eredtilde}
\end{eqnarray}
where
\begin{eqnarray}
\widetilde{w}(\bbox{k}_{1f},\bbox{k}_{1i};P) &=&
\frac{1}{g_0(\bbox{k}_{1f};P)}
\langle \widetilde{G}_0 \widetilde{W} \widetilde{G}_0 \rangle
(\bbox{k}_{1f},\bbox{k}_{1i};P)
\frac{1}{g_0(\bbox{k}_{1i};P)}, \\
\widetilde{\psi}(\bbox{k}_1;E) &=&
\frac{1}{\sqrt{k_1^+k_2^+}} 
\int \frac{dk^-_1}{2(2\pi)} \widetilde{G}_0(k_1;P) \Gamma_{\text{MGF}}(k_1;P),
\end{eqnarray}
and the modified kernel $\widetilde{W}$ is given by
\begin{eqnarray}
\widetilde{W} &=&
K_{\text{ladder}} +
K_{\text{ladder}} 
\left[ \widetilde{G}_0 - \widehat{\widetilde{G}}_0 \right]
\widetilde{W}, \\
\widehat{\widetilde{G}}_0(k_{1f},k_{1i};P)
&=& 
\widetilde{G}_0(k_{1f};P)
\frac{\delta^{(2,+)}(\bbox{k}_{1f}-\bbox{k}_{1i})}{g_0(\bbox{k}_{1f};P)}
\widetilde{G}_0(k_{1i};P).
\end{eqnarray}
It is a feature of the light front that $\widetilde{g}_0=g_0$, so that
the uncrossed approximation only affects the potential, and
Eq.~(\ref{wc:eqn:3eredtilde}) has the same form as Eq.~(\ref{wc:eqn:3ered}).
In the equal-time calculation \cite{Phillips:1996eb}
$\widetilde{g}_0\neq g_0$, so the approximation changes both the Green's
function as well as the potential.

We now expand $\widetilde{W}$ in powers of the coupling constant, and
keep only the lowest order term, $K_{\text{ladder}}$.  According to
Eq.~(\ref{wc:convredtopot}), the light-front potential that corresponds to
this truncation of the kernel is the modified-Green's-function (MGF)
potential $V_{\text{MGF}}$,
\begin{eqnarray}
V_{\text{MGF}}(\bbox{k}_{1f},\bbox{k}_{1i};P) &=& \frac{1}{E}
\frac{1}{g_0(\bbox{k}_{1f};P)}
\langle \widetilde{G}_0 K_{\text{ladder}} \widetilde{G}_0 \rangle
(\bbox{k}_{1f},\bbox{k}_{1i};P)
\frac{1}{g_0(\bbox{k}_{1i};P)}.
\end{eqnarray}
The one-boson-exchange kernel $K_{\text{ladder}}$ is given by the
Feynman diagram, so
\begin{eqnarray}
&&K_{\text{ladder}}(k_{1f},k_{1i};P) \nonumber\\&& \qquad
= \frac{(iM)^2}{(k_{1f}-k_{1i})^2-\mu^2+i\eta} \nonumber \\
&& \qquad =
\left( \frac{1}{k_{1f}^+-k_{1i}^+} \right)
\frac{M^2}{(k_{1f}^--k_{1i}^-) - \mbox{Sign}(k_{1f}^+-k_{1i}^+)
\omega^-(\bbox{k}_{1f}-\bbox{p}_{1i}},
\end{eqnarray}
where the light-front energy of the meson is given by
\begin{eqnarray}
\omega^-(\bbox{q}) &=& \frac{\mu^2 - {\bf q}_\perp^2}{|q^+|} - i \eta.
\end{eqnarray}

By examining the locations of all the poles in the $k^-$ integrals for 
$V_{\text{MGF}}$, we find the integrals are non-vanishing only when
both $x$ and $y$ are between $0$ and $1$.  The sign functions in the
denominator of $K_{\text{ladder}}$ naturally divide $V_{\text{MGF}}$
into two parts, one for $x<y$ and the other for $x>y$.  The integrals in
$V_{\text{MGF}}$ are straightforward, but quite lengthy and tedious.
Therefore, we show only the final answer,
\begin{eqnarray}
&&V_{\text{MGF}}(\bbox{k}_{1f},\bbox{k}_{1i};P) \nonumber\\
&& \qquad = \left(\frac{M}{E}\right)^2 \left[
\frac{\theta(x-y)}{|x-y|} \left( \frac{1}{D_1} +
\frac{N_{i,21}+N_{f,12}}{2D_1^2} +
\frac{N_{i,21} N_{f,12}}{2D_1^3}
\right)
\right.\nonumber \\ & & \qquad
\phantom{
= \left(\frac{M}{E}\right)^2 \left[
\right. } \left.+
\frac{\theta(y-x)}{|y-x|} \left( \frac{1}{D_2} +
\frac{N_{f,21}+N_{i,12}}{2D_2^2} + 
\frac{N_{i,12} N_{f,21}}{2D_2^3}
\right) \right], \label{wc:mgfpot}
\end{eqnarray}
where
\begin{eqnarray}
N_{f,12} &=& \frac{k_{1f}^+}{k_{2f}^+}(E - k_{1f}^- - k_{2f}^-), \\
N_{f,21} &=& \frac{k_{2f}^+}{k_{1f}^+}(E - k_{1f}^- - k_{2f}^-), \\
D_1 &=& E - k_{1i}^- - k_{2f}^- - \omega^-(\bbox{k}_{1f}-\bbox{k}_{2i}), \\
D_2 &=& E - k_{1f}^- - k_{2i}^- - \omega^-(\bbox{k}_{1i}-\bbox{k}_{2f}).
\end{eqnarray}
The expressions for $N_{i,12}$ and $N_{i,21}$ are obtained by replacing
$f$ with $i$ in $N_{f,12}$ and $N_{f,21}$.

What is the physical interpretation of this modified-Green's-function
potential?  The first term multiplying each $\theta$ function gives the
OBE potential we derived before from the perturbation theory.  There the
$D$ in the denominators corresponds to one meson exchange.  The second
and third terms multiplying the $\theta$ functions, with $D^2$ and $D^3$
in the denominators appear to be effective two- and three-meson-exchange
terms.  Since time-ordered perturbation theory does not apply to the
modified Bethe-Salpeter equation that we use, the exact nature of these
terms is not easy to understand.  However, it is clear that these terms
increase the strength of the potential, and should mimic the
higher-order diagrams that are not being included explicitly.

The only dependence on the direction of the perpendicular components of
$k_f$ and $k_i$ comes from the $D$'s.  This allows the azimuthal-angle
integration of $V_{\text{MGF}}$ to be done easily, as shown in
Appendix~\ref{a:ints:angint}.

\section{Results for the Wick-Cutkosky Model}

For our numerical work, we pick the meson mass to be $0.15$ times that
of the nucleon, so $\mu = 0.15 M$.  This is chosen so that our ground
state can be considered a toy model of deuterium, and also to facilitate
comparison with the results of Nieuwenhuis and Tjon
\cite{Nieuwenhuis:1996mc}, Phillips and Afnan \cite{Phillips:1996ed},
and Schoonderwoerd, Bakker, and Karmanov \cite{Schoonderwoerd:1998pk}.
Nieuwenhuis and Tjon used the Feynman-Schwinger
representation (FSR) of the two-particle Green's
function \cite{Simonov:1993kp} in the quenched approximation without the
mass and vertex renormalization terms \cite{Nieuwenhuis:1996mc}.  Their
result is to be considered the full solution that the Bethe-Salpeter and
Hamiltonian equations approximate.  For the Bethe-Salpeter equation,
computation of the bound-state energies for models similar to ours have
been done for the ladder \cite{Schwartz:1965} and ladder plus crossed
\cite{LevineWright3} kernels over 30 years ago.  More recent results are
found in Refs.~\cite{Phillips:1996ed,Theussl:1999xq}, where the
solutions are compared to those given by the FSR approach.

\subsection{Ground State Results} \label{wc:ch:wcgroundres}

We want to examine the ground state energies obtained for each of the
light-front potentials derived in this chapter, and see how they compare
the energies calculated in other approaches. Our main interest is in
finding out how well the light-front potentials approximate the
Bethe-Salpeter results, and also in how well both the perturbative and
non-perturbative potential agree with the Feynman-Schwinger
representation approach.

Consider first how the light-front Hamiltonian approach fits in with the
other approaches.  As discussed in section~\ref{wc:tbe:sbSec}, different
light-front potentials can be derived from Bethe-Salpeter equations with
different kernels.  We have mentioned that the OBE+TBE:SB potential
should approximate the ladder Bethe-Salpeter equation, and similarly the
OBE+TBE potential should approximate the Bethe-Salpeter equation when
the ladder plus crossed kernel is used.  The best that these truncated
Hamiltonians can do is approximate their respective Bethe-Salpeter
equations.

\begin{figure}
\begin{center}
\epsfig{angle=0,width=5.0in,height=4.0in,file=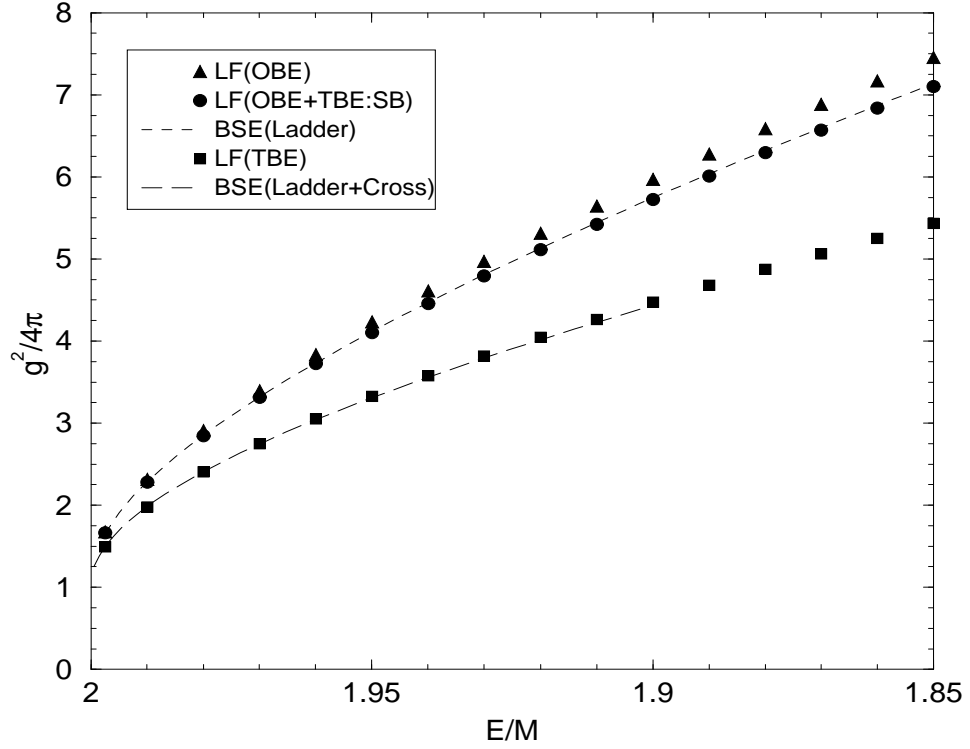}
\caption
[The spectra calculated using the OBE,
OBE+TBE:SB and OBE+TBE potentials are displayed in this figure, along
with the Bethe-Salpeter spectra for the ladder and 
ladder plus crossed kernels.]
{The spectra calculated using the OBE,
OBE+TBE:SB, and OBE+TBE potentials are displayed in this figure, along
with the spectra for the Bethe-Salpeter equation for the ladder and
ladder plus crossed kernels. The curves for the ladder 
(ladder plus crossed) Bethe-Salpeter equation and the light-front
OBE+TBE:SB (OBE+TBE) potentials are very close to each other, almost
indistinguishable in this figure.  $E$ is the energy of the ground state
of two nucleons, and $M$ is the mass of the nucleons.  The meson mass is
$\mu = 0.15 M$.  The binding energy $B$ is related to the bound state energy
via $E=2M-B$.
\label{wc:bslffig}}
\end{center}
\end{figure}

With this in mind, we evaluate the coupling constant versus bound-state
energy curves (which we will call the spectrum) for the Hamiltonian
equation with the OBE potential, the OBE+TBE:SB potential, and the
OBE+TBE potential.  For the range of values we use here, we find
numerical errors in the value of $g^2$ are less than 2\%.
Our results (without error bars) are plotted along with the results
obtained with the ladder BSE \cite{Phillips:1996ed}, and ladder plus
crossed BSE \cite{dandata} in Fig.~\ref{wc:bslffig}.  We note that the
OBE+TBE:SB potential agrees well with the ladder BSE, and the OBE+TBE
potential agrees with the ladder plus crossed BSE.  This is the best
that a Hamiltonian can do, so this result is interpreted as evidence
that, in general, the higher-order diagrams are very small for the
ground state on the light front.

\begin{figure}
\begin{center}
\epsfig{angle=0,width=5.0in,height=4.0in,file=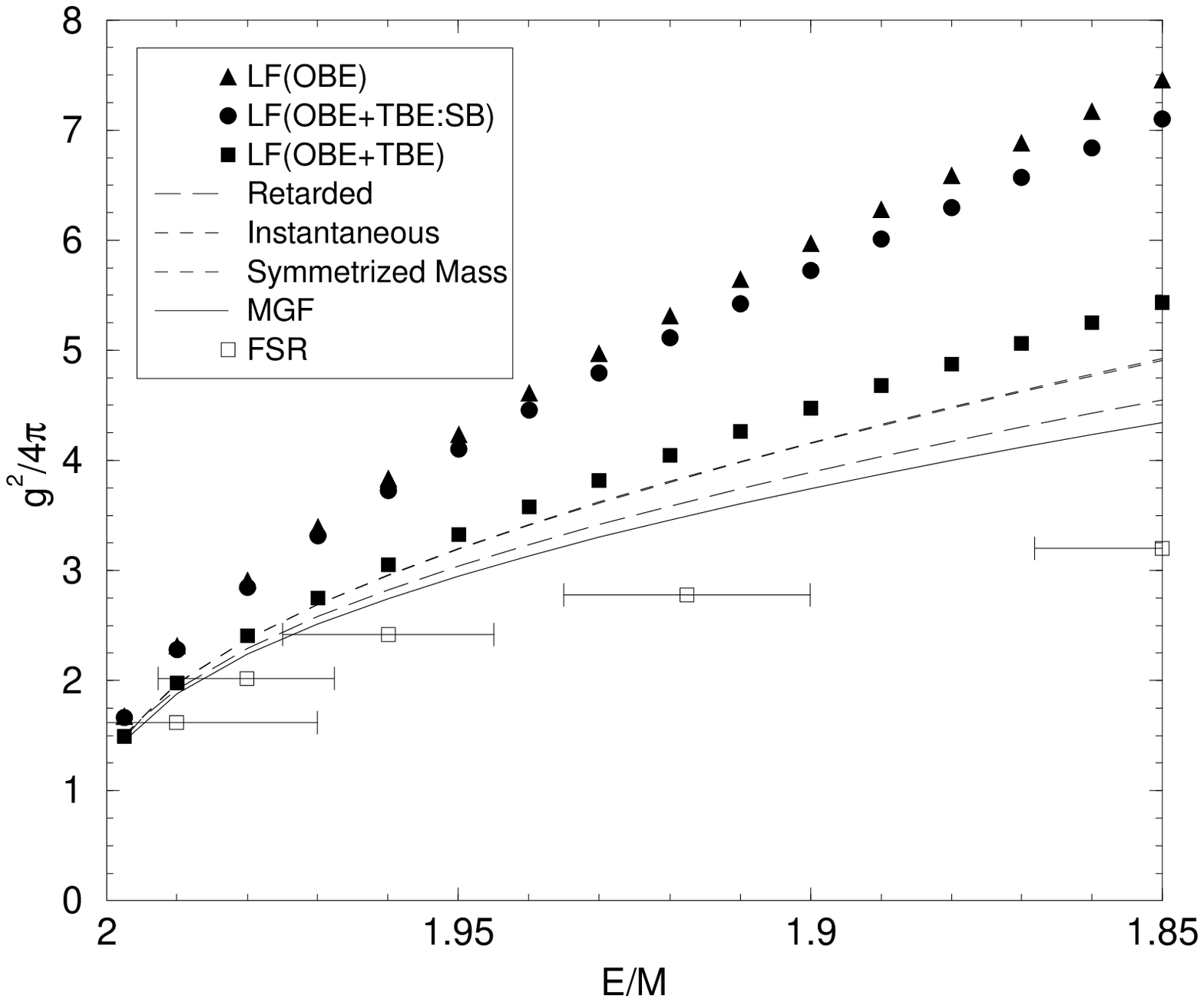}
\caption[The spectra of all three of the perturbative light-front
potentials 
(OBE, OBE+TBE:SB, OBE+TBE)
and the four non-perturbative light-front potentials
(modified-Green's-function, retarded, instantaneous, symmetrized mass)
are plotted here along with the spectra for the full solution (FSR).
The instantaneous curve lies very close to the symmetrized mass curve,
hence the same line style is used for both.]
{The spectra for the seven light-front potentials (OBE, OBE+TBE:SB,
OBE+TBE, retarded, instantaneous, symmetrized mass,
modified-Green's-function) are plotted here along with the spectra for
the full solution (FSR).  The instantaneous curve lies very close to the
symmetrized mass curve, hence the same line style is used for both.
\label{wc:sumfig}}
\end{center}
\end{figure}

If all one wanted was a way to approximate the spectra for
Bethe-Salpeter equations with different kernels, one could just use the
truncated potentials that the LFTOPT provide.  However, the true goal is
to approximate the spectra for the full ground state, which in this
model is given by the FSR approach \cite{Nieuwenhuis:1996mc}.  We expect
that the non-perturbative potentials should give a better approximation
of the full solution than the perturbative potentials, since the
non-perturbative potentials attempt to incorporate physics from
higher-order diagrams, although this is not immediately clear by looking
at the forms of the potentials used. We plot the results for all of the
light-front potentials described in this chapter, the three truncated
potentials (OBE, OBE+TBE:SB, and OBE+TBE) and the four non-perturbative
potentials (symmetrized mass, retarded, instantaneous, and modified
Green's function) along with the results for the full theory in
Fig.~\ref{wc:sumfig}.  For deeply-bound states, there is considerable
disagreement between the perturbative results and the full results,
while the non-perturbative results are in better agreement, with the
modified-Green's-function (MGF) potential achieving the closest
agreement.  For lightly-bound states, the results for all of the
potentials appear converge to each other, close to the full result.

For the modified-Green's-function potential, only the first term of the
expansion in $g^2$ was kept.  In principle, higher-order terms could
be calculated.  However, since there was fairly good agreement between
the OBE potential and the ladder Bethe-Salpeter equation, it is expected
that the MGF potential will give results that are close to the
ladder Bethe-Salpeter equation using the modified-Green's-function.

\subsection{Excited State Results} \label{wc:ch:wccalcres}

Now, we are ready to address the problem of rotational invariance of the
light-front potentials, and see how the bound states are affected by the
breaking of rotational invariance. We expect a splitting of the
bound-state energies for different $m$ states, and we also expect that
the states will not be pure angular momentum states. In order to
examine these effects, we must have excited states, preferably those
with non-zero angular momentum.

Furthermore, since the results of the previous section show that the
ground-state energies calculated with the OBE+TBE potentials are in
excellent agreement with the Bethe-Salpeter results, we expect that the
excited states calculated with the light-front potentials will at least
be similar to the Bethe-Salpeter results. This would mean that the
excited states should {\em almost} be angular momentum eigenstates and
could be classified easily.

First, we consider the solution for the excited states of the
Bethe-Salpeter equation. The technology for doing these bound-state
Bethe-Salpeter equation calculations was developed over 30 years ago
\cite{LevineWright1,LevineWright2}. Since the eigenstates of the ladder
Bethe-Salpeter equation are also eigenstates of the total angular
momentum, there is exact degeneracy in the energies of the different $m$
states for the same angular momentum. For the range of parameters used
in this study, we find that the numerical errors in $g^2$ are less than
0.5\%.
The numerical errors are largest for the most deeply-bound states
($E\approx 1.85 M$) with the largest coupling constants
($g^2/4\pi \approx 50$).

\begin{figure}
\begin{center}
\epsfig{angle=0,width=5.0in,height=4.0in,file=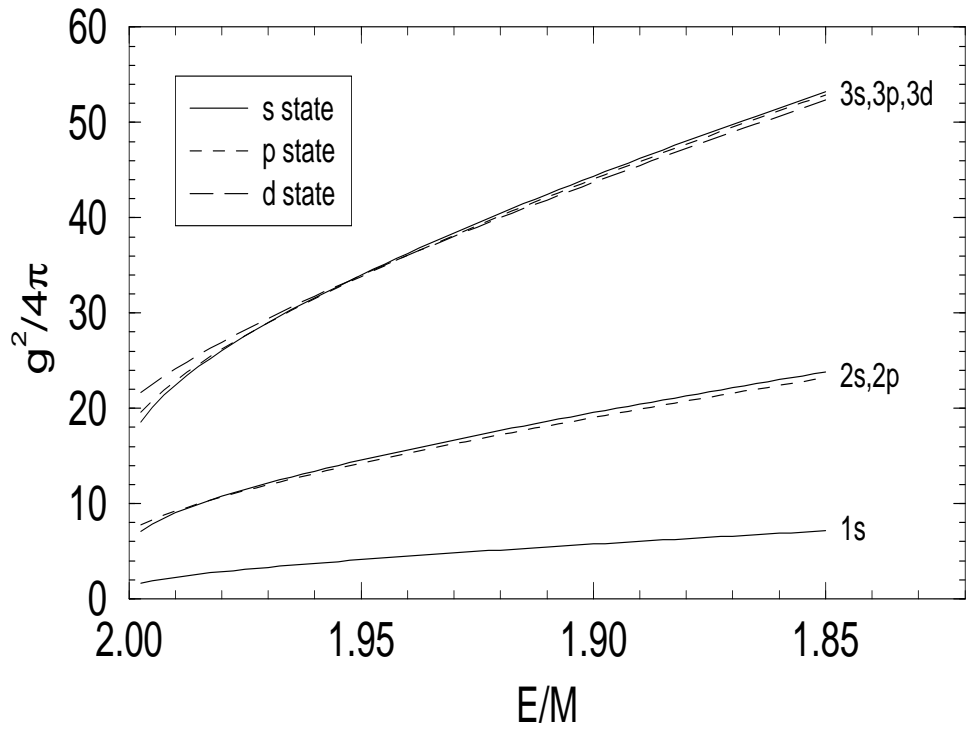}  
\caption[The first three energy bands for the ladder
Bethe-Salpeter equation.]
{The first three energy bands for the ladder
Bethe-Salpeter equation. $E$ is the energy of the bound state of two
nucleons, and $M$ is the mass of each of the two nucleons.  The mass of
the meson is $\mu = 0.15 M$.
\label{wc:bsfig}}
\end{center}
\end{figure}

The solutions to the ladder Bethe-Salpeter equation form bands where
excited states with different orbital angular momentum are approximately
degenerate with each other, as shown in Fig.~\ref{wc:bsfig}.  This
approximate degeneracy is due to the fact that when $\mu \rightarrow 0$,
this model can be shown to have the same degeneracies as the
non-relativistic hydrogen atom \cite{Wick:1954eu,Cutkosky:1954ru}. Thus,
in that limit, all states with the same principal quantum number have
the same energy.  When $\mu \neq 0$, that degeneracy is broken, but only
slightly, as we can see from Fig.~\ref{wc:bsfig}. Because of this, we
will label our states using atomic spectroscopy notation.

Next, we consider the two light-front Schr\"odinger equations given by
Eqs.~(\ref{wc:obeham}) and (\ref{wc:tbeham}).  These equations are
solved numerically (for each parity and several $m$ values) for the
spectrum $g(E)$ for a range of bound-state energies $E$. The symmetries
of the light-front Hamiltonian allow us to classify the states according
to $m$ and the action under parity, so as an example, we calculate the
spectra with even parity and $m=0$. For the range of values we use, we
find the numerical errors in $g^2$ are less than 2\%.
The errors are largest for the most deeply bound states with the largest
coupling constants, as was the case for the Bethe-Salpeter equation.

\begin{figure}
\begin{center}
\epsfig{angle=0,width=5.0in,height=5.0in,file=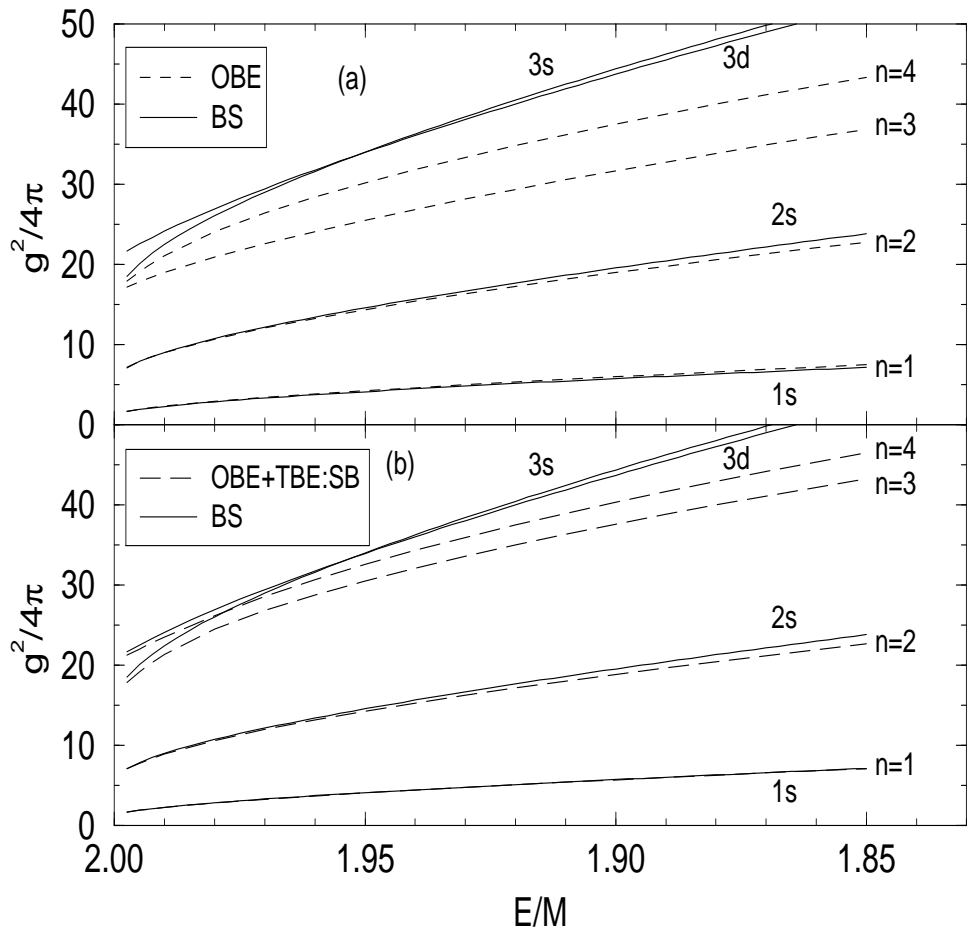}
\caption[Spectra for the four lowest wave functions of even parity and
$m=0$ (a) for the OBE equation (short-dashed line) and the
Bethe-Salpeter equation (solid line), (b) for the OBE+TBE:SB equation
(long-dashed line) and the Bethe-Salpeter equation (solid line).]
{Spectra for the four lowest wave functions of even parity and
$m=0$ (a) for the OBE equation (short-dashed line) and the
Bethe-Salpeter equation (solid line), (b) for the OBE+TBE:SB equation
(long-dashed line) and the Bethe-Salpeter equation (solid line).  The
curves are labeled by $n$, which indicates that the curve belongs to the
$n^{\text{th}}$ eigenvector. The curve for the $1s$ state for the
Bethe-Salpeter equation and the curve for the first OBE+TBE:SB wave
function are very close together, but distinct.
\label{wc:plotraw}}
\end{center}
\end{figure}

We plot both the OBE spectra [in Fig.~\ref{wc:plotraw}(a)] and the
OBE+TBE:SB spectra [in Fig.~\ref{wc:plotraw}(b)] along with the
Bethe-Salpeter spectra for the even parity and $m=0$ states. Based on
the energies, we see that the $n=1$ state is expected to be the $1s$
state, the $n=2$ state the $2s$ state, and the $n=3$ and $n=4$ states
the $3s$ or $3d$ states respectively.  We see approximate agreement
between both of
the truncated Hamiltonian results and the Bethe-Salpeter results for
lightly-bound systems (where $E \approx 2M$).

Our states are not manifest eigenstates of total angular momentum $J^2$.
However, if our approximation of the Hamiltonian potential was good
enough, we would be able to identify the $n=3$ and $n=4$ states of the
light-front Hamiltonian calculation with $3s$ and $3d$ states of the
Bethe-Salpeter equation unambiguously.  This would determine the angular
momentum of the light-front Schr\"odinger equation wave functions.  From
Fig.~\ref{wc:plotraw}, we see both that the addition of the TBE:SB potential
brings the $n=3$ and $n=4$ states closer to the Bethe-Salpeter results,
and that in the lightly-bound region the $n=3$ and $n=4$ states can be
identified as $3s$ and $3d$ states.  For more deeply-bound states, this
identification cannot be made.

An alternative approach to assigning angular momenta labels would be to 
perform a partial-wave decomposition of the wave functions in the real
angular momentum basis.  The wave functions could then be labeled by the
angular momentum component which is dominant.  In light-front dynamics, 
this is difficult to do since the perpendicular components of the real
angular momentum operator ($J_x$ and $J_y$) are dynamical, which makes
the angular momentum operator as complicated as the Hamiltonian.  We are
encouraged to look for an alternative operator which is easier to use,
yet approximates the behavior of the real angular momentum.  We shall
use the ``relative angular momentum'', defined by
$\bbox{L}=\bbox{x}\times\bbox{k}$, where the vectors are equal-time
three-vectors.  The relative momentum $\bbox{k}$ is defined by 
Eq.~(\ref{wc:eteq}) and $\bbox{x}$ is canonically conjugate to
$\bbox{k}$. The ``relative angular momentum'' $\bbox{L}$ is used to help
analyze our solutions, and is not the same as the real angular momentum
that can be derived from the Lagrangian using the energy-momentum
tensor.  However, we will see that the ``relative angular momentum''
gives results that are expected from the real angular momentum, so our
use of the ``relative angular momentum'' in place of the real one
appears to be justified.  

A partial-wave decomposition is performed on the wave functions
(represented in the relative momentum basis) to obtain the radial wave
functions $R_{l,m}$ for all ``relative angular momentum'' states
$Y_{l,m}$. Since the potential is not manifestly rotationally invariant
when written in terms of the relative momentum, our wave functions will
have support from many different partial waves.  Not all partial waves
are allowed; only those partial-wave states with the same $J_3$ and
parity quantum numbers as the wave function give non-vanishing radial
wave functions.  We have 
\begin{eqnarray}
\langle \bbox{k} | \psi^n_{m,p} \rangle &=& \sum_{l=m}^\infty 
Y_l^m(\theta,\phi) R^{n,p}_{l,m}(k).
\end{eqnarray}
We define the fraction of the wave function with ``relative angular
momentum'' $l$ as
\begin{eqnarray}
f^n_l &=& \int_0^\infty dk\, k^2 \left|R^{n,p}_{l,m}(k)\right|^2.
\end{eqnarray}
The fractions $f^n_l$ are a measure of the amount of ``relative angular
momentum'' state $l$ in the $n^{\text{th}}$ eigenfunction.

\begin{figure}
\begin{center}
\epsfig{angle=0,width=5.0in,height=6.0in,file=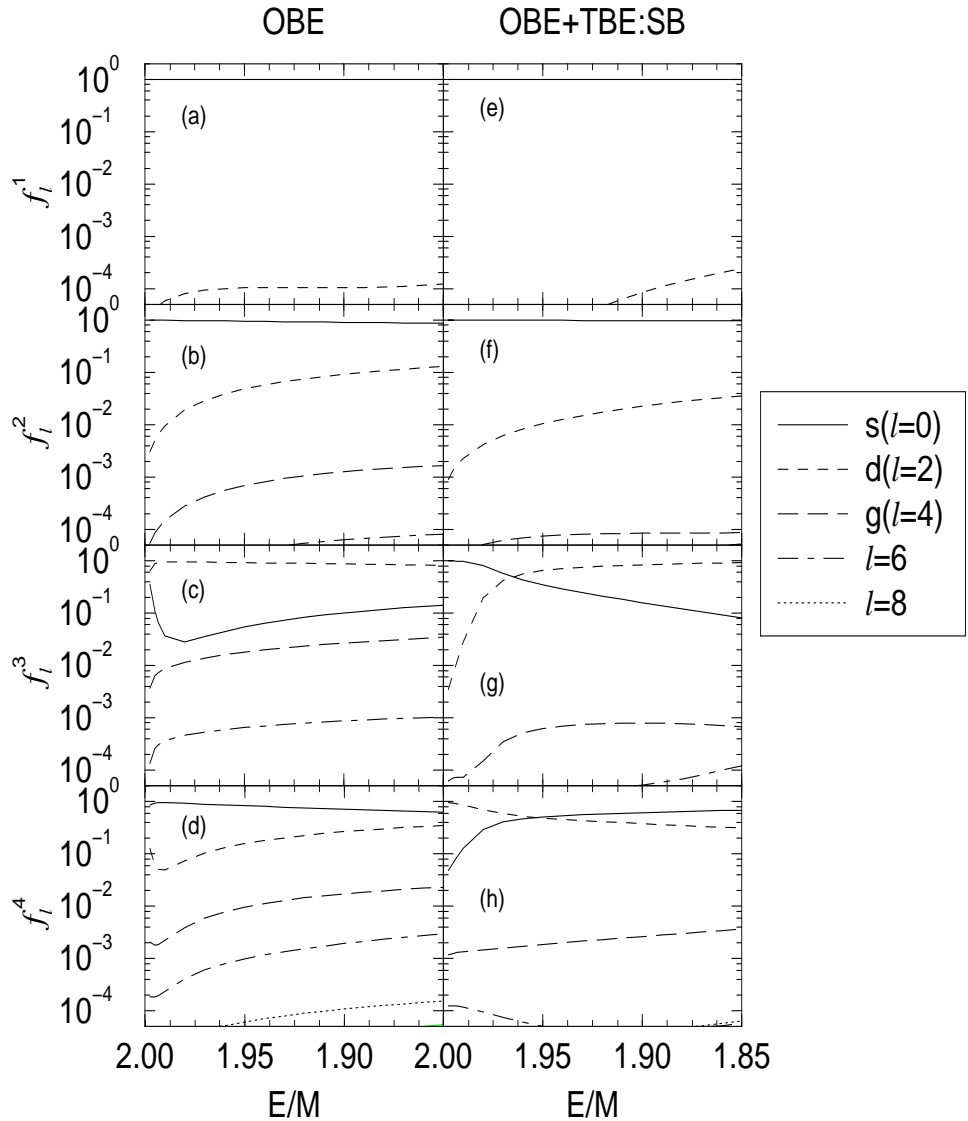}  
\caption
[The fractions of each angular momentum state in several wave functions
is plotted as a function of the binding energy $E/M$. The results for
the OBE truncation are shown along with the OBE+TBE:SB truncation. The
OBE wave functions, in general, have more support from a larger number
of partial waves.]
{The fractions of each angular momentum state in several wave functions
is plotted as a function of the binding energy $E/M$. The wave functions
chosen are the four lowest coupling constant wave functions with $m=0$
and even parity. The results for the OBE truncation are shown on the
left, (a-d), and the OBE+TBE:SB truncation are shown on the right,
(e-h).  Note that the OBE wave functions, in general, have more support
from a larger number of partial waves.
\label{wc:fracfig}}
\end{center}
\end{figure}

To illustrate these ``relative angular momentum'' fractions, we perform
this analysis on the four lowest coupling constant wave functions with
even parity and $m=0$ for both the OBE and OBE+TBE:SB truncations, the same
as in Fig.~\ref{wc:plotraw}.  We show $f^n_l$ as a function of $E$ in
Fig.~\ref{wc:fracfig}.  (The plots for the $m=\pm1,\pm2$ states show
similar behavior, with different $l$ values.)  Note that more higher
``relative angular momentum'' states contribute for deeper bound
states.  This is the same behavior that would be expected if the real
angular momentum was used to perform the partial wave decomposition.
Recall that the real angular momentum will commute with the full
potential if no truncation is made.  However, the potentials we use are
truncated, and neglecting the higher-order terms breaks the rotational
invariance of the potential.  Also, both the binding and the importance
of the higher-order diagrams increase with the coupling constant.  Thus,
both truncations we consider will break rotational invariance more when
the states are more deeply bound.  In the absence of having the real
angular momentum fractions, we will use the ``relative angular
momentum'' fractions.

Examination of the $f^n_l$ curves in Fig.~\ref{wc:fracfig} shows us
that, as postulated earlier, the $n=1$ and $n=2$ states shown in
Figs.~\ref{wc:fracfig}(a,e) and \ref{wc:fracfig}(b,f) are predominately
$s$-wave states, which we label the $1s$ and $2s$ states respectively.
For the OBE truncation, the $n=3$ state shown in
Fig.~\ref{wc:fracfig}(c) is predominately $d$-wave (labeled the $3d$
state) and the $n=4$ state shown in Fig.~\ref{wc:fracfig}(d) is
predominately $s$-wave (labeled the $3s$ state), with little mixing
between the two states.  For the OBE+TBE:SB truncation, the $n=3$ and $n=4$
states are mixtures of both the $3s$ and $3d$ states, as seen in
Figs.~\ref{wc:fracfig}(g,h). In fact, we see a level crossing between
the $n=3,4$ states in that the $n=3$ state is predominantly $3d$ for
lightly-bound systems, but as the binding increases, the $n=3$ state
becomes predominantly $3s$.  Again, this type of mixing would also be
found if the real angular momentum was used.

If this $3s$-$3d$ mixing is ignored, then Fig.~\ref{wc:fracfig} shows
clearly that when the TBE:SB is included the amount of ``relative angular
momentum'' states mixed in actually {\em decreases} as compared to the
OBE results.  This implies that the wave functions of the OBE+TBE:SB
equation are better ``relative angular momentum'' states than the wave
functions of the OBE equation.  Thus the TBE:SB potential restores some
rotational invariance to the OBE calculation.

We attempt to classify the wave functions as states with definite
angular momentum.  If a wave function has the most support from the
``relative angular momentum''  state $l$, we classify that state as
having angular momentum $l$.  This procedure designates the first two
wave functions for both the OBE and the OBE+TBE:SB cases as $s$-wave
states, which is clearly the right thing to do.  For the third and
fourth wave functions of the OBE+TBE:SB equation, there are points where
the fraction of the $s$-wave state equals the fraction of the $d$-wave
state.  Here it no longer clear that such a state should be assigned a
definite ``relative angular momentum'' value; we do so regardless and
analyze the consequences later.

\begin{figure}
\begin{center}
\epsfig{angle=0,width=5.0in,height=4.0in,file=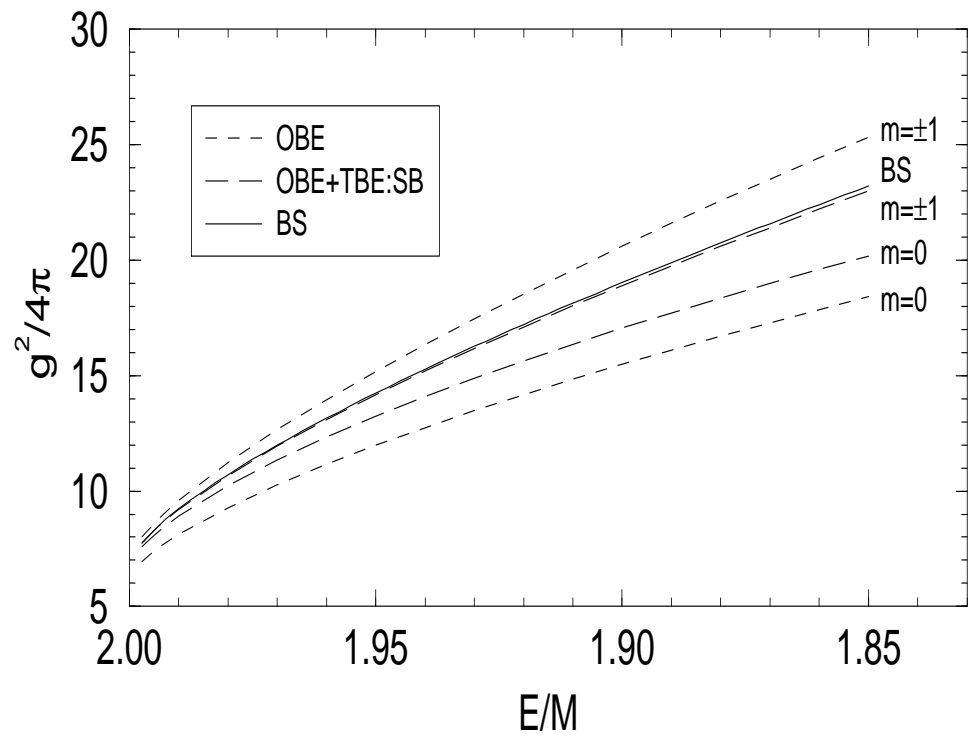}  
\caption{The spectra for the first $p$-wave state, the $2p$ state. The
Bethe-Salpeter (BS) result is plotted with a solid line, the OBE results
with the short-dashed lines, and the OBE+TBE:SB with the long-dashed
lines.
\label{wc:2pfig}}
\end{center}
\end{figure}

\begin{figure}
\begin{center}
\epsfig{angle=0,width=5.0in,height=4.0in,file=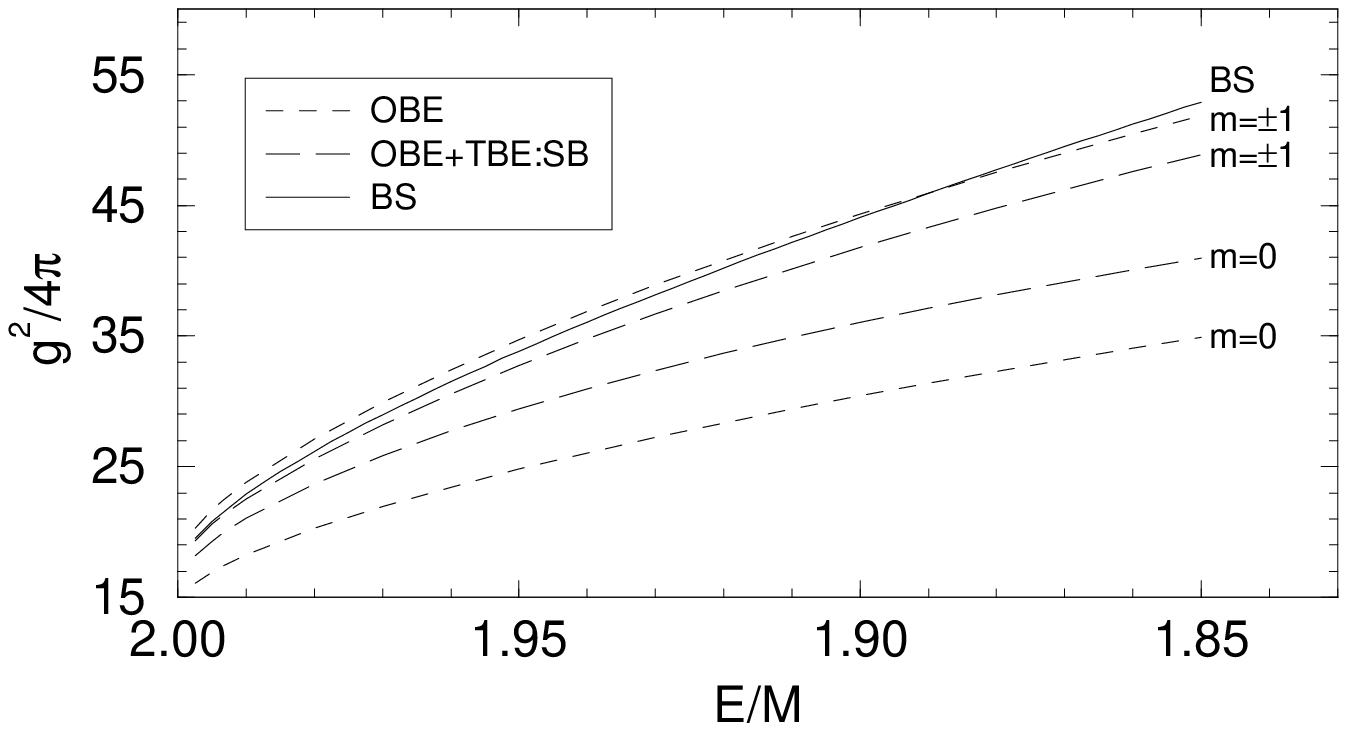}   
\caption{The spectra for the second $p$-wave state, the $3p$ state. The
Bethe-Salpeter (BS) result is plotted with a solid line, the OBE results
with the short-dashed lines, and the OBE+TBE:SB with the long-dashed
lines. 
\label{wc:3pfig}}
\end{center}
\end{figure}

\begin{figure}
\begin{center}
\epsfig{angle=0,width=5.0in,height=6.5in,file=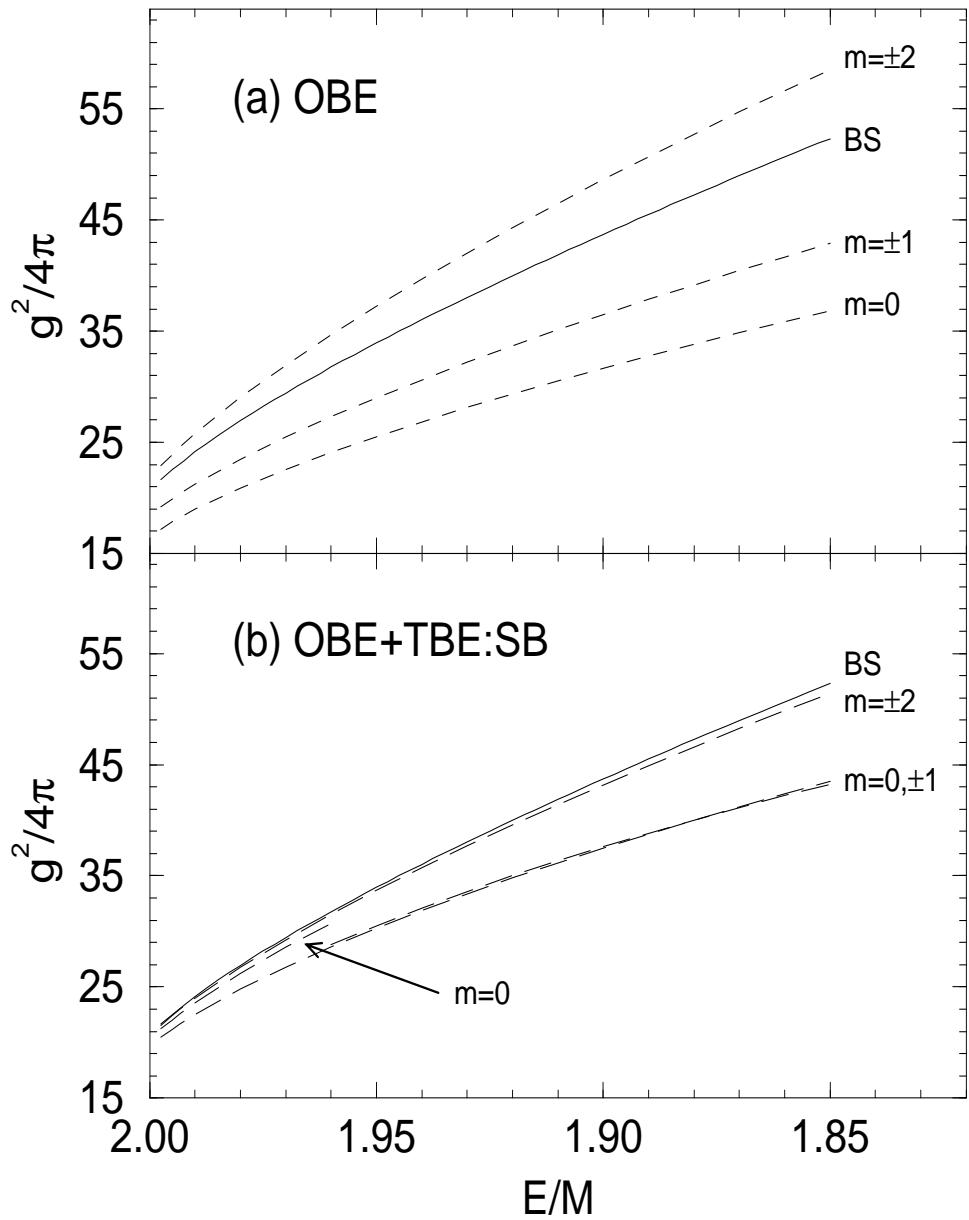}
\caption[
The first $d$-wave state, the $3d$ state. The Bethe-Salpeter (BS) result
is plotted with a solid line, the OBE results with short-dashed lines,
and the OBE+TBE:SB results with long-dashed lines.
]
{The first $d$-wave state, the $3d$ state. The Bethe-Salpeter (BS) result
is plotted with a solid line. In (a) the OBE results are the
short-dashed lines, and in (b) the OBE+TBE:SB results are the long-dashed
lines. Note that there are two separate $m=0$ curves when the OBE+TBE:SB
are used, one extending a quarter of the way over, and the other hiding
under the $m=\pm1$ curve with the OBE+TBE:SB.
\label{wc:3dfig}}
\end{center}
\end{figure}

We now examine the breaking of rotational invariance of the two
truncations based on the comparison of states with different $m$
projections of the same angular momentum.  Since the $s$-wave state has
only $m=0$, there is nothing to analyse in that case.  Further
discussion of the ground-state $s$-wave appears in
Ref.~\cite{Cooke:2000ef}. In Figs.~\ref{wc:2pfig}, \ref{wc:3pfig}, and
\ref{wc:3dfig}, we plot the Bethe-Salpeter bound-state spectra along
with the spectra for the states constructed with the OBE and the OBE+TBE:SB
potentials.  The different $m$ states for the Bethe-Salpeter equation
are exactly degenerate as a result of the rotational invariance of the
equation. The curves from the Hamiltonian theory do not exhibit the
exact degeneracy in $m$; the degeneracy is broken whether the OBE or
OBE+TBE:SB potentials are used.  However, we see that the degeneracy is
partially restored when the TBE:SB potential is included, in that for a
given binding energy the spread of the coupling constants is always
smaller.

In Figs.~\ref{wc:2pfig} and \ref{wc:3pfig}, we plot the spectra for the
$2p$ and $3p$ states, respectively.  In both figures, we see that the
$m=0$ and $m=\pm1$ curves move closer together after addition of the TBE:SB
potential.  However, the average of the two curves for each case does
not move much.  In fact, for the $3p$ state in Fig.~\ref{wc:3pfig}, the
$m=\pm1$ curve moves farther away from the ladder Bethe-Salpeter curve
after addition of the TBE:SB.  We also note that in Fig.~\ref{wc:3pfig} the
spread of the spectra is larger than in Fig.~\ref{wc:2pfig} for both the
OBE and the OBE+TBE:SB light-front Schr\"odinger equations.  We attribute
this to the neglect of the higher-order graphs.  Since the states with
different $m$ values would be degenerate if all the higher-order graphs
were included, not including them causes a breaking of the degeneracy.
Because the coupling constant is larger for the $3p$ states than for the
$2p$ states with the same binding energy, the omission of the
higher-order graphs causes a larger breaking of the degeneracy for the
$3p$ states. 

We also consider the first $d$-wave states.  We see in
Fig.~\ref{wc:3dfig} that, as in Figs.~\ref{wc:2pfig} and \ref{wc:3pfig},
the states with different $m$ values move together after addition of
TBE:SB. The effect of the level crossing of the $3s$ and $3d$ states shown
in Figs.~\ref{wc:fracfig}(g) and \ref{wc:fracfig}(h) is seen here in
that there are two disjoint curves for $m=0$.  When the level crossing
occurs, at $E/M \approx 1.96$, the wave function which originally was
the $3s$ state becomes the $3d$ state and vice versa.

There are some problems that this level crossing causes.  In the
intermediate region where the $3s$ and $3d$ states are about equally
mixed, it is probably not physically sensible to call the state a $3s$
or $3d$ state. However, plotting the curves gives us an indication of
what the wave functions are doing, and for this case we see that the
$m=0$ curve is always bounded by the $m=\pm1$ and $m=\pm2$ curves.  A
more restrictive classification scheme would leave a gap (in $E/M$)
between the two $m=0$ curves and it would not be so clear that the
bounding of the $m=0$ curve which we observe here occurs. 

Finally, we note that the spread of the curves in the $3d$ case is
approximately the same as the spread in the $3p$ case for each
truncation. This tells us that the higher-order graphs have the same
qualitative effect in both states.

\section{Conclusions} \label{wc:conclusions}

We have examined several aspects of bound states in a massive
Wick-Cutkosky model using several light-front potentials. First, we used
three different truncations of the effective potential derived from the
perturbative field theory and four approximations that attempt to
incorporate non-perturbative physics.  For each of these potentials, we
calculate the coupling constant that gives the ground-state for a given
bound-state energy (the spectrum), and compare to the spectra from
different approaches found in the literature.  We find fairly good
agreement between all the methods for lightly-bound systems.

For the full range of binding energies studied, the results for
calculations including one- and two-boson-exchange potentials agree with
the Bethe-Salpeter equation using the physically equivalent truncation
of the kernel  within the numerical errors.  (This is a consequence of
examining the ground state.  For the excited states more higher-order
light-front time-ordered graphs are required to get the same level of
agreement \cite{Cooke:1999yi}.)  The agreement for the case with the
stretched box diagrams (OBE+TBE:SB) has been shown previously by Sales
{\it et al.} \cite{Sales:1999ec}; the result with the crossed-box
contribution (OBE+TBE) is new.  This excellent agreement with the
Bethe-Salpeter results has an undesirable consequence: The BSE results
are known to be a poor approximation of the full solution for deeply
bound systems, so the truncated Hamiltonian approach cannot provide a
good approximation to the full solution in that regime.

The non-perturbative potentials based on physical considerations give a
better approximation of the full solution than the potentials obtained
from LFTOPT.  For all binding energies, the modified-Green's-function
potential achieves the closest agreement with the full solution of all
the potentials considered here.  However, there is still considerable
disagreement between approximate potentials and the full result for
deeply bound states.  We interpret this as an indication that the
approximations, while incorporating some non-perturbative physics, do
not go far enough.  In the weakly-bound regime, which is of relevance
for deuteron calculations, the spectra for all of the potentials are
close together, indicating that light-front dynamics provides a good
description of lightly-bound systems.

After examining the ground state thoroughly, we considered the low-lying
excited states of the Wick-Cutkosky model. We considered two truncations
of the light-front Hamiltonian, the OBE and OBE+TBE:SB truncations.
Using these truncations, a ``relative angular momentum'' operator was
used to study the partial-wave decompositions of the bound-state wave
functions. We found that the ``relative angular momentum'' operator
acting on those states yield behavior similar to that expected from the
real angular momentum.  This result encourages the use of this operator
to classify the states according to their angular momentum values $l$,
and to study the degeneracy of the spectra for states with the same $l$
value but with different projections $m$. We found less breaking of the
degeneracy when the OBE+TBE:SB potential was used than when the OBE
potential was used. Both of these findings indicate that the OBE+TBE:SB
truncation of the Hamiltonian breaks rotational invariance less than the
OBE truncation alone. 

However, there is still some discrepancy between our truncated
Hamiltonian spectra and the ladder Bethe-Salpeter spectra.  In general,
our OBE+TBE:SB results for the $p$- and $d$-wave states show deeper
binding than the Bethe-Salpeter results.  Not surprisingly, this
disagreement is largest for the most deeply bound-states where the
coupling is largest. This difference would be removed if all the
higher-order pieces of the potential were included.  These higher-order
pieces are also needed for the full restoration of rotational
invariance.

\chapter{Realistic Nuclear Models} \label{ch:pionly}

In this chapter, we start with an effective nuclear Lagrangian
\cite{Miller:1997cr,Miller:1999ap} for nucleons and six mesons which
incorporates a
chiral symmetry. We derive the light-front Hamiltonian from the
Lagrangian and address some of the difficulties encountered. Then the
Hamiltonian is used to construct the Feynman rules for nucleon-nucleon
potentials. The issues of the breaking of rotational invariance and
chiral symmetry by certain subsets of the potentials are considered and
addressed. We use the Feynman rules and the symmetry considerations to
construct new
one-meson-exchange (OME) and
two-meson-exchange (TME) 
light-front nucleon-nucleon (LFNN) potentials.

The potentials are used in numeric calculations for two families of
nuclear models. For our first set of calculations, we assume that the
pion is the only meson that interacts with the nucleons, giving us a
special case of the full nuclear model. This pion-only model is inspired
by the non-relativistic one-pion-exchange model used by Friar, Gibson,
and Payne \cite{Friar:1984wi}. We generalize their model to form the
basis of our pion-only light-front model, which includes relativity
automatically. 
 
Our second set of calculations uses the full model with all six mesons
($\pi$, $\sigma$, $\rho$, $\omega$, $\eta$, and $\delta$) for the
interaction. This model is an extension of the light-front
model used by Miller and Machleidt \cite{Miller:1999ap}. A new feature
of this model is that light-front energy dependent denominators
are used in the potentials; the denominators used in
Ref.~\cite{Miller:1999ap} are energy independent.

\section{Model and Formalism}

Our starting point is a nuclear Lagrangian \cite{Miller:1997cr}
which incorporates a non-linear chiral model for the pions. The
Lagrangian is based on the linear representations of chiral symmetry
used by Gursey \cite{Gursey:1960yy}. It is invariant (in the limit
where $m_\pi\rightarrow0$) under chiral transformations. 

The model prescribes the use of nucleons $\psi'$ (or $\psi$, the free
nucleon field, as discussed in Section~\ref{sec:elemferm}) and six
mesons: the $\pi$, $\delta$ (also known as the $a_0(980)$), $\sigma$
(also known as the $f_0(400-1200)$), $\eta$, $\rho$, and $\omega$
mesons. The coupling of each meson to the nucleon is governed by the
combination of the meson's spin and isospin. The $\pi$ and $\eta$ are
pseudoscalars, the
$\rho$ and $\omega$ are vectors, and the $\delta$ and $\sigma$ are
scalars. Under isospin transformations, the $\pi$, $\rho$, and
$\delta$ are isovector particles while the $\eta$, $\omega$, and
$\sigma$ are isoscalar particles.

The use of scalar mesons is meant as a simple representation of part of
the two-pion-exchange potential which causes much of the medium
range attraction between nucleons \cite{Machleidt:1987hj,Machleidt:1989}.
It can also be interpreted as the effect of fundamental scalar mesons
\cite{Black:1998wt,Black:1999dx,Black:2000qq}.

The Lagrangian ${\mathcal L}$ is based on the one used in
Refs.~\cite{Miller:1997xh,Miller:1997cr,Miller:1999ap}. It is given by
\begin{eqnarray}
{\mathcal L} &=&
-\frac{1}{4} \bbox{\rho}^{\mu\nu} \cdot \bbox{\rho}_{\mu\nu} 
+ \frac{m_\rho^2}{2}\bbox{\rho}^\mu \cdot \bbox{\rho}_\mu
-\frac{1}{4} \omega^{\mu\nu}\omega_{\mu\nu} +
\frac{m_\omega^2}{2}\omega^\mu \omega _\mu
\nonumber \\ & &
+ \frac{1}{4}f^2 \mbox{Tr} (\partial_\mu U \, \partial^\mu U^\dagger)
+\frac{1}{4}m_\pi^2f^2 \, \mbox{Tr}(U +U^\dagger-2)
\nonumber \\ & &
+\frac{1}{2} (\partial_\mu \sigma \partial^\mu \sigma -m_\sigma^2 \sigma^2) 
+\frac{1}{2} (\partial_\mu \bbox{\delta} \cdot 
              \partial^\mu \bbox{\delta}       -m_\delta^2 \bbox{\delta}^2) 
+\frac{1}{2} (\partial_\mu \eta   \partial^\mu \eta   -m_\eta^2   \eta^2  ) 
\nonumber\\&&
+\overline{\psi}' \Big[
\gamma^\mu
(i\partial_\mu
-g_\rho   \bbox{\rho  }_\mu\cdot\bbox{\tau}
-g_\omega \omega_\mu
) - U (M
+g_\sigma\sigma
+g_\delta\bbox{\delta}\cdot\bbox{\tau}
+ig_\eta\gamma_5\eta)
\Big] \psi' \label{eq:mainNNlag},
\end{eqnarray}
where the bare masses of the nucleon and the mesons are given by $M$ and
$m_\alpha$ where $\alpha=\pi,\eta,\sigma,\delta,\rho,\omega$. We have
defined $V^{\mu\nu}\equiv\partial^\mu{}V^\nu-\partial^\nu{}V^\mu$ for
$V=\rho,\omega$. The unitary 
matrix $U$ can be chosen to have one of the three forms $U_i$:
\begin{eqnarray}
U_1\equiv e^{i  \gamma_5 \bbox{\tau\cdot\pi}/f},\quad
U_2\equiv \frac
{1+i\gamma_5\bbox{\tau}\cdot\bbox{\pi}/2f}
{1-i\gamma_5\bbox{\tau}\cdot\bbox{\pi}/2f}, \quad
U_3\equiv \sqrt{1-\pi^2/f^2}+i\gamma_5\bbox{\tau\cdot\pi}/f, \label{us}
\end{eqnarray}
which correspond to different definitions of the fields. Note that
each of these definitions can be expanded to give
\begin{eqnarray}
U &=& 1 + i \gamma_5 \frac{\bbox{\tau}\cdot\bbox{\pi}}{f}
- \frac{\pi^2}{2f^2} + {\mathcal O}
\left(\frac{\pi^3}{f^3}\right) \label{pi:pertU}.
\end{eqnarray}
In this work, we consider at most two meson exchange potentials, so we
consider $U$ to be defined by Eq.~(\ref{pi:pertU}).

In the limit where $m_\pi\rightarrow0$, this Lagrangian, 
is invariant under the chiral transformation 
\begin{eqnarray}
\psi^\prime\to e^{i \gamma_5 \bbox{\tau}\cdot\bbox{a}}\psi^\prime,\qquad
U\to e^{-i \gamma_5 \bbox{\tau}\cdot\bbox{ a}} \;U\; 
e^{-i \gamma_5 \bbox{\tau}\cdot \bbox{a}}.
\label{chiral}\end{eqnarray}
In this model the other mesons are not affected by the transformation
because they are not chiral partners of the pion. This is in contrast to
the Lagrangian given in Refs.~\cite{Miller:1997xh,Miller:1997cr}, where
the mass and scalar interaction terms for the nucleon were written as
$MU+g_s\phi$ instead of $U(M+g_s\phi)$.

\section{Non-interacting Nucleon-Nucleon Theory}

The light-front Hamiltonian is derived from this Lagrangian using the
same approach used in section~\ref{wc:ch:wcformalism},
the approach used by Miller \cite{Miller:1997cr} and others
\cite{Chang:1973xt,Chang:1973qi,Yan:1973qf,Yan:1973qg}. The basic idea
is to write the light-front Hamiltonian ($P^-$) as the sum of a free,
non-interacting part and a term containing the interactions. We consider
the free part first.

\subsection{Free Field Expansions}

The solutions for the free fields are similar to those obtained by using
equal-time dynamics. In fact, the solutions are formally related by a
change of variable, and so the most obvious difference between the two is
due to the Jacobian. The field equations have the general form
(when Lorentz, spinor, and isospin indices are suppressed) of
\begin{eqnarray}
\alpha(x) &=& \int \frac{d^2k_\perp dk^+ \theta(k^+)}{(2\pi)^{3/2} \sqrt{2
k^+}} \left[
a_\alpha        (\bbox{k}) e^{-i k^\mu x_\mu} +
a_\alpha^\dagger(\bbox{k}) e^{+i k^\mu x_\mu} \right],
\end{eqnarray}
where $\alpha=\pi,\eta,\sigma,\delta,\rho,\omega,\psi$. Note that in the
exponentials, 
\begin{eqnarray}
k^\mu x_\mu &=& \frac{1}{2} \left( k^+ x^- + k^- x^+ \right) -
\bbox{k}_\perp \cdot \bbox{x}_\perp.
\end{eqnarray}
In particular, the solutions for all the mesons and the nucleon field
are
\begin{eqnarray}
\bbox{\pi}(x) &=& \int \frac{d^2k_\perp dk^+ \theta(k^+)}{(2\pi)^{3/2} \sqrt{2
k^+}} \left[
\bbox{a}_\pi        (\bbox{k}) e^{-i k^\mu x_\mu} +
\bbox{a}_\pi^\dagger(\bbox{k}) e^{+i k^\mu x_\mu} \right], \\
\eta(x) &=& \int \frac{d^2k_\perp dk^+ \theta(k^+)}{(2\pi)^{3/2} \sqrt{2
k^+}} \left[
a_\eta        (\bbox{k}) e^{-i k^\mu x_\mu} +
a_\eta^\dagger(\bbox{k}) e^{+i k^\mu x_\mu} \right], \\
\bbox{\delta}(x) &=& \int \frac{d^2k_\perp dk^+
\theta(k^+)}{(2\pi)^{3/2} \sqrt{2 k^+}} \left[
\bbox{a}_\delta        (\bbox{k}) e^{-i k^\mu x_\mu} +
\bbox{a}_\delta^\dagger(\bbox{k}) e^{+i k^\mu x_\mu} \right], \\
\sigma(x) &=& \int \frac{d^2k_\perp dk^+ \theta(k^+)}{(2\pi)^{3/2} \sqrt{2
k^+}} \left[
a_\sigma        (\bbox{k}) e^{-i k^\mu x_\mu} +
a_\sigma^\dagger(\bbox{k}) e^{+i k^\mu x_\mu} \right], \\
\bbox{\rho}^\mu(x) &=& \int \frac{d^2k_\perp dk^+
\theta(k^+)}{(2\pi)^{3/2} \sqrt{2 k^+}}
\sum_{s=1,3} \epsilon^\mu(\bbox{k},s) \left[
\bbox{a}_\rho        (\bbox{k},s) e^{-i k^\mu x_\mu} +
\bbox{a}_\rho^\dagger(\bbox{k},s) e^{+i k^\mu x_\mu} \right], \\
\omega^\mu(x) &=& \int \frac{d^2k_\perp dk^+ \theta(k^+)}{(2\pi)^{3/2}
\sqrt{2 k^+}} \sum_{s=1,3} \epsilon^\mu(\bbox{k},s) \left[
a_\omega        (\bbox{k},s) e^{-i k^\mu x_\mu} +
a_\omega^\dagger(\bbox{k},s) e^{+i k^\mu x_\mu} \right], \\
\psi(x) &=& \sqrt{2M}
\int \frac{d^2k_\perp dk^+ \theta(k^+)}{(2\pi)^{3/2} \sqrt{2
k^+}} \nonumber \\
&& \qquad \qquad \times 
\sum_{\lambda=+,-}
\sum_{t_3=+,-}
 \!\! \left[
u(\bbox{k},\lambda) b        (\bbox{k}) e^{-i k^\mu x_\mu} +
v(\bbox{k},\lambda) d^\dagger(\bbox{k}) e^{+i k^\mu x_\mu} \right]
\chi_{t_3}.
\end{eqnarray}
The polarization vectors are the usual ones. The most general of the
commutation relations is
\begin{eqnarray}
\left[ a_{\alpha,i}(\bbox{k},s),a_{\beta,j}^\dagger(\bbox{k}',s')
\right] &=&
\delta_{\alpha,\beta} \delta_{i,j} \delta_{s,s'}
\delta^{(2,+)}(\bbox{k}-\bbox{k}'),
\end{eqnarray}
where $\alpha$, $i$, and $s$ denote the meson type, isospin, and
spin, and $\delta^{(2,+)}$ is defined by Eq.~(\ref{wc:eq:deltwoplus}).
The anti-commutation relations are
\begin{eqnarray}
\left\{ b(\bbox{k},\lambda),b^\dagger(\bbox{k}',\lambda') \right\}  =
\left\{ d(\bbox{k},\lambda),d^\dagger(\bbox{k}',\lambda') \right\} &=&
\delta_{\lambda,\lambda'} \delta^{(2,+)}(\bbox{k}-\bbox{k}').
\end{eqnarray}
All other (anti-)commutation relations vanish. The spinors are
normalized so that
$\overline{u}(\bbox{p},\lambda')u(\bbox{p},\lambda)=\delta_{\lambda'\lambda}$.
For more information on the definition of the spinors, see
Appendix~\ref{app:spinors}.

\subsection{Non-interacting Hamiltonians}

The general form of the non-interacting Hamiltonian for each meson is 
\begin{eqnarray}
P^-_0(\alpha) &=& \int d^2k_\perp dk^+ \theta(k^+)
a_\alpha^\dagger(\bbox{k}) a_\alpha(\bbox{k})
\frac{k_\perp^2 + m_\alpha^2}{k^+}.
\end{eqnarray}
For the vector mesons ($\rho$ and $\omega$), there is an implicit sum
over the meson spins. Explicitly, this means that for vector mesons
$a_V^\dagger(\bbox{k})a_V(\bbox{k})\rightarrow\sum_{s=1,3}a_V^\dagger(\bbox{k},s)a_V(\bbox{k},s)$.
Likewise, the sum over the isospin of the isovector mesons ($\pi$,
$\delta$, and $\rho$) is implicit. The sum over isospin can be made
explicit by writing
$a_I^\dagger(\bbox{k})a_I(\bbox{k})\rightarrow\sum_{i=1,3}a_{I,i}^\dagger(\bbox{k})a_{I,i}(\bbox{k})$.
The non-interacting Hamiltonian for the nucleons has a similar form as
well, 
\begin{eqnarray}
P^-_0(\psi) &=& \int d^2k_\perp dk^+ \theta(k^+)
\left[ \sum_{\lambda=+,-}
b^\dagger(\bbox{k},\lambda) b(\bbox{k},\lambda) +
d^\dagger(\bbox{k},\lambda) d(\bbox{k},\lambda) \right]
\frac{k_\perp^2 + M^2}{k^+}.
\end{eqnarray}
These equations are what one expects, since a free particle with momenta
$k_\perp$ and $k^+$ has light-front energy
$k^-=\frac{k_\perp^2+m^2}{k^+}$.

\section{Interacting Nucleon-Nucleon Theory}

The interaction Hamiltonians are derived from the Lagrangian in
Eq.~(\ref{eq:mainNNlag}) using the techniques presented in
chapter~\ref{wc:ch:wcmodel}. However, there are some additional
complications due to the structure of the interactions.

One complication 
is that the chiral coupling of the pion field to the nucleons through
the $U$ matrix generates vertices with any number of pions. This is
addressed simply by expanding the $U$ matrix in powers of $\frac{1}{f}$,
and considering the interaction Hamiltonians order by order.

Another complication is due to the fact that both the vector mesons and
the fermions have components which depend on other components of the
field
\cite{Miller:1997cr,Soper:1971sr,Kogut:1970xa,Bjorken:1971ah,Soper:1971wn}.
Vector meson fields have four components, but only three degrees of
freedom. Likewise, fermion fields have four spinor components, but only
two degrees of freedom. When the dependent components are expressed
explicitly in terms of the independent components, we obtain new
(effective) interaction Hamiltonians for instantaneous vector mesons and
fermions. A complete derivation is given by Miller in
Ref.~\cite{Miller:1997cr}, we illustrate only the main points of the
derivation here. 

\subsection{Expanding the Pion Interaction}

We start by Taylor-expanding $U$ in powers of $\frac{1}{f}$,
after which the derivation of the
one-meson-interaction Hamiltonian ${P'}^-_{I,1}$ is 
straightforward. (The prime indicates that it is in terms of $\psi'$,
not $\psi$. We derive the expressions for $P^-_{I,1}$ in the
section~\ref{sec:elemferm}.) The result is
\begin{eqnarray}
{P'}^-_{I,1} &=&
\int d^2x_\perp dx^- \overline{\psi}'(x)
\Big[
  g_\rho   \gamma^\mu   \rho_{\mu,i}(x) \tau_i
+ g_\omega \gamma^\mu \omega_\mu    (x)
+ g_\delta            \delta_i      (x) \tau_i
+ g_\sigma            \sigma        (x)
\nonumber\\&&
\phantom{\int d^2x_\perp dx^- \overline{\psi}(x)\Big[}\,\,
+ g_\pi  (i\gamma_5) \tau_i \pi_i(x)
+ g_\chi (i\gamma_5)        \chi (x)
\Big] \psi'(x). \label{eq:pm1Primespecific}
\end{eqnarray}
We have defined a dimensionless coupling constant
$g_\pi\equiv\frac{M}{f}$. To save space and to generalize, we define
\begin{eqnarray}
\Gamma_\alpha &=& \left\{\begin{array}{cl}
i\gamma^5  & \mbox{ if $\alpha$ is a pseudoscalar meson ($\pi$, $\eta$)} \\
1          & \mbox{ if $\alpha$ is a scalar meson ($\delta$, $\sigma$)} \\
\gamma^\mu & \mbox{ if $\alpha$ is a vector meson ($\rho$, $\omega$)} \\
\end{array} \right. \\
T_\alpha &=& \left\{\begin{array}{cl}
\tau_i & \mbox{ if $\alpha$ is an isovector meson ($\pi$, $\delta$, $\rho$)} \\
1      & \mbox{ if $\alpha$ is an isoscalar meson ($\eta$, $\sigma$, $\omega$)}
\\
\end{array} \right.
\end{eqnarray}
and denote the meson fields by $\Phi_\alpha$. This allows us to write
\begin{eqnarray}
P^-_{I,1} &=&
\sum_{\alpha=\pi,\eta,\sigma,\delta,\rho,\omega}
\int d^2x_\perp dx^- \overline{\psi}'(x)
g_\alpha \Gamma_\alpha T_\alpha \Phi_\alpha(x)
\psi'(x) \label{eq:pm1Prime},
\end{eqnarray}
where the appropriate sums over the meson indices are implicit.

The next step is to derive the two-meson-interaction Hamiltonian which
arises from chiral symmetry, ${P'}^-_{I,2c}$:
\begin{eqnarray}
{P'}^-_{I,2c} &=&
\int d^2x_\perp dx^- \overline{\psi}'(x)\Big[
- \frac{g^2_\pi}{2M}                 \tau_i \tau_j \pi_i(x) \pi_j(x)
\nonumber\\&&\phantom{\int d^2x_\perp dx^- \overline{\psi}(x)\Big[}\,\,
+ \frac{g_\pi g_\phi}{M} (i\gamma^5) \tau_i        \pi_i(x) \phi (x)
\nonumber\\&&\phantom{\int d^2x_\perp dx^- \overline{\psi}(x)\Big[}\,\,
- \frac{g_\pi g_\chi}{M}             \tau_i        \pi_i(x) \chi (x)
\Big] \psi'(x) \\
&=&
\sum_{\alpha=\pi,\eta,\sigma,\delta} \frac{g_\pi g_\alpha}{M} s_\alpha
\int d^2x_\perp dx^- \overline{\psi}'(x) 
\left[\Gamma_\pi    T_\pi    \Phi_\pi   (x) \right]
\left[\Gamma_\alpha T_\alpha \Phi_\alpha(x) \right]
\psi'(x) \label{eq:rawtme2},
\end{eqnarray}
where $s_\alpha$ is a symmetry factor, equal to $\frac{1}{2}$ when
$\alpha=\pi$, and $1$ otherwise. When the contact interaction is used to
calculate diagrams, an additional factor is picked up for the $\pi\pi$
contact term (due to indistinguishability) which cancels the
symmetry factor $s_\pi$. Note that this contact interaction
involves only scalar and pseudoscalar mesons.

\subsection{Elimination of Dependent Fermion Components}
\label{sec:elemferm}
\doquote
{It does not do to leave a dragon out of your calculations, if you live
near him.}
{J.R.R.~Tolkien} 

We are now ready to express the dependent components of $\psi'$ in terms
of the independent components, and address the problem of instantaneous
nucleons. The generation of instantaneous interactions is a general
feature of theories with interacting fermions in light-front
dynamics. This allows us to use a simplified model to demonstrate how
these instantaneous nucleons arise. In particular, we want to postpone
the discussion of the complication introduced by the vector mesons
until the next section. To this end, we choose to remove all mesons
except the $\sigma$ from the Lagrangian given in
Eq.~(\ref{eq:mainNNlag}). (The $\sigma$ is chosen since it has the simplest
coupling to the nucleon.) The equation of motion for the nucleons is then
\begin{eqnarray}
i \NEG\partial \psi' &=& ( M + g_\sigma\sigma ) \psi'.
\label{eq:eomNucleon} 
\end{eqnarray}
Applying the projection operators $\Lambda_\pm=\frac{1}{2}\gamma^0\gamma^\pm$
(defined in Appendix~\ref{ch:notation}) to Eq.~(\ref{eq:eomNucleon})
splits it into two equations,
\begin{eqnarray}
i \partial^- \psi_+' &=& [ \bbox{\alpha}_\perp \cdot \bbox{p}_\perp +
\beta (M + g_\sigma\sigma ) ] \psi_-' \label{eomNucleonP}, \\
i \partial^+ \psi_-' &=& [ \bbox{\alpha}_\perp \cdot \bbox{p}_\perp +
\beta (M + g_\sigma\sigma ) ] \psi_+' \label{eomNucleonM},
\end{eqnarray}
where $\psi_\pm'=\Lambda_\pm\psi'$. This split is useful because in
Eq.~(\ref{eq:eomNucleon}), all four components of the nucleon field are
interrelated, while in Eqs.~(\ref{eomNucleonP}) and (\ref{eomNucleonM}),
the two components of $\psi'_+$ are related to the two components of
$\psi'_-$, and vice versa.

First, notice that Eq.~(\ref{eomNucleonP}) involves $\partial^-$, a
dynamic operator in light-front dynamics. Dynamic operators should be
avoided since they involve the interaction, and are therefore
complicated. We use Eq.~(\ref{eomNucleonP}) to avoid that complication and
relate the components of $\psi$. Secondly, to
keep the relation as simple as possible, we do not attempt to invert the
spinor matrix on the right side of Eq.~(\ref{eomNucleonP}). Requiring that
the equation for the dependent components be both a kinematic equation
and simple equation forces us to choose $\psi_+'$ as the independent
components. The dependent components, $\psi_-'$, are defined by
\begin{eqnarray}
\psi_-' &=& \frac{1}{p^+} \left[ \bbox{\alpha}_\perp \cdot
\bbox{p}_\perp + \beta (M + g_\sigma \sigma ) \right] \psi_+'.
\end{eqnarray}
Notice that the dependent components consist of a non-interacting part
and an interacting part. We separate these parts by defining $\psi$
(without a prime) to be the free nucleon field, and $\xi$ to be the part
of $\psi'$ that is due to interactions. So
\begin{eqnarray}
\psi' &=& \psi + \xi \label{eq:psippsi},
\end{eqnarray}
where
\begin{eqnarray}
\xi = \xi_- 
&=& \frac{1}{p^+} \beta ( g_\sigma \sigma ) \psi_+
 = \frac{\gamma^+}{2p^+} ( g_\sigma \sigma ) \psi_+.
\end{eqnarray}
This allows us to write
\begin{eqnarray}
\psi'
&=& \psi + \frac{\gamma^+}{2p^+} ( g_\sigma \sigma ) \psi_+ \\
&=& \psi + \frac{\gamma^+}{2p^+} ( g_\sigma \sigma ) \psi'
\label{eq:psipSigma}
\end{eqnarray}
The last equation follows is obtained from noting that $(\gamma^+)^2=0$,
which implies that $\gamma^+\psi'=\gamma^+\psi_+$. 

Plugging Eq.~(\ref{eq:psipSigma}) into Eq.~(\ref{eq:pm1Primespecific})
and removing all mesons except the $\sigma$ meson, we obtain
\begin{eqnarray}
{P'}^-_{I,1} &=& P^-_{I,1} + P^-_{I,2}, \\
P^-_{I,1} &=&
\int d^2x_\perp dx^- \overline{\psi}(x) g_\sigma \sigma(x) \psi(x),
\\
P^-_{I,2} &=&
\int d^2x_\perp dx^-
\overline{\psi}(x) \left[
g_\sigma \sigma(x)
\frac{\gamma^+}{2p^+}
g_\sigma \sigma(x)
\right] \psi(x)
\Big] \label{eq:sigmaHamWinst}.
\end{eqnarray}

We interpret the $\frac{\gamma^+}{2p^+}$ factor as a
type of nucleon propagator that joins any two meson interactions
(although having two of these propagators adjacent to each other causes
the interaction to vanish since $(\gamma^+)^2$). Because this propagator
does not allow for an
energy denominator (as it is already between two potentials),
$\frac{\gamma^+}{2p^+}$ is called an {\it instantaneous} propagator.

Thus, when constructing the diagrams for the light-front potentials,
we must also include instantaneous propagators for the nucleons in
addition to the usual propagators. (This is one of the dragons mentioned
earlier.)

\subsection{Elimination of Dependent Vector Meson Components}

Like the nucleons, the vector mesons have a dependent component that 
contains interactions and must be eliminated. This process is complicated
somewhat by the fact that the dependent nucleon components must be eliminated
at the same time. 
The salient points of the combined elimination of the dependent nucleon 
and vector meson components are discussed in detail by Miller
\cite{Miller:1997cr}.

The result is that the vector meson field must be redefined and 
an instantaneous vector meson propagator is generated
in addition to an instantaneous nucleon propagator. However, when the
nucleon-nucleon potential is calculated, the redefinition of the vector
meson field exactly cancels the contribution of the instantaneous
vector meson. The result is that the potentials can formally be
calculated using the original vector meson field.

In this work, we use that result to simplify our derivations of
nucleon-nucleon potentials by formally using the na\"{\i}ive form of the
vector meson field. Thus, we find an interaction Hamiltonian similar
to the one shown in Eq.~(\ref{eq:sigmaHamWinst}),
\begin{eqnarray}
{P'}^-_{I,1} &=& P^-_{I,1} + P^-_{I,2}, \\ 
P^-_{I,1} &=& 
\sum_{\alpha=\pi,\eta,\sigma,\delta,\rho,\omega}
\int d^2x_\perp dx^- \overline{\psi}(x)
g_\alpha \Gamma_\alpha T_\alpha \Phi_\alpha(x)
\psi(x), \label{eq:rawome1} \\
P^-_{I,2} &=&
\sum_{\alpha_1,\alpha_2=\pi,\eta,\sigma,\delta,\rho,\omega}
\int d^2x_\perp dx^-
\overline{\psi}(x)
\left[g_{\alpha_1} \Gamma_{\alpha_1} T_{\alpha_1} \Phi_{\alpha_1}(x)\right]
\frac{\gamma^+}{2p^+}
\nonumber\\&&\qquad\qquad\qquad\qquad\qquad\qquad\qquad
\left[g_{\alpha_2} \Gamma_{\alpha_2} T_{\alpha_2} \Phi_{\alpha_2}(x)\right]
\psi(x) \label{eq:rawtme1}.
\end{eqnarray}
We may continue to interpret $\frac{\gamma^+}{2p^+}$ as an instantaneous
nucleon propagator, since in the derivation in Ref.~\cite{Miller:1997cr}
it is clear that the potential vanishes when there are two adjacent 
instantaneous propagators. In Refs.~\cite{Miller:1997cr,Miller:1999ap},
sign of the coupling of the vector mesons in the equations equivalent to
Eq.~(\ref{eq:rawtme1}) has the wrong sign; the coupling of the mesons in
Eq.~(\ref{eq:rawtme1}) must be the same as in Eq.~(\ref{eq:rawome1}).

Also note that in principle the same prescription has to be applied to
the contact interaction, although to the order of two mesons, this
simply has the effect of removing the primes. We find that
\begin{eqnarray}
P^-_{I,2c} &=&
\sum_{\alpha=\pi,\eta,\sigma,\delta} \frac{s_\alpha}{M}
\int d^2x_\perp dx^- \overline{\psi}(x) 
\left[g_\pi    \Gamma_\pi    T_\pi    \Phi_\pi   (x) \right]
\left[g_\alpha \Gamma_\alpha T_\alpha \Phi_\alpha(x) \right]
\psi(x) \label{eq:mediumtme2}.
\end{eqnarray}
Note that $P^-_{I,2c}$ and $P^-_{I,2}$ have forms that are very
similar. In fact, we can obtain $P^-_{I,2c}$ from $P^-_{I,2}$ by making
the following changes:
\begin{enumerate}
\item Replace $\frac{\gamma^+}{2p^+}$ with $\frac{s_\alpha}{M}$.
\item Replace $\alpha_1$ with $\pi$.
\item Restrict the sum on $\alpha_2$ to $\pi,\eta,\sigma,\delta$.
\end{enumerate}

\subsection{Interaction Hamiltonians in Momentum Space}

We look at the matrix element of the
interaction Hamiltonians between initial and final states given by
$|k_i,\lambda_i,\tau_i\rangle=b^\dagger(k_i,\lambda_i)\chi_{\tau_i}|0\rangle$
and 
$\langle k_f,\lambda_f,\tau_f|=\langle0|b(k_f,\lambda_f)\chi^\dagger_{\tau_f}$
where both the bra and the ket have units of $[M^{-3/2}]$.
We find that
\begin{eqnarray}
P^-_{I,1}(f,i)
&=&
\sum_{\alpha=\pi,\eta,\sigma,\delta,\rho,\omega}
\frac{g_\alpha 2M}{2(2\pi)^3 \sqrt{k_f^+ k_i^+}} 
\int \frac{d^2q_\perp dq^+ \theta(q^+)}{ \sqrt{2(2\pi)^3} \sqrt{q^+}}
\nonumber\\&&
\overline{u}(\bbox{k}_f,\lambda_f) \Gamma_\alpha u(\bbox{k}_i,\lambda_i)
\chi^\dagger_{\tau_f} T_\alpha \chi_{\tau_i}
\nonumber\\&&
\int d^2x_\perp dx^- 
e^{+i k_f^\mu x_\mu}e^{-i k_i^\mu x_\mu} \left[
a_\alpha        (\bbox{q}) e^{-i q^\mu x_\mu} +
a_\alpha^\dagger(\bbox{q}) e^{+i q^\mu x_\mu} \right] F_\alpha(\bbox{q}),
\end{eqnarray}
where $\bbox{q}=(q^+,\bbox{q}_\perp)$
and there are implicit
$\theta$-functions on $q^+$ for each particle to ensure that the
light-front energy is positive. These are occasionally suppressed to
simplify the equations. We include a meson-nucleon form factor
$F_\alpha$ to phenomenologically account for the fact that the mesons
and nucleons are composite objects. 

Now, we evaluate $P^-$ at $x^+=0$ and use
\begin{eqnarray}
\int d^2x_\perp dx^- e^{i (k_f-k_i-q)^\mu x_\mu}
&=& 2(2\pi)^3 \delta^{(2,+)}(k_f-k_i-q),
\end{eqnarray}
to write
\begin{eqnarray}
P^-_{I,1}(f,i)
&=& 
\sum_{\alpha=\pi,\eta,\sigma,\delta,\rho,\omega}
g_\alpha 2M \frac{
\overline{u}(\bbox{k}_f,\lambda_f) \Gamma_\alpha u(\bbox{k}_i,\lambda_i)
}{\sqrt{2(2\pi)^3} \sqrt{k_f^+ k_i^+ q^+}} 
\chi^\dagger_{\tau_f} T_\alpha \chi_{\tau_i}
\nonumber\\&&
\left[
a_\alpha        (\bbox{q}) \theta(  k_f^+-k_i^+ ) +
a_\alpha^\dagger(\bbox{q}) \theta(-(k_f^+-k_i^+)) \right] F_\alpha(\bbox{q}).
\end{eqnarray}
We define
$\bbox{q}\equiv\mbox{sign}(k_f^+-k_i^+)(\bbox{k}_f-\bbox{k}_i)$, which
ensures that the $q^+$ of the meson is positive.

Next we consider the two-meson interactions. First take the interaction
with the instantaneous propagator given by Eq.~(\ref{eq:rawtme1}). Plugging
in the field definitions gives
\begin{eqnarray}
P^-_{I,2}(f,i)
&=& 
\sum_{\alpha_1,\alpha_2=\pi,\eta,\sigma,\delta,\rho,\omega}
\frac{g_{\alpha_1} g_{\alpha_2}  2M}{2(2\pi)^3 \sqrt{k_f^+ k_i^+}} 
\int \frac{d^2q_{1\perp} dq_1^+ \theta(q_1^+)}{(2\pi)^{3/2} \sqrt{2 q_1^+}}
\int \frac{d^2q_{2\perp} dq_2^+ \theta(q_2^+)}{(2\pi)^{3/2} \sqrt{2 q_2^+}}
\nonumber\\&&
\frac{\theta(k_m^+)}{2k_m^+}
\overline{u}(\bbox{k}_f,\lambda_f) 
\Gamma_{\alpha_2}
\gamma^+
\Gamma_{\alpha_1}
          u (\bbox{k}_i,\lambda_i)
\chi^\dagger_{\tau_f} T_{\alpha_2} T_{\alpha_1} \chi_{\tau_i}
\nonumber\\&&
\int d^2x_\perp dx^- 
e^{+i k_f^\mu x_\mu}e^{-i k_i^\mu x_\mu} \left[
a_{\alpha_2}        (\bbox{q}_2) e^{-i q_2^\mu x_\mu} +
a_{\alpha_2}^\dagger(\bbox{q}_2) e^{+i q_2^\mu x_\mu} \right]
F_{\alpha_2} (\bbox{q}_2)
\nonumber\\&&
\left[
a_{\alpha_1}        (\bbox{q}_1) e^{-i q_1^\mu x_\mu} +
a_{\alpha_1}^\dagger(\bbox{q}_1) e^{+i q_1^\mu x_\mu} \right]
F_{\alpha_1} (\bbox{q}_1)\\
&=& 
\sum_{\alpha_1,\alpha_2=\pi,\eta,\sigma,\delta,\rho,\omega}
g_{\alpha_1} g_{\alpha_2}  2M 
\frac{\overline{u}(\bbox{k}_f,\lambda_f) \Gamma_{\alpha_2}
\gamma^+
\Gamma_{\alpha_1}
u (\bbox{k}_i,\lambda_i)}{2(2\pi)^3 \sqrt{k_f^+ k_i^+}}
\nonumber\\&&
\left[ \chi^\dagger_{\tau_f} T_{\alpha_2} T_{\alpha_1} \chi_{\tau_i} \right]
\int \frac{d^2k_{m\perp} dk_m^+ 
\theta(k_m^+)}{2k_m^+\sqrt{q_1^+ q_2^+}}
\nonumber\\&& \left[
a_{\alpha_2}        (k_f-k_m) \theta(k_f^+-k_m^+) +
a_{\alpha_2}^\dagger(k_m-k_f) \theta(k_m^+-k_f^+) \right] 
F_{\alpha_2} (\bbox{q}_2)
\nonumber\\&&\left[
a_{\alpha_1}        (k_m-k_i) \theta(k_m^+-k_i^+) +
a_{\alpha_1}^\dagger(k_i-k_m) \theta(k_i^+-k_m^+) \right] 
F_{\alpha_1} (\bbox{q}_1) \label{eq:instantham}
\end{eqnarray}
Note that the momenta $q_1$ and $q_2$ are implicitly functions of the
momenta $k_f$, $k_m$, and $k_i$.

We could also write out the contact interaction given by
Eq.~(\ref{eq:mediumtme2}) in momentum space, but it is related to
Eq.~(\ref{eq:instantham}) by replacing $\frac{\theta(p^+)\gamma^+}{2p^+}$
with $\frac{s_\alpha}{M}$ and restricting the allowed values of the
$\alpha$'s.

\section{Feynman Rules for Nucleon-Nucleon Potentials}

Now that we have the one-meson- and two-meson-exchange expressed in
momentum space, we are now ready to write out the Feynman rules for
diagrams in our model. For simplicity, the only diagrams considered are
those where a meson emitted by one nucleon is absorbed by the other
nucleon. Since we are only interested in the two-nucleon to two-nucleon
potentials, we follow the same approach as outlined in
section~\ref{wc:uxsec} to derive the rules. We denote a ``normal''
nucleon propagator by a solid line with an arrow, an instantaneous
nucleon propagator by a solid line with an stroke across it, mesons
of all type by a dashed line, and energy denominator terms by a
vertical, light, dotted line.
\begin{enumerate}
\item Overall factor of 
$\frac{4M^2\delta^{\perp,+}(P_f-P_i)}{2(2\pi)^3\sqrt{k^+_{1f}k^+_{2f}k^+_{1i}k^+_{2i}}}$.
\item Usual light-front rules for $p_\perp$ and $p^+$ momentum
conservation.
\item Factor of $\frac{\theta(q_i)}{q_i}$ for each internal line,
including any instantaneous nucleon lines.
\item
A factor of $\frac{1}{P^--\sum_iq_i^-}$ for each energy denominator.
\item Each meson connects the two nucleons, and each end of the meson
line has a factor of $g_\alpha\Gamma_\alpha T_\alpha F_\alpha(q)$. The
indices of the isospin factors on each end of the meson are contracted
together. The Lorentz indices of the gamma matrices are contracted with
$-g^{\mu\nu}$ for the vector mesons.
\item For each contact vertex, multiply by a factor of $\frac{1}{M}$. If
the vertex is a $\pi-\pi$ vertex, symmetrize the $T_\pi
T_\pi=\tau_i\tau_j$ by replacing it with $\delta_{i,j}$.
\item Factor of
$\frac{\NEG{k}+M}{2M}=\sum_{\lambda}u(k,\lambda)\overline{u}(k,\lambda)$
for each propagating nucleon and $\frac{\gamma^+}{2}$ for an
instantaneous nucleon.
\item Integrate with $4M^2 \int \frac{d^2k_\perp dk^+}{2(2\pi)^3}$ over
any internal momentum loops.
\item Put the spinor factors for nucleon 1 and 2 between
$\overline{u}u$'s and the isospin factors between the initial and final
state isospin.
\end{enumerate}

From this list, it is useful to summarize what needs to be done to
convert a graph with an instantaneous nucleon to one with a contact
interaction: 
\begin{enumerate}
\item Replace $\frac{\theta(k^+)\gamma^+}{k^+}$ with $\frac{1}{M}$.
\item If both mesons are pions, replace $T_i T_j$ with $\delta_{i,j}$.
\end{enumerate}

These rules make it easy to write down what various potentials are.

\section{Nucleon-Nucleon Perturbative Potentials}

The meson exchange potentials have the same basic form as in
Section~\ref{wc:uxsec}, however we must include the contact interaction
and instantaneous nucleon propagators for the nuclear model used
here. First, we discuss how to include the contact diagrams from the
standpoint of the Bethe-Salpeter equation to maintain chiral symmetry,
then we begin to calculate the light-front potentials.

\subsection{The Bethe-Salpeter Equation} \label{nt:bseequiv}

\begin{figure}
\begin{center}
\epsfig{angle=0,width=5.0in,height=0.8in,file=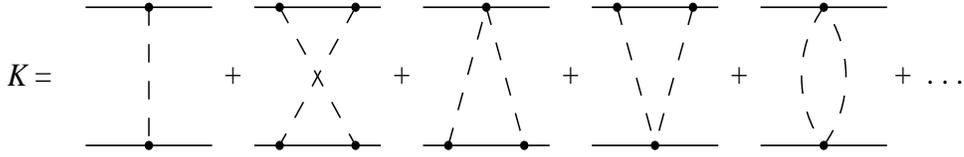}
\caption{The first several terms of the full kernel for the
Bethe-Salpeter equation of the nuclear model with chiral symmetry.
\label{fig:nt.fullbseker}}
\end{center}
\end{figure}

The kernel of the Bethe-Salpeter equation
\cite{Nambu:1950vt,Schwinger:1951ex,%
Schwinger:1951hq,Gell-Mann:1951rw,Salpeter:1951sz} for this nuclear
model is richer than the one presented in section~\ref{wc:bseequiv} for
the Wick-Cutkosky model. This is due mainly to the presence of the
contact interactions which are generated by the chirally invariant
coupling of the pion to the nucleon. Several of the lowest-order pieces
of the full kernel $K$ are shown in Fig.~\ref{fig:nt.fullbseker}.
(Note that for Feynman diagrams, it is useful to combine the ``normal''
nucleon propagator with the instantaneous nucleon propagator
\cite{Schoonderwoerd:1998qj,Schoonderwoerd:1998iy,Schoonderwoerd:1998jm},
and denote the combination with a solid line.)

As discussed in section~\ref{wc:bseequiv}, each of these Feynman
diagrams is covariant. This means that we may choose any of the diagrams
from $K$ to construct a new kernel $K'$, and the infinite series of
potential diagrams physically equivalent (Sec.~\ref{wc:bseequivcomp}) to
$K'$ will also be covariant. 

In practice, this means that when deciding which two-meson-exchange
potentials to include for calculating the deuteron wave function, we
may neglect the crossed diagrams. Although including only the box and
contact two-meson-exchange diagrams may affect the exact binding energy
calculated, we should find a partial restoration of rotational
invariance. We reiterate that the focus of this work is on understanding
the effects of the breaking of rotational invariance and how to restore
it; our goal is not precise agreement with experimental results.

\subsubsection{Chiral Symmetry}

We also want to keep the potentials chirally symmetric as well. Whereas
Lorentz symmetry is maintained by using a kernel with any Feynman
diagrams (with potentially arbitrary coefficients), chiral symmetry
relates the strength of the $\pi\pi$ contact interaction to the strength
of the pion-nucleon coupling. 

\begin{figure}
\begin{center}
\epsfig{angle=0,width=3.0in,height=0.75in,file=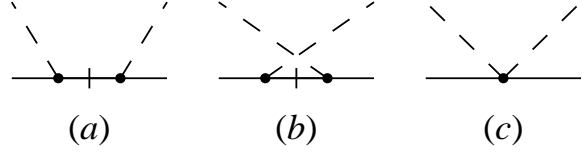}
\caption{The non-vanishing diagrams for pion-nucleon scattering at
threshold: (a) ${\mathcal M}_U$, (b) ${\mathcal M}_X$, and (c)
${\mathcal M}_C$. The mesons here are pions.
\label{fig:nt.chiralcan}}
\end{center}
\end{figure}

Chiral symmetry tells us that for pion-nucleon scattering at threshold,
the time-ordered graphs approximately cancel
\cite{Miller:1997cr}. Furthermore, upon closer examination, we find that
all the light-front time-ordered graphs for the scattering amplitude
vanish except for the two graphs with instantaneous nucleons and the
contact graph. These graphs are shown in
Fig.~\ref{fig:nt.chiralcan}. Using the Feynman rules, and denoting the
nucleon momentum by $k$ and the pion momentum by $q$, we find that
\begin{eqnarray}
{\mathcal M}_U &=& C \frac{\tau_i\tau_j}{2(k^++q^+)}u(k')\gamma^+u(k), \\
{\mathcal M}_X &=& C \frac{\tau_j\tau_i}{2(k^+-q^+)}u(k')\gamma^+u(k), \\
{\mathcal M}_C &=& C \frac{-\delta_{i,j}}{M}        u(k')        u(k),
\end{eqnarray}
where the factors common to all amplitudes are denoted by $C$.

For threshold scattering, we take $k=k'=M$ and $q=q'=m_\pi$. In that
limit, we find
\begin{eqnarray}
{\mathcal M}_U &=& C'\frac{ \delta_{i,j}+i\epsilon_{i,j,k}\tau_k}{2(M+m_\pi)},\\
{\mathcal M}_X &=& C'\frac{ \delta_{i,j}-i\epsilon_{i,j,k}\tau_k}{2(M-m_\pi)},\\
{\mathcal M}_C &=& C'\frac{-\delta_{i,j}}{M},     
\end{eqnarray}
where $C'=C \overline{u}(k')u(k)$. In the limit that $m_\pi\rightarrow
0$, the sum of these three terms vanishes. The term in these equations
proportional to $\tau_k$ is the famous Weinberg-Tomazowa term
\cite{Brandsden:1973,Adler:1968}.

The fact that the amplitudes cancel only when the contact interaction is
included demonstrates that chiral symmetry can have a significant effect
on calculations. In terms of two-pion-exchange potentials, this result
means that the contact potentials cancel strongly with both the iterated
box potentials and the crossed potentials. This serves to reduce the
strength of the total two-pion-exchange potential, which should lead to
more stable results.

However, since we do not use the crossed graphs for the nucleon-nucleon
potential, we must come up with a prescription which divides the contact
interactions into two parts which cancel the box and crossed diagrams
separately. We do this by formally defining two new contact
interactions, so that 
\begin{eqnarray}
{\mathcal M}_{C_U} &\equiv& \frac{M}{2(M+m_\pi)} {\mathcal M}_C, \\
{\mathcal M}_{C_X} &\equiv& {\mathcal M}_C - {\mathcal M}_{C_U}.
\end{eqnarray}
With these definitions, we find that at threshold and in the chiral
limit,
\begin{eqnarray}
{\mathcal M}_U + {\mathcal M}_{C_U} &=& 0, \\
{\mathcal M}_X + {\mathcal M}_{C_X} &=& 0.
\end{eqnarray}
This indicates that we can incorporate approximate chiral symmetry
without including crossed graphs simply by weighting each graph with a
contact interaction by a factor of $\frac{M}{2(M+m_\pi)}$.

\subsection{OME Potential}

\begin{figure}
\begin{center}
\epsfig{angle=0,width=5.0in,height=1.1in,file=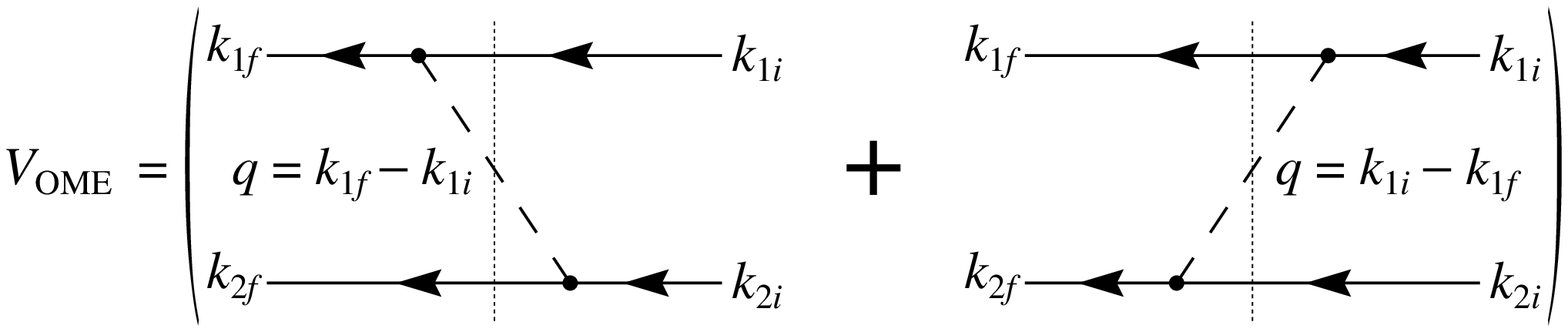}
\caption{The two diagrams which contribute to the OME potential for each
meson.
\label{fig:obepot}}
\end{center}
\end{figure}

The one meson exchange (OME) potential connects an initial state with
two nucleons to a final state with two nucleons, has one meson in
the intermediate state, and has the meson emitted and absorbed by
different nucleons. With these restrictions, along with the fact that in
light-front dynamics the interaction does not allow for particles to be
created from the vacuum, we find that there each meson has only two
diagrams for the OME potential. These diagrams are shown in
Fig.~\ref{fig:obepot}.

Note that chiral symmetry does not impose restrictions when considering
the OME potential alone. It does affect the TME potentials considered,
as discussed in the following section.

The Feynman rules derived in the previous section are used to derive the
potential for these diagrams. We factor out an overall factor of 
$\frac{4M^2\delta(k_{1f}+k_{2f}-k_{1i}-k_{2i})}{2\sqrt{k_{1f}^+k_{1i}^+k_{2f}^+k_{2i}^+}}$
that is common to all the two-nucleon potentials and suppress it from
now on.  Then we get
\begin{eqnarray}
V_{\text{OME},\alpha}
&=& 
\frac{g_\alpha^2 \bbox{T}_{\alpha,1}\cdot\bbox{T}_{\alpha,2}}{(2\pi)^3}
\overline{u}(\bbox{k}_{1f},\lambda_{1f})\Gamma_\alpha u(\bbox{k}_{1i},\lambda_{1i})
\overline{u}(\bbox{k}_{2f},\lambda_{2f})\Gamma_\alpha u(\bbox{k}_{2i},\lambda_{2i})
F_\alpha^2(\bbox{q})
\nonumber\\&&
\left[
\frac{\theta(k_{1f}^+-k_{1i}^+)}{
(k_{1f}^+-k_{1i}^+)(P^--k_{1i}^--k_{2f}^-)
 -m_\alpha^2-(k_{1i,\perp}-k_{1f,\perp})^2} \right.\nonumber\\&&\left.
+
\frac{\theta(k_{1i}^+-k_{1f}^+)}{
(k_{1i}^+-k_{1f}^+)(P^--k_{1f}^--k_{2i}^-)
 -m_\alpha^2-(k_{1i,\perp}-k_{1f,\perp})^2}
\right] \label{nuc:eq:obepotraw}.
\end{eqnarray}

To simplify the potential, we define
\begin{eqnarray}
a &\equiv&
\left[ \theta(k_{1f}^+-k_{1i}^+)\times
(k_{1f}^+-k_{1i}^+)(P^--k_{1i}^--k_{2f}^-) \right.+
\nonumber\\&&
\left.
\theta(k_{1i}^+-k_{1f}^+)\times(k_{1i}^+-k_{1f}^+)(P^--k_{1f}^--k_{2i}^-)
\right]
-k_{i,\perp}^2-k_{f,\perp}^2, \\
b &\equiv& 2 k_{i,\perp}k_{f,\perp},
\end{eqnarray}
so
\begin{eqnarray}
V_{\text{OME},\alpha} 
&=& \frac{g_\alpha^2 \bbox{T}_{\alpha,1}\cdot\bbox{T}_{\alpha,2}}{(2\pi)^3}
F_\alpha(\bbox{q})^2 \frac{
\overline{u}(\bbox{k}_{1f},\lambda_{1f})\Gamma_\alpha u(\bbox{k}_{1i},\lambda_{1i})
\overline{u}(\bbox{k}_{2f},\lambda_{2f})\Gamma_\alpha u(\bbox{k}_{2i},\lambda_{2i})
}{\left[a -       m_\alpha^2 + b \cos(\phi_f-\phi_i)\right]}.
\end{eqnarray}

Now we consider the precise form to use for the meson-nucleon form
factor. We assume that it has a $n$-pole form \cite{Machleidt:1987hj},
so that the denominator of the meson-nucleon form factor has the
same form as the denominator of the potential in the scattering
regime. In particular, $\Lambda_\alpha$ playing the role of $m_\alpha$
for the form factor. For simplicity, we choose that the
denominator of the form factor has the same form of the denominator of
the potential. Thus,
\begin{eqnarray}
F_\alpha(\bbox{q}) &=& \left(\frac{\Lambda_\alpha^2 - m_\alpha^2}
{a - \Lambda_\pi^2 + b \cos(\phi_f-\phi_i)} \right)^{n_\alpha}.
\label{eq:formfacexp}
\end{eqnarray}

Inserting the explicit expression of for the meson-nucleon form factor
into the potential results in
\begin{eqnarray}
V_{\text{OME},\alpha}
&=&
\frac{g_\alpha^2 \bbox{T}_{\alpha,1}\cdot\bbox{T}_{\alpha,2}}{(2\pi)^3}
(\Lambda_\alpha^2-m_\alpha^2)^{2n_\alpha} \nonumber\\&&
\frac{
\overline{u}(\bbox{k}_{1f},\lambda_{1f})\Gamma_\alpha u(\bbox{k}_{1i},\lambda_{1i})
\overline{u}(\bbox{k}_{2f},\lambda_{2f})\Gamma_\alpha u(\bbox{k}_{2i},\lambda_{2i})
}{
\left[a -       m_\alpha^2 + b \cos(\phi_f-\phi_i)\right]
\left[a - \Lambda_\alpha^2 + b \cos(\phi_f-\phi_i)\right]^{2n_\alpha}
}.
\end{eqnarray}

Now, since the light-front potentials have exact rotational invariance
about the $z$-axis, they conserve the $J_z$ quantum number $m$. Thus,
these potentials connect only states with the same value of $m$. This
means that, in general, light-front potentials may be written as
\begin{eqnarray}
V(\phi_f,\phi_i) &=& \sum_{m=-\infty}^\infty e^{i m (\phi_i-\phi_f)} V^m,
\end{eqnarray}
where $V^m$ is the potential in the magnetic quantum number basis. This
relation can be inverted to obtain $V^m$ in terms of $V(\phi_f,\phi_i)$,
\begin{eqnarray}
V^m_{\text{OME},\alpha} &=&
\frac{g_\pi^2 \bbox{T}_{\alpha,1}\cdot\bbox{T}_{\alpha,2}}{(2\pi)^3}
(\Lambda_\alpha^2-m_\alpha^2)^{2n_\alpha}
\nonumber\\&&
\int \frac{d\phi_f}{2\pi} e^{i m \phi_f}
\frac{
\left[
\overline{u}(\bbox{k}_{1f},\lambda_{1f})\Gamma_\alpha u(\bbox{k}_{1i},\lambda_{1i})
\overline{u}(\bbox{k}_{2f},\lambda_{2f})\Gamma_\alpha u(\bbox{k}_{2i},\lambda_{2i})
\right]_{\phi_i=0}
}{
\left[a -       m_\alpha^2 + b \cos(\phi_f)\right]
\left[a - \Lambda_\alpha^2 + b \cos(\phi_f)\right]^{2n_\alpha}
}.
\end{eqnarray}

To perform the integration, we need an expression for $\phi_f$
dependence the $\overline{u}u$ matrix elements.  These are calculated in
Appendix~\ref{app:spinors}. Summarizing, we can write
\begin{eqnarray}
\overline{u}(\bbox{k}_{1f},\lambda_{1f})\Gamma_\alpha u(\bbox{k}_{1i},\lambda_{1i})
\overline{u}(\bbox{k}_{2f},\lambda_{2f})\Gamma_\alpha u(\bbox{k}_{2i},\lambda_{2i})
&=& \sum_j C_j(\Gamma_\alpha,\Gamma_\alpha) e^{ij\phi_f},
\end{eqnarray}
where the $C$ depends implicitly on all the variables on the left-hand
side except $\phi_f$ and $\phi_i$.

Thus,
\begin{eqnarray}
V^m_{\text{OME},\alpha} &=&
g_\alpha^2 (\bbox{T}_{\alpha,1}\cdot\bbox{T}_{\alpha,2})
\frac{(\Lambda_\alpha^2-m_\alpha^2)^{2n_\alpha}}{(2\pi)^3}
\sum_j C_j(\Gamma_\alpha,\Gamma_\alpha) \nonumber\\&&
\int \frac{d\phi_f}{2\pi}
\frac{e^{i(m+j)\phi_f}}{
\left[a -       m_\alpha^2 + b \cos(\phi_f)\right]
\left[a - \Lambda_\alpha^2 + b \cos(\phi_f)\right]^{2n_\alpha}}.
\end{eqnarray}
The cosine integral is denoted by $I$, and is calculated in
Appendix~\ref{theintegral}. Substituting $I$ for the integral, we obtain
\begin{eqnarray}
V^m_{\text{OME},\alpha} &=& g_\alpha^2
(\bbox{T}_{\alpha,1}\cdot\bbox{T}_{\alpha,2})
\frac{(\Lambda_\alpha^2-m_\alpha^2)^{2n_\alpha}}{(2\pi)^3} \nonumber\\&&
\sum_j C_j(\Gamma_\alpha,\Gamma_\alpha)
I(m+j,a-m_\alpha^2,1,a-\Lambda_\alpha^2,2n_\alpha,b) \label{eq:omefinished}.
\end{eqnarray}
All of the terms in Eq.~(\ref{eq:omefinished}) are known, and the
potential can now be calculated numerically.

\subsection{TME Potentials}

\begin{figure}[!htp]
\begin{center}
\epsfig{angle=0,width=6in,height=7.0in,file=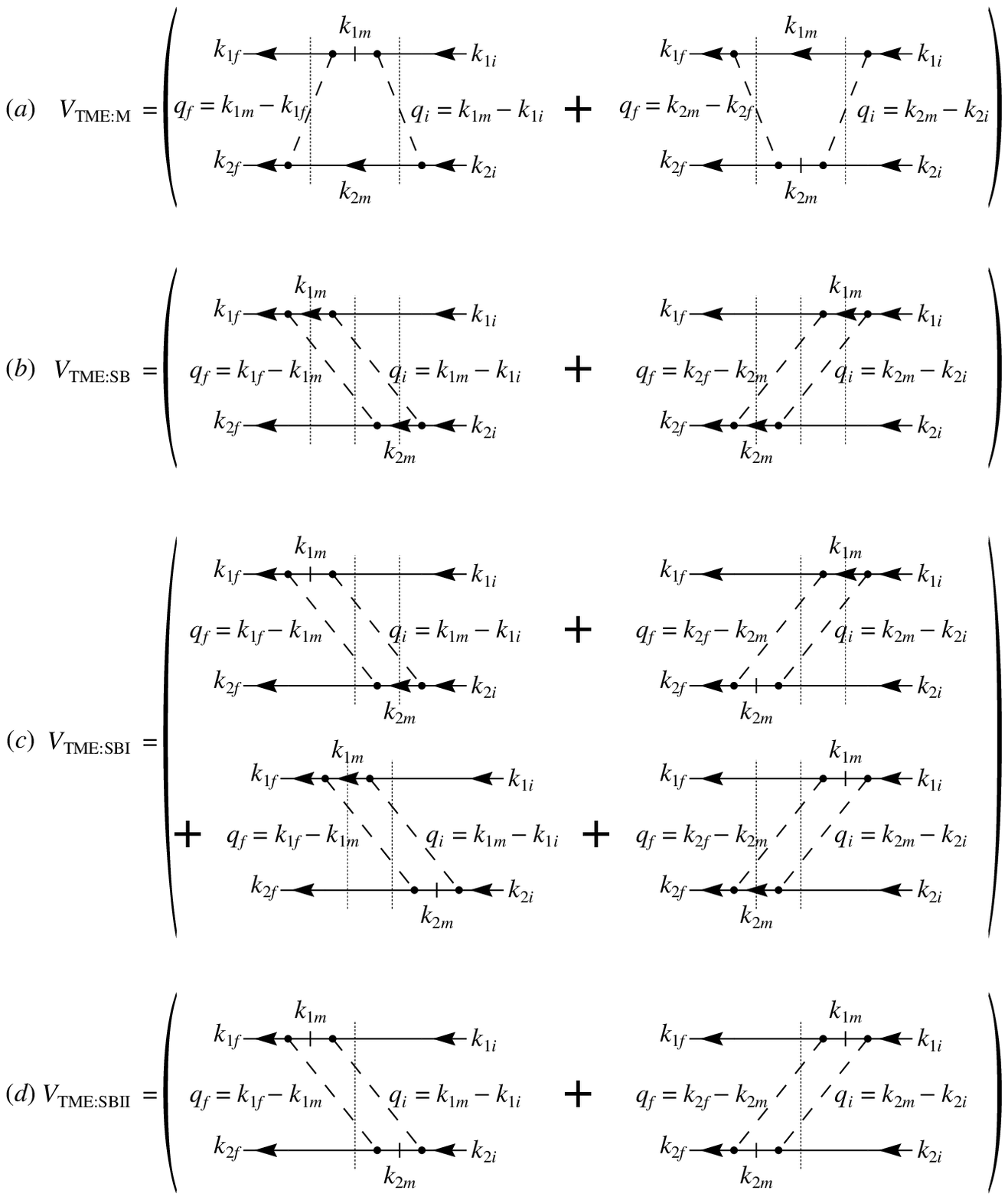}
\caption[The non-contact TME potentials:
the Mesa potential, 
the stretched box potential, 
the stretched instantaneous potential, and
the stretched double instantaneous potential.]
{The TME potentials for
(a) $V_{\text{TME:M}}$ (the Mesa potential), 
(b) $V_{\text{TME:SB}}$ (the stretched box potential), 
(c) $V_{\text{TME:SBI}}$ (the stretched instantaneous potential), and
(d) $V_{\text{TME:SBII}}$ (the stretched double instantaneous potential).
Note that the graphs on the right side are obtained from the graphs on the
left side by relabeling $1\leftrightarrow 2$.
\label{fig:tbepot}}
\end{center}
\end{figure}

\begin{figure}[!htp]
\begin{center}
\epsfig{angle=0,width=5in,height=6.75in,file=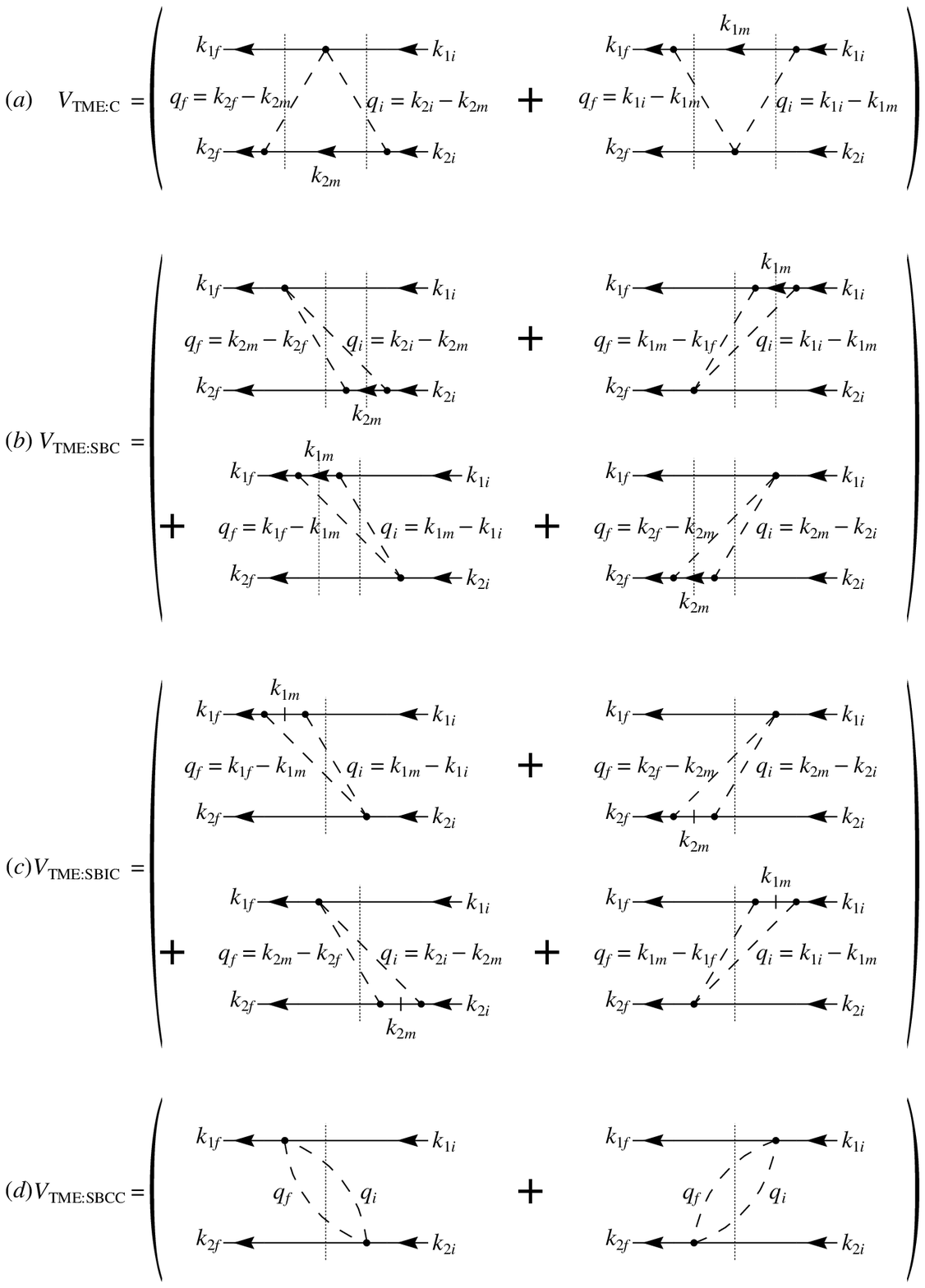}
\caption[The contact TME potentials:
the contact potential, 
the stretched contact potential, 
the stretched instantaneous contact potential, and
the stretched double contact potential.]
{The TME potentials that include the contact interaction for
(a) $V_{\text{TME:C}}$ (the contact potential), 
(b) $V_{\text{TME:SBC}}$ (the stretched contact potential), 
(c) $V_{\text{TME:SBIC}}$ (the stretched instantaneous contact potential), and
(d) $V_{\text{TME:SBCC}}$ (the stretched double contact potential).
Note that the graphs on the right side are obtained from the graphs on the
left side by relabeling $1\leftrightarrow 2$.
\label{fig:tbecontpot}}
\end{center}
\end{figure}

The two-meson-exchange (TME) potentials consider here are the box
diagrams (see Fig.~\ref{fig:tbepot}) and the contact diagrams (see
Fig.~\ref{fig:tbecontpot}). We do not consider the crossed diagrams
because they are not needed to restore rotational invariance, as shown
in Section~\ref{nt:bseequiv}. We derive a few TME potentials to
illustrate the general form that they take.

\subsubsection{Stretched Box Potential}

We use the Feynman rules to write the stretched-box potential shown in
Fig.~\ref{fig:tbepot}(b),
\begin{eqnarray}
V^{\alpha_f,\alpha_i}_{\text{TME:SB}}
&=& \frac{g^2_{\alpha_f} g^2_{\alpha_i} 4M^2
[\bbox{T}_{\alpha_f,1}\cdot\bbox{T}_{\alpha_f,2}]
[\bbox{T}_{\alpha_i,1}\cdot\bbox{T}_{\alpha_i,2}]}{(2\pi)^3}
\int \frac{d^2k_{2m\perp} dk_{2m}^+}{2(2\pi)^3 k_{1m}^+k_{2m}^+}
\nonumber\\&&
{}_{1f}\langle\left( \Gamma_{\alpha_f} \frac{\NEG{k}_{1m}+M}{2M}
\Gamma_{\alpha_i} \right)\rangle_{1m} \times
{}_{2f}\langle\left( \Gamma_{\alpha_f} \frac{\NEG{k}_{2m}+M}{2M}
\Gamma_{\alpha_i} \right)\rangle_{2m}
\nonumber\\&&
\theta(k_{1m}^+)\theta(k_{2m}^+)
\theta(k_{2i}^+-k_{2m}^+)\theta(k_{2m}^+-k_{2f}^+)
\nonumber\\&&
\frac{F_{\alpha_f}(\bbox{q}_f)^2}{
(k_{2m}^+-k_{2f}^+)(P^--k_{1m}^--k_{2f}^-) - m_{\alpha_f}^2
- (\bbox{k}_{1f\perp}-\bbox{k}_{1m\perp})^2}
\nonumber\\&&
\frac{1}{P^- - k_{1i}^- - k_{2f}^- - q_f^- - q_i^-}
\nonumber\\&&
\frac{F_{\alpha_i}(\bbox{q}_i)^2}{
(k_{2i}^+-k_{2m}^+)(P^--k_{1i}^--k_{2m}^-) - m_{\alpha_i}^2
- (\bbox{k}_{1m\perp}-\bbox{k}_{1i\perp})^2 }
\nonumber\\&&
+ \{ 1 \leftrightarrow 2 \}
\label{eqn:strechedfermionbox}.
\end{eqnarray}
To compress notation, we defined
${}_{1f}\langle\equiv\overline{u}(\bbox{k}_{1f},\lambda_{1f})$,
$\rangle_{1i}\equiv{}u(\bbox{k}_{1i},\lambda_{1i})$, and so on.
The symbol $\{ 1\leftrightarrow 2\}$ means that all labels 1 are
replaced with 2 and vice versa. This is a way of explicitly stating the
potential is invariant under exchange of nucleons 1 and 2.

We also used the following relation to simplify
Eq.~(\ref{eqn:strechedfermionbox}): 
\begin{eqnarray}
&&\sum_{\tau_{2m}}
\chi^\dagger_{\tau_{2f}} T_{\alpha_f,j} \chi_{\tau_{2m}}
\chi^\dagger_{\tau_{2m}} T_{\alpha_i,i} \chi_{\tau_{2i}}
\chi^\dagger_{\tau_{1f}} T_{\alpha_f,j} \chi_{\tau_{1m}} 
\chi^\dagger_{\tau_{1m}} T_{\alpha_i,i} \chi_{\tau_{1i}} 
\nonumber\\
&=&
\langle \tau_f | 
[\bbox{T}_{\alpha_f,1}\cdot\bbox{T}_{\alpha_f,2}]
[\bbox{T}_{\alpha_i,1}\cdot\bbox{T}_{\alpha_i,2}]
| \tau_i \rangle \nonumber \\
&=&
[\bbox{T}_{\alpha_f,1}\cdot\bbox{T}_{\alpha_f,2}]
[\bbox{T}_{\alpha_i,1}\cdot\bbox{T}_{\alpha_i,2}].
\end{eqnarray}

\subsubsection{Mesa Potential}

We consider the Mesa potential next, $V_{\text{TME:M}}$, shown in
Fig.~\ref{fig:tbepot}(a). This is one of the potentials which is due to
the instantaneous nucleon, a feature not found in the Wick-Cutkosky
model. Using the Feynman rules results in
\begin{eqnarray}
V^{\alpha_f,\alpha_i}_{\text{TME:M}}
&=& \frac{g^2_{\alpha_f} g^2_{\alpha_i} 4M^2
[\bbox{T}_{\alpha_f,1}\cdot\bbox{T}_{\alpha_f,2}]
[\bbox{T}_{\alpha_i,1}\cdot\bbox{T}_{\alpha_i,2}]}{(2\pi)^3}
\int \frac{d^2k_{2m\perp} dk_{2m}^+}{2(2\pi)^3 k_{1m}^+k_{2m}^+}
\nonumber\\&&
{}_{1f}\langle\left( \Gamma_{\alpha_f} \frac{\gamma^+}{2}
\Gamma_{\alpha_i} \right)\rangle_{1m} \times
{}_{2f}\langle\left( \Gamma_{\alpha_f} \frac{\NEG{k}_{2m}+M}{2M}
\Gamma_{\alpha_i} \right)\rangle_{2m}
\nonumber\\&&
\theta(k_{1m}^+)\theta(k_{2m}^+)
\theta(k_{2i}^+-k_{2m}^+)\theta(k_{2m}^+-k_{2f}^+)
\nonumber\\&&
\frac{F_{\alpha_f}(\bbox{q}_f)^2}{
(k_{2f}^+-k_{2m}^+)(P^--k_{1f}^--k_{2m}^-) - m_{\alpha_f}^2
- (\bbox{k}_{1f\perp}-\bbox{k}_{1m\perp})^2}
\frac{1}{2M}
\nonumber\\&&
\frac{F_{\alpha_i}(\bbox{q}_i)^2}{
(k_{2i}^+-k_{2m}^+)(P^--k_{1i}^--k_{2m}^-) - m_{\alpha_i}^2
- (\bbox{k}_{1m\perp}-\bbox{k}_{1i\perp})^2 }
\nonumber\\&&
+ \{ 1 \leftrightarrow 2 \}
\label{eqn:mesapot}.
\end{eqnarray}

\subsubsection{Contact Potential}

The last potential which we calculate explicitly is the contact
potential $V_{\text{TME:C}}$, shown in Fig.~\ref{fig:tbecontpot}(a). The
rules relate this potential to $V_{\text{TME:M}}$ in a simple
fashion. If both the initial and final mesons are not pions, then the
potential is
\begin{eqnarray}
V^{\alpha_f,\alpha_i}_{\text{TME:C}}
&=& \frac{g^2_{\alpha_f} g^2_{\alpha_i} 4M^2
[\bbox{T}_{\alpha_f,1}\cdot\bbox{T}_{\alpha_f,2}]
[\bbox{T}_{\alpha_i,1}\cdot\bbox{T}_{\alpha_i,2}]}{(2\pi)^3}
\int \frac{d^2k_{2m\perp} dk_{2m}^+}{2(2\pi)^3 k_{2m}^+M}
\nonumber\\&&
{}_{1f}\langle\left( \Gamma_{\alpha_f} \frac{1}{2}
\Gamma_{\alpha_i} \right)\rangle_{1m} \times
{}_{2f}\langle\left( \Gamma_{\alpha_f} \frac{\NEG{k}_{2m}+M}{2M}
\Gamma_{\alpha_i} \right)\rangle_{2m}
\nonumber\\&&
\theta(k_{2m}^+)
\theta(k_{2i}^+-k_{2m}^+)\theta(k_{2m}^+-k_{2f}^+)
\nonumber\\&&
\frac{F_{\alpha_f}(\bbox{q}_f)^2}{
(k_{2f}^+-k_{2m}^+)(P^--k_{1f}^--k_{2m}^-) - m_{\alpha_f}^2
- (\bbox{k}_{1f\perp}-\bbox{k}_{1m\perp})^2}
\frac{1}{2M}
\nonumber\\&&
\frac{F_{\alpha_i}(\bbox{q}_i)^2}{
(k_{2i}^+-k_{2m}^+)(P^--k_{1i}^--k_{2m}^-) - m_{\alpha_i}^2
- (\bbox{k}_{1m\perp}-\bbox{k}_{1i\perp})^2 }
\nonumber\\&&
+ \{ 1 \leftrightarrow 2 \}
\label{eqn:contactpot}.
\end{eqnarray}
To get the contact potential for the pions, the following change has to
be made:
\begin{eqnarray}
[\bbox{T}_{\alpha_f,1}\cdot\bbox{T}_{\alpha_f,2}]
[\bbox{T}_{\alpha_i,1}\cdot\bbox{T}_{\alpha_i,2}]
&\rightarrow&
\tau_2^2 = 3.
\end{eqnarray}
This is due the additional symmetry that the two pion vertex has.

Before this potential is used in calculations, it must be multiplied by
a factor of $\frac{M}{2(M+m_\pi)}$ (as discussed in
Section~\ref{nt:bseequiv}) if the crossed diagrams are not included.

\subsubsection{Common Features and Simplification}

Each potential in Figs.~\ref{fig:tbepot} and \ref{fig:tbecontpot} can be
written schematically in the following, general form:
\begin{eqnarray}
V_{\text{TME}}
&=& \frac{g^4 4M^2 T^4(\Lambda^2-m^2)^{4n_\alpha}}{(2\pi)^3}
\int \frac{d^2k_{2m\perp} dk_{2m}^+}{2(2\pi)^3}
f(k_{1m}^+,k_{2m}^+,q_f^+,q_i^+)
\nonumber\\&&
{}_{1f}\langle\left( \Gamma_{\alpha_f} {\mathcal M}_1
\Gamma_{\alpha_i} \right)\rangle_{1m} \times
{}_{2f}\langle\left( \Gamma_{\alpha_f} {\mathcal M}_2
\Gamma_{\alpha_i} \right)\rangle_{2m}
\nonumber\\&&
\left[a_f -       m_{\alpha_f}^2 + b_f \cos(\phi_f-\phi_m) \right]^{-n_f}
\left[a_f - \Lambda_{\alpha_f}^2 + b_f \cos(\phi_f-\phi_m) \right]^{-2n_{\alpha_f}}
\nonumber\\&&
\left[a_m + b_{mf} \cos(\phi_f-\phi_m)
          + b_{mi} \cos(\phi_m-\phi_i) \right]^{-n_m}
\nonumber\\&&
\left[a_i -       m_{\alpha_i}^2 + b_i \cos(\phi_m-\phi_i) \right]^{-n_i}
\left[a_i - \Lambda_{\alpha_i}^2 + b_i \cos(\phi_m-\phi_i) \right]^{-2n_{\alpha_i}},
\end{eqnarray}
where the expression for $F_\alpha$ in Eq.~(\ref{eq:formfacexp}) was used.

The TME potentials, like the OME potential, have exact rotational
invariance about the $z$-axis, and thus conserve $m$.  We project the
potential onto states of definite $m$ by setting $\phi_i$ to zero and
integrating the potential with $\int d\phi_f e^{im\phi_f}/2\pi$.

Now we have to obtain the $\overline{u}u$ matrix elements explicitly to
perform the azimuthal integrations. Appendix~\ref{app:spinors} shows that
we can write
\begin{eqnarray}
&&\left[ 
{}_{1f}\langle\left( \Gamma_{\alpha_f} {\mathcal M}_1
\Gamma_{\alpha_i} \right)\rangle_{1m} \times
{}_{2f}\langle\left( \Gamma_{\alpha_f} {\mathcal M}_2
\Gamma_{\alpha_i} \right)\rangle_{2m}
\right]_{\phi_i=0} \nonumber \\
&&\qquad= \sum_{j,k}
C_{j,k}(
\Gamma_{\alpha_f},{\mathcal M}_1,\Gamma_{\alpha_i};
\Gamma_{\alpha_f},{\mathcal M}_2,\Gamma_{\alpha_i})
e^{i j (\phi_f-\phi_m)}
e^{i k \phi_m}.
\end{eqnarray}
Note that the $C$ is implicitly a function of the momentum and helicity
of the initial and final states, but is independent of the azimuthal
angles $\phi_f$ and $\phi_i$. 

Then, after a change of variables,
$\phi_m\rightarrow\phi'_i$,
$\phi_f-\phi_m\rightarrow\phi'_f$, and
$\phi_f\rightarrow\phi'_f+\phi'_i$, the potential can be written as
\begin{eqnarray}
V_{\text{TME}}
&=& \frac{g^4 4M^2 T^4(\Lambda^2-m^2)^{4n_\alpha}}{(2\pi)^3}
\sum_{j,k}
\int \frac{dk_{2m\perp} \, k_{2m\perp} \, dk_{2m}^+}{2(2\pi)^2}
f(k_{1m}^+,k_{2m}^+,q_f^+,q_i^+)
\nonumber\\&&
C_{j,k}(
\Gamma_{\alpha_f},{\mathcal M}_1,\Gamma_{\alpha_i};
\Gamma_{\alpha_f},{\mathcal M}_2,\Gamma_{\alpha_i})
\nonumber\\&&
\int_0^{2\pi} \frac{d\phi'_f}{2\pi}
\int_0^{2\pi} \frac{d\phi'_i}{2\pi}
e^{i (m+j) \phi'_f}
e^{i (m+k) \phi'_i}
\nonumber\\&&
\left[a_f -       m_{\alpha_f}^2 + b_f    \cos\phi'_f \right]^{-n_f}
\left[a_f - \Lambda_{\alpha_f}^2 + b_f    \cos\phi'_f \right]^{-2n_{\alpha_f}}
\nonumber\\&&
\left[a_m                 + b_{mf} \cos\phi'_f
                          + b_{mi} \cos\phi'_i \right]^{-n_m}
\nonumber\\&&
\left[a_i -       m_{\alpha_i}^2 + b_i    \cos\phi'_i \right]^{-n_i}
\left[a_i - \Lambda_{\alpha_i}^2 + b_i    \cos\phi'_i \right]^{-2n_{\alpha_i}}.
\end{eqnarray}
The azimuthal-angle integrals taken together are denoted as $I$, and the
method for calculating the $\phi'_m$ and $\phi'_f$ integrals is
discussed in Appendix~\ref{theintegral}. Summarizing, we can write 
\begin{eqnarray}
V_{\text{TME}}
&=& \frac{g^4 4M^2 T^4(\Lambda^2-m^2)^{4n_\alpha}}{(2\pi)^3}
\sum_{j,k}
\int \frac{dk_{2m\perp} \, k_{2m\perp} \, dk_{2m}^+}{2(2\pi)^2}
f(k_{1m}^+,k_{2m}^+,q_f^+,q_i^+)
\nonumber\\&&
C_{j,k}(
\Gamma_{\alpha_f},{\mathcal M}_1,\Gamma_{\alpha_i};
\Gamma_{\alpha_f},{\mathcal M}_2,\Gamma_{\alpha_i})
\nonumber\\&&
I(
a_f-m_{\alpha_f}^2, a_f-\Lambda_{\alpha_f}^2, b_f, n_f, 2n_{\alpha_f}, m+j;
\nonumber\\&&\phantom{I( }
a_m, b_{mf}, b_{mi}, n_m;
\nonumber\\&&\phantom{I( }
a_i-m_{\alpha_i}^2, a_i-\Lambda_{\alpha_i}^2, b_i, n_i, 2n_{\alpha_i}, m+k).
\end{eqnarray}
This potential is evaluated using the numerical integration techniques
discussed in Appendix~\ref{ch:integrations}.

\subsection{Further Development of the Light-front Schr\"odinger
Equation} \label{sec:furtherlfnn}

The potentials derived here possess a high degree of symmetry. To solve
the light-front Schr\"odinger equation efficiently, these symmetries
should be explicitly exploited, as was done in
section~\ref{wc:sec:furtherdev}. In addition to the invariance of the
potentials
under parity, there are additional symmetries due to the conservation
of nucleon helicity and invariance under time reversal
\cite{Machleidt:1987hj}. 

We follow Machleidt's approach for taking advantage of the symmetry of
rotationally invariant potentials with helicity
\cite{Machleidt:1989}. However, since the light-front potentials derived
here do not have full rotational invariance, Machleidt's approach must
be modified. The symmetry properties of helicity matrix elements are
rederived in Appendix~\ref{app:rotmat} without assuming full rotational
invariance of the potentials. These results allow a modified version of
Machleidt's
approach to be combined with the exploitation of parity (using the
transformation from light-front coordinates to equal-time coordinates)
discussed in section~\ref{wc:sec:furtherdev}. 
In particular, the potentials are initially calculated in the
$|p_{\text{ET}},\theta,M,\lambda_1,\lambda_2\rangle$ basis, although the
relations in Appendix~\ref{app:rot:summ} are used to transform to the 
$|p_{\text{ET}},J,M,L,S\rangle$ basis, which is useful to solve for the
wave functions. Note that the potentials connect states with different
$J$ values in general.
Once the symmetries have been explicitly expressed, we can discretize
the Schr\"odinger equation as done in Appendix~\ref{ch:numericaleq}.

Note that we apply the transformations to the potential in order to
simplify the calculation of the wave functions. Once the wave functions
are
obtained, we may apply the inverse of the transformations to the wave
functions to express them in the helicity basis
($|\bbox{p}_\perp,p^+,\lambda_1,\lambda_2\rangle$ ) 
or, by using Eq.~(\ref{eq:rot:spinhel}), the Bjorken and Drell spin
basis
($|\bbox{p}_\perp,p^+,m_1,m_2\rangle$).

\section{Results for the Nucleon-Nucleon Potentials}

The next step towards numerically calculating the bound states for these
potentials is to choose the parameters (meson masses, coupling
constants, etc.) for the potentials. We consider two models: a pion-only
model and a light-front nucleon-nucleon model which incorporates
all six mesons used in the Refs.~\cite{Miller:1997cr,Miller:1999ap}.

\subsection{Pion-only Results}

As a starting point, we look for inspiration from the non-relativistic
one-pion-exchange (OPE) model used by Friar, Gibson, and Payne (FGP)
\cite{Friar:1984wi}. They use the conventional OPE potential along with
a $n$-pole pion-nucleon form factor to regulate the high-momentum
(short-range) part of the potential. By tuning the form factor
parameters to reproduce the physical binding energy of the deuteron,
they find that the other the deuteron observables are also reproduced
fairly accurately. This leads to the conclusion that this OPE potential
model is a good starting point for investigating the deuteron
observables.

The basic parameters of the FGP model are the mass of the nucleon
($M=938.958$~MeV), the mass of the pion ($m_\pi=138.03$~MeV), and the
pion decay constant ($f_0^2=0.079$), which corresponds to a coupling
constant of $\frac{g_\pi^2}{4\pi}=\frac{4M^2f_0^2}{m_\pi^2}=14.6228$ in
our formalism.  The family of $n$-pole pion-nucleon form factors they
consider have the form
\begin{eqnarray}
F(q)&=&\left( \frac
{\Lambda^2-m_\pi^2}
{\Lambda^2+\bbox{q}^2} \right)^n.
\end{eqnarray}
The parameter $\Lambda$ is fit to reproduce the deuteron binding energy
for a given value of $n$. 

We can obtain the FGP model from our light-front nucleon-nucleon
potential by considering only pion exchanges and performing a
non-relativistic reduction.
To understand our results, it is important to first consider how well
the light-front potential and wave functions approximate the
non-relativistic potential and wave functions. 

It is useful to start by reviewing some of the properties of the
deuteron. First, the deuteron is very lightly bound. Second, although a
majority of the deuteron wave function resides in the non-relativistic
regime, it has a high-momentum tail that falls off rather slowly
\cite{Machleidt:1987hj}. Recalling our
experience with the Wick-Cutkosky model, where we found that mass
splitting between states with different $m$ values is small for lightly
bound states, it may appear plausible that masses of the $m=0$ and $m=1$
states of the deuteron should be approximately the same when calculated
with
the light-front OPE potential. However, the high-momentum tail of the
deuteron enhances the effects of the potential's relativistic
components. Since the breaking of rotational invariance is a
relativistic effect, as shown in Eq.~(\ref{wc:OBEpotet}), this implies
that mass splitting may be large for the deuteron calculated with the
light-front potential.

To clearly understand how a large mass splitting might arise in the
light-front pion-only model, we take a step back and consider a scalar
version of the pion-only model, in which we assume that the pion has a
scalar coupling to the nucleon. This allows for a more direct comparison
with the Wick-Cutkosky model, since the main difference between the
scalar-pion-only potential and the Wick-Cutkosky potential 
is the pion-nucleon form
factor. Since the denominator of the pion-nucleon form factor has the
same form as
the denominator of the potential, the form factors do not significantly
change the rotational properties of the scalar-pion-only
potential. Another difference is that factors of $\overline{u}u$ appear
in the numerator of the scalar-pion-only potential, but the effect of
these terms is small. 

In Table~\ref{pi:tab:checkscalarpionbe}, we show the binding energies
for deuterons calculated with the scalar-pion-only model. Two
pion-nucleon form
factors are considered, the first one has $\Lambda=1.0$~GeV, for which
the coupling constant was fit to give the correct binding energy for the
non-relativistic potential, and the second form factor uses
$\Lambda=1.915$~GeV, which was fit to give the correct binding energy for
the light-front potential. Those form factors are used to calculate the
binding energies for the non-relativistic and light-front potentials, as
well as the instantaneous and retarded potentials, which are
relativistic and defined by analogy with Eqs.~(\ref{wc:instpot}) and
(\ref{wc:retpot}). We find that the binding energies for all of the
potentials have the same order of magnitude, and that the light-front
potentials have consistently lower binding energies, which confirms
the behavior observed in Fig.~\ref{wc:sumfig}. In addition, the binding
energies
of the $m=0$ and $m=1$ light-front potentials are essentially
degenerate for both form factors.

\begin{table}
\caption{
Binding energies for deuterons calculated with several potentials and
two different pion-nucleon form factors for the scalar-pion-only
model. The $n$
parameter of the form factor is 1, the coupling constant is
$\frac{g^2}{4\pi}=0.424$, and the pion mass is used as the mass of the
exchanged meson. 
\label{pi:tab:checkscalarpionbe}}
\begin{center}
\begin{tabular}{c|c|c} \hline \hline 
Potential
& $\Lambda=1.0$~GeV  
& $\Lambda=1.915$~GeV 
\\ \hline
Non-relativistic   & 2.2236~MeV  & 4.2539~MeV \\ 
Instantaneous      & 2.1581~MeV  & 4.0862~MeV \\ 
Retarded           & 2.3111~MeV  & 4.5224~MeV \\ 
Light-front, $m=0$ & 1.2027~MeV  & 2.2296~MeV \\ 
Light-front, $m=1$ & 1.2027~MeV  & 2.2294~MeV \\ 
\hline \hline
\end{tabular} 
\end{center}
\end{table}

Now we ready to consider the (pseudoscalar) pion-only model. The only
difference between this potential and the scalar-pion-only potential is
that the numerator contains factors of $\overline{u}i\gamma^5u$ instead
of $\overline{u}u$. Although this is formally a small change, it has a
large
effect on the binding energies. The pseudoscalar coupling generates a
tensor force, which is more sensitive to the relativistic components of
the wave function than the scalar force. In general, this means that the
differences in binding energies obtained with different potentials, such
as those shown in Table~\ref{pi:tab:checkscalarpionbe}, will be larger
for the pion-only potential. 

In Table~\ref{pi:tab:checkpionbe}, we show the binding energies for
deuterons calculated with the pion-only model. Two pion-nucleon form
factors are
considered, the first one has $\Lambda=1.01$~GeV, which was fit for the
non-relativistic pion-only model \cite{Friar:1984wi}, and the second
form factor uses $\Lambda=1.9$~GeV, which was fit to give the most
reasonable binding
energy for the light-front potentials. We find that the binding energies
vary greatly depending on which potential is used. In fact, the
light-front potentials do not bind the deuteron with the first form
factor, and with the second form factor, the mass splitting between the
$m=0$ and $m=1$ states is very large. These facts indicate that the
deuteron wave functions are very sensitive to subtle changes in the
relativistic structure of the pion-only potentials.  

\begin{table}
\caption{
Binding energies for deuterons calculated with several potentials and
two different pion-nucleon form factors for the (pseudoscalar)
pion-only model. The
$n$ parameter of the form factor is 1, the coupling constant
is $\frac{g^2}{4\pi}=14.6$, and the pion mass is used as the mass of the
exchanged meson. The negative binding energy for the light-front
potentials in the first column indicates that the states are not bound. 
\label{pi:tab:checkpionbe}}
\begin{center}
\begin{tabular}{c|c|c } \hline \hline 
Potential
& $\Lambda=1.01$~GeV 
& $\Lambda=1.9$~GeV  
\\ \hline
Non-relativistic   &  2.2244~MeV  &  227.019~MeV  \\
Instantaneous      &  0.0146~MeV  &   28.192~MeV  \\
Retarded           &  1.3590~MeV  &  130.472~MeV  \\
Light-front, $m=0$ & -0.0269~MeV  &    0.788~MeV  \\
Light-front, $m=1$ & -0.0246~MeV  &    8.856~MeV  \\
\hline \hline
\end{tabular} 
\end{center}
\end{table}

Now that we understand that relativistic effects are important for the
OPE potential, we can start to analyze the TPE potential. 
We reiterate here that the main objective of this work is to investigate
this breaking of rotational invariance and how it is restored. Obtaining
results which are in agreement with experimental data is of lesser
importance; we must have states that transform correctly under rotations
before we can address the experimental data. Na\"{\i}vely, one might think
that we can choose any set of two-pion-exchange graphs which is a
truncation of
a rotationally-invariant infinite series of graphs. In particular, one
might think that using the box graphs shown in Fig.~\ref{fig:tbepot}
(the two-pion-exchange graphs without the graphs mandated by chiral
symmetry, denoted as the non-chiral-two-pion-exchange (ncTPE) potential)
would be adequate for our analysis of rotational invariance.

Another choice of a set of two-pion-exchange graphs is sum of the box
graphs shown in Fig.~\ref{fig:tbepot} and the contact graphs shown in 
Fig.~\ref{fig:tbecontpot}. To incorporate chiral symmetry as accurately
as possible without including the crossed two-pion-exchange potentials,
we weight the contact vertex with a factor of $\frac{M}{2(M+m_\pi)}$, as
explained in Section~\ref{nt:bseequiv}. Note that in the sum, which
we call the two-pion-exchange (TPE) potential, chiral symmetry provides
a cancelation between the box diagrams and the contact diagrams. This
indicates that the results obtained with the TPE potential will be more
stable than those obtained with the ncTPE potential. 

To check this stability and the restoration of rotational invariance of
the pion-only model, we
calculate the energy of the deuteron using the OPE, OPE+TPE, and
OPE+ncTPE potentials. We verify that the results are independent of the
choice of the pion-nucleon form factor by considering three different
choices for the form factor.

For the first form factor, we choose $n=1$ and find that
$\Lambda=1.9$~GeV gives a reasonable range of binding energies. We use
the form factor to calculate the lowest energy bound state with
arbitrary total angular momentum. In addition, we take the $J=1$
component of the potential and use it to perform the same
calculation. We expect to obtain similar results since the deuteron is a
spin one object.

The results for the first pion-nucleon form factor are shown in
Table~\ref{pi:tab:Lam19n1}. Note that the results do not depend much on
whether a restriction to spin one is applied. This shows that 
even though the eigenstates of the light-front Hamiltonian are not in
principle eigenstates of the angular momentum, numerically they are {\em
almost} angular momentum eigenstates.
However, this does not mean that rotational invariance is unbroken.
Rotations relate the different $m$ states, and from the mass splittings,
we see that the states do not transform correctly.

Table~\ref{pi:tab:Lam19n1} also shows that both the mass splitting and
the difference in the percentage of the D-state wave function decrease
when TME diagrams are included. In addition, the wave functions are
almost completely in the $J=1$ state. As for the D-state probability, we
first note that it increases with the binding energy. It is consistent
(given the range of values it and the binding energy take) with the
value of 6\%
reported in Ref.~\cite{Friar:1984wi} for the FGP model.

Another thing to note in Table~\ref{pi:tab:Lam19n1} are the effects of
using the ncTPE potential is used. As mentioned earlier, this potential
is greater in magnitude than the TPE potential and should have a larger
effect on the binding energy. In fact, the effect is so large that it
serves to unbind the deuteron. (Strictly speaking, this indicates only
that the binding energy is very small or zero; the error in the binding
energy calculation increases as the binding energy approaches zero.)
Because the ncTPE potential has such a large effect by itself, it is
impossible to determine what effect it has on the rotational properties
of the state. Only the TPE potential can be used to analyze the
restoration of the state's rotational invariance.

\begin{table}
\caption
[
The values of the binding energy, percentage of
the wave function in the D state, and the percentage of the wave function
in the $J=1$ state for the $m=0$ and $m=1$ states for different
potentials.
The pion-nucleon form factor uses $n=1$ and $\Lambda=1.9$~GeV.
]
{
The values of the binding energy, percentage of
the wave function in the D state, and the percentage of the wave function
in the $J=1$ state for the $m=0$ and $m=1$ states for different
potentials.
The values obtained for the full potential (which connects states of
arbitrary total angular momentum) are shown along with the values
obtained for potentials restricted to the $J=1$ sector which are shown
in parentheses.
The pion-nucleon form factor uses $n=1$ and $\Lambda=1.9$~GeV.
\label{pi:tab:Lam19n1}}
\begin{center}
\begin{tabular}{c|c|c|c|c|c|c|c} \hline \hline 
Potential &
\multicolumn{3}{c|}{Binding Energy (MeV)} &
\multicolumn{2}{c|}{D state (\%)} &
\multicolumn{2}{c}{$J=1$ (\%)} \\ \hline
& m=0 & m=1 & Diff & m=0 & m=1 & m=0 & m=1 \\ \hline
\begin{tabular}{c} OPE\end{tabular} &		       	       
\!\!\!\!\begin{tabular}{c} -0.7884 \\(-0.7129) \end{tabular}\!\!\!\! &
\!\!\!\!\begin{tabular}{c} -8.8561 \\(-7.6970) \end{tabular}\!\!\!\! &
\!\!\!\!\begin{tabular}{c}  8.0677 \\ (6.9841) \end{tabular}\!\!\!\! &
\!\!\!\!\begin{tabular}{c}  4.09   \\ (3.97)   \end{tabular}\!\!\!\! &
\!\!\!\!\begin{tabular}{c} 12.27   \\(11.99)   \end{tabular}\!\!\!\! &
\!\!\!\!\begin{tabular}{c} 99.99   \\(100)     \end{tabular}\!\!\!\! &
\!\!\!\!\begin{tabular}{c} 99.78   \\(100)     \end{tabular}\!\!\!\!
\\ \hline						       	       
\begin{tabular}{c} OPE \\ +TPE \end{tabular} &		       	       
\!\!\!\!\begin{tabular}{c} -0.6845 \\(-0.5886) \end{tabular}\!\!\!\! &
\!\!\!\!\begin{tabular}{c} -2.5606 \\(-2.0644) \end{tabular}\!\!\!\! &
\!\!\!\!\begin{tabular}{c}  1.8761 \\(1.4758)  \end{tabular}\!\!\!\! &
\!\!\!\!\begin{tabular}{c}  3.98   \\(3.79)    \end{tabular}\!\!\!\! &
\!\!\!\!\begin{tabular}{c}  8.08   \\(7.61)    \end{tabular}\!\!\!\! &
\!\!\!\!\begin{tabular}{c} 99.98   \\(100)     \end{tabular}\!\!\!\! &
\!\!\!\!\begin{tabular}{c} 99.88   \\(100)     \end{tabular}\!\!\!\!
\\ \hline						       	       
\begin{tabular}{c} OPE \\ +ncTPE \end{tabular} &      	       
\!\!\!\!\begin{tabular}{c}  0.0107 \\(0.0140)  \end{tabular}\!\!\!\! &
\!\!\!\!\begin{tabular}{c}  0.0087 \\(0.0155)  \end{tabular}\!\!\!\! &
\!\!\!\!\begin{tabular}{c}  0.0020 \\(-0.0015) \end{tabular}\!\!\!\! &
\!\!\!\!\begin{tabular}{c}  0.15   \\(0.11)    \end{tabular}\!\!\!\! &
\!\!\!\!\begin{tabular}{c}  0.66   \\(0.58)    \end{tabular}\!\!\!\! &
\!\!\!\!\begin{tabular}{c}100.00   \\(100)     \end{tabular}\!\!\!\! &
\!\!\!\!\begin{tabular}{c}100.00   \\(100)     \end{tabular}\!\!\!\!
\\ \hline \hline
\end{tabular} 
\end{center}
\end{table}

To make sure that the results found are independent of the pion-nucleon
form factor, we performed the same calculation with $n=1$ and
$\Lambda=2.1$~GeV (shown in Table~\ref{pi:tab:Lam21n1}) and with 
$n=2$ and $\Lambda=2.9$~GeV (shown in Table~\ref{pi:tab:Lam29n2}). The
results in these two tables are qualitatively the same as the results in
Table~\ref{pi:tab:Lam19n1}, which demonstrates that these results are
robust.

\begin{table}
\caption{
A pion-nucleon form factor with $n=1$ and $\Lambda=2.1$~GeV is used to
calculate the same quantities as shown in Table~\ref{pi:tab:Lam19n1}.
\label{pi:tab:Lam21n1}}
\begin{center}
\begin{tabular}{c|c|c|c|c|c|c|c} \hline \hline 
Potential &
\multicolumn{3}{c|}{Binding Energy (MeV)} &
\multicolumn{2}{c|}{D state (\%)} &
\multicolumn{2}{c}{$J=1$ (\%)} \\ \hline
& m=0 & m=1 & Diff & m=0 & m=1 & m=0 & m=1 \\ \hline
\begin{tabular}{c}OPE\end{tabular} &
\!\!\!\!\begin{tabular}{c} -1.7357 \\ (-1.6103) \end{tabular}\!\!\!\! &
\!\!\!\!\begin{tabular}{c}-14.9116 \\(-13.2111) \end{tabular}\!\!\!\! &
\!\!\!\!\begin{tabular}{c} 13.1759 \\ (11.6008) \end{tabular}\!\!\!\! &
\!\!\!\!\begin{tabular}{c}  5.60   \\  (5.49)   \end{tabular}\!\!\!\! &
\!\!\!\!\begin{tabular}{c} 14.30   \\ (14.14)   \end{tabular}\!\!\!\! &
\!\!\!\!\begin{tabular}{c} 99.98   \\ (100)     \end{tabular}\!\!\!\! &
\!\!\!\!\begin{tabular}{c} 99.71   \\ (100)     \end{tabular}\!\!\!\!
\\ \hline
\begin{tabular}{c}OPE\\+TPE\end{tabular} &
\!\!\!\!\begin{tabular}{c} -1.6296 \\ (-1.4551) \end{tabular}\!\!\!\! &
\!\!\!\!\begin{tabular}{c} -4.5609 \\ (-3.8203) \end{tabular}\!\!\!\! &
\!\!\!\!\begin{tabular}{c}  2.9313 \\  (2.3652) \end{tabular}\!\!\!\! &
\!\!\!\!\begin{tabular}{c}   5.62  \\  (5.43)   \end{tabular}\!\!\!\! &
\!\!\!\!\begin{tabular}{c}   9.89  \\  (9.53)   \end{tabular}\!\!\!\! &
\!\!\!\!\begin{tabular}{c}  99.96  \\ (100)     \end{tabular}\!\!\!\! &
\!\!\!\!\begin{tabular}{c}  99.85  \\ (100)     \end{tabular}\!\!\!\!
\\ \hline
\begin{tabular}{c}OPE\\+ncTPE\end{tabular} &
\!\!\!\!\begin{tabular}{c} -0.0219 \\ (-0.0045) \end{tabular}\!\!\!\! &
\!\!\!\!\begin{tabular}{c} -0.0180 \\  (0.0056) \end{tabular}\!\!\!\! &
\!\!\!\!\begin{tabular}{c} -0.0039 \\ (-0.0101) \end{tabular}\!\!\!\! &
\!\!\!\!\begin{tabular}{c}  0.69   \\  (0.38)   \end{tabular}\!\!\!\! &
\!\!\!\!\begin{tabular}{c}  1.33   \\  (0.86)   \end{tabular}\!\!\!\! &
\!\!\!\!\begin{tabular}{c} 99.99   \\ (100)     \end{tabular}\!\!\!\! &
\!\!\!\!\begin{tabular}{c} 99.99   \\ (100)     \end{tabular}\!\!\!\!
\\ \hline\hline
\end{tabular} 
\end{center}
\end{table}

\begin{table}
\caption{
A pion-nucleon form factor with $n=2$ and $\Lambda=2.9$~GeV is used to
calculate the same quantities as shown in Table~\ref{pi:tab:Lam19n1}.
\label{pi:tab:Lam29n2}}
\begin{center}
\begin{tabular}{c|c|c|c|c|c|c|c} \hline \hline 
Potential &
\multicolumn{3}{c|}{Binding Energy (MeV)} &
\multicolumn{2}{c|}{D state (\%)} &
\multicolumn{2}{c}{$J=1$ (\%)} \\ \hline
& m=0 & m=1 & Diff & m=0 & m=1 & m=0 & m=1 \\ \hline
\begin{tabular}{c}OPE\end{tabular} &
\!\!\!\!\begin{tabular}{c}  -1.2060 \\ (-1.1068) \end{tabular}\!\!\!\! &
\!\!\!\!\begin{tabular}{c} -10.7746 \\ (-9.4180) \end{tabular}\!\!\!\! &
\!\!\!\!\begin{tabular}{c}   9.5686 \\  (8.3112) \end{tabular}\!\!\!\! &
\!\!\!\!\begin{tabular}{c}   4.86   \\  (4.74)   \end{tabular}\!\!\!\! &
\!\!\!\!\begin{tabular}{c}  13.06   \\ (12.82)   \end{tabular}\!\!\!\! &
\!\!\!\!\begin{tabular}{c}  99.98   \\ (100)     \end{tabular}\!\!\!\! &
\!\!\!\!\begin{tabular}{c}  99.75   \\ (100)     \end{tabular}\!\!\!\!
\\ \hline
\begin{tabular}{c}OPE\\+TPE\end{tabular} &
\!\!\!\!\begin{tabular}{c} -1.0781 \\ (-0.9494) \end{tabular}\!\!\!\! &
\!\!\!\!\begin{tabular}{c} -3.3983 \\ (-2.7863) \end{tabular}\!\!\!\! &
\!\!\!\!\begin{tabular}{c}  2.3202 \\ (1.8369)  \end{tabular}\!\!\!\! &
\!\!\!\!\begin{tabular}{c}  4.78   \\ (4.59)    \end{tabular}\!\!\!\! &
\!\!\!\!\begin{tabular}{c}  8.95   \\ (8.52)    \end{tabular}\!\!\!\! &
\!\!\!\!\begin{tabular}{c}  99.97  \\ (100)     \end{tabular}\!\!\!\! &
\!\!\!\!\begin{tabular}{c}  99.87  \\ (100)     \end{tabular}\!\!\!\!
\\ \hline
\begin{tabular}{c}OPE\\+ncTPE\end{tabular} &
\!\!\!\!\begin{tabular}{c} -0.0002 \\  (0.0069) \end{tabular}\!\!\!\! &
\!\!\!\!\begin{tabular}{c} -0.0028 \\  (0.0106) \end{tabular}\!\!\!\! &
\!\!\!\!\begin{tabular}{c}  0.0026 \\ (-0.0037) \end{tabular}\!\!\!\! &
\!\!\!\!\begin{tabular}{c}  0.30   \\  (0.20)   \end{tabular}\!\!\!\! &
\!\!\!\!\begin{tabular}{c}  0.90   \\  (0.68)   \end{tabular}\!\!\!\! &
\!\!\!\!\begin{tabular}{c} 100.00  \\  (100)    \end{tabular}\!\!\!\! &
\!\!\!\!\begin{tabular}{c} 99.99   \\  (100)    \end{tabular}\!\!\!\!
\\ \hline \hline
\end{tabular} 
\end{center}
\end{table}

\begin{figure}
\begin{center}
\epsfig{angle=0,width=5.0in,height=4.0in,file=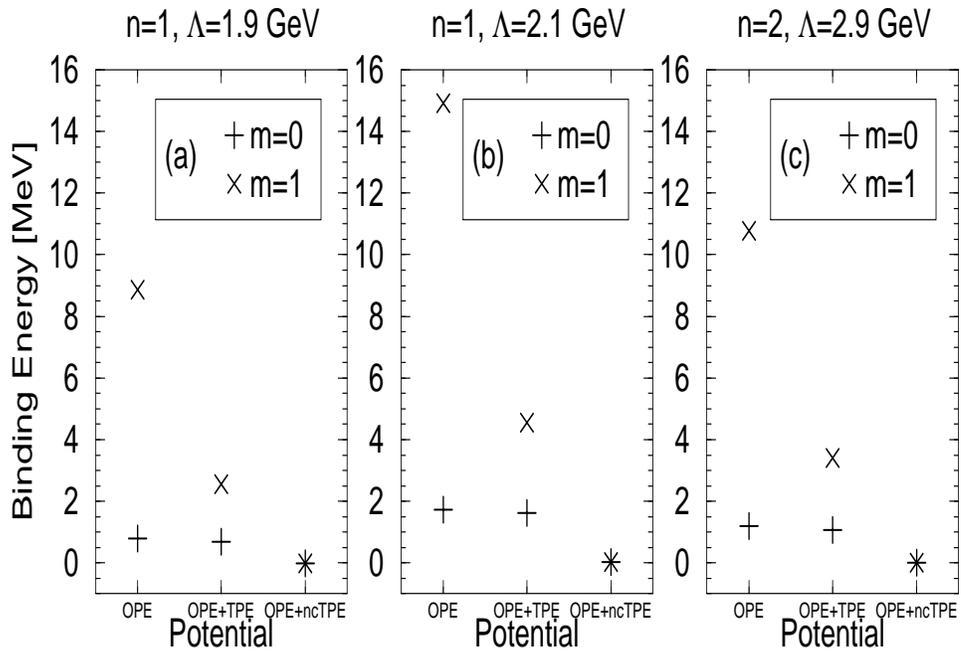}
\caption
[The values of the binding energy for the $m=0$ and $m=1$ states
for different pion-only light-front potentials. Three different
pion-nucleon form factors are used.]
{The values of the binding energy for the $m=0$ and $m=1$ states
for different pion-only light-front potentials. Three different
pion-nucleon form factors are used: 
(a) $n=1$ and $\Lambda=1.9$~GeV,
(b) $n=1$ and $\Lambda=2.1$~GeV, and
(c) $n=2$ and $\Lambda=2.9$~GeV.
\label{fig:BEforMsPi}}
\end{center}
\end{figure}

The deuteron masses for the different $m$ states, as shown in
Tables~\ref{pi:tab:Lam19n1}, \ref{pi:tab:Lam21n1}, and
\ref{pi:tab:Lam29n2}, are shown in Fig.~\ref{fig:BEforMsPi} as a
function of which potential was used for the calculation. It is easy to
see that the relatively large mass splitting obtained with the OPE
potential is significantly reduced by including the TPE potential. The
apparent further reduction of the splitting by neglecting the chiral
part of the TPE potential (leaving the ncTPE potential) is an
artifact of the deuteron becoming unbound. Since the unbound states form
a continuum, the lowest energy unbound states will always be essentially
degenerate, regardless of how badly rotational invariance is broken.

\subsection{Light-Front Nucleon-Nucleon Model Results}
\label{nnresults}

Having addressed the simple pion-only model in the previous section, we
are now ready to consider the full nuclear model where the
nucleon-nucleon interaction is mediated by all the six mesons shown in
Table~\ref{nn:tab:mesparams}.
For numerical work, use the parameters for the light-front
nucleon-nucleon (LFNN) potential from the work of Miller and Machleidt
\cite{Miller:1999ap}. Those parameters were fit for a potential that used
a retarded propagator for the energy in the potentials. Since the
potentials used in this paper have energy dependent denominators, the
parameters must be modified somewhat. We choose to vary the coupling
constant for the $\sigma$ meson. The parameters are given in
table~\ref{nn:tab:mesparams}.

\begin{table}
\caption
[The parameters for the $\pi$, $\eta$, $\rho$, $\omega$,
$\delta$, and $\sigma$ mesons.
The $\sigma$ meson is also known as
$f_0(400-1200)$, and the $\delta$ meson is the $a_0(980)$.]
{The parameters for the $\pi$, $\eta$, $\rho$, $\omega$,
$\delta$, and $\sigma$ mesons from Ref.~\cite{Miller:1999ap}.
For the meson type, ``iv'' and ``is'' stand
for isovector and isoscalar, while ``ps'', ``v'', and ``s'' stand for
pseudoscalar, vector, and scalar.
The $\sigma$ meson is also known as
$f_0(400-1200)$, and the $\delta$ meson is the $a_0(980)$.
\label{nn:tab:mesparams}}
\begin{center}
\begin{tabular}{c|c|c|c|c|c|c} \hline \hline 
Meson     & type  & mass [MeV]  & $\Lambda$ [GeV] & $n$ & $\frac{g^2}{4\pi}$ &
$\frac{f}{g}$ \\ \hline
$\pi$     & iv,ps & 138.04 & 1.2  & 1 &                 14.0    &     \\
$\eta$    & is,ps & 547.5  & 1.5  & 1 &                  3.0    &     \\
$\rho$    & iv,v  & 769.   & 1.85 & 2 &                  0.9    & 6.1 \\
$\omega$  & is,v  & 782.   & 1.85 & 2 &                 24.5    & 0.0 \\
$\delta$  & iv,s  & 983.   & 2.0  & 1 &                  2.0723 &     \\
$\sigma$  & is,s  & 550.   & 2.0  & 1 & $f_\sigma\times$ 8.9602 &     \\
\hline\hline
\end{tabular} 
\end{center}
\end{table}

As with all the other deuteron models presented in this paper, 
the light-front one-meson-exchange (OME) potential breaks 
rotational invariance and causes a mass splitting of the deuteron states
with different magnetic quantum numbers. We expect that the
splitting will be removed somewhat by
including higher order potentials.

The first step is to determine which two-meson-exchange potentials
to use. One choice is to use only the two-pion-exchange potentials, TPE
and ncTPE, as defined in the previous section. However, we expect to get
better results using the two-meson-exchange diagrams
generated by all the available mesons, including the contact diagrams
for the pions, which we denote as the two-meson-exchange (TME) potential.
In addition, we can also investigate the effect of leaving out the
contact potentials for the pions, resulting in the non-chiral 
two-meson-exchange (ncTME) potential.

We do not include diagrams with a contact interaction between the nucleon,
a pion, and another meson. This is because, as mentioned in
section~\ref{nt:bseequiv}, the infinite series of the box diagrams is
rotationally invariant and the contact diagrams are not needed to
achieve rotational invariance. Furthermore, they are not required to
control the convergence of the series, since there is no strong
cancellation between the contact diagram and the instantaneous diagrams.

The first step in analyzing the bound states is to determine what range of
$f_\sigma$ gives reasonable results. We iteratively solve for the binding
energy of the deuteron, varying $f_\sigma$ until the binding energy matches
the physical value of the binding energy, for each of the potentials. The
results are shown in Table~\ref{nn:tab:fixbseJ1}. We find that a value of 
$f_\sigma$ in the range 1.2 to 1.3 will give reasonable results for the
binding energy. Note that D-state probability (about $3\%$) is lower in
this model than for the energy-independent light-front used in
Ref.~\cite{Miller:1999ap}, were a value of $4.5\%$ is found. This is
expected since the $f_\sigma$ is greater than 1 in this model, meaning
that the scalar interaction is strengthened relative to the tensor
interaction, leading to a decrease in amount of the D-state present.

\begin{table}
\caption{The values of $f_\sigma$ required to give the physical value of
the deuteron mass for a given potential and state with $J_z=m$.
The percent of the wave function in the D-state and in the $J=1$ state
are also shown.
\label{nn:tab:fixbseJ1}}
\begin{center}
\begin{tabular}{c|c|c|c|c|c|c|c} \hline \hline 
Potential &
\multicolumn{3}{c|}{$f_\sigma$} &
\multicolumn{2}{c|}{\% D state} &
\multicolumn{2}{c}{\% $J=1$} \\ \hline
& m=0 & m=1 & Diff & m=0 & m=1 & m=0 & m=1  \\ \hline
OME &
1.2407  &  1.2125  &   0.0282 & 2.87 & 3.55 & 99.99 & 99.97 \\ \hline
\begin{tabular}{c}OME\\+TPE\end{tabular} &
1.2829  &  1.2819  &   0.0010 & 2.96 & 3.23 & 99.99 & 99.97 \\ \hline
\begin{tabular}{c}OME\\+TME\end{tabular} &
1.2968  &  1.3079  &  -0.0111 & 2.95 & 3.28 & 99.99 & 99.96 \\ \hline
\begin{tabular}{c}OME\\+ncTPE\end{tabular} &
1.3064  &  1.3121  &  -0.0057 & 2.99 & 3.16 & 99.98 & 99.97 \\ \hline
\begin{tabular}{c}OME\\+ncTME\end{tabular} &
1.3198  &  1.3397  &  -0.0199 & 2.98 & 3.21 & 99.98 & 99.96 \\ \hline
 \hline
\end{tabular} 
\end{center}
\end{table}

We choose two values of $f_\sigma$, one from the low end of the range (1.22)
and one from the high end (1.2815) for our investigations. Using two values
helps ensure that our results are robust.

\begin{figure}
\begin{center}
\epsfig{angle=0,width=5.0in,height=5.0in,file=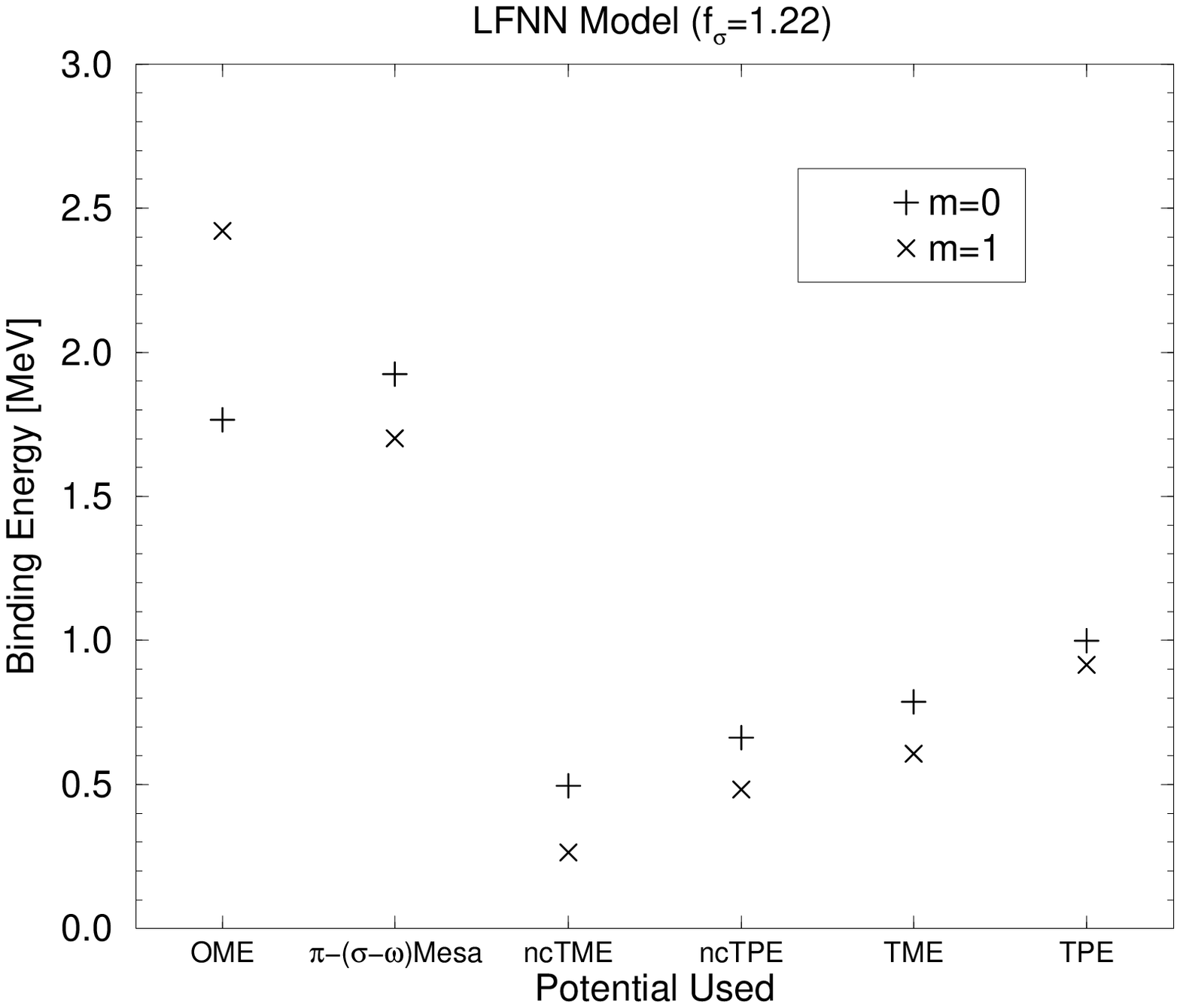}
\caption{The values of the binding energy for the $m=0$ and $m=1$ states
for different nucleon-nucleon light-front potentials. The $\sigma$
coupling constant factor is $f_\sigma=1.22$.
\label{fig:BEforMsAll.1.22}}
\end{center}
\end{figure}

First, we examine the bound states for $f_\sigma=1.22$.
The results for several different choices of the TME potentials 
are shown in Table~\ref{nn:tab:sig122AllJ} and the binding
energies are plotted in Fig.~\ref{fig:BEforMsAll.1.22}. In addition to
the two-meson-exchange potentials mentioned above, we also consider the
$\pi$-$\sigma$ plus $\pi$-$\omega$ Mesa potential.  The reason for
considering this potential is that Carbonell, Desplanques, Karmanov, and
Mathiot \cite{Carbonell:1998rj} have shown that it helps restore
rotational invariance of the deuteron.

In particular, they have used manifestly covariant light-front dynamics 
to analyze the deuteron. They start with a deuteron wave function
calculated in equal-time dynamics, then use a light-front 
one-pion-exchange potential (expanded to lowest order in powers of
$\frac{1}{M}$) to calculate the perturbative corrections to the wave
function. They find that the resulting wave function has an unphysical
dependence on the orientation of the light-front plane, which would
manifest itself as a breaking of rotational invariance in our
formalism. They also use the $\pi$-$\sigma$ and $\pi$-$\omega$ Mesa
potentials (expanded to lowest order in powers of $\frac{1}{M}$) to
calculate the correction to the wave function. When the wave function
corrections are combined, they find that the directional dependence of
the longest-range part of the deuteron wave function cancels exactly.

This implies that for our model, using the $\pi$-$\sigma$ plus
$\pi$-$\omega$ Mesa potential (which we denote by
$\pi$-($\sigma$-$\omega$)) should partially restore the rotational
invariance of the deuteron, assuming that the breaking of rotational
invariance is due primarily to the one-pion-exchange potential. Note that
since we solve for the deuteron wave function self-consistently and to
all orders for our potentials, we do not expect to find exactly the same
result as Ref.~\cite{Carbonell:1998rj}. 

The first thing to notice about the data in
Table~\ref{nn:tab:sig122AllJ} is that the results are essentially the
same regardless of if arbitrary angular momentum or is used or if the
potential is restricted to the $J=1$ sector. This was also seen in the
previous section for the pion-only model. It means that the wave
functions are numerically approximate to angular momentum eigenstates.

Next we notice the splittings between masses and D state percentages for
the $m=0$ and $m=1$ states. The dependence on $m$ implies that the
states do not transform
correctly under rotations. All of the two-meson-exchange potentials used
reduce the splittings by similar amounts, by about 60\%
for the binding energy and by about 70\%
for the percent D state. Note also that the mass splittings for the
pion-only model were much larger. 

Examining the effects of the individual two-meson-exchange potentials,
we see that $\pi$-($\sigma$-$\omega$)) potential does reduce the mass
splitting, but it does not fully remove it. This is expected since the
OME potential includes more than just the pion potential, and the
potential is relativistic.

Next, we compare the ncTME and ncTPE potentials to the TME and TPE
potentials. The non-chiral potentials reduce the binding energy more
than the chiral potentials, as we expected from our experience from the
pion-only model. However, unlike for the pion-only model, we find that
the chiral and non-chiral potential have fairly similar effects.

Finally, notice that the mass splitting for the TPE potential is much
smaller than for the other two-meson-exchange potentials. By itself,
this does not imply that the rotational properties of the deuteron
calculated with that potential are significantly better than those from
other two-meson-exchange potentials. The individual potentials that make
up the TME potential are fairly large in magnitude, but vary in
sign. This means that using any subset of those potentials may result in
either a larger or smaller mass splitting. In this case, it is
smaller. To investigate this further, we examine the currents for
the TME and TPE deuteron wave function in Chapter~\ref{ch:ffdeut}.

\begin{figure}
\begin{center}
\epsfig{angle=0,width=5.0in,height=5.0in,file=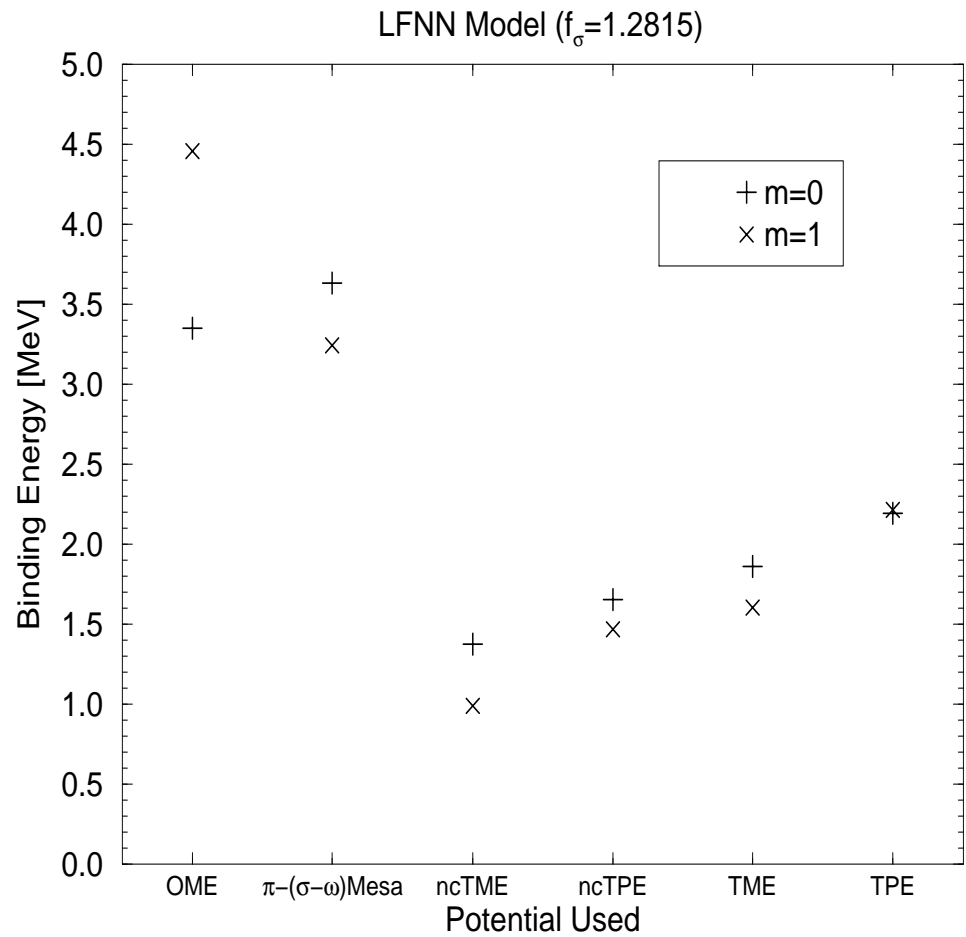}
\caption{The values of the binding energy for the $m=0$ and $m=1$ states
for different nucleon-nucleon light-front potentials. The $\sigma$
coupling constant factor is $f_\sigma=1.2815$.
\label{fig:BEforMsAll.1.28}}
\end{center}
\end{figure}

To verify that our results are independent of the value of $f_\sigma$, we
recalculate the deuteron properties for each of the potentials with
$f_\sigma=1.2815$. The results are summarized in
Table~\ref{nn:tab:sig128AllJ}, and the binding energies are shown in
Fig.~\ref{fig:BEforMsAll.1.28}. The change in $f_\sigma$ increases the
binding of the states, but the rest of the results are qualitatively the
same.

\begin{table}
\caption[The values of the binding energy, percentage of the wave
function in the D state, and the percentage of the wave function in the
$J=1$ state for the $m=0$ and $m=1$ states for different potentials.
The $\sigma$ coupling constant factor is $f_\sigma=1.22$.]
{The values of the binding energy, percentage of the wave
function in the D state, and the percentage of the wave function in the
$J=1$ state for the $m=0$ and $m=1$ states for different potentials.
The values obtained for the full potential (which connects states of
arbitrary total angular momentum) are shown along with the values
obtained for potentials restricted to the $J=1$ sector which are shown
in parentheses. The $\sigma$ coupling constant factor is
$f_\sigma=1.22$.
\label{nn:tab:sig122AllJ}}
\begin{center}
\begin{tabular}{c|c|c|c|c|c|c|c} \hline \hline 
Potential &
\multicolumn{3}{c|}{Binding Energy (MeV)} &
\multicolumn{2}{c|}{D state (\%)} &
\multicolumn{2}{c}{$J=1$ (\%)} \\ \hline
& m=0 & m=1 & Diff & m=0 & m=1 & m=0 & m=1 \\ \hline
OME only & 
\!\!\!\!\begin{tabular}{c} -1.7653 \\ (-1.7141) \end{tabular}\!\!\!\! &
\!\!\!\!\begin{tabular}{c} -2.4200 \\ (-2.2655) \end{tabular}\!\!\!\! &
\!\!\!\!\begin{tabular}{c} 0.6547  \\ (0.5514)  \end{tabular}\!\!\!\! &
\!\!\!\!\begin{tabular}{c} 2.73    \\ (2.72)    \end{tabular}\!\!\!\! &
\!\!\!\!\begin{tabular}{c} 3.61    \\ (3.54)    \end{tabular}\!\!\!\! &
\!\!\!\!\begin{tabular}{c} 99.99   \\ (100)     \end{tabular}\!\!\!\! &
\!\!\!\!\begin{tabular}{c} 99.96   \\ (100)     \end{tabular}\!\!\!\!
\\ \hline
\begin{tabular}{c}OME\\+$\pi$-($\sigma$-$\omega$) Mesa\end{tabular} &
\!\!\!\!\begin{tabular}{c} -1.9236 \\ (-1.8625) \end{tabular}\!\!\!\! &
\!\!\!\!\begin{tabular}{c} -1.7021 \\ (-1.5340) \end{tabular}\!\!\!\! &
\!\!\!\!\begin{tabular}{c} -0.2215 \\ (-0.3285) \end{tabular}\!\!\!\! &
\!\!\!\!\begin{tabular}{c} 2.80    \\ (2.78)    \end{tabular}\!\!\!\! &
\!\!\!\!\begin{tabular}{c} 3.38    \\ (3.25)    \end{tabular}\!\!\!\! &
\!\!\!\!\begin{tabular}{c} 99.99   \\ (100)     \end{tabular}\!\!\!\! &
\!\!\!\!\begin{tabular}{c} 99.96   \\ (100)     \end{tabular}\!\!\!\!
\\ \hline
\begin{tabular}{c}OME\\+ncTME\end{tabular} &
\!\!\!\!\begin{tabular}{c} -0.4948 \\ (-0.4611) \end{tabular}\!\!\!\! &
\!\!\!\!\begin{tabular}{c} -0.2646 \\ (-0.2034) \end{tabular}\!\!\!\! &
\!\!\!\!\begin{tabular}{c} -0.2302 \\ (-0.2577) \end{tabular}\!\!\!\! &
\!\!\!\!\begin{tabular}{c} 1.97    \\ (1.93)    \end{tabular}\!\!\!\! &
\!\!\!\!\begin{tabular}{c} 1.76    \\ (1.59)    \end{tabular}\!\!\!\! &
\!\!\!\!\begin{tabular}{c} 99.99   \\ (100)     \end{tabular}\!\!\!\! &
\!\!\!\!\begin{tabular}{c} 99.98   \\ (100)     \end{tabular}\!\!\!\!
\\ \hline
\begin{tabular}{c}OME\\+ncTPE\end{tabular} &
\!\!\!\!\begin{tabular}{c} -0.6620 \\ (-0.6232) \end{tabular}\!\!\!\! &
\!\!\!\!\begin{tabular}{c} -0.4825 \\ (-0.4132) \end{tabular}\!\!\!\! &
\!\!\!\!\begin{tabular}{c} -0.1795 \\ (-0.2100) \end{tabular}\!\!\!\! &
\!\!\!\!\begin{tabular}{c} 2.16    \\ (2.13)    \end{tabular}\!\!\!\! &
\!\!\!\!\begin{tabular}{c} 2.09    \\ (1.99)    \end{tabular}\!\!\!\! &
\!\!\!\!\begin{tabular}{c} 99.99   \\ (100)     \end{tabular}\!\!\!\! &
\!\!\!\!\begin{tabular}{c} 99.98   \\ (100)     \end{tabular}\!\!\!\!
\\ \hline
\begin{tabular}{c}OME\\+TME\end{tabular} &
\!\!\!\!\begin{tabular}{c} -0.7861 \\ (-0.7460) \end{tabular}\!\!\!\! &
\!\!\!\!\begin{tabular}{c} -0.6060 \\ (-0.5191) \end{tabular}\!\!\!\! &
\!\!\!\!\begin{tabular}{c} -0.1801 \\ (-0.2269) \end{tabular}\!\!\!\! &
\!\!\!\!\begin{tabular}{c} 2.25    \\ (2.22)    \end{tabular}\!\!\!\! &
\!\!\!\!\begin{tabular}{c} 2.31    \\ (2.19)    \end{tabular}\!\!\!\! &
\!\!\!\!\begin{tabular}{c} 99.99   \\ (100)     \end{tabular}\!\!\!\! &
\!\!\!\!\begin{tabular}{c} 99.97   \\ (100)     \end{tabular}\!\!\!\!
\\ \hline
\begin{tabular}{c}OME\\+TPE\end{tabular} &
\!\!\!\!\begin{tabular}{c} -0.9981 \\ (-0.9531) \end{tabular}\!\!\!\! &
\!\!\!\!\begin{tabular}{c} -0.9155 \\ (-0.8253) \end{tabular}\!\!\!\! &
\!\!\!\!\begin{tabular}{c} -0.0826 \\ (-0.1278) \end{tabular}\!\!\!\! &
\!\!\!\!\begin{tabular}{c} 2.42    \\ (2.39)    \end{tabular}\!\!\!\! &
\!\!\!\!\begin{tabular}{c} 2.57    \\ (2.48)    \end{tabular}\!\!\!\! &
\!\!\!\!\begin{tabular}{c} 99.99   \\ (100)     \end{tabular}\!\!\!\! &
\!\!\!\!\begin{tabular}{c} 99.98   \\ (100)     \end{tabular}\!\!\!\!
\\ \hline \hline
\end{tabular} 
\end{center}
\end{table}

\begin{table}
\caption{
The $\sigma$ coupling constant factor used here,
$f_\sigma=1.2815$, 
distinguishes this table from Table~\ref{nn:tab:sig122AllJ}.
\label{nn:tab:sig128AllJ}}
\begin{center}
\begin{tabular}{c|c|c|c|c|c|c|c} \hline \hline 
Potential &
\multicolumn{3}{c|}{Binding Energy (MeV)} &
\multicolumn{2}{c|}{D state (\%)} &
\multicolumn{2}{c}{$J=1$ (\%)} \\ \hline
& m=0 & m=1 & Diff & m=0 & m=1 & m=0 & m=1 \\ \hline
OME only & 
\!\!\!\!\begin{tabular}{c} -3.3500\\(-3.2818) \end{tabular}\!\!\!\! &
\!\!\!\!\begin{tabular}{c} -4.4546\\(-4.2622) \end{tabular}\!\!\!\! &
\!\!\!\!\begin{tabular}{c}  1.1046\\(0.9804)  \end{tabular}\!\!\!\! &
\!\!\!\!\begin{tabular}{c} 3.09   \\(3.06)    \end{tabular}\!\!\!\! &
\!\!\!\!\begin{tabular}{c} 3.97   \\(3.92)    \end{tabular}\!\!\!\! &
\!\!\!\!\begin{tabular}{c} 99.99  \\(100)     \end{tabular}\!\!\!\! &
\!\!\!\!\begin{tabular}{c} 99.96  \\(100)     \end{tabular}\!\!\!\!
\\ \hline
\begin{tabular}{c}OME\\+$\pi$-($\sigma$-$\omega$) Mesa\end{tabular} &
\!\!\!\!\begin{tabular}{c} -3.6331 \\ (-3.5520) \end{tabular}\!\!\!\! &
\!\!\!\!\begin{tabular}{c} -3.2408 \\ (-3.0194) \end{tabular}\!\!\!\! &
\!\!\!\!\begin{tabular}{c} -0.3923 \\ (-0.5326) \end{tabular}\!\!\!\! &
\!\!\!\!\begin{tabular}{c} 3.10    \\ (3.10)    \end{tabular}\!\!\!\! &
\!\!\!\!\begin{tabular}{c} 3.85    \\ (3.77)    \end{tabular}\!\!\!\! &
\!\!\!\!\begin{tabular}{c} 99.99   \\ (100)     \end{tabular}\!\!\!\! &
\!\!\!\!\begin{tabular}{c} 99.95   \\ (100)     \end{tabular}\!\!\!\!
\\ \hline
\begin{tabular}{c}OME\\+ncTME\end{tabular} &
\!\!\!\!\begin{tabular}{c} -1.3766 \\ (-1.3230) \end{tabular}\!\!\!\! &
\!\!\!\!\begin{tabular}{c} -0.9901 \\ (-0.8807) \end{tabular}\!\!\!\! &
\!\!\!\!\begin{tabular}{c} -0.3865 \\ (-0.4423) \end{tabular}\!\!\!\! &
\!\!\!\!\begin{tabular}{c} 2.67    \\ (2.65)    \end{tabular}\!\!\!\! &
\!\!\!\!\begin{tabular}{c} 2.64    \\ (2.54)    \end{tabular}\!\!\!\! &
\!\!\!\!\begin{tabular}{c} 99.99   \\ (100)     \end{tabular}\!\!\!\! &
\!\!\!\!\begin{tabular}{c} 99.97   \\ (100)     \end{tabular}\!\!\!\!
\\ \hline
\begin{tabular}{c}OME\\+ncTPE\end{tabular} &
\!\!\!\!\begin{tabular}{c} -1.6532 \\ (-1.5939) \end{tabular}\!\!\!\! &
\!\!\!\!\begin{tabular}{c} -1.4693 \\ (-1.3578) \end{tabular}\!\!\!\! &
\!\!\!\!\begin{tabular}{c} -0.1839 \\ (-0.2361) \end{tabular}\!\!\!\! &
\!\!\!\!\begin{tabular}{c} 2.81    \\ (2.79)    \end{tabular}\!\!\!\! &
\!\!\!\!\begin{tabular}{c} 2.88    \\ (2.81)    \end{tabular}\!\!\!\! &
\!\!\!\!\begin{tabular}{c} 99.99   \\ (100)     \end{tabular}\!\!\!\! &
\!\!\!\!\begin{tabular}{c} 99.97   \\ (100)     \end{tabular}\!\!\!\!
\\ \hline
\begin{tabular}{c}OME\\+TME\end{tabular} &
\!\!\!\!\begin{tabular}{c} -1.8617 \\ (-1.8021) \end{tabular}\!\!\!\! &
\!\!\!\!\begin{tabular}{c} -1.6032 \\ (-1.4696) \end{tabular}\!\!\!\! &
\!\!\!\!\begin{tabular}{c} -0.2585 \\ (-0.3325) \end{tabular}\!\!\!\! &
\!\!\!\!\begin{tabular}{c} 2.85    \\ (2.83)    \end{tabular}\!\!\!\! &
\!\!\!\!\begin{tabular}{c} 3.05    \\ (2.97)    \end{tabular}\!\!\!\! &
\!\!\!\!\begin{tabular}{c} 99.99   \\ (100)     \end{tabular}\!\!\!\! &
\!\!\!\!\begin{tabular}{c} 99.96   \\ (100)     \end{tabular}\!\!\!\!
\\ \hline
\begin{tabular}{c}OME\\+TPE\end{tabular} &
\!\!\!\!\begin{tabular}{c} -2.1915 \\ (-2.1267) \end{tabular}\!\!\!\! &
\!\!\!\!\begin{tabular}{c} -2.2137 \\ (-2.0831) \end{tabular}\!\!\!\! &
\!\!\!\!\begin{tabular}{c} 0.0222  \\ (-0.0436) \end{tabular}\!\!\!\! &
\!\!\!\!\begin{tabular}{c} 2.95    \\ (2.94)    \end{tabular}\!\!\!\! &
\!\!\!\!\begin{tabular}{c} 3.23    \\ (3.17)    \end{tabular}\!\!\!\! &
\!\!\!\!\begin{tabular}{c} 99.99   \\ (100)     \end{tabular}\!\!\!\! &
\!\!\!\!\begin{tabular}{c} 99.97   \\ (100)     \end{tabular}\!\!\!\!
\\ \hline \hline
\end{tabular} 
\end{center}
\end{table}

\chapter{Form Factors of the Deuteron} \label{ch:ffdeut}
\doquote
{It is theory that decides what can be observed.}
{Albert Einstein}

In Chapter~\ref{ch:pionly}, we considered several different truncations
of the light-front nucleon-nucleon (LFNN) potential and used them to
solve for the deuteron wave function. In this chapter, we use those wave
functions to solve for the deuteron current, which is used to calculate
the deuteron electromagnetic and axial form factors. We can analyze the
rotational properties of the deuteron current like we did for the wave
function. 

In this chapter, we first outline the covariant theory of the
electromagnetic form factors for spin-1 objects, like the deuteron. Then
we recall the features of light-front calculations (including the
breaking of rotational invariance) of the form factors. After that, we
review the covariant and light-front tools for calculating axial form
factors. The formalism is then applied to calculate the electromagnetic
and axial currents and form factors for the light-front deuteron wave
functions.

One notable feature of this calculation is that it is done entirely with
light-front dynamics. The covariant Lagrangian generates light-front
potentials, which generate light-front wave functions, which are used in
a light-front calculation of the deuteron current and form factors. This
is different from other approaches which use deuteron wave functions
calculated from equal-time dynamics, then transformed to the light front
\cite{Carbonell:1994uy,Carbonell:1995yi,Chung:1988my,Frankfurt:1993ut,%
Frederico:1991vb,Kondratyuk:1984kq,Cardarelli:1995yq,Arndt:1999wx}.

\section{Electromagnetic Form Factors}

\subsection{Covariant Theory}

\begin{figure}
\begin{center}
\epsfig{angle=0,width=3.25in,height=1.8in,file=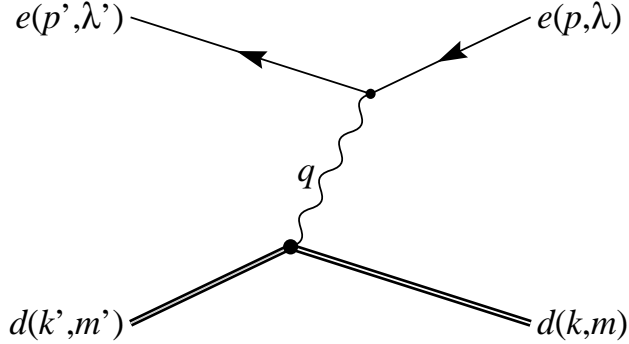}
\caption{The Feynman diagram for one-photon-exchange electron-deuteron
scattering.
\label{form:edscat}}
\end{center}
\end{figure}

In the one-photon-exchange approximation, shown in
Fig.~\ref{form:edscat}, the amplitude of the
scattering process $ed\rightarrow ed$ is just the contraction of the
electron and deuteron currents, multiplied by the photon propagator,
\begin{eqnarray}
\langle p', \lambda' | j_\mu^{e} | p, \lambda \rangle
\frac{1}{q^2}
\langle k', m' | J^\mu_d | k, m \rangle,
\end{eqnarray}
where
\begin{eqnarray}
\langle p', \lambda' | j_\mu^{e} | p, \lambda \rangle
&=&
e \overline{u}(\bbox{p}',\lambda') \gamma_\mu u(\bbox{p},\lambda).
\end{eqnarray}

From Lorentz covariance, current conservation, parity invariance, and time reversal invariance
\cite{Hummel:1990zz,Zuilhof:1980ae,Rupp:1990sg,Garcon:2001sz,Arnold:1981zj},
we infer that the deuteron form factor can be written as
\begin{eqnarray}
\langle k', m' | J^\mu_d | k, m \rangle
&=&
- \frac{e}{2 M_d}
(e^*)^\rho(\bbox{k}',m')
J^\mu_{\rho \sigma}
e^\sigma(\bbox{k},m), \label{eq:deutcurrent}
\end{eqnarray}
where the spin-1 polarization vectors satisfy
\begin{eqnarray}
e_\mu^*(\bbox{k},m) e^\mu(\bbox{k},m') &=& -\delta_{m,m'}, \\
\sum_{m} e^*_\mu(\bbox{k},m) e_\nu(\bbox{k},m) &=& - g_{\mu \nu} +
\frac{k_\mu k_\nu}{M_d^2}, \\
k_\mu e^\mu(\bbox{k},m) &=& 0,
\end{eqnarray}
and the operator $J^\mu_{\rho\sigma}$ is given by
\begin{eqnarray}
J^\mu_{\rho\sigma} &=& (k'_\mu + k_\mu) \left[
g_{\rho \sigma} F_1(q^2) - \frac{q_\rho q_\sigma}{2 M_d^2} F_2(q^2)
\right] - I^{\mu \nu}_{\rho \sigma} q_\nu G_1(q^2),
\end{eqnarray}
where
$I_{\rho\sigma}^{\mu\nu}=g_\rho^\mu{}g_\sigma^\nu-g_\rho^\nu{}g_\sigma^\mu$
is the generator of infinitesimal Lorentz transformations, $q=k'-k$, and
$F_1$, $F_2$, and $G_1$ are functions of $q^2$, the invariant mass of
the photon. Some conventions \cite{Chung:1988my,Frankfurt:1993ut} denote
$G_1$ as $-F_3$.

The $F_1$, $F_2$, and $G_1$ form factors
are related to the deuteron charge, magnetic, and quadrapole form
factors, denoted by $F_C$, $F_M$, and $F_Q$ respectively, 
by \cite{Coester:1975hj,Hummel:1990zz,Rupp:1990sg}
\begin{eqnarray}
F_C &=& F_1 + \frac{2}{3} \eta \left[ F_1 + (1+\eta) F_2 - G_1 \right],
\label{ff:eq:fc} \\
F_M &=& G_1, \\
F_Q &=& F_1 + (1+\eta) F_2 - G_1, \label{ff:eq:fq}
\end{eqnarray}
where
\begin{eqnarray}
\eta &=& \frac{-q^2}{4 M^2_d}. \label{eq:defeta}
\end{eqnarray}
Note that since both the initial and final electron and deuteron are
on-shell, $-q^2>0$. We define $Q^2=-q^2$.

At zero-momentum transfer, the deuteron charge, magnetic, and
quadrapole form factors are simply
\begin{eqnarray}
F_C(0) &=& 1, \\
F_M(0) &=& \frac{M_d}{m_p} \mu_d = 1.71293,\\
F_Q(0) &=& M_d^2 Q_d = 25.8525,
\end{eqnarray}
where $m_p$ is the mass of the proton,
$\mu_d$ is the magnetic moment of the deuteron (in nuclear magnetons
$\mu_N=\frac{e}{2m_p}$), and $Q_d$ is the electric quadrapole moment of
the deuteron (in units of the deuteron mass, although typically it is
measured in $e\cdot$barns). The numerical values of the form factors are
determined by using experimentally measured values
\cite{Garcon:2001sz,Ericson:1983ei,Mohr:2000}.

Another set of form factors are $G_0$, $G_1$, and $G_2$, commonly used
in light-front dynamics
\cite{Cardarelli:1995yq,Chung:1988my,Frankfurt:1993ut}. They are simply
related to the charge, magnetic, and quadrapole form factors,
\begin{eqnarray}
G_0 &=& F_C, \\
G_1 &=& -F_M, \\
G_2 &=& \frac{\sqrt{8}}{3}\eta F_Q.
\end{eqnarray}

The electron-deuteron cross-section is measured to determine the
deuteron form factors for large momentum transfers. The
unpolarized cross-section has the form 
\cite{Phillips:1998uk,Garcon:2001sz}
\begin{eqnarray}
\frac{d\sigma}{d\Omega_e} &=&
\frac{\sigma_{\text{Mott}}}{ 1+\frac{2E_e}{M_d}\sin^2\frac{\theta_e}{2}}
\left(
A(Q^2) + B(Q^2) \tan^2 \left( \frac{\theta_e}{2} \right) \right),
\end{eqnarray}
where 
\begin{eqnarray}
\sigma_{\text{Mott}} &=&
\left( 
\frac{\alpha \cos\frac{\theta_e}{2}}
{2 E_e \sin^2\frac{\theta_e}{2}} \right)^2,
\end{eqnarray}
and $E_e$ is the electron beam energy, $\theta_e$ is the angle by
which the electron scatters, and $\alpha$ is the fine-structure
constant. The structure functions $A$ and $B$ can be expressed in terms
of the charge, magnetic, and quadrapole form factors,
\begin{eqnarray}
A &=& F_c^2 + \frac{8}{9} \eta^2 F_Q^2 + \frac{2}{3}\eta F_M^2, \\
B &=& \frac{4}{3} \eta (1+\eta) F_M^2.
\end{eqnarray}

The three form factors $F_1$, $F_2$, and $G_1$ cannot be determined
uniquely from the unpolarized scattering data since that measurement of
provides only two structure functions. More information is needed, which
can be obtained
from experiments where the polarization of the electron and/or deuteron
is measured \cite{Garcon:2001sz,Arnold:1981zj}. The most commonly
measured quantity is the tensor polarization observable, $T_{20}$,
\begin{eqnarray}
T_{20} &=& -
\frac{\frac{8}{3}\eta F_C F_Q + \frac{8}{9}\eta^2 F_Q^2 + 
\frac{1}{3}\eta F_M^2 \left(1+2(1+\eta)\tan^2\frac{\theta_e}{2} \right)}
{\sqrt{2}\left(A+B \tan^2\frac{\theta_e}{2}\right)}.
\end{eqnarray}
Note that $T_{20}$ depends on the angle $\theta_e$. For ease of
comparison, data for $T_{20}$ is usually presented for
$\theta_e=70^\circ$. Alternatively, one can eliminate the angular
dependence by defining $\widetilde{T}_{20}$, which is $T_{20}$ with
$F_M$ set equal to zero,
\begin{eqnarray}
\widetilde{T}_{20} &=& -
\frac{\frac{8}{3}\eta F_C F_Q + \frac{8}{9}\eta^2 F_Q^2}
{\sqrt{2}\left(F_C^2 + \frac{8}{9} \eta^2 F_Q^2\right)}.
\end{eqnarray}

The extraction of the structure functions from the data is
difficult. In practice, each cross-section measurement is performed at
a different momentum transfer and angle, making it impossible to exactly
disambiguate, for example, $A$ and $B$. One needs to interpolate between
values of $B$ to calculate values for $A$, and vice versa.
The Jefferson Lab t$_{20}$ Collaboration used a self-consistent method
for obtaining the structure functions from the scattering data
\cite{Abbott:2000ak}. They consider a
variety of theoretical models for the deuteron form factors, then fit
the parameters of the models to the measured cross-sections. After
each model is fit, it is used as the interpolation function 
to disentangle the structure functions. This two-step process helps
minimize model-dependent effects in the values of $A$ and $B$.

\subsection{Light Front Calculation} \label{sec:nplfcalcem}

Light-front dynamics is particularly well suited to calculating form
factors. One reason is that the generators of boosts in the one, two,
and plus directions are kinematic, so that wave functions calculated
with a truncated potential transform correctly under boosts. This
feature is especially important for form factors at high momentum
transfer since the wave functions must undergo a large boost.

\begin{figure}
\begin{center}
\epsfig{angle=0,width=5.75in,height=1.0in,file=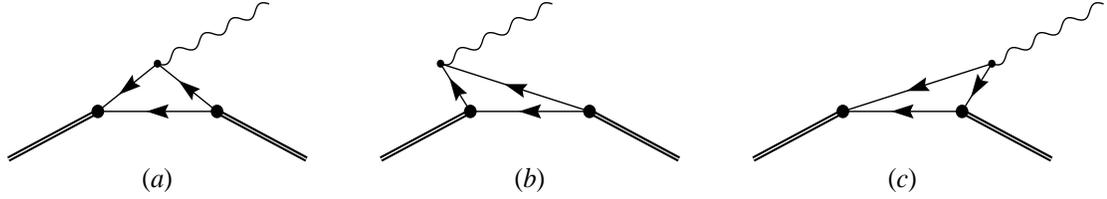}
\caption{The lowest-order graphs which contribute to the deuteron
current matrix element. 
\label{form:deut.cur}}
\end{center}
\end{figure}

Another, more subtle, reason for using the light front is that many of
the graphs which contribute to the current vanish identically. For
example, the three lowest-order graphs for the current are shown in
Fig.~\ref{form:deut.cur}. The double line denotes the deuteron, and the
vertex of the deuteron lines and the nucleon lines represents the
deuteron wave function. The graph labeled (a) does not vanish and is
calculated in section~\ref{ff:impapp}.

Fig.~\ref{form:deut.cur}(b) vanishes in light-front dynamics. To see
why, we first note three facts: the plus component of each particle in
light-front dynamics is non-negative (for massive particles it must be
positive), the plus component of the momentum is conserved, and the plus
momentum of the vacuum is zero. Combining these facts, we find that any
vertex which has particles on one side and vacuum on the other must
vanish. In other words, the vacuum is trivially empty, and no graphs
couple to it.

For Fig.~\ref{form:deut.cur}(c), the coupling of the photon to
the nucleon goes like
$\overline{u}_{\text{LF}}\gamma^\mu{}v_{\text{LF}}$, where the
light-front spinors are defined in Appendix~\ref{app:spinors}. For
$\mu=+$, this matrix element is suppressed maximally, and thus $J^+$ is
the ``good'' component of the current
\cite{Frankfurt:1993ut,Dashen:1966,Kondratyuk:1984kq}.
We calculate only $J^+$, since it is the most stable.

We do not consider the contribution of higher-order graphs to the deuteron current, such as graphs where the photon couples to a meson or a nucleon while a meson is present. The omission of the meson-exchange currents does not affect the rotational properties of the current, although it does affect the overall values of the deuteron form factors. This is acceptable since we are only concerned with the rotational properties in this work, not in the detailed results.

The neglect of the graphs where the photon couples to a nucleon while a meson is present may affect the rotational properties of the current. This is because the deuteron current shown in Fig.~\ref{form:deut.cur}(a) is not formally conserved \cite{Phillips:1998uk}, but current conservation is necessary (although not sufficient) for the current to have the correct properties under rotation. To construct the conserved current operator associated with a given wave function, the current must include diagrams that are related to the potential used to calculate the wave function. We expect that these diagrams are small since they contain meson propagators, and that neglecting them should not significantly affect the conservation of the deuteron current or the rotational properties of the current.

\subsubsection{Symmetries of the Electromagnetic Current} 

Now, we use symmetries to relate the components of
$\langle k', m'|J^+(q)| k,m\rangle$. Although in light-front dynamics
some generators of Lorentz transformations are dynamic, such as the
generators of rotations from the $x$-$y$ plane into the $x$ direction
are dynamic, we are free to boost in the plane 
perpendicular to the $z$-axis, boost in the plus direction, and rotate
about the $z$-axis, since the generators of those transformations are
kinematic. In addition, we will find how the states and the
current operator transforms under parity and time-reversal.

The kinematic generators allow us to choose which frame to evaluate the
current in. We choose the Breit frame \cite{Frankfurt:1993ut}, where
$q^+=q^-=q_\perp^y=0$ and $q_\perp^x=Q$. We also choose the plus
momentum of the deuteron to be $M_d$, since $M_d$ is value of the plus
momentum in deuteron's rest frame. To simplify notation, we define the
matrix elements of $J^+$ as 
\begin{eqnarray}
I^+_{m',m}(Q)
&=& 
\left\langle
 \frac{\bbox{q}_\perp}{2}, m' \right| J^+(Q) \left|
-\frac{\bbox{q}_\perp}{2}, m 
\right\rangle. \label{eq:ff:defimm}
\end{eqnarray}
This quantity is represented with $I_{m'm}$ instead of $J_{m'm}$ because
$J_{m'm}$ is used to represent the matrix elements of $J$ using the
instant-form spin basis. This distinction is discussed later in this
section.

Next, we consider rotation about the $z$-axis by $\pi$ ($R_z(\pi)$),
parity ($\Pi$), and time-reversal ($\Theta$). First note that
the current operator transforms as
\begin{eqnarray}
R_z(\pi) J^\pm R_z(\pi) &=& J^\pm, \\
\Pi J^\pm \Pi &=& J^\mp, \\
\Theta J^\pm \Theta &=& J^\mp.
\end{eqnarray}
Obviously, we must use both parity and time reversal to retain
the plus component of the current. Now look at how the deuteron states
transform:
\begin{eqnarray}
R_z(\pi)  | -\bbox{k}_\perp, m \rangle
      &=& | +\bbox{k}_\perp, m \rangle (-1)^m, \\
\Theta\Pi | -\bbox{k}_\perp, m \rangle
      &=& | -\bbox{k}_\perp, -m \rangle (-1)^m.
\end{eqnarray}
Although the symmetry operators can introduce extra, constant phase
factors, they are omitted here since they cancel when calculating matrix
elements. These results make it easy to find what each of these
symmetry operators does to the matrix element. Under rotations a
rotation of $\pi$ about the $z$-axis, the current matrix element
transforms to
\begin{eqnarray}
I^+_{m',m}(Q) &\rightarrow& (-1)^{m-m'} \left\langle
-\frac{\bbox{q}_\perp}{2}, m' \right| J^+(Q) \left|
+\frac{\bbox{q}_\perp}{2}, m  \right\rangle. \label{eq:ff:currot}
\end{eqnarray}
Under parity followed by time reversal, the matrix element becomes
\begin{eqnarray}
I^+_{m',m}(Q) &\rightarrow& (-1)^{m-m'} \left\langle
-\frac{\bbox{q}_\perp}{2}, -m \right| J^+(Q) \left|
+\frac{\bbox{q}_\perp}{2}, -m' \right\rangle. \label{eq:ff:curpith} \\
\end{eqnarray}
Since $J^\mu$ is Hermitian, we can also take the complex conjugate of
the matrix element to get
\begin{eqnarray}
I^+_{m',m}(Q) &\rightarrow&  \left\langle
-\frac{\bbox{q}_\perp}{2}, m \right| J^+(Q) \left|
+\frac{\bbox{q}_\perp}{2}, m' \right\rangle. \label{eq:ff:curcc}
\end{eqnarray}

The appropriate combinations of
Eqs.~(\ref{eq:ff:currot}-\ref{eq:ff:curcc}) give \cite{Chung:1988my}
\begin{eqnarray}
I^+_{m',m}(Q)
&=& (-1)^{m-m'} I^+_{-m',-m }(Q), \label{eq:ff:blah1} \\
&=& (-1)^{m-m'} I^+_{ m , m'}(Q). \label{eq:ff:blah2}
\end{eqnarray}
The same relations also apply for $J^+_{m'm}$ matrix elements.
Eqs.~(\ref{eq:ff:blah1}) and (\ref{eq:ff:blah2})
imply that of the nine possible matrix elements of
$J^+$, there are only four independent components. We choose those
components to be $I^+_{11}$, $I^+_{10}$, $I^+_{1-1}$, and $I^+_{00}$. It
is helpful to express the matrix elements in a matrix to see the
symmetry properties explicitly.
\begin{eqnarray}
I^+_{m',m} &=&
\left(\begin{array}{ccc}
 I^+_{1 1} &  I^+_{10} & I^+_{1-1} \\
-I^+_{1 0} &  I^+_{00} & I^+_{1 0} \\
 I^+_{1-1} & -I^+_{10} & I^+_{1 1} \\
\end{array}\right). \label{eq:deutcurlf}
\end{eqnarray}

\subsubsection{Rotational Invariance and the Angular Condition}
\label{sec:rotinv}

This is as far as we can go with light-front dynamics, but there
should be an additional redundancy in our matrix elements. We have
derived four independent components, whereas in a fully covariant
framework there are only three form factors. The resolution of this
conflict is that full rotational invariance imposes an
{\em angular condition} on the light-front matrix elements.

The angular condition can be found by using the deuteron polarization
vectors
for the Breit frame to calculate the current given in
Eq.~(\ref{eq:deutcurrent}) in terms of $F_1$, $F_2$, and $G_1$.
By comparing that result with Eq.~(\ref{eq:deutcurlf}), we obtain
linear relations between the current matrix elements ($I^+_{11}$,
$I^+_{10}$, $I^+_{1-1}$, and $I^+_{00}$) and the form factors ($F_1$,
$F_2$, and $G_1$) and also find that in general there is a deviation
from the angular condition, which we denote with $\Delta$, given by
\cite{Cardarelli:1995yq}
\begin{eqnarray}
\Delta &=& -I^+_{00} + (1+2\eta) I^+_{11} + I^+_{1-1} - 2\sqrt{2\eta}
I^+_{10},
\end{eqnarray}
where $\eta$ is defined by Eq.~(\ref{eq:defeta}). Since $\Delta$
vanishes when the deuteron current transforms correctly
under rotations, we interpret $\Delta$ as a measure of the extent to
which the current matrix elements transform incorrectly.

The form factors are overdetermined by the current matrix elements, which means there are many different ways to express the form factors in terms of the current matrix elements. In order to make a definite prescription, one the current matrix elements (or a linear combination of matrix elements) is identified as ``bad'', and the angular condition is used to eliminate it from the expressions for the form factors. This classification of matrix elements as ``good'' or ``bad'' is similar to the one made for choosing which component of the current to use.

When $\Delta$ is zero, all the prescriptions must be equivalent, while a non-zero $\Delta$ means that the form factors depend on which prescription is chosen. Since $\Delta$ is non-zero in general, it is important to choose the best prescription to obtain the form factor. 

We consider four different prescriptions \cite{Cardarelli:1995yq} in this work.
This allows us to study how sensitively the form factors depend on the
prescription used. First, Grach and
Kondratyuk (GK) derive a relation in which $I^+_{00}$ is considered
bad. Using the angular condition to eliminate the $I^+_{00}$ for the
equations, they obtain
\cite{Grach:1984hd}
\begin{eqnarray}
G_{0,GK} &=& \frac{1}{3} \left[ (3-2\eta) I^+_{11}
+ 2\sqrt{2\eta} I^+_{10} + I^+_{1-1} \right], \\
G_{1,GK} &=& 2 \left( I^+_{11} - \frac{1}{\sqrt{2\eta}} I^+_{10} \right), \\
G_{2,GK} &=& \frac{2\sqrt{2}}{3} \left( -\eta I^+_{11}
+ \frac{2\eta} I^+_{10} - I^+_{1-1} \right).
\end{eqnarray}
Brodsky and Hiller (BH) use a prescription where $I^+_{11}$ is bad,
which results in \cite{Brodsky:1992px}
\begin{eqnarray}
G_{0,BH} &=& \frac{1}{3(1+2\eta)} \left[ (3-2\eta) I^+_{00}
+ 8\sqrt{2\eta} I^+_{10} + 2(2\eta-1) I^+_{1-1} \right], \\
G_{1,BH} &=& \frac{2}{1+2\eta} \left[ I^+_{00} - I^+_{1-1} 
+ (2\eta-1) \frac{I^+_{10}}{\sqrt{2\eta}} \right], \\
G_{2,BH} &=& \frac{2\sqrt{2}}{3(1+2\eta)} \left[ \sqrt{2\eta} I^+_{10}
- \eta I^+_{00} - (1+\eta) I^+_{1-1} \right].
\end{eqnarray}

Frankfurt, Frederico, and Strikman (FFS) start by analyzing the deuteron
current using an equal-time spin basis, where the current matrix
elements
are $J^+_{m'm}$ \cite{Frankfurt:1993ut}. Their prescription uses the
Cartesian basis to find that $J^+_{zz}$ is the bad current. Upon
transformation to the spherical basis and use of the Melosh
transformation \cite{Kondratyuk:1984kq,Melosh:1974cu} to relate
$J^+_{m'm}$ to $I^+_{m'm}$, they find 
\begin{eqnarray}
G_{0,FFS} &=& \frac{1}{3(1+\eta)} \left[
(2\eta+3) I^+_{11} + 2 \sqrt{2\eta} I^+_{10} - \eta I^+_{00}
+ (2\eta + 1) I^+_{1-1} \right], \\
G_{1,FFS} &=& \frac{1}{1+\eta} \left[ I^+_{11} + I^+_{00} - I^+_{1-1}
-\frac{2(1-\eta)}{\sqrt{2\eta}} I^+_{10} \right], \\
G_{2,FFS} &=& \frac{\sqrt{2}}{3(1+\eta)} \left[ -\eta I^+_{11}
+ 2\sqrt{2\eta} I^+_{10} - \eta I^+_{00} - (\eta+2) I^+_{1-1} \right].
\end{eqnarray}
Chung, Polyzou, Coester, and Keister (CCKP) choose the canonical
expressions for the form factors $G_0$, $G_1$, and $G_2$ in terms of the
equal-time current as the starting point
\cite{Chung:1988my,Coester:1975hj}. They use rotations and the Melosh
transformation to express the equal-time current matrix elements in
terms of the light-front current matrix elements, and find that
\begin{eqnarray}
G_{0,CCKP} &=& \frac{1}{6(1+\eta)} \left[
(3-2\eta) (I^+_{11} + I^+_{00} ) 
+ 10 \sqrt{2\eta} I^+_{10} + (4\eta-1) I^+_{1-1} \right], \\
G_{1,CCKP} &=& G_{1,FFS}, \label{eq:cckpffs1} \\
G_{2,CCKP} &=& G_{2,FFS}. \label{eq:cckpffs2}
\end{eqnarray}

Note that all the prescriptions reviewed here are related by the angular condition. When $\Delta$ is zero, each prescription gives the same results.

\subsection{Impulse Approximation on the Light Front} \label{ff:impapp}

We now are ready to relate the deuteron wave function to the current
expressed in Eq.~(\ref{eq:ff:defimm}). The deuteron wave functions
solved in Chapter~\ref{ch:pionly} are used. We use the representation of
the wave function in the spin $|\bbox{p}_\perp,p^+,m_1,m_2\rangle$
basis, as discussed in Sec.~\ref{sec:furtherlfnn}. Note that these spins
are expressed in the usual Bjorken and Drell representation
\cite{Bjorken:1964}.
By inserting a two complete sets of states into Eq.~(\ref{eq:ff:defimm})
and making the momentum of
particle 1 explicit, we get
\begin{eqnarray}
J^+_{m',m}(q) &=& \int d^2p_\perp dp^+
\sum_{m'_1,m_1,m_2}
\left\langle \frac{\bbox{q}_\perp}{2},m' \left.|
p^+,\bbox{p}_\perp+\frac{\bbox{q}_\perp}{2}, m'_1, m_2
\right. \right\rangle
\nonumber \\ && \qquad 2
\left\langle p^+,\bbox{p}_\perp+\frac{\bbox{q}_\perp}{2}, m'_1 \left|
 J_{(S)}^+(\bbox{q}_\perp)
\left| p^+,\bbox{p}_\perp-\frac{\bbox{q}_\perp}{2}, m_1 
\right.
\right.
\right\rangle
\nonumber \\ && \qquad
\left\langle p^+,\bbox{p}_\perp-\frac{\bbox{q}_\perp}{2}, m_1,
m_2 \left| -\frac{\bbox{q}_\perp}{2}, m
\right.\right\rangle, \label{eq:ff:curwf}
\end{eqnarray}
where the spin directions of particles 1 and 2 are labeled $m_1$ and
$m_2$, respectively. Also, since the deuteron is an isoscalar
combination of nucleons, the isovector component of the nucleon current
does not contribute and the isoscalar nucleon current is the same for
both nucleons. This allows us to simply double the isoscalar current of
particle 1 instead of using the isoscalar currents of both particle 1
and 2.

We want to boost the deuteron wave functions to the rest frame, since
that is where the wave functions are calculated. Note that the boosts in
the $x$-$y$ plane 
transforms a general vector $(\bbox{k}_\perp,k^+)$ by
$\bbox{k}_\perp\rightarrow\bbox{k}_\perp-\bbox{q}_\perp\frac{k^+}{q^+}$,
leaving $k^+$ unchanged. This means that the boost that puts the
deuteron in its rest frame transforms the individual nucleon momentum by
\begin{eqnarray}
\bbox{p}_\perp-\frac{\bbox{q}_\perp}{2} \rightarrow 
\bbox{p}_\perp-(1-x)\frac{\bbox{q}_\perp}{2},
\end{eqnarray}
where we use the Bjorken $x$-variable $x=\frac{p^+}{M_d}$.  Note that
the spin labels do not change.  This is because the boost is in the
perpendicular direction, so the spin labels (defined to point in the $z$
direction) are not affected.

This allows Eq.~(\ref{eq:ff:curwf}) to be written as \cite{Arndt:1999wx}
\begin{eqnarray}
J^+_{m',m}(q) &=& \int d^2p_\perp dp^+
\sum_{m'_1,m_1,m_2}
\left\langle m' \left| p^+,\bbox{p}_\perp+(1-x)\frac{\bbox{q}_\perp}{2},
m'_1, m_2 \right.\right\rangle
\nonumber \\ && \qquad
2
\left\langle p^+,\bbox{p}_\perp+\frac{\bbox{q}_\perp}{2}, m'_1 \left|
 J_{(S)}^+(\bbox{q}_\perp)
\left| p^+,\bbox{p}_\perp-\frac{\bbox{q}_\perp}{2}, m_1 
\right.\right.\right\rangle
\nonumber \\ && \qquad
\left\langle\left. p^+,\bbox{p}_\perp-(1-x)\frac{\bbox{q}_\perp}{2}, m_1,
m_2 \right| m \right\rangle. \label{eq:ff:deutcurwf}
\end{eqnarray}
We have dropped explicit mention of the deuteron's momentum.

Before we continue, we note that we can write the nucleon current matrix
elements using the $u$ spinors,
\begin{eqnarray}
&&\left\langle p^+,\bbox{p}_\perp+\frac{\bbox{q}_\perp}{2}, m'_1 \left|
 J_{(S)}^+(\bbox{q}_\perp)
\left| p^+,\bbox{p}_\perp-\frac{\bbox{q}_\perp}{2}, m_1 
\right.\right.\right\rangle
\nonumber\\&&\qquad=
\overline{u}_{\text{BD}}\left(p^+,\bbox{p}_\perp+\frac{\bbox{q}_\perp}{2},
m'_1 \right)
 J_{(S)}^+(\bbox{q}_\perp)
u_{\text{BD}}\left(p^+,\bbox{p}_\perp-\frac{\bbox{q}_\perp}{2},
m_1\right),
\end{eqnarray}
where we have used the ``BD'' subscript to denote these as Bjorken and
Drell spinors. Since the nucleons are on-shell, as they must be in a
Hamiltonian theory, we interpret 
$u_{\text{BD}}\left(p^+,\bbox{p}_\perp,m\right)$ as 
$u_{\text{BD}}\left(
p_x=p_{\perp,x},
p_y=p_{\perp,y},
p_z=\frac{p^+-p^-}{2}
\right)$, where $p^-=\frac{M^2+p_\perp^2}{p^+}$.

However, the matrix elements can be calculated easily if we convert the
Bjorken and Drell spinors to light-front spinors. This transformation is
formally accomplished with the Melosh transformation $R_M$, given by
\cite{Chung:1988my,Frankfurt:1993ut},
\begin{eqnarray}
R_M &=& \frac{p^+ + M - i
\bbox{\sigma}\cdot(\bbox{n}\times\bbox{p}_\perp)}{\sqrt{(p^++M)^2+p_\perp^2}},
\end{eqnarray}
where $\bbox{n}$ points in the direction of the light front. Note that
the Melosh transformation is unitary.

This transformation converts a Bjorken and Drell spinor to a light-front
spinor in the following manner,
\begin{eqnarray}
u_{\text{LF}}(p^+,\bbox{p}_\perp,m) &=& \sum_{m'}(R_M^\dagger)_{m,m'}
u_{\text{BD}}(p^+,\bbox{p}_\perp,m').
\end{eqnarray}
Using this transformation, we rewrite Eq.~(\ref{eq:ff:deutcurwf}) as
\begin{eqnarray}
J^+_{m',m}(q) &=& \int d^2p_\perp dp^+
\sum_{m'_1,m_1,m_2}
\left[
\sum_{m^{\prime\prime}_1}
\left\langle m' \left| p^+,\bbox{p}_\perp+(1-x)\frac{\bbox{q}_\perp}{2},
m''_1, m_2 \right.\right\rangle
(R_M)_{m^{\prime\prime}_1,m'_1}
\right]
\nonumber \\ && \qquad
2
\overline{u}_{\text{LF}}\left( p^+,
\bbox{p}_\perp+\frac{\bbox{q}_\perp}{2}, m'_1 \right)
 J_{(S)}^+(\bbox{q}_\perp)
u_{\text{LF}}\left(p^+,\bbox{p}_\perp-\frac{\bbox{q}_\perp}{2}, m_1 \right)
\nonumber \\ && \qquad
\left[
\sum_{m^{\prime\prime\prime}_1}
(R_M^\dagger)_{m^{\prime\prime\prime}_1,m_1}
\left\langle\left. p^+,\bbox{p}_\perp-(1-x)\frac{\bbox{q}_\perp}{2},
m^{\prime\prime\prime}_1, m_2 \right| m \right\rangle
\right]
\label{eq:ff:deutcurwf2}.
\end{eqnarray}

\subsection{The Nucleon Form Factors}

From Lorentz covariance, current conservation, parity invariance, and time reversal invariance
\cite{Zuilhof:1980ae,Rupp:1990sg}, the isoscalar part of the nucleon
current can be expressed as
\begin{eqnarray}
J_{(S)}^\mu(q) &=& \gamma^\mu F^{(S)}_1(q)
+ i \frac{\sigma^{\mu\nu} q_\nu}{2M}  F^{(S)}_2(q).
\end{eqnarray}

Taking the matrix elements of $J^+$ with the light-front spinors gives
\begin{eqnarray}
\overline{u}(k', \lambda') J_{(S)}^+(\bbox{q}_\perp) u(k, \lambda)
&=& F^{(S)}_1(q)
\overline{u}_{\text{LF}}(k', \lambda') \gamma^+ u_{\text{LF}}(k, \lambda)
\nonumber\\&&
+ F^{(S)}_2(q)
\frac{q_\perp^i}{2M}
\overline{u}_{\text{LF}}(k',\lambda') \gamma^+ \gamma_\perp^i
u_{\text{LF}}(k,\lambda),
\label{fermcurrent}
\end{eqnarray}
where we have parameterized
\begin{eqnarray}
{k'}^+ &=& k^+, \\
\bbox{k}_\perp  &=& \bbox{p}_\perp + \frac{\bbox{q}_\perp}{2},\\
\bbox{k}'_\perp &=& \bbox{p}_\perp - \frac{\bbox{q}_\perp}{2}.
\end{eqnarray}
We use the representation of the spinors given in
Appendix~\ref{app:spinors} to simplify the spinor matrix elements
appearing in Eq.~\ref{fermcurrent}. First consider
\begin{eqnarray}
\overline{u}_{\text{LF}}(k', \lambda') \gamma^+ u_{\text{LF}}(k, \lambda)
&=&
\frac{1}{M k^+}
\chi^\dagger_{\text{LF},\lambda'}
\left[ ( M - \bbox{\alpha}^\perp\cdot{\bbox{k}'}^\perp ) \Lambda_+ 
+ k^+ \Lambda_- \right]
\gamma^+
\nonumber\\&& \qquad\qquad\times
\left[ \Lambda_- 
( M + \bbox{\alpha}^\perp\cdot\bbox{k}^\perp )
+ \Lambda_+ k^+ \right] \chi_{\text{LF},\lambda}, \\
&=& \frac{k^+}{M} \delta_{\lambda'\lambda}.
\end{eqnarray}
Next,
\begin{eqnarray}
\overline{u}_{\text{LF}}(k', \lambda') \gamma^+ \gamma_\perp^-
u_{\text{LF}}(k, \lambda)
&=&
\frac{1}{M k^+}
\chi^\dagger_{\text{LF},\lambda'}
\left[ ( M - \bbox{\alpha}^\perp\cdot{\bbox{k}'}^\perp ) \Lambda_+ 
+ k^+ \Lambda_- \right]
\gamma^+ \gamma^i
\nonumber\\&& \qquad\qquad\times
\left[ \Lambda_- 
( M + \bbox{\alpha}^\perp\cdot\bbox{k}^\perp )
+ \Lambda_+ k^+ \right] \chi_{\text{LF},\lambda}, \\
&=& \frac{k^+}{M}
\chi^\dagger_{\text{LF},\lambda'} \gamma^+ \gamma_\perp^i
\chi_{\text{LF},\lambda}, \\
&=& - i \epsilon^{ij3} 
\frac{k^+}{M} \chi^\dagger_{\lambda'} \sigma^j \chi_\lambda.
\end{eqnarray}
Combining these, we find \cite{Frankfurt:1993ut,Kondratyuk:1984kq}
\begin{eqnarray}
\langle k', \lambda' | J_{(S)}^+(\bbox{q}_\perp) | k, \lambda \rangle
&=& \frac{k^+}{M}
\chi_{\lambda'}^\dagger \left[
F^{(S)}_1(q) - \frac{1}{2M} F^{(S)}_2(q)
\left( i q^i_\perp \epsilon^{ij3} \sigma^j_\perp \right)
\right]
\chi_\lambda \label{eq:ff:nuccurrent}
\end{eqnarray}

We can rewrite Eq.~(\ref{eq:ff:nuccurrent}) as
\begin{eqnarray}
\langle k', \lambda' | J_{(S)}^+(\bbox{q}_\perp) | k, \lambda \rangle
&=& 
F^{(S)}_1(q)
\langle k', \lambda' | J_{(1S)}^+(\bbox{q}_\perp) | k, \lambda \rangle
+ \nonumber\\&&
F^{(S)}_2(q)
\langle k', \lambda' | J_{(2S)}^+(\bbox{q}_\perp) | k, \lambda \rangle,
\end{eqnarray}
which, when inserted into Eq.~(\ref{eq:ff:deutcurwf2}), gives
\begin{eqnarray}
I^+_{m',m}(q) &=&
F^{(S)}_1(q) I^+_{(1)m',m}(q) +
F^{(S)}_2(q) I^+_{(2)m',m}(q),
\end{eqnarray}
where we define
\begin{eqnarray}
I^+_{(i)m',m}(q) &=& \int d^2p_\perp dp^+
\sum_{m'_1,m_1,m_2}
\langle m' | p^+,\bbox{p}_\perp+(1-x)\frac{\bbox{q}_\perp}{2},
m'_1, m_2 \rangle
\nonumber \\ && \qquad
2
\langle p^+,\bbox{p}_\perp+\frac{\bbox{q}_\perp}{2}, m'_1 |
 J_{(iS)}^+(\bbox{q}_\perp)
| p^+,\bbox{p}_\perp-\frac{\bbox{q}_\perp}{2}, m_1 \rangle
\nonumber \\ && \qquad
\langle p^+,\bbox{p}_\perp-(1-x)\frac{\bbox{q}_\perp}{2}, m_1,
m_2 | m \rangle, \label{eq:ff:i12}
\end{eqnarray}
for $i=1,2$. Note that both $J^+_{(1)m',m}(q)$ and $J^+_{(2)m',m}(q)$
must satisfy the same equations as $J^+_{m',m}(q)$ does. In particular,
this means that the angular condition should apply to $J^+_{(1)m',m}(q)$
and $J^+_{(2)m',m}(q)$ independently, so we consider the deviation from
the angular condition for each.

\begin{figure}
\begin{center}
\epsfig{angle=0,width=5.5in,height=5.5in,file=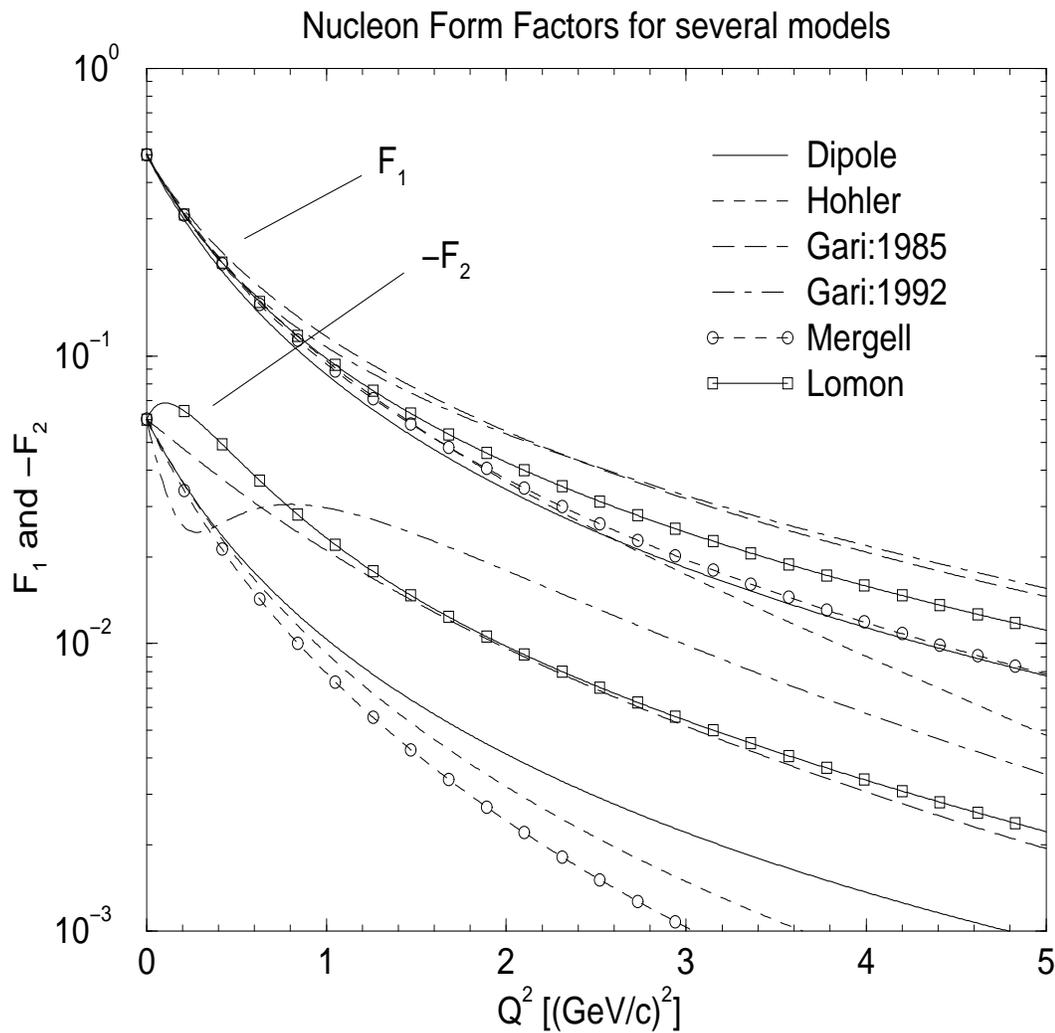}
\caption[The $F_1$ and $-F_2$ isoscalar nucleon form factors for six
different models: the dipole model, Hohler, Gari:1985, Gari:1992,
Mergell, and Lomon.]
{The $F_1$ and $-F_2$ isoscalar nucleon form factors for six
different models: the dipole model,
Hohler \cite{Hohler:1976ax},
Gari:1985 \cite{Gari:1985ia},
Gari:1992 \cite{Gari:1992qw,Gari:1992},
Mergell \cite{Mergell:1996bf}, and
Lomon \cite{Lomon:2001ga}.
\label{form:diag:diffNucFF}}
\end{center}
\end{figure}

There are many parameterizations of the isoscalar nucleon form factors
$F^{(S)}_1(q)$ and $F^{(S)}_2(q)$. Since the measurement of the
electron-nucleon cross section is difficult, the data have large errors
and are consistent with several different models of the nucleon form
factors. Some of the models representative of those proposed in the
literature are: the dipole
model, fit 8.2 of Hohler \cite{Hohler:1976ax}, Gari:1985 \cite{Gari:1985ia},
model 3 of Gari:1992 \cite{Gari:1992qw,Gari:1992}, best fit for the
multiplicative parameterization of Mergell \cite{Mergell:1996bf}, and
model DR-GK(1) of Lomon \cite{Lomon:2001ga}. The $F^{(S)}_1(q)$ and
$F^{(S)}_2(q)$ form factors for each of these models are shown in
Fig.~\ref{form:diag:diffNucFF}.

We can relate the isovector form factors to $G_{Ep}$, $G_{Mp}$,
$G_{En}$, and $G_{Mn}$, the proton electric, proton
magnetic, neutron electric, and neutron magnetic form factors,
respectively, with \cite{deJager:1999ce}
\begin{eqnarray}
F^{(S)}_1 &=& 
\frac{ G_{Ep}+G_{En}+ \tau \left(G_{Mp} + G_{Mn} \right)}{2(1+\tau)}, \\
F^{(S)}_2 &=& 
\frac{-G_{Ep}-G_{En}+ G_{Mp} + G_{Mn}}{2(1+\tau)},
\end{eqnarray}
where $\tau\equiv\frac{Q^2}{4M}$. The value of $\tau$ is approximately 1
at a momentum transfer of about 5~GeV$^2$, the upper range of momentum
transfers that we consider. Since the overall magnitudes of the form
factors are similar at this momentum transfer, it is important to
measure each of the form factors with the same accuracy and cover the
same range of momentum transfers. Currently, the most poorly known form
factor is $G_{En}$, both in terms of the magnitude of the error and in
the number of data points \cite{Lomon:2001ga}.

In Section~\ref{ch:ff:results}, we will find that for momentum transfers
greater than about 2~GeV$^2$, the spread in the values of the deuteron
form factors due to the breaking of rotational invariance on the light
front is smaller than the spread in values due to using the various
nucleon form factors. It is uncertainty of the nucleon form factors, not
the use of the light front, that limits the accuracy of the deuteron
form factors at large momentum transfers. Only more accurate
measurements of the nucleon form factors, especially $G_{En}$, will
allow for more accurate deuteron form factor calculations. 

\section{Axial Form Factors} \label{sec:axialstuff}

The formalism used for the axial current and form factor is very similar
to that used for the electromagnetic current and form factor. Thus, most
of the discussion from the previous section carries over here with only
slight modifications. We highlight only the differences.

The derivation of the symmetries of the axial current matrix elements is
almost the same as in Section~\ref{sec:nplfcalcem}, with the
exception that under parity, the axial current picks up a negative sign
\begin{eqnarray}
\Pi J_5^\pm \Pi &=& -J_5^\mp.
\end{eqnarray}
When this is propagated through the algebra, we find that
\begin{eqnarray}
I^+_{(5)m',m}(Q)
&=& -(-1)^{m-m'} I^+_{(5)-m',-m }(Q), \label{eq:ff:blah1axial} \\
&=& (-1)^{m-m'} I^+_{(5) m , m'}(Q). \label{eq:ff:blah2axial}
\end{eqnarray}
Eqs.~(\ref{eq:ff:blah1axial}) and (\ref{eq:ff:blah2axial})
imply that of the nine possible matrix elements of
$J_5^+$, there are only two independent components. We choose those
components to be $I^+_{(5)11}$, and $I^+_{(5)10}$. It
is helpful to express the matrix elements in a matrix to see the
symmetry properties explicitly.
\begin{eqnarray}
I^+_{(5)m',m} &=&
\left(\begin{array}{ccc}
 I^+_{(5)1 1} &  I^+_{(5)10} &  0            \\
-I^+_{(5)1 0} &  0           & -I^+_{(5)1 0} \\
 0            &  I^+_{(5)10} & -I^+_{(5)1 1} \\
\end{array}\right). \label{eq:deutcurlfaxial}
\end{eqnarray}

We have derived two independent components, but an analysis of the
covariant theory shows that only one deuteron form factor ($F_A$)
contributes for the plus component of the axial current 
\cite{Frederico:1991vb}. This implies that the requirement of full
rotational invariance imposes an angular condition on the light-front
axial current matrix elements. 
The deviation from the angular condition, denoted by $\Delta$, given by
\cite{Frederico:1991vb},
\begin{eqnarray}
\Delta &=& \frac{\sqrt{2\eta}}{2} I^+_{(5)11} - I^+_{(5)10}.
\end{eqnarray}

Since the deuteron axial form factor is overdetermined by the current
matrix elements, we need to classify the current matrix elements as
either ``good'' or ``bad'' to eliminate ambiguity. We consider two such
choices.

Frankfurt, Frederico, and Strikman (FFS) find that the $J^+_{(5)zz}$ is
the bad matrix element \cite{Frankfurt:1993ut}. After transforming to
the spherical basis and using the Melosh transformation, they find that
\begin{eqnarray}
F_A &=& \frac{1}{2(1+\eta)} \left( I^+_{(5)11} + \sqrt{2\eta}
I^+_{(5)10} \right).
\end{eqnarray}
Frederico, Henley, and Miller (FHM) use the behavior of the matrix
elements in the non-relativistic limit to determine that the bad element
is $I^+_{(5)10}$ \cite{Frederico:1991vb}. This means that
\begin{eqnarray}
F_A &=& \frac{1}{2} I^+_{(5)11}.
\end{eqnarray}

The current matrix elements are calculated using the nucleon
axial current. The general form of nucleon axial current is given by
\cite{Frederico:1991vb},
\begin{eqnarray}
J_{(5)n}^\mu &=& \gamma^\mu \gamma^5 F_A^n + q^\mu \gamma^5 F_P^n.
\label{eq:nucaxialff}
\end{eqnarray}
Since we choose to $\mu=+$ and work in the Breit frame, where $q^+=0$,
Eq.~(\ref{eq:nucaxialff}) reduces to
\begin{eqnarray}
J^+_{(5)n} &=& \gamma^+ \gamma^5 F_A^n. \label{eq:nucaxialffplus}
\end{eqnarray}
The light-front spinor matrix elements of Eq.~(\ref{eq:nucaxialffplus})
can be computed using expressions given in
Appendix~\ref{app:spinors}. The result is
\begin{eqnarray}
\langle k', \lambda' | J_{(5)n}^+(\bbox{q}_\perp) | k, \lambda \rangle
&=& \frac{k^+}{M} F^n_A(q) 
\chi_{\lambda'} \sigma^3 \chi_\lambda \label{eq:ff:nucaxialcurrent}.
\end{eqnarray}

We consider only one model for the nucleon axial form factor since the
deuteron axial current has such a simple dependence on it. We choose to
use the dipole model
\begin{eqnarray}
F_A(Q^2) &=& \frac{F_A(0)}{\left(1+\frac{Q^2}{M_A^2}\right)^2},
\end{eqnarray}
where $M_A$ is the axial mass. For our calculations, we use the value
for the axial mass determined by Liesenfeld {\it et
al}.~\cite{Liesenfeld:1999mv}.

\section{Results for the Form Factors} \label{ch:ff:results}

We use the deuteron wave functions obtained for the light-front
nucleon-nucleon potential in Chapter~\ref{ch:pionly} to calculate the
deuteron currents and form factors. This gives a solution where
light-front dynamics is used consistently throughout. For the
potential, we choose the light-front nucleon-nucleon potential with
$f_\sigma=1.2815$. We have verified that the results do not change
significantly when $f_\sigma=1.22$ is used. 

\begin{figure}
\begin{center}
\epsfig{angle=0,width=5.5in,height=5.5in,file=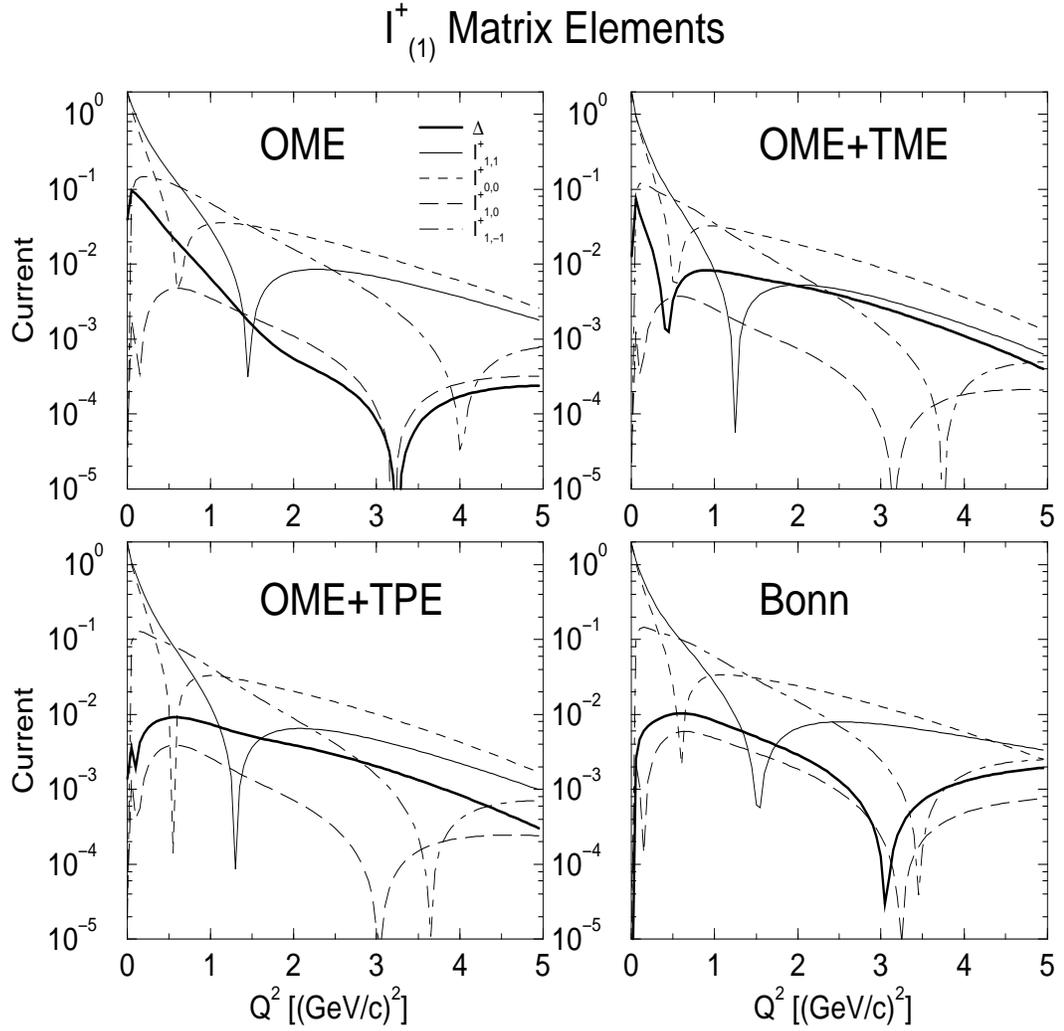}
\caption{The matrix elements of $I^+_{(1)m'm}$, the component of the
electromagnetic current which multiplies the nucleon $F_1$ form factor,
calculated with the wave function from the OME, OME+TME, OME+TPE, and
Bonn potentials.
\label{form:diag:j1}}
\end{center}
\end{figure}

Figure~\ref{form:diag:j1} show the currents and the associated angular
condition for $I^+_{(1)}$, given by Eq.~(\ref{eq:ff:i12}), for several
different deuteron wave functions. Results are shown for the wave
function from the OME, OME+TME, and OME+TPE potentials (calculated in 
Chapter~\ref{ch:pionly}), and the parameterization of the deuteron wave
function for the energy independent Bonn potential
\cite{Machleidt:1987hj}. The current matrix elements (but not $\Delta$)
are approximately the same regardless of which wave function is used. This
consistency is important, since it verifies that the gross features of
all the models are the same.

We find that $\Delta$ for $I^+_{(1)}$ is much smaller than the largest
matrix elements when using the OME wave function. This means that the
matrix elements of the $I^+_{(1)}$ current operator
transform very well under rotations. This is
somewhat surprising, since we found earlier that the binding energies
for the OME wave functions have a large splitting, indicating that OME
wave functions transforms poorly under rotations. 

To understand this apparent discrepancy, note that some diagrams are not included in the calculation of the current although they are required for manifest current conservation. This problem is addressed in Section~\ref{sec:nplfcalcem}. The missing graphs should have a small effect on the rotational properties of the current, but they may have a large effect on $\Delta$. In particular, since the OME+TME current is missing more graphs than the OME current, it is reasonable that the OME+TME current transforms more poorly under rotations than the OME current.

Comparing the current calculated with the OME wave function to those
calculated with other potentials, we find that for momentum transfers of
more than 1~GeV$^2$ the OME  $I^+_{(1)}$ current has the best
transformation properties under rotation of all the $I^+_{(1)}$
currents shown.

For smaller momentum transfers, the transformation properties of the Bonn
and OME+TME wave functions are the best. This is expected, since in the
limit of no momentum transfer, the current $I^+_{(1)m'm}$ is simply the
overlap of deuteron wave functions, $\langle m'|m\rangle$. If the
initial and final states have the same mass, the matrix element is
simply $\delta_{m'm}$, which satisfies the angular condition. However,
if the states do not have the same mass (which implies that $m'\neq m$),
there will be a non-zero overlap between the two states, which violates
the angular condition. Since the masses of the deuteron states are
exactly the same for the Bonn wave function, and approximately the same
for the OME+TPE wave function, they have a small $\Delta$ at low
momentum transfer. However, the OME wave functions, having the
largest mass splitting, have the largest $\Delta$ at low momentum
transfers.

\begin{figure}
\begin{center}
\epsfig{angle=0,width=5.5in,height=5.5in,file=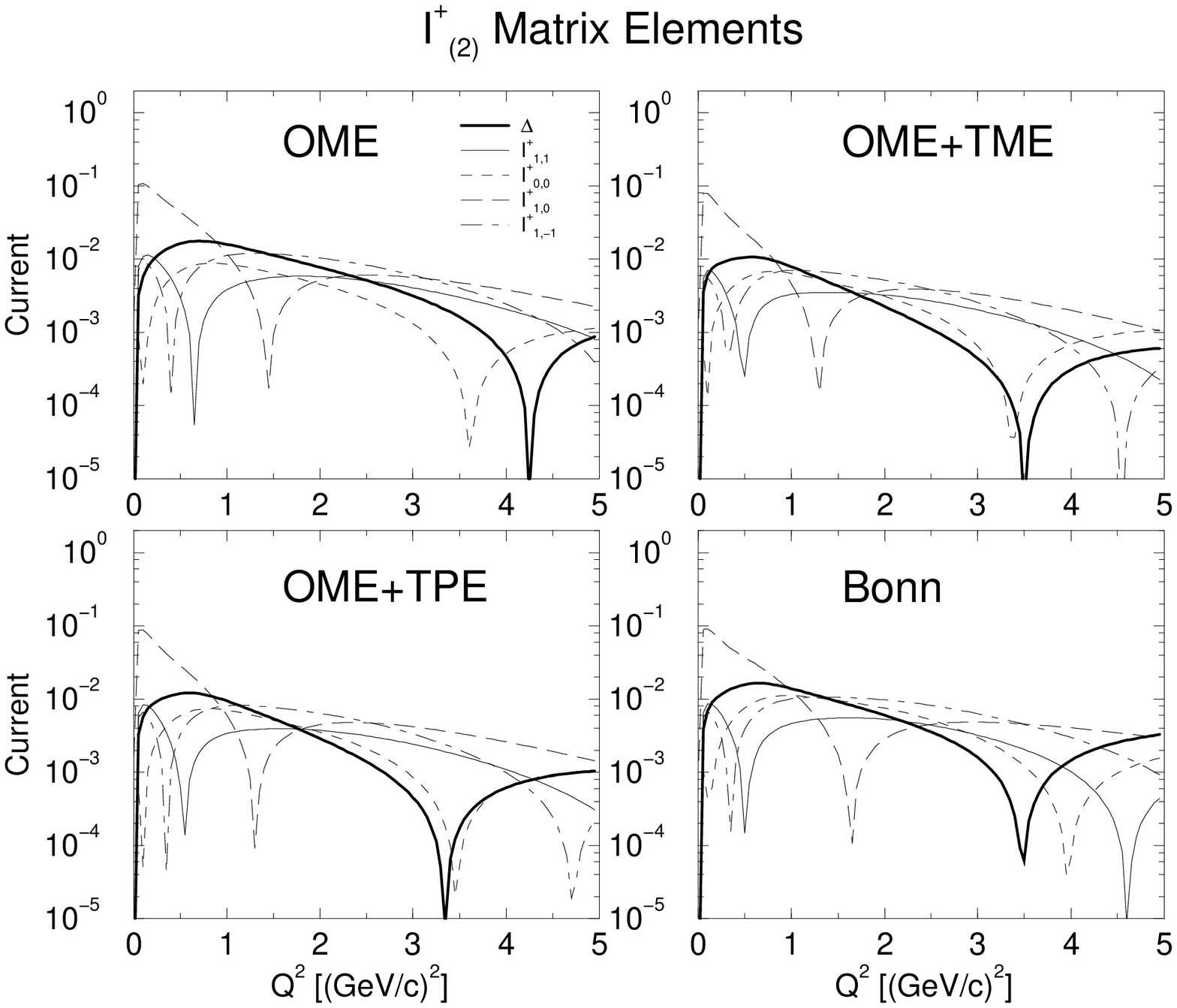}
\caption{The matrix elements of $I^+_{(2)m'm}$, the component of the
electromagnetic current which multiplies the nucleon $F_2$ form factor,
calculated with the wave function from the OME, OME+TME, OME+TPE, and
Bonn potentials.
\label{form:diag:j2}}
\end{center}
\end{figure}

\begin{figure}
\begin{center}
\epsfig{angle=0,width=5.5in,height=5.5in,file=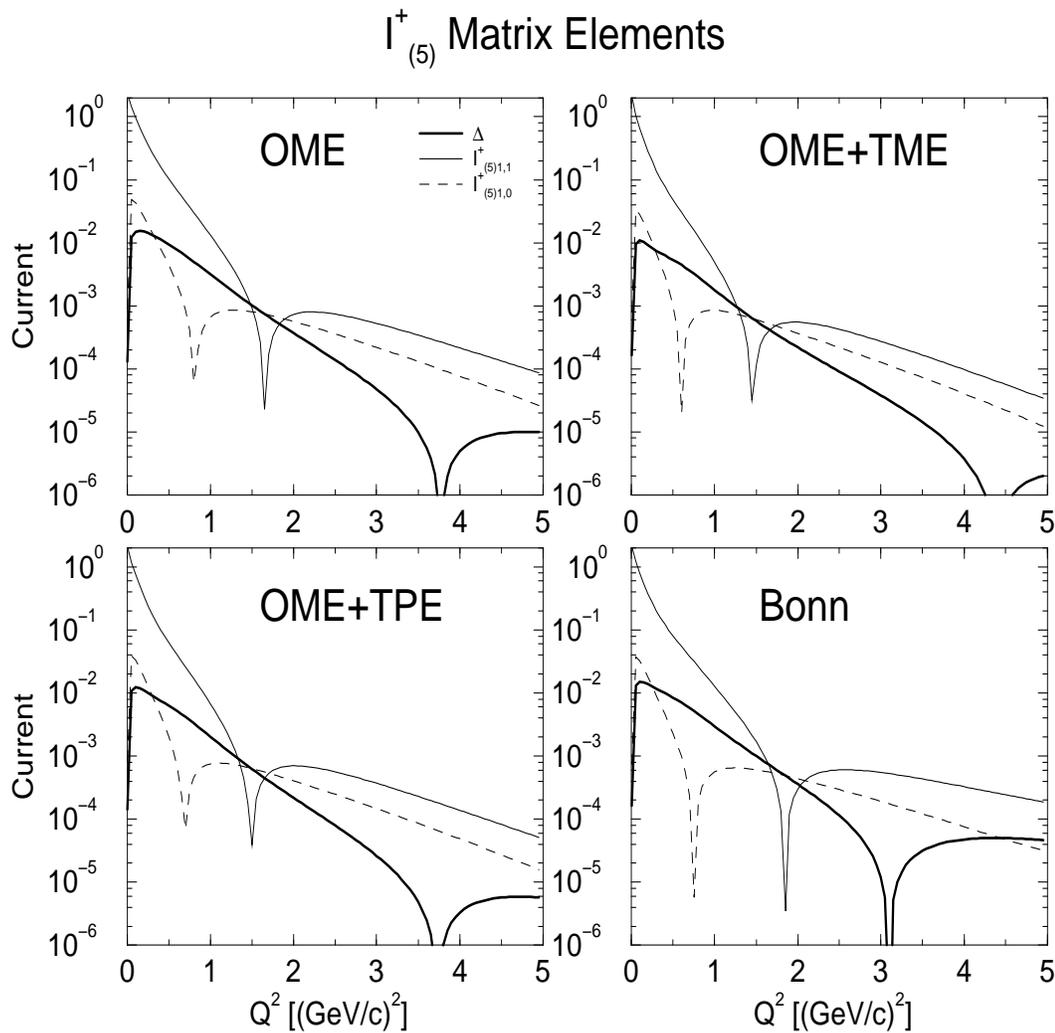}
\caption{The matrix elements of $I^+_{(5)m'm}$, the deuteron axial
current including the nucleon axial form factor, calculated with the
wave function from the OME, OME+TME, OME+TPE, and Bonn potentials.
\label{form:diag:j5}}
\end{center}
\end{figure}

Figures~\ref{form:diag:j2} and \ref{form:diag:j5} show the current
matrix elements and the angular condition for $I^+_{(2)}$ and
$I^+_{(5)}$, respectively. The general features of these figures are the
same as in Figure~\ref{form:diag:j1}, with one important exception. In
both figures, the $\Delta$ for the
OME wave function has about the same magnitude as the $\Delta$'s for the
other wave functions. This means that the rotational properties
of $I^+_{(2)}$ and $I^+_{(5)}$ currents are approximately the same
regardless of which wave function is used. This is a surprising result
since it indicates that the rotational properties of the current depend
more on how the current is constructed than on which wave function is
used.

In Fig.~\ref{form:diag:j2}, we find that the magnitude of $\Delta$ is
almost the same as the magnitude of the largest matrix element of
$I^+_{(1)}$. This means there is a large deviation from the angular
condition, and that form factors calculated with this current may depend
strongly on which matrix element is chosen as ``bad''. We show below
that this is not the case for the electromagnetic form factors.

We find that $\Delta$ is much smaller than the largest matrix element of
the axial currents shown in Fig.~\ref{form:diag:j5} for most values of
momentum transfer. This means that the deuteron axial form factor will
be essentially independent of which matrix element is chosen as ``bad'',
except for within the range of 1.5 to 2~GeV$^2$.

\begin{figure}
\begin{center}
\epsfig{angle=0,width=5.5in,height=5.5in,file=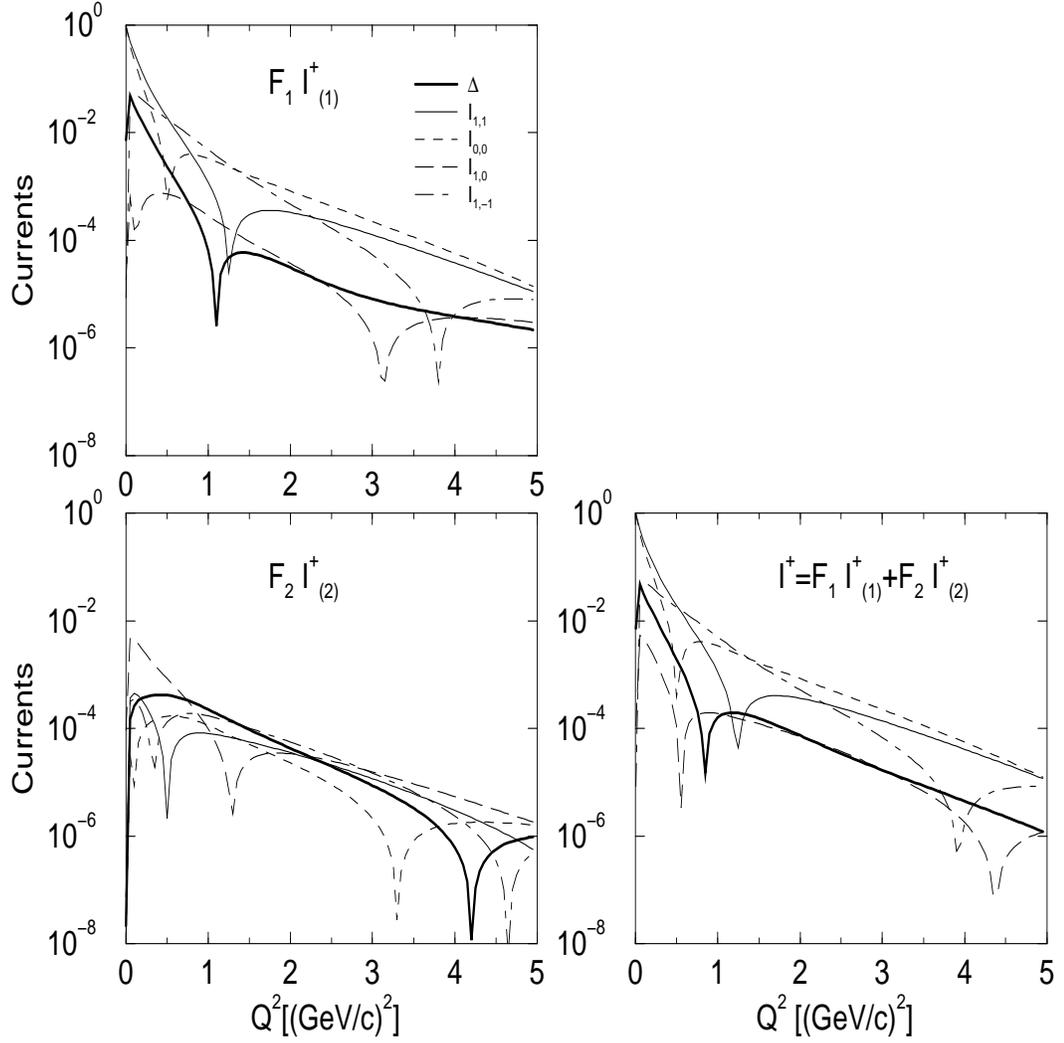}
\caption[
The matrix elements for $F_1I^+_{(1)m'm}$, $F_2I^+_{(2)m'm}$, and
$I^+_{m'm}$ calculated with the OME wave functions. The Gari:1985
nucleon isoscalar form factors are used for $F_1$ and $F_2$.
]{
The matrix elements for $F_1I^+_{(1)m'm}$, $F_2I^+_{(2)m'm}$, and
$I^+_{m'm}$ calculated with the OME wave functions. The Gari:1985
nucleon isoscalar form factors are used for $F_1$ and $F_2$
\cite{Gari:1985ia}.
\label{form:diag:gk1985f1f2both}}
\end{center}
\end{figure}

Now we combine the two parts of the electromagnetic current, 
$I^+_{(1)}$ and $I^+_{(2)}$, with the nucleon form factors 
$F_1$ and $F_2$ to get the total current.
Figure \ref{form:diag:gk1985f1f2both} shows the currents for
$F_1I^+_{(1)}$ and $F_2I^+_{(2)}$, as well as the sum, $I^+$.
The Gari:1985 nucleon form factors are used \cite{Gari:1985ia}.
We find that $F_1I^+_{(1)}$ gives the largest contribution to the total
current, and because $\Delta$ is small for $I^+_{(1)}$, $\Delta$ is also
small for the total current, meaning that the total current transforms well
under rotations. Thus, in spite of the fact that $\Delta$ is
approximately the same size as the current matrix elements for
$I^+_{(2)}$, the deuteron electromagnetic form factors should not
depend too strongly on the choice of the ``bad'' matrix element. This is
especially true for the form factors calculated with the OME wave
function. 

\begin{figure}
\begin{center}
\epsfig{angle=0,width=5.5in,height=5.5in,file=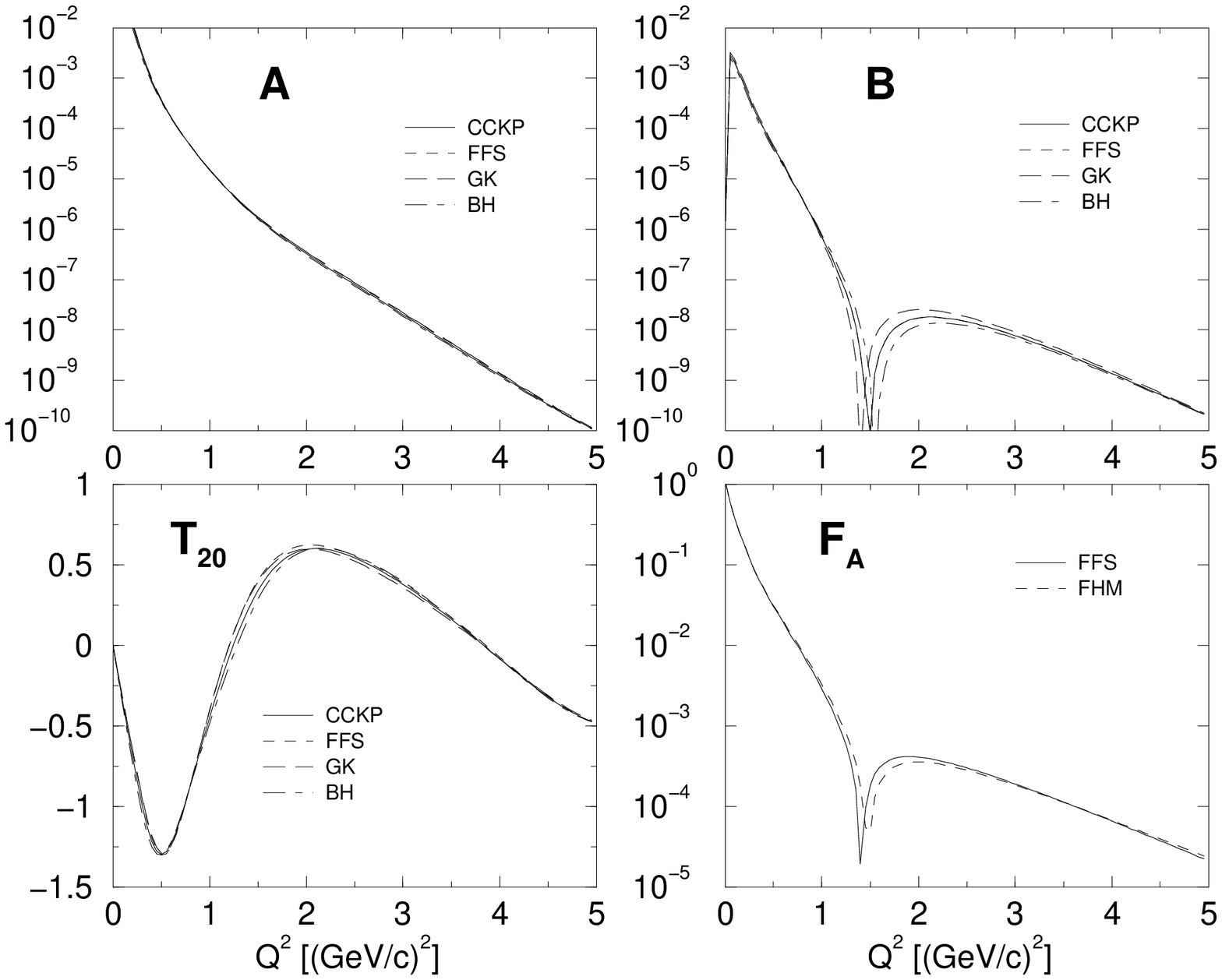}
\caption[
The form factors $A$, $B$, $T_{20}$, and $F_A$ calculated using
the various choices of the ``bad'' matrix element.
The definitions of the ``bad'' matrix elements are given in
Sections~\ref{sec:rotinv} and \ref{sec:axialstuff}.
The
OME wave function is used, along with the Lomon nucleon form factors
for the electromagnetic form factors, and the
Liesenfeld nucleon form factor for the axial
form factor.
]{
The form factors $A$, $B$, $T_{20}$, and $F_A$ calculated using
the various choices of the ``bad'' matrix element.
The definitions of the ``bad'' matrix elements are given in
Sections~ \ref{sec:rotinv} and \ref{sec:axialstuff}.
The
OME wave function is used, along with the Lomon nucleon form factors
\cite{Lomon:2001ga} for the electromagnetic form factors, and the
Liesenfeld nucleon form factor \cite{Liesenfeld:1999mv} for the axial
form factor.
\label{form:diag:ome.allbad}}
\end{center}
\end{figure}

We calculate the form factors $A$, $B$, $T_{20}$, and $F_A$ using the
OME wave function, and show the results in
Fig.~\ref{form:diag:ome.allbad}. In general, the form factors do not
depend strongly on which matrix element is chosen as ``bad'', in
agreement what what we predicted in the previous paragraph. The only
exception is for the $B$ form factor, and to a lesser extent the $F_A$
form factor, near where they crosses zero. This is not too surprising,
since a small constant shift in any function near a zero-crossing has a
large effect in a logarithmic plot. Also, we note that the FFS and CCKP
choices of the ``bad'' matrix element give the same value for $B$. This
is a consequence of Eqs.~(\ref{eq:cckpffs1}) and (\ref{eq:cckpffs2}).

\begin{figure}
\begin{center}
\epsfig{angle=0,width=5.5in,height=5.5in,file=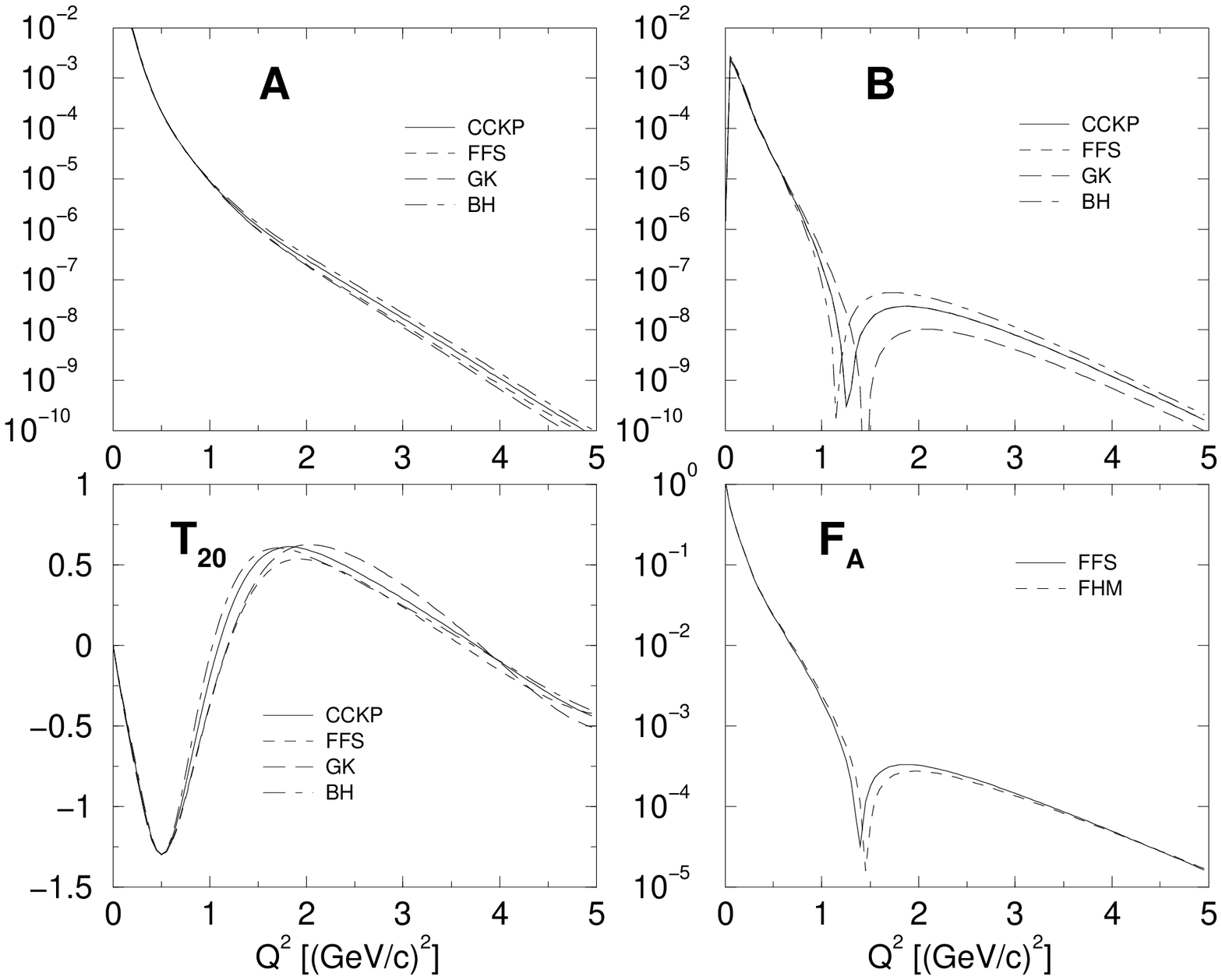}
\caption[
The form factors $A$, $B$, $T_{20}$, and $F_A$ calculated using
the various choices of the ``bad'' matrix element.
The definitions of the ``bad'' matrix elements are given in
Sections~\ref{sec:rotinv} and \ref{sec:axialstuff}.
The
OME+TME wave function is used, along with the Lomon nucleon form factors
for the electromagnetic form factors, and the
Liesenfeld nucleon form factor for the axial
form factor.
]{
The form factors $A$, $B$, $T_{20}$, and $F_A$ calculated using
the various choices of the ``bad'' matrix element.
The definitions of the ``bad'' matrix elements are given in
Sections~\ref{sec:rotinv} and \ref{sec:axialstuff}.
The
OME+TME wave function is used, along with the Lomon nucleon form factors
\cite{Lomon:2001ga} for the electromagnetic form factors, and the
Liesenfeld nucleon form factor \cite{Liesenfeld:1999mv} for the axial
form factor.
\label{form:diag:tme.allbad}}
\end{center}
\end{figure}

We also use the OME+TME wave function to calculate the form factors $A$,
$B$, $T_{20}$, and $F_A$, which we show in
Fig.~\ref{form:diag:tme.allbad}. We argued earlier that the these
electromagnetic form factors depend more strongly on which matrix
element is chosen as ``bad'' that those calculated with the OME wave
function, and that dependence is clear in this figure. At low momentum
transfers, the dependence on the change is fairly small, but as the
momentum transfer increases, so does the dependence. The axial form
factor is not affected as strongly, primarily because each wave function
generates an axial current which  violates the angular condition by
approximately the same amount.

\begin{figure}
\begin{center}
\epsfig{angle=0,width=5.5in,height=5.5in,file=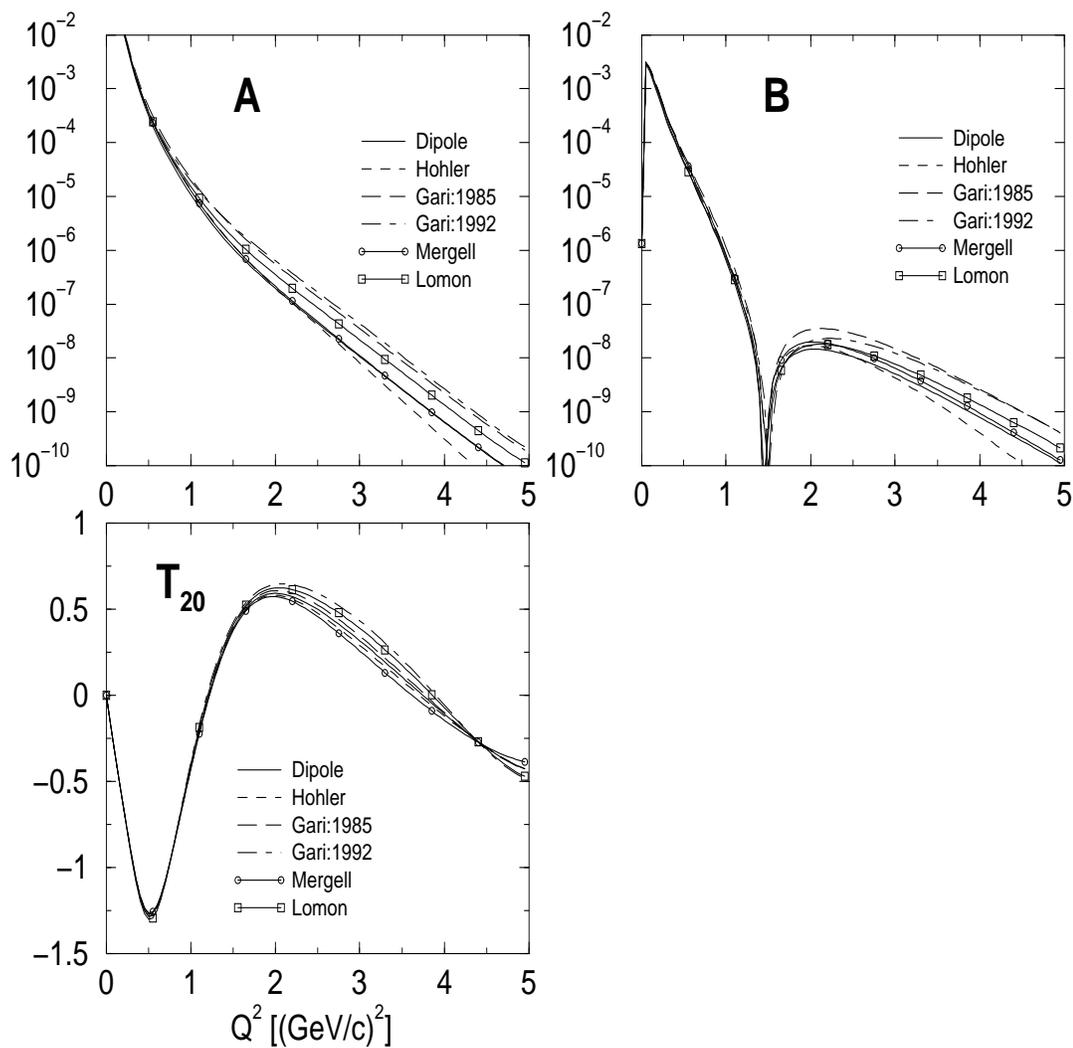}
\caption{
The electromagnetic form factors $A$, $B$, and $T_{20}$ calculated using
the various choices of the nucleon isoscalar form factors. The
OME wave function is used, along with the FFS choice of the ``bad''
deuteron current matrix. The axial form factor is not shown since its
dependence on different form factors is trivial.
\label{form:diag:ome.allff}}
\end{center}
\end{figure}

Since there are many different models of the nucleon electromagnetic
form factors, we calculate the deuteron electromagnetic form factors
using each of them to see what effect the differences have. The results
are shown in Fig.~\ref{form:diag:ome.allff}. At low momentum transfers,
all the nucleon form factors give close to the same results. However,
when the momentum transfers is large, we find a large spread in the
values due to nucleon form factors. In fact, this spread in $A$ and $B$
is larger than the spread of values obtained from using different
``bad'' matrix elements with the OME+TME wave functions. In other words,
in order to obtain accurate results for $A$ and $B$ at momentum
transfers over 2~GeV$^2$, it is more important to determine which
nucleon form factor to use than when ``bad'' matrix to use. The spread
in $T_{20}$ is about the same size regardless of whether the different
``bad'' matrix elements or the different nucleon form factors are used.

\begin{figure}
\begin{center}
\epsfig{angle=0,width=5.5in,height=5.5in,file=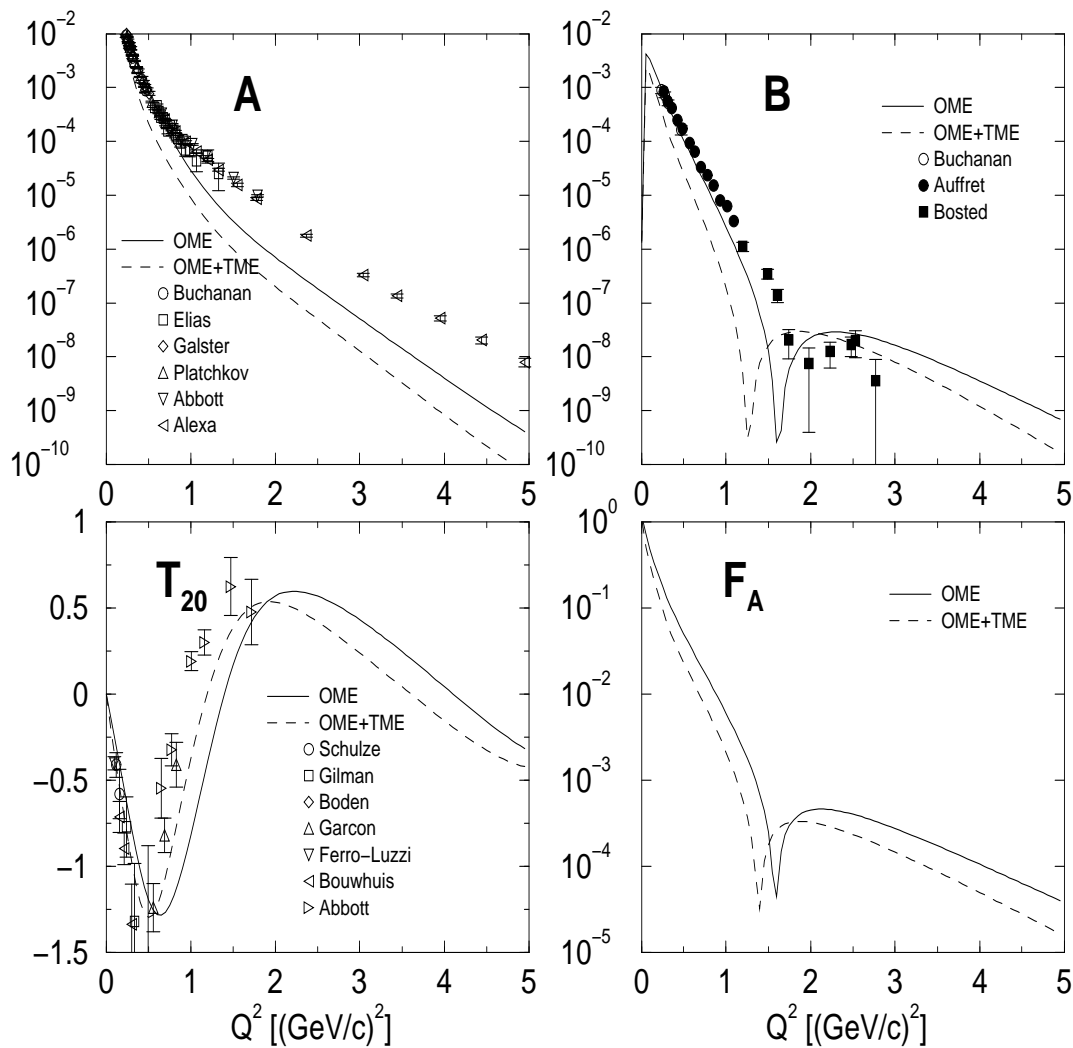}
\caption{The $A$, $B$, $T_{20}$, and $F_A$ form factors for the OME and
OME+TME potentials, along with data. See the accompanying text for an
explanation of the data.
\label{form:diag:OME.expt2}}
\end{center}
\end{figure}

Finally, in Fig.~\ref{form:diag:OME.expt2}, we compare the $A$, $B$,
$T_{20}$, and $F_A$ form factors for the OME and OME+TME wave functions
to experimental data. The ``bad'' component was chosen according to FFS,
and the nucleon form factors of Lomon were used for $A$, $B$, and $T_{20}$,
while the Liesenfeld axial nucleon form factor was used for $F_A$. The
data for $A$ is from: Buchanan {\it et al.} \cite{Buchanan:1965},
Elias {\it et al.} \cite{Elias:1969},
Galster {\it et al.} \cite{Galster:1971kv},
Platchkov {\it et al.} \cite{Platchkov:1990ch},
Abbott {\it et al.} \cite{Abbott:1998sp}, and
Alexa {\it et al.} \cite{Alexa:1999fe};
the data for $B$ is from: Buchanan {\it et al.} \cite{Buchanan:1965},
Auffret {\it et al.} \cite{Auffret:1985}, and
Bosted {\it et al.} \cite{Bosted:1990hy};
and the data for $T_{20}$ is from:
Schulze {\it et al.} \cite{Schulze:1984ms}, 
Gilman {\it et al.} \cite{Gilman:1990vg}, 
Boden {\it et al.} \cite{Boden:1991un}, 
Garcon {\it et al.} \cite{Garcon:1994vm}, 
Ferro-Luzzi {\it et al.} \cite{Ferro-Luzzi:1996dg}, 
Bouwhuis {\it et al.} \cite{Bouwhuis:1998jj}, and
Abbott {\it et al.} \cite{Abbott:2000fg}.

There is a rather large difference between the form factors calculated
with the OME and OME+TME wave functions. This difference is due
primarily to the fact that the OME wave functions are more deeply bound
than the OME+TME wave functions, and it can be reduced by choosing a
different sigma coupling constant $f_\sigma$ for the OME and OME+TME
potentials. However, for our analysis of rotational invariance, it is
important to keep $f_\sigma$ fixed.

The difference between the calculated form factors and the data is also
quite large. This is not unexpected, since in our model of the current,
meson exchange currents are not included. It is known that these have a
large effect on the form factors at large momentum transfers
\cite{Phillips:1998uk,Wiringa:1995wb,Schiavilla:1991ug}. Including these
effects would bring the form factors into better agreement with the
data. However, we emphasize again that agreement with the data is not a
priority of this work. Our goal is to gain a better understanding of the
breaking of rotational invariance by the light front, and how to restore
that invariance. Only after we have that understanding can we pursue
accurate calculation of the form factors with light-front dynamics.

\chapter{Conclusions} \label{ch:conclusions}
\doquote
{If a thing can not go on forever, it will come to an end.}
{Herbert Stein}

The issue of rotational invariance in light-front dynamics must be
addressed before one attempt to use light-front dynamics for
high-precision calculations. In this work, we have sought to
find ways to quantify the level to which rotational invariance is
broken. In addition, we have investigated potentials of different
orders, since we know that if all orders of the potential were included,
rotational invariance would be restored. We have used light-front
dynamics to obtain one-meson-exchange (OME) and two-meson-exchange (TME)
potentials for several different nuclear theory Lagrangians.

In Chapter~\ref{wc:ch:wcmodel}, the massive Wick-Cutkosky model is
used to investigate the breaking of rotational invariance in deeply
bound states. We studied the pseudo angular momentum decomposition of
each wave function into different eigenstates of pseudo angular
momentum. In general, we found that most
states are approximately angular momentum eigenstates. That is, a
partial-wave decomposition of the wave functions showes that most states
have one dominant (highest percentage) angular momentum
component. Including the TBE potentials increases the fraction of the
dominant component in most states. We also showed that the
spectra generated by the Bethe-Salpeter equation using the ladder and
crossed ladder are well approximated with the spectra generated by the
matching OME+TME potentials. In addition, we found that the spectra for
the full ground state calculation cannot be reproduced accurately for
deeply bound states using either the light-front Hamiltonian or the
Bethe-Salpeter equation.

In Chapter~\ref{ch:pionly}, a model Lagrangian for nuclear physics which
includes chiral symmetry was used to derive two models, a light-front
pion-only model, and a new light-front nucleon-nucleon potential. Those
potentials are then used to calculate the binding energy and wave
function for the $m=0$ and $m=1$ states of the deuteron. We find that
for both models, the splitting between the $m=0$ and $m=1$ states was
smaller for the OME+TME potential as compared to the OME potential
gives. We also find that the effects of chiral symmetry must be included
to obtain sensible results for the OPE+TPE pion-only model. 

In Chapter~\ref{ch:ffdeut}, the wave functions obtained in
Chapter~\ref{ch:pionly} are used to calculate the form factors of the
deuteron using only light-front dynamics throughout.
In light-front dynamics, there are four independent components
of the deuteron current. However, the requirement of rotational
invariance introduces an angular condition that the four components must
satisfy, reducing the number of physically independent components to
three. The deviation of the calculated current components from the
angular condition is denoted by $\Delta$. We find that $\Delta$ is very
small for the deuteron wave functions calculated with the OME
potential. This is an important result, since it means that although
{\em in principle} the light-front calculation of the deuteron current
does not transform correctly under rotations, {\em in practice} it does
quite well. The smallness of $\Delta$ means that any reasonable
prescription for eliminating the dependent component of the current gives
essentially the same results; the uncertainty introduced by the various
nucleon form factors is greater for $A$ and $B$, and about the same size
for $T_{20}$.

We also found that $\Delta$ is significantly larger when the TME
potentials are used. Since the previous results in
Chapters~\ref{wc:ch:wcmodel} and \ref{ch:pionly} indicate that the
rotational properties of the TME wave function are better that for the
OME wave function, we interpret the increase in $\Delta$ as an
indication that extra diagrams need to be included in the current
calculation to restore rotational invariance. That is, having a wave
function with good rotational properties is not sufficient to obtain
matrix elements of the current operator with good rotational properties.

\section*{Future Extensions}
\doquote
{I never see what has been done; I only see what remains to be done.}
{Marie Curie} 

This work can be extended in many ways. Perhaps the most important
addition is to add the higher-order graphs to the deuteron current 
that are required for current conservation. This may improve the 
quality of the deuteron form factors when the wave function is 
calculated using the two-meson-exchange potentials.

Another obvious extension is to include the crossed TME diagrams. This
would have two effects. Recall that we included the chiral contact
diagrams using a prescription to maintain approximate chiral
symmetry. When the crossed diagrams are included, we simply include the
full contact potentials which included chiral symmetry without
approximation. The second effect is that the crossed diagrams are needed
in principle to obtain an accurate solution for the deuteron. However,
these effects are not needed to study the breaking of rotational
invariance. 

The contribution to the deuteron form factors from the meson form
factors is another effect that could be investigated. This is possible
since our potential is derived from a Lagrangian, and we can calculate
the mesonic components of the deuteron wave function. Obviously, the
contribution to the form factor due to the mesons must be taken into
account when attempting a high precision calculation. However, including
these components, like including the crossed diagrams, does not affect
the overall rotational invariance properties of the deuteron form
factors.

\def\bibquote{\doquote
{One could not be a successful scientist without realizing that, in
contrast to the popular conception supported by newspapers and mothers
of scientists, a goodly number of scientists are not only narrow-minded
and dull, but also just stupid.}{James Watson}}

\bibliographystyle{h-physrev3}
\bibliography{thesis}

\appendix
\chapter{Notation, Conventions, and Useful Relations} \label{ch:notation}
\doquote
{The best way to be boring is to leave nothing out.}
{Voltaire}

For a general four-vector $a$ with components $(a^0,a^1,a^2,a^3)$ in the
equal-time basis, we define the light-front variables 
\begin{eqnarray}
a^\pm          &=& a^0 \pm a^3, \\
\boldsymbol{a}_\perp &=& (a^1,a^2),
\end{eqnarray}
so the 4-vector $a^\mu$ can be denoted in the light-front basis as
\begin{eqnarray}
a&=&(a^+,a^-,\boldsymbol{a}_\perp).
\end{eqnarray}
Using this, we find that the scalar product is
\begin{eqnarray}
a \cdot b &=& a^\mu b_\mu
= \frac{1}{2}\left(a^+ b^- + a^- b^+ \right) - \boldsymbol{a}_\perp \cdot 
\boldsymbol{b}_\perp.
\end{eqnarray}
This defines $g_{\mu\nu}$, with $g_{+-}=g_{-+}=1/2$, $g_{11}=g_{22}=-1$,
and all other elements of $g$ vanish.  The elements of $g^{\mu\nu}$ are
obtained from the condition that $g^{\mu\nu}$ is the inverse of
$g_{\mu\nu}$, so
$g^{\alpha\beta}g_{\beta\lambda}=\delta^\alpha_\lambda$.  Its elements
are the same as those of $g_{\mu\nu}$, except that
$g^{-+}=g^{+-}=2$. Thus,
\begin{eqnarray}
a^\pm &=& 2 a_\mp.
\end{eqnarray}
and the partial derivatives are similarly given by
\begin{eqnarray}
\partial^\pm &=& 2 \partial_\mp = 2 \frac{\partial}{\partial x^\mp}.
\end{eqnarray}

To find the physical consequences of this coordinate system, consider
the commutation relations $[p^\mu,x^\nu] = i g^{\mu\nu}$, which yield
\begin{eqnarray}
\,[       p^\pm     ,      x^\mp      ] &=& 2i, \\
\,[ \boldsymbol{p}_\perp^i,\boldsymbol{x}_\perp^j ] &=& -i \delta_{i,j},
\end{eqnarray}
with the other commutators equal to zero.  This means that
$\boldsymbol{x}_\perp^i$ is
canonically conjugate to $\boldsymbol{p}_\perp^i$, and $x^\pm$ is
conjugate to
$p^\mp$.  Since $x^+$ plays the role
of time (the light-front time) in light-front dynamics, and $p^-$ is
canonically conjugate to $x^+$, this means that $p^-$ is the
light-front energy and that the light-front Hamiltonian is given by
$P^-$.

In any Hamiltonian theory, particles have the an energy defined
by the on-shell constraint $k^2=m^2$.  This implies that the light-front
energy of a particle is
\begin{eqnarray}
k^- &=& \frac{m^2 +\boldsymbol{k}_\perp^2}{k^+}.
\end{eqnarray}
The independent components of the momentum can be written as a
light-front three-vector $\boldsymbol{k}_{\text{LF}}$, denoted by
\begin{eqnarray}
\boldsymbol{k}_{\text{LF}} &=& (k^+,\boldsymbol{k}_\perp).
\end{eqnarray}

For dealing with spin, we require the Pauli sigma matrices, which are
\begin{eqnarray}
(\sigma^1,\sigma^2,\sigma^3) &=&
\left(
\left(\begin{array}{cc} 0 &  1 \\ 1 &  0 \end{array} \right),
\left(\begin{array}{cc} 0 & -i \\ i &  0 \end{array} \right),
\left(\begin{array}{cc} 1 &  0 \\ 0 & -1 \end{array} \right)
\right) \label{sigmamat}.
\end{eqnarray}

The Bjorken and Drell convention \cite{Bjorken:1964} for the gamma
matrices is used in this work. They specify that 
\begin{eqnarray}
\gamma^0 = \beta &=&
\left(\begin{array}{cc} 1 & 0 \\ 0 & -1 \end{array} \right), \\
\bbox{\gamma} = \beta \bbox{\alpha} &=&
\left(\begin{array}{cc} 0 & \sigma \\ -\sigma & 0 \end{array} \right), \\
\gamma^5 = i \gamma^0 \gamma^1 \gamma^2 \gamma^3 &=&
\left(\begin{array}{cc} 0 & 1 \\ 1 & 0 \end{array} \right).
\end{eqnarray}

The spin matrices $S^i$ then are
\begin{eqnarray}
S^i &=& \frac{1}{2} \Sigma^i = -\frac{1}{2} \gamma^5 \gamma^i, \\
\Sigma^i &=& 
\left(\begin{array}{cc} \sigma^i & 0 \\ 0 & -\sigma^i
\end{array} \right).
\end{eqnarray}
Using $\bbox{\Sigma}$, we can express the helicity operator as
$H=\widehat{p}\cdot\bbox{\Sigma}$, which has eigenvalues $\pm 1$. This
is useful since the helicity is invariant under rotations, a property
that is used in Appendix~\ref{app:rotmat}.

It is useful to define the spinor projection operators $\Lambda_\pm$ by
\begin{eqnarray}
\Lambda_\pm &=& \frac{1}{4} \gamma^\mp \gamma^\pm = \frac{1}{2} \gamma^0
\gamma^\pm = \frac{1}{2} (I\pm \alpha^3).
\end{eqnarray}
These satisfy the requirements for projection operators,
\begin{eqnarray}
\Lambda_+ + \Lambda_- &=& 1, \\
(\Lambda_\pm)^2 &=& \Lambda_\pm, \\
\Lambda_\pm \Lambda_\mp &=& 0.
\end{eqnarray}

We summarize the effect these projection operators have on the gamma
matrices: 
\begin{eqnarray}
\Lambda_\pm \gamma^0 &=& \gamma^0 \Lambda_\mp, \\
\Lambda_\pm \gamma^\pm &=& 0 = \gamma^\pm \Lambda_\mp, \\
\Lambda_\pm \gamma^\mp &=& \gamma^\mp = \gamma^\mp \Lambda_\mp, \\
\Lambda_\pm \gamma^\perp &=& \gamma^\perp \Lambda_\pm,
\end{eqnarray}
and under conjugation,
\begin{eqnarray}
\gamma^0 \Lambda_\pm^\dagger \gamma^0 &=& \Lambda_\mp.
\end{eqnarray}

\chapter{Conversion to Matrix Form} \label{ch:numericaleq}
\doquote
{Let us work without theorizing. It's the only way to make life endurable.}
{Voltaire} 

To solve for the bound-state wave function for the Wick-Cutkosky model
numerically, the light-front
Schr\"{o}dinger equation given in Eq.~(\ref{wc:fullse6}) must be
discretized and cast in matrix form. The solution of the light-front
nucleon-nucleon model proceeds along similar lines. The equation is first 
symmetrized to get
\begin{eqnarray}
-\int_0^\infty dp_{\text{ET}} \int_0^{\pi/2} d\theta_p
V^S_{p,m}(k_{\text{ET}},\theta_k;p_{\text{ET}},\theta_p)
\psi^S_{p,m}(p_{\text{ET}},\theta_p)
&=&
\psi^S_{p,m}(k_{\text{ET}},\theta_k),
\label{a:nint:fullse7}
\end{eqnarray}
where
\begin{eqnarray}
V^S_{p,m}(k_{\text{ET}},\theta_k;p_{\text{ET}},\theta_p)
\!&=&\!\!
B(k_{\text{ET}})
A(k_{\text{ET}},\theta_k)
V_{p,m}(k_{\text{ET}},\theta_k;p_{\text{ET}},\theta_p)
A(p_{\text{ET}},\theta_p)
B(p_{\text{ET}}),
\\
\psi^S_{p,m}(k_{\text{ET}},\theta_k)
&=&
B(k_{\text{ET}})^{-1} A(k_{\text{ET}},\theta_k)
\psi_{p,m}(k_{\text{ET}},\theta_k), \\
A(k_{\text{ET}},\theta_k) &=&
\sqrt{\frac{2 k_1^+ k_2^+  k^2_{\text{ET}} \sin\theta_k}{(k^0)}}, \\
B(k_{\text{ET}}) &=& \sqrt{\frac{-1}{E^2 - 4 (k^0)^2}}.
\end{eqnarray}

Before discretizing the integrals, note that we may write
\cite{Cooke:2000ef} 
\begin{eqnarray}
\int_0^\infty dp \, f(p) &=& 
a \int_0^1 du \left(f(au) + \frac{f(a/u)}{u^2} \right). \label{a:nint:inttrick}
\end{eqnarray}
Using this trick, the integral in Eq.~\ref{a:nint:fullse7} over
$p_{\text{ET}}$ can be written as an integral over a finite range.
Since we are concerned with a bound state, the wave function is
damped for large momenta, and the second term of
Eq.~(\ref{a:nint:inttrick}) converges as $u$ approaches zero. Another
choice that can be used is \cite{Machleidt:1987hj,Machleidt:1989}
\begin{eqnarray}
\int_0^\infty dp \, f(p) &=& 
\frac{a\pi}{4} \int_{-1}^1
\frac{dx}{\cos^2\left(\frac{\pi(x+1)}{4}\right)}
f\left(a \tan\left(\frac{\pi(x+1)}{4}\right)
\right). \label{a:nint:inttrick2}
\end{eqnarray}

After using Eq.~(\ref{a:nint:inttrick}) or Eq.~(\ref{a:nint:inttrick2}),
the integrals in Eq.~\ref{a:nint:fullse7} are over a
finite range and can be discretized using Gauss-Legendre quadrature.
The specific routines for the quadrature are
given by {\it Numerical Recipes in C} \cite{Press:1992}.  This
conversion gives a
matrix equation that approximates the original Eq.~\ref{a:nint:fullse7},
\begin{eqnarray}
-V^S_{p,m}(g(E),E) \psi^S_{p,m} &=& \psi^S_{p,m},
\label{a:nint:fullse8}
\end{eqnarray}
where the explicit dependence of $V^S_{p,m}$ 
on the binding energy $E$ and the coupling constant $g$ is shown.  This
equation must be solved self-consistently for the spectrum $g(E)$.

The approach we use is to first solve for the spectrum for the OBE
potential.  The eigenvalue equation
\begin{eqnarray}
V^S_{p,m,\text{OBE}}(E)
\psi^S_{p,m}
&=& \alpha
\psi^S_{p,m},
\end{eqnarray}
where
\begin{eqnarray}
\alpha &=& \frac{-1}{g_{\text{OBE}}(E)^2}.
\end{eqnarray}
The ground-state wave function is the eigenvector that corresponds to the
smallest eigenvalue $\alpha$.  We calculate the wave function and the
smallest coupling constant using EISPACK
\cite{eispack:smith,eispack:garbow} routines for a 
range of energies to map out the spectrum.

Using the coupling constant for the OBE potential as a starting point,
we can use Eq.~\ref{a:nint:fullse8} for higher-order potentials that include
$N$ meson exchanges.  For a
given energy, the coupling constant $g(E)$ is initially chosen as
$g_{\text{OBE}}(E)$, then we take the equation
\begin{eqnarray}
\left[
\sum_{n=1}^N g(E)^{2n} V^S_{p,m,n\text{BE}}(E)
\right]
\psi^S_{p,m}
&=& \beta
\psi^S_{p,m},
\end{eqnarray}
where $V_{n\text{BE}}$ is the $n$-boson-exchange potential, and solve
it as an eigenvalue equation for $\beta$.  The coupling constant $g(E)$
is varied until the the lowest eigenvalue is $\beta=-1$, at which point
$g(E)$ is the correct value of the spectrum corresponding to the
ground-state wave function $\psi^S_{p,m}$.

\chapter{Azimuthal-angle and Loop Integrals}
\label{ch:integrations}

We outline the methods used to perform the azimuthal-angle integrations
for OBE, TME, and MGF potentials, and the loop integrals for the TME
potentials. Although these methods are specifically for the
Wick-Cutkosky potentials (presented in Chapter~\ref{wc:ch:wcmodel}),
they can easily be generalized for the light-front nucleon-nucleon
potentials (presented in Chapter~\ref{ch:pionly}) by using the integrals
discussed in Appendix~\ref{theintegral}.

\section{Azimuthal-angle Integration of the OBE and MGF Potentials}
\label{a:ints:angint}
\doquote
{Never express yourself more clearly than you are able to think.}
{Niels Bohr}

In this section, we evaluate the azimuthal-angle integration of the OBE
potential in Eq.~(\ref{wc:OBEpot2}) and the first term in MGF potential in
Eq.~(\ref{wc:mgfpot}), using the prescription for azimuthal-angle
integration given in Eq.~(\ref{wc:angintlf}).  One of the integrals is
easily done since since the potential is independent of the azimuthal
angle between the two perpendicular momenta, so
\begin{eqnarray}
V(k^+,k_\perp;p^+,p_\perp) &\propto&
\left[
 \theta(x-y) \int_0^{2\pi} \frac{d\phi}{A_1 + B \cos\phi} + \right. 
\nonumber\\
&& \phantom{\left[\right.}\left.
\theta(y-x) \int_0^{2\pi} \frac{d\phi}{A_2 + B \cos\phi}
\right], \label{a:ints:obeangint}
\end{eqnarray}
where
\begin{eqnarray}
A_1 &=& (k_1^+-p_1^+)(E-p^-_1-k^-_2)-\mu^2-p_\perp^2-k_\perp^2, \\
A_2 &=& (p_1^+-k_1^+)(E-k^-_1-p^-_2)-\mu^2-p_\perp^2-k_\perp^2, \\
B   &=& 2k_\perp p_\perp.
\end{eqnarray}
The integrals in Eq.~\ref{a:ints:obeangint} are easily done to give
\begin{eqnarray}
\int_0^{2\pi} \frac{d\phi}{A + B \cos\phi} &=&
\frac{-2\pi}{\sqrt{A^2-B^2}}, \label{a:ints:int1}
\end{eqnarray}
which is negative since $A<0$. Using this, the azimuthal-angle-averaged
OBE potential is given by
\begin{eqnarray}
V_{\text{OBE}}(k^+,k_\perp;p^+,p_\perp) &=& -2\pi
\left(\frac{M}{E}\right)^2 E \left[
 \frac{\theta(x-y)}{\sqrt{A_1^2-B^2}}
+\frac{\theta(y-x)}{\sqrt{A_2^2-B^2}}
\right].
\end{eqnarray}
It is straightforward to rewrite this equation for the potential in terms
of the equal-time coordinates.

When the other terms in the MGF potential are azimuthal-angle averaged, 
integrations similar to the one given in Eq.~\ref{a:ints:int1} are encountered,
with the denominator squared or cubed.  We note that
\begin{eqnarray}
\int_0^{2\pi} \frac{d\phi}{(A + B \cos\phi)^2} &=&
\frac{-2\pi}{\sqrt{A^2-B^2}} \, \frac{A}{A^2-B^2}, \\
\int_0^{2\pi} \frac{d\phi}{(A + B \cos\phi)^3} &=&
\frac{-2\pi}{\sqrt{A^2-B^2}} \, \frac{2A^2+B^2}{(A^2-B^2)^2},
\end{eqnarray}
so the azimuthal-angle-averaged MGF potential is given by
\begin{eqnarray}
&&V_{\text{MGF}}(\boldsymbol{k}_1,\boldsymbol{p}_1;P) \nonumber\\
&& \qquad=-2\pi \left(\frac{M}{E}\right)^2 \left[
\frac{\theta(x-y)}{\sqrt{A_1^2-B^2}} \left( 1 +
\frac{N_{p,21}+N_{k,12}}{2D_{1,2}} +
\frac{N_{p,21} N_{k,12}}{2D_{1,3}}
\right)
\right.\nonumber \\ & &
\phantom{
 -2\pi \left(\frac{M}{E}\right)^2 \left[
\right. } \left.+
\frac{\theta(y-x)}{\sqrt{A_2^2-B^2}} \left( 1 +
\frac{N_{k,21}+N_{p,12}}{2D_{2,2}} + 
\frac{N_{p,12} N_{k,21}}{2D_{2,3}}
\right) \right],
\end{eqnarray}
where
\begin{eqnarray}
D_{i,2} &\equiv& \frac{A^2-B^2}{A}, \\
D_{i,3} &\equiv& \frac{(A^2-B^2)^2}{2A^2+B^2},
\end{eqnarray}
and $i=1,2$.

\section{Azimuthal-angle and Loop Integration of the TBE
Potentials} 
\label{a:ints:loopints}
\doquote
{A child of five would understand this. Send someone to fetch a child of
five.}
{Groucho Marx} 

As in Section~\ref{a:ints:angint}, we want the azimuthal-angle integrals
of the TBE potentials given in
Eqs.~(\ref{wc:TBESBpot}-\ref{wc:TBEZXpot}).  For these potentials, there
is also a loop integral that has to be done. We start by analyzing the
equations schematically.  Each of the terms in the TBE potentials can be
written in the following form,
\begin{eqnarray}
&&V_{\text{TBE}}(k^+,k_\perp;p^+,p_\perp) \nonumber\\
&&\qquad=
\int_0^\infty \frac{q_\perp dq_\perp}{2(2\pi)^3} \int_0^1 dz \,
J(k^+,q^+,p^+)
I(k^+,k_\perp,q^+,q_\perp,p^+,p_\perp), \\
&&I(k^+,k_\perp,q^+,q_\perp,p^+,p_\perp) \nonumber\\
&&\qquad= 
\int_0^{2\pi} d\phi_q d\phi_p
\frac{1}{A_1+B_1\cos\phi_q}
\nonumber \\ & &\qquad\qquad \times
\frac{1}{A_2+B_2\cos\phi_q+C_2\cos\phi_p+D_2\cos(\phi_p-\phi_q)}
\nonumber \\ & &\qquad\qquad \times
\frac{1}{A_3+B_3\cos\phi_q+C_3\cos\phi_p+D_3\cos(\phi_p-\phi_q)},
\end{eqnarray}
where the $A$'s, $B$'s, $C$'s, and $D$'s may have dependence on $k^+$,
$k_\perp$, $p^+$, $p_\perp$, $q^+=zE$, and $q_\perp$; they are
independent of the azimuthal angles.  These functions can be easily
determined for  each potential by examining the forms of the original
equations.  The rotational invariance of the potential about the
$z$-axis allows the $\phi_k$ integration to be done trivially.

In the integrand of $I$, only the last two terms depend on $\phi_p$.  To
emphasize this, we write
\begin{eqnarray}
&&I(k^+,k_\perp,q^+,q_\perp,p^+,p_\perp) \nonumber \\
&&\qquad=
\int_0^{2\pi} d\phi_q
\frac{I_2(k^+,k_\perp,\boldsymbol{q},p^+,p_\perp)}
{(A_1+B_1\cos\phi_q)(A_2+B_2\cos\phi_q)(A_3+B_3\cos\phi_q)},
\\
&&I_2(k^+,k_\perp,\boldsymbol{q},p^+,p_\perp) \nonumber \\
&&\qquad= 
\int_0^{2\pi} 
\frac{d\phi_p}{(1+a_2\cos\phi_p+b_2\sin\phi_p)(1+a_3\cos\phi_p+b_3\sin\phi_p)},
\end{eqnarray}
where, for $i=2,3$,
\begin{eqnarray}
a_i &=& \frac{C_i+D_i \cos\phi_q}{A_i+B_i \cos\phi_q}, \\
b_i &=& \frac{    D_i \sin\phi_q}{A_i+B_i \cos\phi_q}.
\end{eqnarray}
The integral in $I_2$ is evaluated to obtain
\begin{eqnarray}
&&I_2(k^+,k_\perp,\boldsymbol{q},p^+,p_\perp) \nonumber\\
&&\qquad= 
\frac{2\pi}{(a_2-a_3)^2 + (b_2-b_3)^2 - (a_2 b_3 - a_3 b_2)^2} \nonumber
\\ & & \qquad \times
\left(
\frac{a_2(a_2-a_3)+b_2(b_2-b_3)}{\sqrt{1-a_2^2-b_2^2}}
+
\frac{a_3(a_3-a_2)+b_3(b_3-b_2)}{\sqrt{1-a_3^2-b_3^2}}
\right).
\end{eqnarray}

The remaining three-dimensional loop integral in $V_{\text{TBE}}$ on
$\boldsymbol{q}$ is done using numeric techniques.  The trick introduced in
Appendix \ref{ch:numericaleq} is used to convert the semi-infinite
$q_\perp$
integration into an integration on a compact range.  Before doing the
$z$ integral, the range of integration is limited by using the step
functions.  Gauss-Legendre quadrature, given by {\it Numerical Recipes
in C} \cite{Press:1992},  is used to evaluate all the integrals. 

Since each of the parts of the full TBE potential (TBE:SB, TBE:SX,
\ldots) given in Chapter~\ref{wc:ch:wcmodel}
should be Hermitian and invariant under interchange of particle
1 and 2, these invariances can be used as a self-consistency check.
Each matrix element is calculated twice, first by using the
straightforward approach, then particle labels 1 and 2 are interchanged
and it is calculated again.  The results are compared, and if they
differ by an unacceptable amount, the number of quadrature points is
increased and the element is recalculated.  In order to get the
numerical accuracy of the potentials correct to within 1\%, 
we start with ten points for the $q_\perp$ integral, six points for the
$\phi_q$ integral, and three points for the $z$ integral, resulting in a
three-dimensional integral using 180 points.

\chapter{Cosine Integrals}
\label{theintegral}
\doquote
{According to my point of view, logic and aesthetics cannot be in
conflict with one another.  Perhaps there is something lacking in my
logical reasoning.}
{M.C.~Escher}

We wish to solve
\begin{eqnarray}
I&=&\int_0^{2\pi} \frac{d\phi}{2\pi} e^{i \phi m} 
\prod_{j=1}^N \frac{1}{(a_j + b_j \cos \phi)^{n_j}},
\end{eqnarray}
where $m$ is an integer, $n_j$ are a positive integers, and $a_j$ and
$b_j$ real numbers. Note that if one of the $b_j$'s is zero, then we can
reduce the integral to one with $N-1$ sets of parameters.  We therefore
assume that $b_j\neq0$.

Some things to note about this integral:
\begin{enumerate}
\item{} \label{shifty} 
Since the range of integration is $2\pi$, and all the functions in the
integrand are periodic in that range, the range of integration can be
shifted by an arbitrary amount.
\item{} \label{swapy} 
When the variable of integration is changed from $\phi$ to $-\phi$,
the limits of integration can be rewritten as
\begin{eqnarray}
\int_0^{2\pi} d\phi \rightarrow 
\int_0^{-2\pi} d(-\phi) =
\int_{-2\pi}^0 d\phi =
\int_0^{2\pi} d\phi. 
\end{eqnarray}
Here we have used item number \ref{shifty}.  Since $\cos(-z)=\cos(z)$,
the only thing that changes in the integrand is the sign in the exponential.
Hence, $I(m)=I(-m)$, or the integral is only a function of the
magnitude of $m$.
\item{} The integral is real, since $I(m)^*=I(-m)=I(m)$.
\end{enumerate}

In addition,
\begin{eqnarray}
\frac{d^m}{d a^m} \frac{1}{a+b\cos \phi}
&=&
(-1)^m m! \frac{1}{(a+b\cos \phi)^{m+1}},
\end{eqnarray}
so that
\begin{eqnarray}
&&\int_0^{2\pi} \frac{d\phi}{2\pi} e^{i \phi m}
\prod_{j=1}^N \frac{1}{(a_j + b_j \cos \phi)^{n_j}} \nonumber\\
&&\qquad =
\prod_{l=1}^N \frac{(-1)^{(n_l-1)}}{(n_l-1)!} \frac{d^{n_l-1}}{d a_l^{n_l-1}}
\int_0^{2\pi} \frac{d\phi}{2\pi} e^{i \phi m} 
\prod_{j=1}^N \frac{1}{a_j + b_j \cos \phi}. \label{a:ci:firstint}
\end{eqnarray}
Since $b_j\neq0$, we can factor out a
$\prod_{j=1}^N\frac{1}{b_j^{n_j}}$, and we relabel
$\frac{a_j}{b_j}\rightarrow a_j$. Then, the non-trivial integral is
\begin{eqnarray}
I(\{a_i\},m,N)
&=&
\int_0^{2\pi} \frac{d\phi}{2\pi} e^{i \phi |m|} 
\prod_{j=1}^N \frac{1}{a_j + \cos \phi} \nonumber \\
&=&
\int_0^{2\pi} \frac{d\phi}{2\pi} e^{i \phi |m|} 
\prod_{j=1}^N \frac{2 e^{i \phi}}
{e^{2 i \phi} + 2a_j e^{i \phi} + 1}.
\end{eqnarray}

Now we let $z=e^{i\phi}$, so $d\phi e^{i\phi}= dz/i$, which allows the
integral to be converted to a contour integral over the contour $C$, a
unit circle in the complex plane. Then,
\begin{eqnarray}
I(\{a_i\},m,N) 
&=&
\int_C \frac{dz}{2\pi i} z^{|m|-1}
\prod_{j=1}^N \frac{2 z}{z^2 + 2a_j z + 1} \nonumber \\
&=&
\int_C \frac{dz}{2\pi i} z^{|m|-1}
\prod_{j=1}^N \frac{2 z}{(z-z_{j+})(z-z_{j-})},
\end{eqnarray}
where
\begin{eqnarray}
z_{j\pm} = \left\{
\begin{array}{ll}
-a_j \pm \mbox{sign}(a_j) \sqrt{a_j^2 -  1 }
& \mbox{if } |a_j| > 1 \\
-a_j \mp i\sqrt{ 1  - a_j^2} = -e^{\pm i \phi_i)} 
& \mbox{if } |a_j|\leq 1 \\ 
\end{array} \right\}.
\end{eqnarray}
Here $\phi_i=\arccos \sqrt{1-a_i^2}$.

We will only consider the case where $|a_j|>1$. In this case, only the
only poles that contribute are the ones at $z=z_{j+}$.  The contour
integral is done by inspection to obtain
\begin{eqnarray}
I(\{a_i\},m,N) 
&=&
\sum_{k=1}^N
\int_{z_{k+}} \frac{dz}{2\pi i}
\frac{2 z^{|m|}}{(z-z_{k+})(z-z_{k-})}
\prod_{j=1,j\neq k}^N
\frac{2 z}{z^2 + 2 a_j z + 1} \\
&=&
\sum_{k=1}^N \frac{2 z_{k+}^{|m|}}{z_{k+}-z_{k-}}
\prod_{j=1,j\neq k}^N
\frac{2 z_{k+}}
{z_{k+}^2 + 2 a_j z_{k+} + 1}.
\end{eqnarray}
Note that
\begin{eqnarray}
z_{k\pm}^2 + 2 a_j z_{k\pm} + 1
&=& z_{k\pm}^2 + 2 a_k z_{k\pm} + 1
+ 2 (a_j -a_k) z_{k\pm}
= 2 (a_j -a_k) z_{k\pm}, \\
z_{k+}-z_{k-} &=& 
2\, \mbox{sign}(a_k) \sqrt{a_k^2-1}.
\end{eqnarray}
We define $c_k$ and $d_k$ by
\begin{eqnarray}
c_k &\equiv& \frac{z_{k+}-z_{k-}}{2}, \\
k_k &\equiv& z_{k+}^{|m|}.
\end{eqnarray}
With the appropriate replacements, we obtain
\begin{eqnarray}
I(\{a_i\},m,N) 
&=& \sum_{k=1}^N \frac{d_k}{c_k} \prod_{j=1,j\neq k}^N \frac{1}{a_j-a_k}.
\end{eqnarray}

Putting the $b_j$'s back in using the $a_i\rightarrow\frac{a_i}{b_i}$,
we get
\begin{eqnarray}
c_k &=& \mbox{sign}(a_k) \sqrt{a_k^2-b_k^2}, \\
d_k &=& \left(\frac{c_k-a_k}{b_k}\right)^{|m|}, \\
\int_0^{2\pi} \frac{d\phi}{2\pi} e^{i \phi m} 
\prod_{j=1}^N \frac{1}{a_j + b_j \cos\phi}
&=&
\sum_{k=1}^N \frac{d_k}{c_k}
\prod_{j=1,j\neq k}^N \frac{1}{a_j-a_k \frac{b_j}{b_k}}.
\end{eqnarray}
This expression is used in Eq.~(\ref{a:ci:firstint}) get
\begin{eqnarray}
&&\int_0^{2\pi} \frac{d\phi}{2\pi} e^{i \phi m} 
\prod_{j=1}^N \frac{1}{(a_j + b_j \cos\phi)^{n_j}} \nonumber\\
&& \qquad =
\prod_{l=1}^N \frac{(-1)^{(n_l-1)}}{(n_l-1)!} \frac{d^{n_l-1}}{d a_l^{n_l-1}}
\sum_{k=1}^N \frac{d_k}{c_k}
\prod_{j=1,j\neq k}^N \frac{1}{a_j-a_k \frac{b_j}{b_k}} \label{eqn:firstcosint}.
\end{eqnarray}

Specificly, for $N=1$, we can drop the subscripts.  For $n=1$,
\begin{eqnarray}
\int_0^{2\pi} \frac{d\phi}{2\pi} e^{i \phi m} 
\frac{1}{a + b \cos\phi}
&=& \frac{d}{c},
\end{eqnarray}
and for $n=2$,
\begin{eqnarray}
\int_0^{2\pi} \frac{d\phi}{2\pi} e^{i \phi m} 
\frac{1}{(a + b \cos\phi)^2}
&=& - \frac{d}{d a} \frac{d}{c}
= f_2 \frac{d}{c} \\
f_2 &=& \frac{a+|m| \, c}{c^2}.
\end{eqnarray}
Since $I(a,b,m)$ is like an eigenfunction of $\frac{d}{da}$, (but not
exactly, since the ``eigenvalue'' is a function of $a$), we an use
recursion to find that for general $n>2$,
\begin{eqnarray}
\int_0^{2\pi} \frac{d\phi}{2\pi} e^{i \phi m} 
\frac{1}{(a + b \cos\phi)^n}
&=& f_n I(a,b,m), \\
f_n &=& \frac{1}{n-1}\left( f_2 f_{n-1} - \frac{d}{da} f_{n-1} \right),
\end{eqnarray}
Using this we get
\begin{eqnarray}
f_1 &=& 1, \\
f_2 &=& \frac{a+|m| \, c}{c^2}, \\
f_3 &=& \frac{3 a^2 + c^2(m^2-1)+3a|m|c}{2c^4}.
\end{eqnarray}
The higher-order $f_i$'s can easily be found from analytic iteration
using a program like {\it Mathematica}, but apparently they do not have
a nice analytic form. We note that in general, $f_i$ is multiplied by a
factor of $c^{-2(i-1)}$.

We now check the equation for situations that may be numerically
unstable. First note that as $c\rightarrow0$, the $f$'s are singular,
which is due to the form of the original integral. Since the singular
part is multiplicative, it can easily be treated.

Instabilities appear when $b$ is very small. in which case we have to
take care of $d$ carefully as it appears to be singular. Let
$b=2a\delta$, so we can write
\begin{eqnarray*}
d_k &=& \left(\frac{\sqrt{1-4\delta^2}-1}{2\delta}\right)^{|m|} 
\approx \left[
-\delta
 ( 1 + \delta^2
 ( 1 + \delta^2
 ( 2 + \delta^2
 ( 5 + \delta^2
 (14 + \delta^2 \ldots ))))) \right]^{|m|}.
\end{eqnarray*}
This expression demonstrates that $d_k$ is not go as $b^{-|m|}$, but as
$b^{+|m|}$ as $b\rightarrow0$.

Now we analyze what happens where there is more than one term in
the denominator. We note first that
\begin{eqnarray}
\left[ \left( \frac{d}{dx} \right)^n f(x)g(x) \right]
&=& \sum_{m=0}^n \frac{n!}{m!(n-m)!}
\left[\left( \frac{d}{dx} \right)^m     f(x)\right]
\left[\left( \frac{d}{dx} \right)^{n-m} g(x)\right].
\end{eqnarray}
This means that Eq.~(\ref{eqn:firstcosint}) can be written as
\begin{eqnarray}
&&\int_0^{2\pi} \frac{d\phi}{2\pi} e^{i \phi m} 
\prod_{j=1}^N \frac{1}{(a_j + b_j \cos\phi)^{n_j}} \nonumber\\
&&\qquad =
\sum_{k=1}^N
\left[\prod_{l=1}^N \sum_{i_l=0}^{n_l-1} \right]
\left[
\frac{(-1)^{(n_l-1-i_l)}}{(n_l-1-i_l)!} \frac{d^{n_l-1-i_l}}{d a_l^{n_l-1-i_l}}
\left\{ \frac{d_k}{c_k} \right\}_{\text{on}}
\right]
\nonumber\\&&\qquad\qquad \times
\left[
\frac{(-1)^{(i_l)}}{(i_l)!} \frac{d^{i_l}}{d a_l^{i_l}}
\left\{
\prod_{j=1,j\neq k}^N \frac{1}{a_j-a_k \frac{b_j}{b_k}}
\right\}_{\text{on}}
\right].
\end{eqnarray}
The product can be broken into two parts, one where $l=k$ and another
where $l\neq k$.
\begin{eqnarray}
&&\int_0^{2\pi} \frac{d\phi}{2\pi} e^{i \phi m} 
\prod_{j=1}^N \frac{1}{(a_j + b_j \cos\phi)^{n_j}} \nonumber\\
&&\qquad=
\sum_{k=1}^N
\left[\prod_{l=1,l\neq k}^N \sum_{i_l=0}^{n_l-1} \right]
\left[
\frac{(-1)^{(n_l-1-i_l)}}{(n_l-1-i_l)!} \frac{d^{n_l-1-i_l}}{d a_l^{n_l-1-i_l}}
\left\{ \frac{d_k}{c_k} \right\}_{\text{on}}
\right]
\nonumber\\&&\qquad\qquad \times
\left[
\frac{(-1)^{(i_l)}}{(i_l)!} \frac{d^{i_l}}{d a_l^{i_l}}
\left\{
\prod_{j=1,j\neq k}^N \frac{1}{a_j-a_k \frac{b_j}{b_k}}
\right\}_{\text{on}}
\right]
\nonumber\\&&\qquad
\times 
\left[\sum_{i_k=0}^{n_k-1} \right]
\left[
\frac{(-1)^{(n_k-1-i_k)}}{(n_k-1-i_k)!} \frac{d^{n_k-1-i_k}}{d a_k^{n_k-1-i_k}}
\left\{ \frac{d_k}{c_k} \right\}_{\text{on}}
\right]
\nonumber\\&&\qquad\qquad \times
\left[
\frac{(-1)^{(i_k)}}{(i_k)!} \frac{d^{i_k}}{d a_k^{i_k}}
\left\{
\prod_{j=1,j\neq k}^N \frac{1}{a_j-a_k \frac{b_j}{b_k}}
\right\}_{\text{on}}
\right].
\end{eqnarray}
Now we note that $\frac{d}{da_l}\frac{d_k}{c_k}=0$, and so on, to get
\begin{eqnarray}
&&\int_0^{2\pi} \frac{d\phi}{2\pi} e^{i \phi m} 
\prod_{j=1}^N \frac{1}{(a_j + b_j \cos\phi)^{n_j}} \nonumber\\
&&\qquad=
\sum_{k=1}^N
\prod_{l=1,l\neq k}^N 
\left[
\frac{(-1)^{(n_l-1)}}{(n_l-1)!} \frac{d^{n_l-1}}{d a_l^{n_l-1}}
\left\{
\prod_{j=1,j\neq k}^N \frac{1}{a_j-a_k \frac{b_j}{b_k}}
\right\}_{\text{on}}
\right]
\nonumber\\&&\qquad\qquad\times 
\frac{d_k}{c_k} \sum_{i_k=0}^{n_k-1} f_{n_k-i_k}
\left[
\frac{(-1)^{i_k}}{i_k!} \frac{d^{i_k}}{d a_k^{i_k}}
\left\{
\prod_{j=1,j\neq k}^N \frac{1}{a_j-a_k \frac{b_j}{b_k}}
\right\}_{\text{on}}
\right]. \label{eq:cibigthing}
\end{eqnarray}
The first derivative term in Eq.~(\ref{eq:cibigthing}) can be written as
\begin{eqnarray}
&&\int_0^{2\pi} \frac{d\phi}{2\pi} e^{i \phi m} 
\prod_{j=1}^N \frac{1}{(a_j + b_j \cos\phi)^{n_j}} \nonumber\\
&&\qquad=
\sum_{k=1}^N \frac{d_k}{c_k} \sum_{i_k=0}^{n_k-1} f_{n_k-i_k}
\frac{(-1)^{i_k}}{i_k!} \frac{d^{i_k}}{d a_k^{i_k}}
\prod_{j=1,j\neq k}^N \frac{1}{\left(a_j-a_k
\frac{b_j}{b_k}\right)^{n_j}}.
\label{eq:ci:firstderterm}
\end{eqnarray}

Now, we consider the situation where all the $b_i$ have the same
value,$b$. In the case where $k=N$, the complicated part of
Eq.~(\ref{eq:ci:firstderterm}) is 
\begin{eqnarray}
h_N(i_N,N)
&\equiv& \frac{1}{i_N!} \frac{d^{i_N}}{d a_N^{i_N}}
\prod_{j=1}^{N-1} \frac{1}{(a_j-a_N)^{n_j}} \\
&=& \sum_{i_{N-1}=0}^{i_N} h_{N-1}(i_{N-1},N-1)
\frac{1}{(i_N-i_{N-1})!} 
\nonumber\\&&\qquad\times
\frac{d^{(i_N-i_{N-1})}}{d a_N^{(i_N-i_{N-1})}}
\frac{1}{(a_{N-1}-a_N)^{n_{N-1}}} \\
&=& \sum_{i_{N-1}=0}^{i_N} h_{N-1}(i_{N-1},N-1)
\frac{(i_N-i_{N-1}+  n_{N-1}-1)!}
     {(i_N-i_{N-1})!(n_{N-1}-1)!}
\nonumber\\&&\qquad\times
\frac{1}{(a_{N-1}-a_N)^{n_{N-1}+i_N-i_{N-1}}} \\
&=& \prod_{l=3}^N
\left[
\sum_{i_{l-1}=0}^{i_l}
\frac{(i_l-i_{l-1}+  n_{l-1}-1)!}
     {(i_l-i_{l-1})!(n_{l-1}-1)!} \frac{1}{(a_{l-1}-a_N)^{n_{l-1}+i_l-i_{l-1}}}
\right]
\\&&\times
\frac{(i_2+n_1-1)!}{i_2!(n_1-1)!} \frac{1}{(a_1-a_N)^{n_1+i_2}}.
\end{eqnarray}
From this, we can cyclically permute (or shuffle in any
order) the labels to get a similar expression for $k\neq N$.

For definiteness, we consider a few special cases. If $n_j=1$ for all
$j$ except $N$, we find that
\begin{eqnarray}
h_2(i,1) &=& \frac{\delta_{i,0}}{(a_2-a_1)^{n_2}}, \\
h_2(i,2) &=& \frac{1}{(a_1-a_2)^{i+1}},
\end{eqnarray}
and
\begin{eqnarray}
h_3(i,1) &=& \frac{\delta_{i,0}}{(a_2-a_1)(a_3-a_1)^{n_3}}, \\
h_3(i,2) &=& \frac{\delta_{i,0}}{(a_1-a_2)(a_3-a_2)^{n_3}}, \\
h_3(i,3) &=& \sum_{j=0}^i \frac{1}{(a_1-a_3)^{j+1}(a_2-a_3)^{i-j+1}}.
\end{eqnarray}
Using these we can write
\begin{eqnarray}
g_k &=& \sum_{i_k=0}^{n_k-1} (-1)^{i_k} f_{n_k-i_k} h_N(i_k,k), \\
\int_0^{2\pi} \frac{d\phi}{2\pi} e^{i \phi m} 
\prod_{j=1}^N \frac{1}{(a_j + b_j \cos\phi)^{n_j}}
&=& \sum_{k=1}^N \frac{d_k}{c_k} g_k.
\end{eqnarray}

\section*{Approximations}

For $N=2$, consider what happens when $a_1 \approx a_2$. We can write
$a_2 = a$, $a_1 = a - \delta$, so that
\begin{eqnarray}
&&I(a_1,a_2,b,n_1,n_2,m)\nonumber\\
&&\qquad= \int_0^{2\pi} \frac{d\phi}{2\pi} e^{i \phi m} 
\frac{1}{(a_2 + b \cos\phi - \delta)^{n_1}}
\frac{1}{(a_2 + b \cos\phi)^{n_2}} \\
&&\qquad= \int_0^{2\pi} \frac{d\phi}{2\pi} e^{i \phi m} 
\frac{1}{(a_2 + b \cos\phi)^{n_1+n_2}}
\sum_{j=0}^\infty \frac{(n_1-1+j)!}{(n_1-1)!j!}
\frac{\delta^j}{(a+b\cos\phi)^j} \\
&&\qquad= \left[ \sum_{j=0}^\infty (a_2-a_1)^j \frac{(n_1-1+j)!}{(n_1-1)!j!}
f_{n_1+n_2+j} \right] \frac{d_2}{c_2}.
\end{eqnarray}

For $N=3$, we could have $a_2\approx a_3$ (or equivalently
$a_1\approx a_3$). In this case,
\begin{eqnarray}
&&I(a_1,a_2,a_3,b,1,1,n,m) \nonumber\\
&&\qquad= \sum_{j=0}^\infty (a_3-a_2)^j
\int_0^{2\pi} \frac{d\phi}{2\pi} e^{i \phi m} 
\frac{1}{ a_1 + b \cos\phi}
\frac{1}{(a_3 + b \cos\phi)^{n+j+1}} \\
&&\qquad= \sum_{j=0}^\infty (a_3-a_2)^j I(a_1,a_3,b,1,n+j+1,m).
\end{eqnarray}
When $a_1\approx a_2$ but $a_1\not\approx a_3$, we get
\begin{eqnarray}
I(a_1,a_2,a_3,b,1,1,n,m)
&=& \sum_{j=0}^\infty (a_2-a_1)^j I(a_1,a_3,b,2+j,n,m).
\end{eqnarray}

\section*{Summary}

We define
\begin{eqnarray}
I(\{a_i\},b,\{n_i\},m) &=&
\int_0^{2\pi} \frac{d\phi}{2\pi} e^{i \phi m} 
\prod_i \frac{1}{(a_i+b\cos\phi)^{n_i}}, \\
c_k &=& a_k \sqrt{1-\left(\frac{b_k}{a_k}\right)^2}, \\
d_k &=& \left(\frac{c_k-a_k}{b_k}\right)^{|m|}.
\end{eqnarray}
We assume that $n_i>0$ and $b_i\neq0$.

For $N=1$, the integral is
\begin{eqnarray}
I(a,b,n,m) &=& f_n \frac{d}{c},
\end{eqnarray}
and for $N=2$, the integral is
\begin{eqnarray}
I(a_1,a_2,b,n_1,n_2,m)
&=&
(-1)^{n_2}
\left( \sum_{j=0}^{n_1-1} f_{n_1-j} \frac{(j+n_2-1)!}{j!(n_2-1)!}
\frac{1}{(a_1-a_2)^{j+n_2}} \right)
\frac{d_1}{c_1} \nonumber\\&&
+
(-1)^{n_1}
\left( \sum_{j=0}^{n_2-1} f_{n_2-j} \frac{(j+n_1-1)!}{j!(n_1-1)!}
\frac{1}{(a_2-a_1)^{j+n_1}} \right)
\frac{d_2}{c_2}.
\end{eqnarray}
For $N=2$, and with $n_1=1$ the integral simplifies to
\begin{eqnarray}
I(a_1,a_2,b,1,n,m)
&=& \frac{1}{(a_2-a_1)^n} \frac{d_1}{c_1}
- \left( \sum_{j=0}^{n-1} f_{n-j} \frac{1}{(a_2-a_1)^{j+1}} \right)
\frac{d_2}{c_2}.
\end{eqnarray}
When $N=3$ and $n_1=n_2=1$ the integral is
\begin{eqnarray}
&&I(a_1,a_2,a_3,b,1,1,n,m) \nonumber\\
&&\qquad=
\frac{1}{(a_2-a_1)(a_3-a_1)^n} \frac{d_1}{c_1} +
\frac{1}{(a_1-a_2)(a_3-a_2)^n} \frac{d_2}{c_2} \nonumber\\&& \qquad\qquad+
\left[ \sum_{i=0}^{n-1} (-1)^i f_{n-i}
\sum_{j=0}^i \frac{1}{(a_1-a_3)^{j+1}(a_2-a_3)^{i-j+1}} \right] \frac{d_3}{c_3}.
\end{eqnarray}

\chapter{Check of the Uncrossed Approximation}
\label{a:uxcheck}
\doquote
{For every problem, there is one solution which is simple, neat and wrong.}
{H.~L.~Mencken} 

In this section, we want to check that how well the approximation
\begin{eqnarray}
K_{\text{cross}} &\approx&  K_{\text{ladder}} G_C K_{\text{ladder}},
\end{eqnarray}
works.  Since we are using a Hamiltonian theory and are interested in the
potentials, we compare the potentials defined by
\begin{eqnarray}
V_{\text{TBE:X}} &=& \frac{1}{E} g_0^{-1}
\langle G_0 K_{\text{cross}} G_0 \rangle g_0^{-1}, \\
V_{\text{TBE:UX}} &=& \frac{1}{E} g_0^{-1}
\langle G_0 K_{\text{ladder}} G_C K_{\text{ladder}} G_0 \rangle g_0^{-1}.
\end{eqnarray}
The notation used here is defined in sections \ref{wc:sec:3dredbasic} and
\ref{wc:sect:3dred}.

The TBE crossed potential (TBE:X) can be written as
\begin{eqnarray}
V_{\text{TBE:X}} &=& 
V_{\text{TBE:SX}} +
V_{\text{TBE:TX}} +
V_{\text{TBE:WX}} +
V_{\text{TBE:ZX}},
\end{eqnarray}
where the potentials on the right-hand side are defined in section
\ref{wc:deftbepots}.  Calculation of the TBE approximate uncrossed
potential (TBE:UX) is straightforward, but tedious.  We find that
\begin{eqnarray}
V_{\text{TBE:UX}}(E;\boldsymbol{k}_1,\boldsymbol{p}_1) &=&
\frac{1}{2} \Big[
V_{\text{TBE:UX1}}(E;\boldsymbol{k}_1,\boldsymbol{p}_1) +
V_{\text{TBE:UX1}}(E;\boldsymbol{p}_1,\boldsymbol{k}_1)
\nonumber \\ 
& &  \phantom{ \frac{1}{2} \Big[} +
V_{\text{TBE:UX2}}(E;\boldsymbol{k}_1,\boldsymbol{p}_1) +
V_{\text{TBE:UX2}}(E;\boldsymbol{p}_1,\boldsymbol{k}_1) \Big],
\end{eqnarray}
where
\begin{eqnarray}
V_{\text{TBE:UX1}}(E;\boldsymbol{k}_1,\boldsymbol{p}_1)
&=&
\left( \frac{M}{E} \right)^4
\int \frac{d^2 q_\perp}{2(2\pi)^3} \left[
\int_0^1 dz \frac{\theta(x-z) \theta(z-y)}{(x-z) z^2 (z-y)} 
\right. \nonumber \\ & & \phantom{
\left( \frac{M}{E} \right)^4
} \left. \times
\frac{1}{E - p_1^- - k_2^- - \omega^-(\boldsymbol{q}_1-\boldsymbol{p}_1)
- \omega^-(\boldsymbol{k}_1-\boldsymbol{q}_1)}
\right. \nonumber \\ & & \phantom{
\left( \frac{M}{E} \right)^4
\int \frac{d^2 q_\perp}{2(2\pi)^3} \left[ \right.} \left. \times
\left(
\frac{1}{E - q_1^- - k_2^- - \omega^-(\boldsymbol{k}_1-\boldsymbol{q}_1)}
\right)^2
 \right] \nonumber \\ & & \phantom{
\left( \frac{M}{E} \right)^4
\int \frac{d^2 q_\perp}{2(2\pi)^3}} 
+ \{ 1\leftrightarrow 2\}, \\
V_{\text{TBE:UX2}}(E;\boldsymbol{k}_1,\boldsymbol{p}_1)
&=&
\left( \frac{M}{E} \right)^4
\int \frac{d^2 q_\perp}{2(2\pi)^3} \left[
\int_0^1 dz \frac{\theta(x-z) \theta(y-z)}{(x-z) z^2 (y-z)} 
\right. \nonumber \\ & & \phantom{
\left( \frac{M}{E} \right)^4
\int \frac{d^2 q_\perp}{2(2\pi)^3} \left[ \right.} \left. \times
\frac{1}{E - q_1^- - p_2^- - \omega^-(\boldsymbol{p}_1-\boldsymbol{q}_1)}
\right. \nonumber \\ & & \phantom{
\left( \frac{M}{E} \right)^4
\int \frac{d^2 q_\perp}{2(2\pi)^3} \left[ \right.} \left. \times
\left(
\frac{1}{E - q_1^- - k_2^- - \omega^-(\boldsymbol{k}_1-\boldsymbol{q}_1)}
\right)^2
 \right] \nonumber \\ & & \phantom{
\left( \frac{M}{E} \right)^4
\int \frac{d^2 q_\perp}{2(2\pi)^3}}
+ \{ 1\leftrightarrow 2\}.
\end{eqnarray}

Now consider the azimuthal-angle and loop integrals for these
potentials.  The approach used is similar to that of
Section~\ref{a:ints:loopints}.  Analyzing the potentials reveals that
each of the terms can be written in the following schematic form,
\begin{eqnarray}
&&V_{\text{TBE:UX,i}}(k^+,k_\perp;p^+,p_\perp) \nonumber\\
&&\qquad=
\int_0^\infty \frac{q_\perp dq_\perp}{2(2\pi)^3} \int_0^1 dz \,
J(k^+,q^+,p^+)
I_i(k^+,k_\perp,q^+,q_\perp,p^+,p_\perp) \\
&&I_i(k^+,k_\perp,q^+,q_\perp,p^+,p_\perp), \nonumber\\
&&\qquad= 
\int_0^{2\pi} 
\frac{d\phi_q d\phi}{A_{1,i} + B_{1,i} \cos \phi_q + C_{1,i} \cos \phi}
\left( \frac{1}{A_2 + B_2 \cos \phi_q} \right)^2,
\end{eqnarray}
where $\phi=-\phi_q+\phi_p$, and the $A$'s, $B$'s, and $C$'s may have
dependence on $k^+$, $k_\perp$, $p^+$, $p_\perp$, $q^+=zE$, and
$q_\perp$; they are independent of the azimuthal angles.  These factors
can be easily determined for each potential by examining the forms of
the original equations.  The rotational invariance of the potential
about the three-axis allows the $\phi_k$ integration to be done
trivially.  The $\phi$ integral is easily done to obtain
\begin{eqnarray}
I(k^+,k_\perp,q^+,q_\perp,p^+,p_\perp) &=& -2\pi
\int_0^{2\pi} d\phi_q
\frac{1}{\sqrt{a_i^2-C_{1,i}^2}}
\left( \frac{1}{A_2 + B_2 \cos \phi_q} \right)^2,
\end{eqnarray}
where
\begin{eqnarray}
a_i &=& A_{1,i} + B_{1,i} \cos \phi_q.
\end{eqnarray}

Further simplification is possible for $V_{\text{TBE:UX2}}$, since for
that potential $B_{1,2}=0$,
\begin{eqnarray}
I(k^+,k_\perp,q^+,q_\perp,p^+,p_\perp) &=& \frac{(2\pi)^2 A_2}{
\sqrt{A_{1,2}^2-C_{1,2}^2}\sqrt{A_2^2-B_2^2}(A_2^2-B_2^2)}.
\end{eqnarray}
The techniques discussed in the Section~\ref{a:ints:loopints} are used
to do the remaining loop integrals.

\begin{figure}[!ht]
\begin{center}
\epsfig{angle=0,width=5.0in,height=4.0in,file=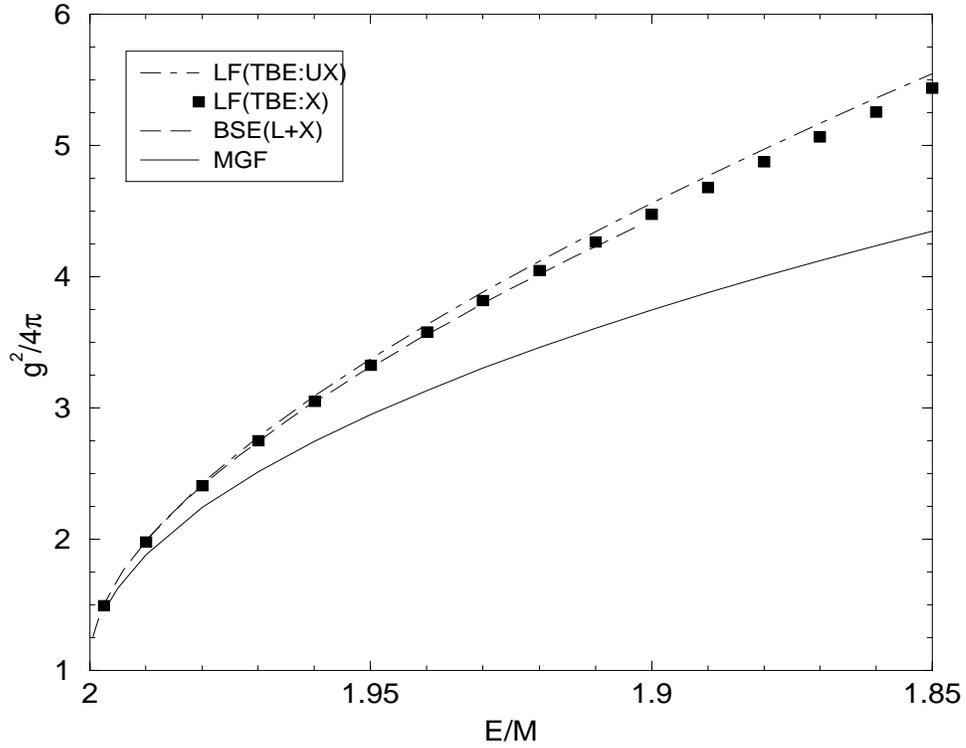}
\caption[
The spectra for the potential derived from the TBE
truncation of the uncrossed approximation, the TBE potential,
the ladder plus crossed Bethe-Salpeter equation, and the
modified-Green's-function potential.  Except for the MGF spectrum, all
the curves lie close to each other.
]
{The spectra for the potential derived from the TBE
truncation of the uncrossed approximation (OBE+TBE:SB+TBE:UX, denoted in
the figure by TBE:UX), the TBE potential (OBE+TBE, denoted by TBE:X),
the ladder plus crossed Bethe-Salpeter equation (BSE(L+X)), and the
modified-Green's-function potential (MGF).  Except for the MGF spectrum,
all the curves lie close to each other.
\label{a:checkux:uxtestfig}}
\end{center}
\end{figure}

The spectra for the OBE+TBE:SB+TBE:UX potential can be calculated and
compared to the OBE+TBE:SB+TBE:X potential (recall that this is the same
as the OBE+TBE potential), the modified-Green's-function potential, and
the ladder plus crossed Bethe-Salpeter equation.  The spectra are
plotted in Fig.~\ref{a:checkux:uxtestfig}.  The spectra for the TBE:UX,
TBE:X and BSE all lie close to each other, which shows that that the
uncrossed approximation is valid.

This shows that the important approximation in the
modified-Green's-function approach is not the uncrossed approximation,
but the addition of the extra interaction terms in
Eq.~(\ref{wc:ladderuncrossBSE}) which serve to mimic the higher order
interactions.

\chapter{Spinors} \label{app:spinors}
\doquote
{Everything should be made as simple as possible, but not simpler.}
{Albert Einstein} 

\section{Definition of the Spinors} \label{sp:spinorapp}

The Dirac equation is
\begin{eqnarray}
(i\NEG\partial - M)\psi(x) = 0.
\end{eqnarray}
We can multiply the Dirac equation by $( i \NEG\partial + M )$ to get
\begin{eqnarray}
0
= ( i \NEG\partial   + M  ) ( i \NEG\partial - M ) \psi(x)
= ( - \NEG\partial^2 + M^2) \psi(x)
= ( -     \partial^2 + M^2) \psi(x).
\end{eqnarray}
This means that for an eigenstate of momentum, $\psi_p(x) \propto e^{\pm
i p_\mu x^\mu}$, with  $p^2=M^2$.  The energy component of $p_\mu$ is
chosen to be positive.

Looking at the positive energy part of $\psi(x)$, (the part where the
sign in the exponent is negative), we let
\begin{eqnarray}
\psi_{p,+}(x) &=& u(\bbox{p}) e^{-ip_\mu x^\mu},
\end{eqnarray}
and for the negative energy part,
\begin{eqnarray}
\psi_{p,-}(x) &=& v(\bbox{p}) e^{+ip_\mu x^\mu}.
\end{eqnarray}
The $u(\bbox{p})$ and $v(\bbox{p})$ are column vectors.  They depend on
only the spatial part of $p$, since the momentum is required to be on
shell.  The column vectors satisfy
\begin{eqnarray}
(\NEG p-M)u(\bbox{p}) &=& (\NEG p+M)v(\bbox{p}) =0, \\
\bar{u}(\bbox{p})(\NEG p-M) &=& \bar{v}(\bbox{p})(\NEG p+M) =0.
\end{eqnarray}
From this point on, we will only be concerned with these momentum space
representations.

To get a Hamiltonian version of the Dirac equation, we multiply by
$\gamma^0$ to get
\begin{eqnarray}
(E - \bbox{\alpha}\cdot\bbox{p} - \beta M)u(\bbox{p}) &=& 0, \\
(E - \bbox{\alpha}\cdot\bbox{p} + \beta M)v(\bbox{p}) &=& 0.
\end{eqnarray}

Using the Bjorken and Drell representation, we can write the Hamiltonian
version of the Dirac equation as 
\begin{eqnarray}
\left(\begin{array}{cc}
E-M & -\bbox{\sigma}\cdot\bbox{p} \\ 
-\bbox{\sigma}\cdot\bbox{p} & E+M
\end{array} \right)
\left(\begin{array}{c} \chi(\bbox{p}) \\ u_B(\bbox{p}) \end{array} \right)
=
\left(\begin{array}{c} 0 \\ 0 \end{array} \right).
\end{eqnarray}
We find that
$u_B(\bbox{p}=\frac{\bbox{\sigma}\cdot\bbox{p}}{E+M}\chi(\bbox{p})$, so
\begin{eqnarray}
u(\bbox{p}) &=& c
\left(\begin{array}{c} 1 \\ \frac{\bbox{\sigma}\cdot\bbox{p}}{E+M}
\end{array} \right) \chi(\bbox{p}), \\
v(\bbox{p}) &=& c
\left(\begin{array}{c} \frac{\bbox{\sigma}\cdot\bbox{p}}{E+M} \\ 1
\end{array} \right) v_A(\bbox{p}),
\end{eqnarray}
where $c$ is the normalization constant.

The normalization we choose is
$\bar{u}(\bbox{p},\lambda')u(\bbox{p},\lambda)=\delta_{\lambda'\lambda}$,
so
\begin{eqnarray}
\delta_{\lambda'\lambda}
&=& \bar{u}(\bbox{p},\lambda')u(\bbox{p},\lambda) \\
&=& |c|^2
\left(\begin{array}{cc} 1 & \frac{\bbox{\sigma}\cdot\bbox{p}}{E+M}
\end{array} \right)
\left(\begin{array}{cc} 1 & 0 \\ 0 & -1\\ \end{array} \right)
\left(\begin{array}{c} 1 \\ \frac{\bbox{\sigma}\cdot\bbox{p}}{E+M}
\end{array} \right)
\chi^\dagger(\bbox{p},\lambda') \chi(\bbox{p},\lambda)
\\
&=& |c|^2
\left( 1 - \frac{\bbox{p}^2}{(E+M)^2} \right)
\chi^\dagger(\bbox{p},\lambda') \chi(\bbox{p},\lambda) \\
&=& |c|^2
\frac{M^2 + p^2 + E^2 + 2 E M + M^2 - p^2}{(E+M)^2}
\chi^\dagger(\bbox{p},\lambda') \chi(\bbox{p},\lambda) \\
&=& |c|^2 \frac{2 M}{E+M}
\chi^\dagger(\bbox{p},\lambda') \chi(\bbox{p},\lambda).
\end{eqnarray}
If we choose also 
$\chi^\dagger(\bbox{p},\lambda') \chi(\bbox{p},\lambda)=1$, then we get
that $c=\sqrt{\frac{W}{2M}}$, where $W=E+M$, so
\begin{eqnarray}
u(\bbox{p}) &=& \sqrt{\frac{W}{2M}}
\left(\begin{array}{c} 1 \\ \frac{\bbox{\sigma}\cdot\bbox{p}}{W}
\end{array} \right) \chi(\bbox{p}), \\
v(\bbox{p}) &=& \sqrt{\frac{W}{2M}}
\left(\begin{array}{c} \frac{\bbox{\sigma}\cdot\bbox{p}}{W} \\ 1
\end{array} \right) v_A(\bbox{p}).
\end{eqnarray}
We can choose any two linearly independent vectors for $\chi(\bbox{p})$
which correspond to the two spins. These different choices lead to the
different definitions of the spinors.

\subsection*{Assorted Spinor Definitions}
\doquote
{One day Alice came to a fork in the road and saw a Cheshire cat in a
tree. `Which road do I take?' she asked. His response was a question:
`Where do you want to go?' `I don't know,' Alice answered. `Then,' said
the cat, `it doesn't matter.' }
{Lewis Carroll}

The spinors used by Bjorken and Drell \cite{Bjorken:1964} are polarized
in the $\widehat{z}$ direction, and so the $\chi$'s are chosen to be
\begin{eqnarray}
\chi_{+1/2} &=& \left(\begin{array}{c} 1 \\ 0 \end{array} \right), \\
\chi_{-1/2} &=& \left(\begin{array}{c} 0 \\ 1 \end{array} \right).
\end{eqnarray}

The light-front spinors are defined to be \cite{Miller:1997cr}
\begin{eqnarray}
u_{\text{LF}}(k,\lambda)
&\equiv& \frac{1}{\sqrt{M k^+}} \left[ M \Lambda_- + (k^+ +
\bbox{\alpha}^\perp\cdot\bbox{k}^\perp ) \Lambda_+ \right]
\chi_{\text{LF},\lambda} \\
&=& \frac{1}{\sqrt{M k^+}} \left[ \Lambda_- 
( M + \bbox{\alpha}^\perp\cdot\bbox{k}^\perp )
+ \Lambda_+ k^+ \right] \chi_{\text{LF},\lambda}, \\
\chi_{\text{LF},\lambda} &\equiv&
\left( \begin{array}{c} \chi_\lambda \\ 0 \end{array} \right),
\end{eqnarray}
where $\chi_\lambda$ is the usual Pauli spinor, and the $\Lambda_\pm$
are the spinor projection operators defined in
Appendix~\ref{ch:notation}. We find that
\begin{eqnarray}
\overline{u}_{\text{LF}}(k,\lambda)
&=& \frac{1}{\sqrt{M k^+}} \chi^\dagger_{\text{LF},\lambda}
\left[ \Lambda_+ M + \Lambda_- (k^+ +
\bbox{\alpha}^\perp\cdot\bbox{k}^\perp ) \right] \\
&=& \frac{1}{\sqrt{M k^+}} \chi^\dagger_{\text{LF},\lambda}
\left[ ( M - \bbox{\alpha}^\perp\cdot\bbox{k}^\perp ) \Lambda_+ 
+ k^+ \Lambda_- \right].
\end{eqnarray}
Note that these spinors are normalized to satisfy
$\overline{u}_{\text{LF}}(k,\lambda')u_{\text{LF}}(k,\lambda)=\delta_{\lambda'\lambda}$.

For helicity spinors, we choose the eigenvectors of the helicity
operator ($\bbox{\sigma}\cdot\widehat{\bbox{p}}$) as the $\chi$'s. In
particular, 
$(\widehat{\bbox{p}}\cdot\bbox{\Sigma})u(\bbox{p},\lambda)=hu(\bbox{p},\lambda)$,
where $h=2\lambda$.  This choice allows us to write
\begin{eqnarray}
u(\bbox{p},\lambda) &=& \sqrt{\frac{W}{2M}}
\left(\begin{array}{c} 1 \\ \frac{hp}{W} \end{array} \right)
\chi_\lambda(\bbox{p})
\end{eqnarray}
where
$(\bbox{\sigma}\cdot\widehat{\bbox{p}})\chi_\lambda(\bbox{p})=h\chi_\lambda(\bbox{p})$.

One useful feature of the helicity spinors is the simplification
of the $\sigma\cdot p$ factor living in the prefactor of $\chi$ in the
definition of $u$.
This is an important feature: all of the directional dependence lives
in the $\chi$ alone.  (For the normal spinors, the $\chi$ are independent
of the momentum, but the prefactor has the dependence.)

Now, we proceed to calculate $\chi_\lambda(\bbox{p})$.  This will be
done by taking two steps: first calculate $\chi_\lambda(\widehat{z})$,
then rotate that to an arbitrary angle.

The first step is easy.  When $\bbox{p}=\widehat{\bbox{z}}$, the
helicity eigenvalue equation is 
\begin{eqnarray}
(\bbox{\sigma}\cdot\widehat{\bbox{z}})\chi_\lambda(\bbox{z})
&=&
\sigma_z \chi_\lambda(\bbox{z}) \\
&=&h \chi_\lambda(\bbox{z}).
\end{eqnarray}
This means that $\chi_\lambda(\bbox{z})$ are just the usual vectors for
the spin-1/2 basis,
\begin{eqnarray}
\chi_{+1/2}(\bbox{z}) &=& \left(\begin{array}{c} 1 \\ 0 \end{array} \right), \\
\chi_{-1/2}(\bbox{z}) &=& \left(\begin{array}{c} 0 \\ 1 \end{array} \right).
\end{eqnarray}

Now for the second step.  To rotate, use the Euler angle representation
of the rotation operator \cite{Sakurai:1994}:
\begin{eqnarray}
R_{\alpha,\beta,\gamma} &=& e^{-i\alpha J_z} e^{-i\beta J_y} e^{-i\gamma J_z}.
\end{eqnarray}
This operator allows us to write
\begin{eqnarray}
u(\bbox{p},\lambda) 
&=& R_{\phi,\theta,\gamma} u(p\times\widehat{\bbox{z}},\lambda) \\
&=& R_{\phi,\theta,\gamma} \sqrt{\frac{W}{2M}}
\left(\begin{array}{c} 1 \\ \frac{hp}{W} \end{array} \right)
\chi_\lambda(\widehat{\bbox{z}}).
\end{eqnarray}
Note that we are free to choose $\gamma$, since all it does is give a
phase to $\chi$.  The traditional approach is to set it equal to
$-\phi$.  This is what Brown and Jackson do \cite{Brown:1976},
and in general it is the
best thing to do, since it gives the correct rotations starting from a
state pointing in an arbitrary direction.  However, since our state just
points in the $z$-direction, this angle just gives a phase.  It will
turn out to be useful to just set $\gamma=0$.  This choice simplifies
the equations we will use later, but be warned that these kets transform
differently from those that Machleidt \cite{Machleidt:1987hj}
and Brown and Jackson use \cite{Brown:1976}, but they are closer to
what Rice and Kim \cite{Rice:1993} use.  Note that for 
totally symmetric potentials, both $\phi$ and $\phi'$ can be set to
zero, so this distinction is a moot point.  However, here (and
in Rice and Kim's work), we cannot set them both to zero, and we find
that our convention is much easier to work with.

So we have
\begin{eqnarray}
u(\bbox{p},\lambda) &=& R_{\phi,\theta,0} \sqrt{\frac{W}{2M}}
\left(\begin{array}{c} 1 \\ \frac{hp}{W} \end{array} \right)
\chi_\lambda(\widehat{\bbox{z}}).
\end{eqnarray}
The operator will not affect the prefactor. Since the spin and the
momentum rotate together, the factor of  $\bbox{\sigma}\cdot\bbox{p}$
remains unchanged.  The rotation operator only hits $\chi$, so that
\begin{eqnarray}
\chi_\lambda(\bbox{p})
&=& R_{\phi,\theta,0} 
\chi_\lambda(\widehat{\bbox{z}}).
\end{eqnarray}
The $\chi$ is an $SU(2)$ spinor, and the $SU(2)$ representation of the
angular momentum operator $J$ is $\sigma/2$.  Note that $\sigma_i^2=1$,
and that in general if $A^2=1$,
\begin{eqnarray}
e^{i\theta A} &=&
\sum_{n=0}^\infty \frac{(-1)^n}{(2n)!}   \theta^{2n} +
\sum_{n=0}^\infty \frac{(-1)^n}{(2n+1)!} \theta^{2n+1} i A \\
&=& \cos\theta + i A \sin\theta.
\end{eqnarray}
Thus,
\begin{eqnarray}
e^{-i\alpha J_z} &=& 
\left(\begin{array}{cc} e^{-i\alpha/2} & 0 \\ 0 & e^{+i\alpha/2}
\end{array} \right), \\
e^{-i\beta J_y} &=& \cos\frac{\beta}{2} - i \sigma_y \sin\frac{\beta}{2}
=
\left(\begin{array}{rr} 
\cos\frac{\beta}{2} & -\sin\frac{\beta}{2} \\
\sin\frac{\beta}{2} &  \cos\frac{\beta}{2} \\
\end{array} \right), \\
e^{-i\theta J_x} &=& \cos\frac{\theta}{2} - i \sigma_x \sin\frac{\theta}{2}
=
\left(\begin{array}{rr} 
   \cos\frac{\theta}{2} & -i \sin\frac{\theta}{2} \\
-i \sin\frac{\theta}{2} &    \cos\frac{\theta}{2} \\
\end{array} \right). \label{eq:expofjx}
\end{eqnarray}
This means that $R$ is
\begin{eqnarray}
R_{\phi,\theta,0} &=&
\left(\begin{array}{rr} 
e^{-i\phi/2} c_2 & - e^{-i\phi/2} s_2 \\
e^{+i\phi/2} s_2 &   e^{+i\phi/2} c_2 \\
\end{array} \right) \label{sp:rsu2},
\end{eqnarray}
where $c_2=\cos\frac{\theta}{2}$ and $s_2=\sin\frac{\theta}{2}$.
This gives
\begin{eqnarray}
\chi_\lambda(\bbox{k}) &=&
R_{\phi,\theta,0} \chi_\lambda(\widehat{\bbox{z}}) =
\chi_\lambda(\widehat{\bbox{p}}) =
\left\{\begin{array}{ll}
\left(\begin{array}{r}  c_2e^{-i\phi/2}\\ s_2e^{+i\phi/2} \end{array} \right)
& \mbox{ if $h=+1$} \\
\left(\begin{array}{r} -s_2e^{-i\phi/2}\\ c_2e^{+i\phi/2} \end{array} \right)
& \mbox{ if $h=-1$} \\
\end{array} \right. \phantom{\left. \right\}} 
\end{eqnarray}

\section{Two Helicity Spinors in the Center of Momentum Frame}

We now address the problem of two particles with spin. We work in the
center-of-momentum frame, and consider the relative momentum between the
particles. 

To start with, consider having the relative momentum pointing in
the $\widehat{\bbox{z}}$ direction. The first particle has momentum
$p\widehat{\bbox{z}}$ and helicity $\lambda_1$.  The helicity spinor for
this particle is just what was discussed in the previous section.

The second particle will have momentum $-p\widehat{\bbox{z}}$ and helicity
$\lambda_2$.  We can relate this to a spinor with the momentum pointing
in the $\widehat{\bbox{z}}$ direction via a rotation,
\begin{eqnarray}
u(-p\widehat{\bbox{z}},\lambda_2)
&=& R_{\phi,\pi,0} u(p\widehat{\bbox{z}},\lambda_2),
\end{eqnarray}
where $\phi$ is a free parameter because it has no meaning when pointing
along the $z$-axis. As argued in the previous section, this can be
recast as
\begin{eqnarray}
\chi_{\lambda_2}(-\widehat{\bbox{z}})
&=& R_{\phi,\pi,0} \chi_{\lambda_2}(\widehat{\bbox{z}}).
\end{eqnarray}
Using Eq.~(\ref{sp:rsu2}), we can write
\begin{eqnarray}
R_{\phi,\pi,0} &=&
\left(\begin{array}{rr} 
0 & - e^{-i\phi/2} \\ e^{+i\phi/2} & 0 \\
\end{array} \right).
\end{eqnarray}
If we choose $\phi=\pi$, this becomes
\begin{eqnarray}
R_{\pi,\pi,0} &=&
\left(\begin{array}{rr} 0 & i \\ i & 0 \\ \end{array} \right) ,
\end{eqnarray}
so that
\begin{eqnarray}
\chi_{\lambda_2}(-\widehat{\bbox{z}})
&=& i \chi_{-\lambda_2}(\widehat{\bbox{z}}).
\end{eqnarray}

When the relative momentum points in an arbitrary direction, we find
that for particle 2,
\begin{eqnarray}
\chi_{\lambda_2}(-\bbox{p})
&=& R_{\phi,\theta,0} \chi_{\lambda_2}(-\widehat{\bbox{z}}) \\
&=& i R_{\phi,\theta,0} \chi_{-\lambda_2}(\widehat{\bbox{z}}) \\
&=& i \chi_{-\lambda_2}(\bbox{p}).
\end{eqnarray}
This is rather simple.

\section{Helicity Spinor Summary}

We have
\begin{eqnarray}
u(\bbox{p},\lambda) &=& \sqrt{\frac{W}{2M}}
\left(\begin{array}{c} 1 \\ f \end{array} \right)
\chi_\lambda(\widehat{\bbox{p}}),
\end{eqnarray}
and
\begin{eqnarray}
\chi_\lambda(\widehat{\bbox{p}}) &=&
\left\{\begin{array}{ll}
\left(\begin{array}{r}  c_2e^{-i\phi/2}\\ s_2e^{+i\phi/2} \end{array} \right)
& \mbox{ if $h=+1$} \\
\left(\begin{array}{r} -s_2e^{-i\phi/2}\\ c_2e^{+i\phi/2} \end{array} \right)
& \mbox{ if $h=-1$} \\
\end{array} \right. \phantom{\left. \right\}} 
\end{eqnarray}
where $c_2=\cos\frac{\theta}{2}$, $s_2=\sin\frac{\theta}{2}$,
$f=\frac{hp}{W}$, and $h=2\lambda$.

When there are two fermions in the center-of-momentum frame, 
we can define $\phi\equiv\phi_1$ and $\theta\equiv\theta_1$ and for
particle two $\phi_2=\pi+\phi$ and $\theta_2=\pi-\theta$. This means
that
\begin{eqnarray}
u(\bbox{p}_i,\lambda_i) &=& \sqrt{\frac{W}{2M}}
\left(\begin{array}{c} 1 \\ f_i \end{array} \right)
\chi_{i,\lambda_i}(\widehat{\bbox{p}}),
\end{eqnarray}
where $i=1,2$ and 
\begin{eqnarray}
\chi_{1,\lambda_1}(\widehat{\bbox{p}})
&=&   \chi_{ \lambda_1}(\widehat{\bbox{p}}), \\
\chi_{2,\lambda_2}(\widehat{\bbox{p}})
&=& i \chi_{-\lambda_2}(\widehat{\bbox{p}}).
\end{eqnarray}

\section{Breakup of Dirac Spinor Matrix Elements}

We are interested in matrix elements with the form
\begin{eqnarray}
\overline{u}(\bbox{p}',\lambda') \Gamma u(\bbox{p},\lambda)
&=& \frac{\sqrt{W'W}}{2M}
\chi_{\lambda'}^\dagger(\bbox{p}') \left[
\left(\begin{array}{cc} 1 & -f' \end{array} \right)
\left(\begin{array}{cc} \Gamma_{AA} & \Gamma_{AB} \\
\Gamma_{BA} & \Gamma_{BB} \end{array} \right)
\left(\begin{array}{c} 1 \\ f \end{array} \right)
\right] \chi_\lambda(\bbox{p}) \nonumber\\
&=&  \frac{\sqrt{W'W}}{2M}\left(
\chi_{\lambda'}^\dagger \Gamma_{AA} \chi_\lambda 
 +  \chi_{\lambda'}^\dagger \Gamma_{AB} \chi_\lambda f
\right. \nonumber \\ && \left.
 -  \chi_{\lambda'}^\dagger \Gamma_{BA} \chi_\lambda f'
 -  \chi_{\lambda'}^\dagger \Gamma_{BB} \chi_\lambda f'f \right).
\end{eqnarray}
The elements of the $\Gamma_{ab}$ matrices will be either the identity
matrix or Pauli sigma matrices.

For example, we have
\begin{eqnarray}
\overline{u}(p',\lambda') u(p,\lambda)
&=& \chi_{\lambda'}^\dagger \chi_\lambda
\frac{\sqrt{W'W}}{2M} \left(1 - f'f \right), \\
\overline{u}(p',\lambda') \gamma^0 u(p,\lambda)
&=& \chi_{\lambda'}^\dagger \chi_\lambda 
\frac{\sqrt{W'W}}{2M} \left(1 + f'f \right), \\
\overline{u}(p',\lambda') \gamma^5 u(p,\lambda)
&=& \chi_{\lambda'}^\dagger \chi_\lambda 
\frac{\sqrt{W'W}}{2M} \left( f - f' \right),
\end{eqnarray}
and some more complicated ones,
\begin{eqnarray}
\overline{u}(p',\lambda') \gamma^i u(p,\lambda)
&=&
\chi_{\lambda'}^\dagger \sigma^i \chi_\lambda 
\frac{\sqrt{W'W}}{2M} \left( f + f' \right), \\
\overline{u}(p',\lambda') \sigma^{\mu\nu} i (p'-p)_\nu  u(p,\lambda)
&=&
2 M \overline{u}(p',\lambda') \gamma^\mu u(p,\lambda)
\nonumber\\&&
- (p'+p)^\mu \overline{u}(p',\lambda') u(p,\lambda).
\end{eqnarray}
We used the Gordon decomposition (Section~\ref{sp:gordon}) to simplify
the last equation. This simplification of the tensor allows us to
only worry about two tricky matrix elements for vector mesons.

With the Gordon decomposition, we can further simplify the vector meson
coupling by using the Dirac equation. Note that the expression for
$\Gamma$ is only valid in this context.
\begin{eqnarray}
\Gamma^\mu_i &=& \left(1+\frac{f}{g}\right) \gamma^\mu - \frac{f}{g}
\frac{1}{2M} (p'_i+p_i)^\mu, \\
U_{\text{v}} &=&
\overline{u}(\bbox{p}'_1,\lambda'_1) \Gamma^\mu_1 u(\bbox{p}_1,\lambda_1)
(-g_{\mu\nu})
\overline{u}(\bbox{p}'_2,\lambda'_2) \Gamma^\nu_2 u(\bbox{p}_2,\lambda_2)
\\
&=&
\left(1+\frac{f}{g}\right)^2
\overline{u}(\bbox{p}'_1,\lambda'_1) \gamma^\mu u(\bbox{p}_1,\lambda_1)
(-g_{\mu\nu})
\overline{u}(\bbox{p}'_2,\lambda'_2) \gamma^\nu u(\bbox{p}_2,\lambda_2)
\nonumber\\&&
+\left(1+\frac{f}{g}\right) \frac{f}{g} \frac{1}{2M}
\overline{u}(\bbox{p}'_1,\lambda'_1) (\NEG{p}'_2+\NEG{p}_2) 
          u (\bbox{p}_1,\lambda_1)
\overline{u}(\bbox{p}'_2,\lambda'_2) u(\bbox{p}_2,\lambda_2)
\nonumber\\&&
+\left(1+\frac{f}{g}\right) \frac{f}{g} \frac{1}{2M}
\overline{u}(\bbox{p}'_1,\lambda'_1) u(\bbox{p}_1,\lambda_1)
\overline{u}(\bbox{p}'_2,\lambda'_2) (\NEG{p}'_1+\NEG{p}_1)
          u (\bbox{p}_2,\lambda_2)
\nonumber\\&&
-\left(\frac{f}{g}\right)^2 (p'_1+p_1)^\mu (p'_2+p_2)_\mu \frac{1}{4M^2}
\overline{u}(\bbox{p}'_1,\lambda'_1) u(\bbox{p}_1,\lambda_1)
\overline{u}(\bbox{p}'_2,\lambda'_2) u(\bbox{p}_2,\lambda_2).
\end{eqnarray}
The Dirac equation can be used to write
$\NEG{p}_2u(p_1,\lambda)=(2E\gamma^0-M)u(p_1,\lambda)$, and also note
that 
\begin{eqnarray}
(p'_1+p_1)^\mu(p'_2+p_2)_\mu
&=& 2 (E'+E)^2 - (p'_1+p_1)^\mu(p'_1+p_1)_\mu \\
&=& 2 (E'+E)^2 - 2 M^2 - 2 E'E + 2 \bbox{p}'\cdot\bbox{p}.
\end{eqnarray}
This means we can write
\begin{eqnarray}
U_{\text{v}} &=&
\left(1+\frac{f}{g}\right)^2
\overline{u}(\bbox{p}'_1,\lambda'_1) \gamma^\mu u(\bbox{p}_1,\lambda_1)
(-g_{\mu\nu})
\overline{u}(\bbox{p}'_2,\lambda'_2) \gamma^\nu u(\bbox{p}_2,\lambda_2)
\nonumber\\&&
-2\left(1+\frac{f}{g}\right) \frac{f}{g} 
\overline{u}(\bbox{p}'_1,\lambda'_1) u(\bbox{p}_1,\lambda_1)
\overline{u}(\bbox{p}'_2,\lambda'_2) u(\bbox{p}_2,\lambda_2)
\nonumber\\&&
+\left(1+\frac{f}{g}\right) \frac{f}{g} \frac{E'+E}{M}
\overline{u}(\bbox{p}'_1,\lambda'_1) \gamma^0 u(\bbox{p}_1,\lambda_1)
\overline{u}(\bbox{p}'_2,\lambda'_2)          u(\bbox{p}_2,\lambda_2)
\nonumber\\&&
+\left(1+\frac{f}{g}\right) \frac{f}{g} \frac{E'+E}{M}
\overline{u}(\bbox{p}'_1,\lambda'_1)          u(\bbox{p}_1,\lambda_1)
\overline{u}(\bbox{p}'_2,\lambda'_2) \gamma^0 u(\bbox{p}_2,\lambda_2)
\nonumber\\&&
-\left(\frac{f}{g}\right)^2 
\frac{(E'+E)^2 - M^2 - E'E + \bbox{p}'\cdot\bbox{p}}{2M^2}
\nonumber\\&&\qquad\times
\overline{u}(\bbox{p}'_1,\lambda'_1) u(\bbox{p}_1,\lambda_1)
\overline{u}(\bbox{p}'_2,\lambda'_2) u(\bbox{p}_2,\lambda_2).
\end{eqnarray}

We are interested in the matrix elements for OME and TBE exchange
potentials. Thus, we want to look at combinations of two $\overline{u}u$
matrix elements.  To save space, we will write
\begin{eqnarray}
\langle \lambda'_1,\lambda'_2 | \lambda_1,\lambda_2 \rangle &=&
\left(\chi_{ \lambda'_1}^\dagger \chi_{ \lambda_1}\right)
\left(\chi_{-\lambda'_2}^\dagger \chi_{-\lambda_2}\right), \\
\langle \lambda'_1,\lambda'_2 | \bbox{\sigma}_1\cdot\bbox{\sigma}_2 
| \lambda_1,\lambda_2 \rangle &=&
\left(\chi_{ \lambda'_1}^\dagger \sigma^i \chi_{ \lambda_1}\right)
\left(\chi_{-\lambda'_2}^\dagger \sigma^i \chi_{-\lambda_2}\right).
\end{eqnarray}

We can write the matrix element for the scalar exchange as
\begin{eqnarray}
U_{\text{s}}&=&
\overline{u}(\bbox{p}'_1,\lambda'_1) u(\bbox{p}_1,\lambda_1) \times
\overline{u}(\bbox{p}'_2,\lambda'_2) u(\bbox{p}_2,\lambda_2) \nonumber\\
&=& 
\frac{W'W}{4M^2} 
\left(1 - \frac{h'_2h_1 p'p}{W'W} \right)
\left(1 - \frac{h'_2h_2 p'p}{W'W} \right)
\langle \lambda'_1,\lambda'_2 | \lambda_1,\lambda_2 \rangle,
\end{eqnarray}
and the matrix element for the pseudoscalar exchange as
\begin{eqnarray}
U_{\text{ps}} &=&
\overline{u}(\bbox{p}'_1,\lambda'_1) i \gamma^5 u(\bbox{p}_1,\lambda_1) \times
\overline{u}(\bbox{p}'_2,\lambda'_2) i \gamma^5 u(\bbox{p}_2,\lambda_2)
\nonumber\\
&=& 
- \frac{W'W}{4M^2} 
\left( \frac{h_1 p}{W} - \frac{h'_1 p'}{W'} \right)
\left( \frac{h_2 p}{W} - \frac{h'_2 p'}{W'} \right)
\langle \lambda'_1,\lambda'_2 | \lambda_1,\lambda_2 \rangle.
\end{eqnarray}

Now for the vector meson. Note that one part is the same as for the
scalar meson. The other part is more complicated, the vector-vector part:
\begin{eqnarray}
U_{\text{vv}} &=&
\overline{u}(\bbox{p}'_1,\lambda'_1) \gamma^\mu u(\bbox{p}_1,\lambda_1)
(-g_{\mu\nu})
\overline{u}(\bbox{p}'_2,\lambda'_2) \gamma^\nu u(\bbox{p}_2,\lambda_2)
\nonumber\\
&=& 
- \frac{W'W}{4M^2} \Big[
\left(1 + \frac{h'_1h_1 p'p}{W'W} \right)
\left(1 + \frac{h'_2h_2 p'p}{W'W} \right)
\langle \lambda'_1,\lambda'_2 | \lambda_1,\lambda_2 \rangle
- \nonumber\\&&
\phantom{- \frac{W'W}{4M^2} \Big[]}
\left( \frac{h_1p}{W} + \frac{h'_1p'}{W'} \right)
\left( \frac{h_2p}{W} + \frac{h'_2p'}{W'} \right)
\langle \lambda'_1,\lambda'_2 | \bbox{\sigma}_1\cdot\bbox{\sigma}_2 
| \lambda_1,\lambda_2 \rangle \Big].
\end{eqnarray}
This means that the full $\overline{u}u$ matrix element for a vector meson with a
non-vanishing tensor coupling is
\begin{eqnarray}
U_{\text{v}}
&=& \left(1+\frac{f}{g}\right)^2 U_{\text{vv}}
+ \frac{f}{g} \left[ 2 + \frac{f}{g} \left( \frac{3}{2} 
+ \frac{-E'E+\bbox{p}'\cdot\bbox{p}}{2M^2} \right) \right] U_{\text{s}} \\
&=& \left(1+\frac{f}{g}\right)^2 U_{\text{vv}}
+ \frac{f}{g} \left[ 2 + \frac{f}{g} \left( \frac{3}{2} 
+ \frac{-E'E+{p'}^3p^3}{2M^2} \right) \right] U_{\text{s}} \nonumber\\&&
+ \left(\frac{f}{g}\right)^2 \frac{p'_\perp p_\perp}{4M^2} U_{\text{s}}
\left( e^{i(\phi'-\phi)}+e^{-i(\phi'-\phi)}\right).
\end{eqnarray}

We can write this out as 
vector-tensor, and tensor-tensor
parts. The vector-vector is just $U_{\text{vv}}$.
\begin{eqnarray}
U_{\text{vt}} &=& \frac{2f}{g} \left[ U_{\text{vv}} + U_{\text{s}} \right] \\
&=& 
- \frac{2f}{g} \frac{W'W}{4M^2} \Big[
2\left(\frac{h'_1h_1 p'p}{W'W} + \frac{h'_2h_2 p'p}{W'W} \right)
\langle \lambda'_1,\lambda'_2 | \lambda_1,\lambda_2 \rangle
- \nonumber\\&&
\phantom{- \frac{W'W}{4M^2} \Big[]}
\left( \frac{h_1p}{W} + \frac{h'_1p'}{W'} \right)
\left( \frac{h_2p}{W} + \frac{h'_2p'}{W'} \right)
\langle \lambda'_1,\lambda'_2 | \bbox{\sigma}_1\cdot\bbox{\sigma}_2 
| \lambda_1,\lambda_2 \rangle \Big], \\
U_{\text{tt}} &=& \left(\frac{f}{g}\right)^2
\left[ U_{\text{vv}} + 
\frac{1}{2} \left( 3 -\frac{E'E}{M^2} 
+ \frac{\bbox{p}'\cdot\bbox{p}}{M^2} \right) U_{\text{s}} \right] \\
&=& 
- \left( \frac{f}{g} \right)^2 \frac{W'W}{4M^2} \Big[
\langle \lambda'_1,\lambda'_2 | \lambda_1,\lambda_2 \rangle
\left\{
\left(1 + \frac{h'_1h_1 p'p}{W'W} \right)
\left(1 + \frac{h'_2h_2 p'p}{W'W} \right)
+ \nonumber \right. \\ && \left. \qquad\qquad
\frac{3M^2 - E'E + \bbox{p}'\cdot\bbox{p}}{2M^2}
\left(1 - \frac{h'_1h_1 p'p}{W'W} \right)
\left(1 - \frac{h'_2h_2 p'p}{W'W} \right)
\right\}
- \nonumber\\&&
\phantom{- \frac{W'W}{4M^2} \Big[]}
\left( \frac{h_1p}{W} + \frac{h'_1p'}{W'} \right)
\left( \frac{h_2p}{W} + \frac{h'_2p'}{W'} \right)
\langle \lambda'_1,\lambda'_2 | \bbox{\sigma}_1\cdot\bbox{\sigma}_2 
| \lambda_1,\lambda_2 \rangle \Big].
\end{eqnarray}

\section{The Gordon Decomposition} \label{sp:gordon}
\doquote
{Nothing shocks me. I'm a scientist.}
{Indiana Jones}

Now, let's look at the Gordon decomposition.  We start by looking at
\begin{eqnarray}
i \sigma^{\mu\nu}
= \frac{-1}{2} \left(
\gamma^\mu \gamma^\nu - \gamma^\nu \gamma^\mu \right)
&=& g^{\mu\nu} - \gamma^\mu \gamma^\nu.
\end{eqnarray}
Then,
\begin{eqnarray}
\overline{u}(p',\lambda') \sigma^{\mu\nu} i (p'-p)_\nu u(p,\lambda)
&=& 
-\overline{u}(p',\lambda') \sigma^{\mu\nu} i p_\nu u(p,\lambda)
-\overline{u}(p',\lambda') i p'_\nu \sigma^{\nu\mu} u(p,\lambda)
\nonumber \\
&=& 
\overline{u}(p',\lambda') \gamma^\mu \NEG{p} u(p,\lambda)
+\overline{u}(p',\lambda') \NEG{p}' \gamma^\mu u(p,\lambda)
\nonumber\\&&
- (p'+p)^\mu \overline{u}(p',\lambda') u(p,\lambda).
\end{eqnarray}
Using the Dirac equation, we see that
\begin{eqnarray}
\overline{u}(p',\lambda') \sigma^{\mu\nu} i (p'-p)_\nu u(p,\lambda)
&=& 
2M \overline{u}(p',\lambda') \gamma^\mu u(p,\lambda)
- (p'+p)^\mu \overline{u}(p',\lambda') u(p,\lambda),
\end{eqnarray}
which is the Gordon identity.

\chapter{Rotation Matrices} \label{app:rotmat}
\doquote
{There comes a time when for every addition of knowledge you forget
something that you knew before. It is of the highest importance,
therefore, not to have useless facts elbowing out the useful ones.}
{Arthur Conan Doyle}

\section{Relations Between Rotation Matrices}

Brown and Jackson give a good introduction to this topic
\cite{Brown:1976}.  We start out with the rotation operator,
parameterized in terms of the Euler angles,
\begin{eqnarray}
R_{\alpha\beta\gamma} &=&
e^{-i\alpha J_z} e^{-i\beta J_y} e^{-i\gamma J_z},
\end{eqnarray}
We can take matrix elements of this, to get
\begin{eqnarray}
\langle J' M' | R_{\alpha\beta\gamma} | J M \rangle
&=& \delta_{J'J} {\mathcal D}^J_{M'M}(\alpha\beta\gamma) \\
&=& \delta_{J'J} e^{-i(\alpha M'+\gamma M)} d^J_{M'M}(\beta).
\end{eqnarray}
Note that the lower indices in $d$ are in the same order as in
${\mathcal D}$.  This is different that the convention in
\cite{Machleidt:1987hj,Brown:1976}, but agrees with
\cite{Rice:1993,Sakurai:1994,Chaichian:1998}.  However, we also define
the matrix elements slightly differently, and the combination of those
two effects makes the calculated Jacobi polynomials the same!

An alternative interpretation of the ${\mathcal D}$'s is that they are
the overlap of $JM\lambda$ kets with $\theta\phi\lambda$ kets. To see
this, generalize Sakurai's approach for kets without spin
\cite{Sakurai:1994}. 
We start out with $|\widehat{z}\lambda\rangle$.  This is a ket for
a particle pointed upwards, with $\lambda$ units of spin in the
$z$-direction.  This ket can be rotated with $R$ to get
\begin{eqnarray}
R_{\phi\theta\gamma} |\widehat{z} \lambda \rangle
&=& |\theta\phi\lambda \rangle.
\end{eqnarray}
Brown and Jackson choose $\gamma=-\phi$, however our convention is to
choose $\gamma=0$ and get 
\begin{eqnarray}
R_{\phi\theta0} |\widehat{z} \lambda \rangle &=& |\theta\phi\lambda \rangle
\end{eqnarray}

Applying a $\langle J' M' \lambda' |$ bra to the previous equation, we
obtain
\begin{eqnarray}
\langle J' M' \lambda' |\theta\phi \lambda \rangle
&=&
\langle J' M' \lambda' | R_{\phi\theta0} |\widehat{z} \lambda \rangle
=
\sum_{J,M,\lambda''}
\langle J' M' \lambda' | R_{\phi\theta0} | JM\lambda'' \rangle
\langle JM\lambda'' |\widehat{z} \lambda \rangle.
\end{eqnarray}
In the last bra-ket, we must have $\lambda=\lambda''$, since the
helicity cannot be changed.
Furthermore, since $\lambda$ is the $J_z$ quantum number for the final
state, we also have $\lambda=M$. These facts allow us to write
\begin{eqnarray}
\langle JM\lambda'' |\widehat{z} \lambda \rangle
&=&
\delta_{\lambda''\lambda} \delta_{M\lambda} \langle JM |\widehat{z} \rangle
=
\delta_{\lambda''\lambda} \delta_{M\lambda} Y_{JM}(0,\phi)
=
\delta_{\lambda''\lambda} \delta_{M\lambda} \sqrt{\frac{2J+1}{4\pi}},
\end{eqnarray}
which means that
\begin{eqnarray}
\langle J M \lambda |\theta \phi \lambda \rangle
&=&
\langle J M | R_{\phi\theta0} | J \lambda \rangle
\sqrt{\frac{2J+1}{4\pi}}
=\sqrt{\frac{2J+1}{4\pi}} {\mathcal D}^J_{M\lambda}(\phi\theta0).
\end{eqnarray}
Since the $\phi$ dependence of the ${\mathcal D}$'s is essentially
trivial, we can replace the $0$ with a $\gamma$ to get
\begin{eqnarray}
{\mathcal D}^J_{M\lambda}(\phi\theta\gamma)
&=&
\sqrt{\frac{4\pi}{2J+1}} 
\langle J M \lambda |\theta \phi \lambda \rangle
e^{-i \gamma \lambda},
\end{eqnarray}
and
\begin{eqnarray}
\langle J M \lambda |\theta \phi \lambda \rangle
&=&
\sqrt{\frac{2J+1}{4\pi}} {\mathcal D}^J_{M\lambda}(\phi \theta 0).
\end{eqnarray}

Using these relations, we find that
\begin{eqnarray}
{{\mathcal D}^J_{M'M}}^*(\alpha\beta\gamma)
&=& {\mathcal D}^J_{MM'}(-\gamma-\beta-\alpha), \\
d^J_{M'M}(\beta) &=& d^J_{MM'}(-\beta), \\
{\mathcal D}^J_{M0}(\phi,\theta,X) &=& \sqrt{\frac{4\pi}{2J+1}} Y_{JM}(\Omega)
= \sqrt{\frac{(J-M)!}{(J+M)!}} P^M_J(\cos\theta) e^{i M \phi}, \\
d^J_{M0}(\beta) &=& \sqrt{\frac{(J-M)!}{(J+M)!}} P^M_J(\cos\beta), \\
d^J_{00}(\beta) &=& P_J(\cos\beta), \\
{\mathcal D}^J_{M'M}(0,0,0) &=& \delta_{M'M} = d^J_{M'M}(0).
\end{eqnarray}

Now, let's use this.  Our first thing to do is look at is the addition
theorem
\begin{eqnarray}
\sum_{M''}
{\mathcal D}^J_{MM''}(\alpha_2,&\beta_2&,\gamma_2)
{\mathcal D}^J_{M''M'}(\alpha_1,\beta_1,\gamma_1) \nonumber\\
&=&
\sum_{M''}
\langle JM  | R_{\alpha_2,\beta_2,\gamma_2} | JM'' \rangle
\langle JM''| R_{\alpha_1,\beta_1,\gamma_1} | JM'  \rangle \\
&=&
\langle JM  | R_{\alpha_2,\beta_2,\gamma_2}
R_{\alpha_1,\beta_1,\gamma_1} | JM'  \rangle \\
&=& \langle JM  | R_{\alpha,\beta,\gamma} | JM'  \rangle \\
&=& {\mathcal D}^J_{MM'}(\alpha,\beta,\gamma).
\end{eqnarray}
We used the group property of the rotations that the product of two
elements gives another element.  The expressions for
$(\alpha,\beta,\gamma)$ in terms of the other angles is 
rather complex in general.

We can also look at the orthogonality relation, considering the
${\mathcal D}$'s as overlaps,
\begin{eqnarray}
\frac{2J+1}{8\pi^2}&& \int_0^{2\pi} d\alpha
\int_0^\pi \sin\beta \, d\beta
\int_0^{2\pi} d\gamma
{{\mathcal D}^{J_1}_{M_1'M_1}}^*(\alpha\beta\gamma)
 {\mathcal D}^{J_2}_{M_2'M_2}   (\alpha\beta\gamma) \nonumber \\
&&=
\frac{2J+1}{8\pi^2}
\sqrt{\frac{4\pi}{2J_1+1}\frac{4\pi}{2J_2+1}}
\int_0^{2\pi} d\alpha
\int_0^\pi \sin\beta \, d\beta
\int_0^{2\pi} d\gamma
\nonumber \\ &&\qquad
\left(
\langle J_1 M_1' M_1 | \alpha \beta M_1 \rangle e^{-i\gamma M_1}
\right)^*
\langle J_2 M_2' M_2 | \alpha \beta M_2 \rangle e^{-i\gamma M_2}
\nonumber \\
&&=
(2J+1) 
\sqrt{\frac{1}{2J_1+1}\frac{1}{2J_2+1}}
\left[ \int_0^{2\pi} \frac{d\gamma}{2\pi} e^{+i\gamma (M_1-M_2)} \right]
\nonumber \\ &&\qquad
\int d\Omega
\langle J_2 M_2' M_2 | \Omega M_2 \rangle 
\langle \Omega M_1 | J_1 M_1' M_1 \rangle  \nonumber \\
&&=
(2J+1) \sqrt{\frac{1}{2J_1+1}\frac{1}{2J_2+1}} \delta_{M_1M_2}
\nonumber \\ &&\qquad
\int d\Omega
\langle J_2 M_2' M_1 | \Omega M_1 \rangle 
\langle \Omega M_1 | J_1 M_1' M_1 \rangle  \nonumber \\
&&=
(2J+1) \sqrt{\frac{1}{2J_1+1}\frac{1}{2J_2+1}} \delta_{M_1M_2}
\langle J_2 M_2' M_1 | J_1 M_1' M_1 \rangle  \nonumber \\
&&= \delta_{M_1M_2} \delta_{J_1J_2} \delta_{M_1'M_2'}.
\end{eqnarray}
So the conclusion is
\begin{eqnarray}
&&\frac{2J+1}{8\pi^2} \int_0^{2\pi} d\alpha
\int_0^\pi \sin\beta \, d\beta
\int_0^{2\pi} d\gamma
{{\mathcal D}^{J_1}_{M_1'M_1}}^*(\alpha\beta\gamma)
 {\mathcal D}^{J_2}_{M_2'M_2}   (\alpha\beta\gamma) \nonumber\\
&& \qquad \qquad = \delta_{M_1M_2} \delta_{M'_1M'_2} \delta_{J_1J_2}. 
\end{eqnarray}

Using this, we find
\begin{eqnarray}
\int d\Omega \, \frac{2J+1}{4\pi} 
{{\mathcal D}^{J_1}_{M_1M'}}^*(\phi,\theta,-\phi)
 {\mathcal D}^{J_2}_{M_2M'}   (\phi,\theta,-\phi)
&=& \delta_{M_1M_2} \delta_{J_1J_2}, \\
\sum_{J,M} \frac{2J+1}{4\pi}
{{\mathcal D}^{J}_{MM'}}^*(\phi',\theta',-\phi')
 {\mathcal D}^{J}_{MM'}   (\phi,\theta,-\phi)
&=& \delta(\Omega-\Omega'), \\
\int_{-1}^1 d\cos\theta \frac{2J+1}{2} 
d^{J_1}_{M'M}(\theta) d^{J_2}_{M'M}(\theta)
&=& \delta_{J_1J_2}, \\
\sum_J \frac{2J+1}{2} d^{J}_{MM'}(\theta') d^{J}_{MM'}(\theta )
&=& \delta(\cos\theta-\cos\theta').
\end{eqnarray}

Now, for two particle state in the center-of-momentum frame, we can
write the ket as $|p\theta\phi\lambda_1\lambda_2\rangle$.  The total
helicity of this state is $\lambda=\lambda_1-\lambda_2$, which is easily
seen if we choose the momentum to be in the $z$-direction.  Since this
ket has total helicity $\lambda$, we can transform it using
\begin{eqnarray}
|pJM\lambda_1\lambda_2\rangle
&=& \int d\Omega \,
|p\theta\phi\lambda_1\lambda_2\rangle
\langle p\theta\phi\lambda_1\lambda_2 | pJM\lambda_1\lambda_2 \rangle \\
&=& \int d\Omega \,
|p\theta\phi\lambda_1\lambda_2\rangle
\langle \theta\phi \lambda | JM \lambda \rangle \\
&=&
\sqrt{\frac{2J+1}{4\pi}} \int d\Omega \,
{{\mathcal D}^J_{M\lambda}}^*(\theta\phi0) \,
|p\theta\phi\lambda_1\lambda_2\rangle,
\label{origtrans}
\end{eqnarray}
where $\lambda = \lambda_1 - \lambda_2$.

We can break apart Eq.~(\ref{origtrans}) as we did for the orthogonality
relations,
\begin{eqnarray}
|pJM\lambda_1\lambda_2\rangle
&=& \sqrt{\frac{2J+1}{4\pi}} \int d\Omega \,
{{\mathcal D}^J_{M\lambda}}^*(\phi, \theta, 0)
|p\theta\phi\lambda_1\lambda_2\rangle \\
&=& \sqrt{\frac{2J+1}{2}} \int_{-1}^1 d\cos\theta \,
d^J_{M \lambda}(\theta) |p\theta M\lambda_1\lambda_2\rangle, \\
|p\theta M\lambda_1\lambda_2\rangle
&=& \frac{1}{\sqrt{2\pi}} \int_0^{2\pi} d\phi \,
e^{i\phi M} |p\theta\phi\lambda_1\lambda_2\rangle,
\end{eqnarray}
and the inverse relations (obtained using the orthogonality relations)
\begin{eqnarray}
|p\theta\phi\lambda_1\lambda_2\rangle
&=& \frac{1}{\sqrt{2\pi}} \sum_M
e^{-i\phi M} 
|p\theta M\lambda_1\lambda_2\rangle \\
&=& \sum_{J,M} \sqrt{\frac{2J+1}{4\pi}}
{\mathcal D}^J_{M\lambda}(\phi, \theta, 0)
|pJM\lambda_1\lambda_2\rangle, \\
|p\theta M\lambda_1\lambda_2\rangle
&=& \sum_J \sqrt{\frac{2J+1}{2}}
d^J_{M \lambda}(\theta) |pJM\lambda_1\lambda_2\rangle.
\end{eqnarray}

\section{Jacobi Polynomials}

\doquote
{Logic, like whiskey, loses its beneficial effect when taken in too
large quantities.}
{Lord Dunsany} 

The Jacobi polynomials are neat guys.  They are defined by
\begin{eqnarray}
\langle J,M' | R_{0\theta0} |J,M\rangle &=& d^J_{M'M}(\theta).
\end{eqnarray}
Note that since the Jacobi polynomials are real,
\begin{eqnarray}
d^J_{M'M}(\theta) =
\left(\langle J,M' | R_{0\theta0} |J,M\rangle\right)^*
&=& \langle J,M | R_{0-\theta0} |J,M' \rangle
= d^J_{MM'}(-\theta).
\end{eqnarray}

There are other symmetries that the Jacobi polynomials obey.  To see
them, we turn to Schwinger's model of angular momentum.  In this model,
we have 
\begin{eqnarray}
|J,M\rangle &=&
\frac{\left(a_+^\dagger\right)^{J+M}}{\sqrt{J+M}}
\frac{\left(a_-^\dagger\right)^{J-M}}{\sqrt{J-M}} |0,0\rangle,
\end{eqnarray}
where  $a_\pm^\dagger$ essentially adds in one half unit of spin, and
changes $M$ by $\pm\frac{1}{2}$.

What does this state do under rotation about the $y$-axis?  Well,
\begin{eqnarray}
R_{0\theta0} |J,M\rangle
&=&
\frac{\left(R_{0\theta0} a_+^\dagger R_{0\theta0}^\dagger \right)^{J+M}}
{\sqrt{J+M}}
\frac{\left(R_{0\theta0} a_-^\dagger R_{0\theta0}^\dagger \right)^{J-M}}
{\sqrt{J-M}}
R_{0\theta0} |0,0\rangle.
\end{eqnarray}
The vacuum is unaffected by rotations.  Now, Sakurai shows that \cite{Sakurai:1994} 
\begin{eqnarray}
R_{0\theta0} a_\pm^\dagger R_{0\theta0}^\dagger
&=& a_\pm^\dagger \cos\frac{\theta}{2} \pm a_\mp^\dagger \sin\frac{\theta}{2}.
\end{eqnarray}
This can be used to get an explicit expression for $d$.

Consider now what happens to the state under a rotation by $\pi$.
We get $a_\pm^\dagger \rightarrow \pm a_\mp^\dagger$.  Thus,
\begin{eqnarray}
R_{0\pi0} |J,M\rangle
&=&
\frac{\left(a_-^\dagger\right)^{J+M}}{\sqrt{J+M}}
\frac{\left(-a_+^\dagger\right)^{J-M}}{\sqrt{J-M}} |0,0\rangle \\
&=& (-1)^{J-M}
\frac{\left(a_+^\dagger\right)^{J-M}}{\sqrt{J-M}}
\frac{\left(a_-^\dagger\right)^{J+M}}{\sqrt{J+M}} |0,0\rangle \\
&=& (-1)^{J-M} |J,-M\rangle,
\end{eqnarray}
which gives a nice relation for Clebsch-Gordon coefficients
\begin{eqnarray}
\langle J_1 M_1 J_2 M_2 | J M \rangle
&=& (-1)^{J_1+J_2+J-M_1-M_2-M}
\langle J_1 -M_1 J_2 -M_2 | J -M \rangle \\
&=& (-1)^{J_1+J_2-J} \langle J_1 -M_1 J_2 -M_2 | J -M \rangle.
\end{eqnarray}

Using this result for the Jacobi polynomials,
\begin{eqnarray}
d^J_{M'M}(\pi-\theta)
&=& \langle J,M' | R_{0\pi-\theta0} |J,M\rangle \\
&=& \langle J,M' | R_{0-\theta0} R_{0\pi0}|J,M\rangle \\
&=& (-1)^{J-M} \langle J,M' | R_{0-\theta0} |J,-M\rangle
= (-1)^{J-M} d^J_{M'-M}(-\theta) \\
&=& (-1)^{J-M'} \langle J,-M' | R_{0-\theta0} |J,M\rangle
= (-1)^{J-M'} d^J_{-M'M}(-\theta).
\end{eqnarray}

We can rotate by a full $2\pi$, and get
\begin{eqnarray}
d^J_{M'M}(2\pi+\theta)
&=& d^J_{M'M}(\theta) \\
&=& \langle J,M' | R_{0\pi0} R_{0\theta0} R_{0\pi0}|J,M\rangle \\
&=& (-1)^{M'-M} \langle J,-M' | R_{0\theta0} |J,-M\rangle
= (-1)^{M'-M} d^J_{-M'-M}(\theta).
\end{eqnarray}

We can also use a trick from Ref.~\cite{Chaichian:1998}.  Note that
\begin{eqnarray}
e^{-i\beta J_y} &=& e^{+i\pi J_x} e^{+i\beta J_y} e^{-i\pi J_x}.
\end{eqnarray}
The $J_x$ exponential is calculated in Eq.~(\ref{eq:expofjx}),
and for an angle of $\pi$ it is
\begin{eqnarray}
e^{-i\pi J_x} &=& i \left(\begin{array}{rr}0&1\\1&0\\\end{array}\right),
\end{eqnarray}
so under this rotation $a_\pm^\dagger \rightarrow i a_\mp^\dagger$, so
\begin{eqnarray}
e^{-i\pi J_x} |J,M\rangle &=& (i)^{J+M+J-M} |J,-M\rangle = (-1)^J |J,-M\rangle.
\end{eqnarray}
Thus,
\begin{eqnarray}
d^J_{M'M}(\theta)
&=& \langle J,M' | e^{+i\beta J_y}
R_{0-\theta0}
e^{-i\beta J_y} |J,M\rangle \\
&=& \langle J,-M' | R_{0-\theta0} |J,-M\rangle \\
&=& d^J_{-M'-M}(-\theta).
\end{eqnarray}

Also, we can use the $x$-rotation to look at
\begin{eqnarray}
d^J_{M'M}(\pi-\theta)
&=& \langle J,M' | R_{0\pi-\theta0} |J,M\rangle \\
&=& \langle J,M' | R_{0-\theta0} R_{0\pi0}|J,M\rangle \\
&=& (-1)^{J-M} \langle J,M' | R_{0-\theta0} |J,-M\rangle
= (-1)^{J-M} d^J_{M'-M}(-\theta) \\
&=& (-1)^{J-M'} \langle J,-M' | R_{0-\theta0} |J,M\rangle
= (-1)^{J-M'} d^J_{-M'M}(-\theta).
\end{eqnarray}

Summarizing what we have so far,
\begin{eqnarray}
d^J_{M'M}(\theta)
&=& d^J_{MM'}(-\theta) \label{jacpoly:ex:mb} \\
&=& d^J_{-M'-M}(-\theta) \label{jacpoly:mm:mmp:mb} \\
&=& (-1)^{M'-M} d^J_{-M'-M}(\theta) \label{jacpoly:mm:mmp}, \\
d^J_{MM'}(\pi-\theta)
&=& (-1)^{J-M } d^J_{M'-M}(-\theta) \label{jacpoly:mm:mb} \\
&=& (-1)^{J-M'} d^J_{-M'M}(-\theta) \label{jacpoly:mmp:mb}.
\end{eqnarray}

What sort of fundamental symmetries do we expect the Jacobi polynomials
to obey?
\begin{enumerate}
\item Under $\theta \rightarrow -\theta$.
Combining Eqs.~(\ref{jacpoly:mm:mmp}) and (\ref{jacpoly:mm:mmp:mb}) gives
\begin{eqnarray}
d^J_{M'M}(\theta) = (-1)^{M'-M} d^J_{M'M}(-\theta) \label{jacpoly:mb}.
\end{eqnarray}

\item Under $m  \leftrightarrow  m'$.
Combining Eqs.~(\ref{jacpoly:ex:mb}), (\ref{jacpoly:mm:mmp:mb}), and
(\ref{jacpoly:mm:mmp}), we get
\begin{eqnarray}
d^J_{M'M}(\theta) &=& (-1)^{M'-M} d^J_{MM'}(\theta) \label{jacpoly:ex}.
\end{eqnarray}

\item Under $m \rightarrow -m$ and $m' \rightarrow -m'$.
Looking at Eq.~(\ref{jacpoly:mm:mmp}), 
\begin{eqnarray}
d^J_{M'M}(\theta) &=& (-1)^{M'-M} d^J_{-M'-M}(\theta) \label{jacpoly:mm:mmp:2}.
\end{eqnarray}

\item Under $\theta \rightarrow \pi\pm\theta$.  Apparently this is
\textit{not} a symmetry that the Jacobi polynomials have.  This must be
accompanied with a $m$ or $m'$ being negated as well.

\item Under $\theta \rightarrow \pi-\theta$ and $m  \rightarrow -m $.
Combining Eqs.~(\ref{jacpoly:mm:mb}) and (\ref{jacpoly:mb}) gives
\begin{eqnarray}
d^J_{M'M}(\pi-\theta)
&=& (-1)^{J-M'} d^J_{M'-M}(\theta) \label{jacpoly:mm:parb}.
\end{eqnarray}

\item Under $\theta \rightarrow \pi-\theta$ and $m' \rightarrow -m'$.
Combining Eqs.~(\ref{jacpoly:mmp:mb}) and (\ref{jacpoly:mb}) gives
\begin{eqnarray}
d^J_{M'M}(\pi-\theta)
&=& (-1)^{J-M} d^J_{-M'M}(\theta) \label{jacpoly:mmp:parb}.
\end{eqnarray}

\end{enumerate}

As a closing note for this section, Sakurai gives an expression for
$d^J_{M'M}$ calculated using Schwinger's angular momentum model
\cite{Sakurai:1994},
\begin{eqnarray}
d^J_{M'M}(\beta)
&=& 
\sqrt{(J+M)!(J-M)!(J+M')!(J-M')!}
\nonumber \\ && \times
\sum_{k=k_{\text{min}}}^{k_{\text{max}}}
(-1)^{k-M+M'}
\frac{s_2^{k          }}{ k!           }
\frac{s_2^{k - M  + M'}}{(k - M  + M')!} \nonumber\\
&&\phantom{\times\sum_{k=k_{\text{min}}}^{k_{\text{max}}}(-1)^{k-M+M'}}
\times 
\frac{c_2^{J + M  - k }}{(J + M  - k )!}
\frac{c_2^{J - M' - k }}{(J - M' - k )!}, \\
k_{\text{min}} &=& \text{Max}(0,M-M'), \\
k_{\text{max}} &=& \text{Min}(J+M,J-M'),
\end{eqnarray}
where $s_2=\sin(\beta/2)$ and $c_2=\cos(\beta/2)$.

\section{Helicity Applied to Potentials with Full Rotational Invariance}

Now, we want to look at matrix elements of the potential
\begin{eqnarray}
\langle \bbox{p}' \lambda_1' \lambda_2' | V |
\bbox{p} \lambda_1 \lambda_2 \rangle
&=&
\sum_{J',M'} \sqrt{\frac{2J'+1}{4\pi}}
{{\mathcal D}^{J'}_{M'\lambda'}}^*(\phi', \theta', 0)
\sum_{J ,M } \sqrt{\frac{2J +1}{4\pi}}
{\mathcal D}^J_{M\lambda}(\phi, \theta, 0) \nonumber \\ && \qquad
\langle p'J'M'\lambda_1'\lambda_2' | V
|pJM\lambda_1\lambda_2\rangle \label{decomp}.
\end{eqnarray}
But if the potential has full rotational invariance, that means that it
is a scalar.  We need to use the Wigner-Eckart theorem, which
states that
\begin{eqnarray}
\langle \alpha' J' M' | T_m^j | \alpha J M \rangle &=&
\langle J j; M m | J j; J' M' \rangle
\frac{\langle \alpha' J' || T^j || \alpha J \rangle}{\sqrt{2j+1}}.
\end{eqnarray}
For a scalar potential, $V=T_0^0$, so
\begin{eqnarray}
\langle \alpha' J' M' | V | \alpha J M \rangle &=&
\delta_{MM'}\delta_{JJ'}
\frac{\langle \alpha' J || V || \alpha J \rangle}{\sqrt{2J+1}}.
\end{eqnarray}
Thus, the potential matrix element is independent of the $M$ value, and
is diagonal in $J$.

This allows us to write
\begin{eqnarray}
\langle \bbox{p}' \lambda_1' \lambda_2' | V |
\bbox{p} \lambda_1 \lambda_2 \rangle
&=&
\sum_J \frac{2J +1}{4\pi}
\left[\sum_M
{{\mathcal D}^{J}_{M\lambda'}}^*(\phi', \theta', 0)
{\mathcal D}^J_{M\lambda}(\phi, \theta, 0) \right]
\nonumber \\ && \qquad
\langle p' J \lambda_1' \lambda_2' | V | p J \lambda_1 \lambda_2 \rangle.
\end{eqnarray}
The $M$ dependence of the matrix element has simply been omitted, since
there is no dependence.

Recall the addition theorem for the rotation matrices,
\begin{eqnarray}
{\mathcal D}^J_{MM'}(\alpha\beta\gamma) &=&
\sum_{M''}
{\mathcal D}^J_{MM''}(\alpha_1\beta_1\gamma_1)
{\mathcal D}^J_{M''M}(\alpha_2\beta_2\gamma_2),
\end{eqnarray}
where 
\begin{eqnarray}
R_{\alpha\beta\gamma}
&=& R_{\alpha_1\beta_1\gamma_1} R_{\alpha_2\beta_2\gamma_2} \\
e^{-i\alpha J_z} e^{-i\beta J_y} e^{-i\gamma M}
&=&
e^{-i\alpha_1 J_z}
e^{-i\beta_1 J_y} e^{-i(\gamma_1+\alpha_2) J_z} e^{-i\beta_2 J_y}
e^{-i\gamma_2 J_z}.
\end{eqnarray}
Using the addition theorem, we get
\begin{eqnarray}
\sum_M
{{\mathcal D}^{J}_{M\lambda'}}^*(\phi', \theta', 0)
{\mathcal D}^J_{M\lambda}(\phi, \theta, 0)
&=& \sum_M
{\mathcal D}^J_{\lambda'M}(0, -\theta', -\phi')
{\mathcal D}^J_{M\lambda }(\phi ,  \theta , 0 ) \\
&=& {\mathcal D}^J_{\lambda'\lambda}(\phi_1'', \theta'', -\phi_2'').
\end{eqnarray}
In general, $\phi_1''\neq\phi_2''$, and $\theta''\neq\theta'-\theta$.
However, we will be able to make some simplifications.

The sum rule allows us to write
\begin{eqnarray}
\langle \bbox{p}' \lambda_1' \lambda_2' | V |
\bbox{p} \lambda_1 \lambda_2 \rangle
&=&
e^{-i\phi_1''\lambda'} e^{i\phi_2''\lambda}
\sum_J \frac{2J +1}{4\pi} d^J_{\lambda'\lambda}(\theta'')
\langle p'J\lambda_1'\lambda_2' | V |pJ\lambda_1\lambda_2\rangle.
\end{eqnarray}
All we are interested is the matrix element buried inside the sum, so we
invert the equation to get
\begin{eqnarray}
\langle p'J\lambda_1'\lambda_2' | V |pJ\lambda_1\lambda_2\rangle
&=&
2\pi \int_{-1}^1 d\cos\theta'' \,
d^J_{\lambda'\lambda}(\theta'')
e^{i\phi_1''\lambda'} e^{-i\phi_2''\lambda}
\langle \bbox{p}' \lambda_1' \lambda_2' | V |
\bbox{p} \lambda_1 \lambda_2 \rangle.
\end{eqnarray}
Note that there are three free parameters in the integrand.  We can
choose $\phi=\phi'=\theta=0$.  This means that $\phi''_1=\phi''_2=0$,
and $\theta''=\theta'$.  Thus,
\begin{eqnarray}
&&\langle p'J\lambda_1'\lambda_2' | V |pJ\lambda_1\lambda_2\rangle\nonumber\\
&&\qquad=
2\pi \int_{-1}^1 d\cos\theta' \,
d^J_{\lambda'\lambda}(\theta')
\langle (p'\sin\theta',0,p'\cos\theta') \lambda_1' \lambda_2' | V |
p\widehat{\bbox{z}} \lambda_1 \lambda_2 \rangle,
\label{tp2jm}
\end{eqnarray}
which is what is advertised in Machleidt's Eq.~(E.41) in
Ref.~\cite{Machleidt:1987hj}.
 
\section{Helicity Applied to Potentials with Cylindrical Invariance}

Now, we want to look at matrix elements of the potential
\begin{eqnarray}
\langle \bbox{p}' \lambda_1' \lambda_2' | V |
\bbox{p} \lambda_1 \lambda_2 \rangle
&=&
\sum_{J',M'} \sqrt{\frac{2J'+1}{4\pi}}
{{\mathcal D}^{J'}_{M'\lambda'}}^*(\phi', \theta', 0)
\sum_{J ,M } \sqrt{\frac{2J +1}{4\pi}}
{\mathcal D}^J_{M\lambda}(\phi, \theta, 0) \nonumber \\ && \qquad
\langle p'J'M'\lambda_1'\lambda_2' | V
|pJM\lambda_1\lambda_2\rangle  \label{decomprotvar}.
\end{eqnarray}
Now we consider that the potential operator is not a scalar, but it
still conserves $M$.  In terms of spherical tensor operators, we can
write
\begin{eqnarray}
V &=& \sum_J V^{(J)}_{M=0}.
\end{eqnarray}
This does not bode well for an approach like the one that was used in
the previous section.

However, we can write
\begin{eqnarray}
\langle \bbox{p}' \lambda_1' \lambda_2' | V |
\bbox{p} \lambda_1 \lambda_2 \rangle
&=&
\sum_{J'} \sqrt{\frac{2J'+1}{4\pi}}
\sum_{J ,M } \sqrt{\frac{2J +1}{4\pi}}
e^{i M(\phi'-\phi)}
d^{J'}_{\lambda'M}(\theta')
d^{J }_{\lambda M}(\theta )
 \nonumber \\ && \qquad
\langle p'J'M\lambda_1'\lambda_2' | V
|pJM\lambda_1\lambda_2\rangle.
\end{eqnarray}
Since the $M$ is all that is conserved, it makes sense to ignore $J$ for
now.  So we change basis to 
\begin{eqnarray}
\langle \bbox{p}' \lambda_1' \lambda_2' | V |
\bbox{p} \lambda_1 \lambda_2 \rangle
&=&
\sum_M \frac{1}{2\pi}
e^{i \Delta\phi M}
\langle p'\theta' M\lambda_1'\lambda_2' | V
|p\theta M\lambda_1\lambda_2\rangle,
\end{eqnarray}
where $\Delta\phi=\phi'-\phi$.  As in the previous section, all we are
interested in is the matrix element buried in the sum.  We invert the
equation to obtain
\begin{eqnarray}
\langle p'\theta' M\lambda_1'\lambda_2' | V
|p\theta M\lambda_1\lambda_2\rangle
&=&
\int_0^{2\pi} d(\Delta\phi) \,
e^{-i \Delta\phi M}
\langle \bbox{p}' \lambda_1' \lambda_2' | V |
\bbox{p} \lambda_1 \lambda_2 \rangle.
\end{eqnarray}
There is one free angle in the integrand.  Setting $\phi=0$ for
simplicity, to get
\begin{eqnarray}
\langle p'\theta' M\lambda_1'\lambda_2' | V
|p\theta M\lambda_1\lambda_2\rangle
&=&
\int_0^{2\pi} d\phi' \, e^{-i \phi' M}
\langle \bbox{p}' \lambda_1' \lambda_2' | V |
p_3, p_\perp \widehat{\bbox{x}} \lambda_1 \lambda_2 \rangle
\label{tp2tm}.
\end{eqnarray}

How does this relate to what we got before, with the $J,M$ matrix
elements?  We need to change basis once more, to get
\begin{eqnarray}
\langle p' J' M \lambda_1'\lambda_2' | V |p J M\lambda_1\lambda_2\rangle 
&=&
\sqrt{\frac{2J'+1}{2}} \int_{-1}^1 d\cos\theta' \,d^{J'}_{M\lambda'}(\theta')
\nonumber\\&&\qquad \times
\sqrt{\frac{2J+1}{2}} \int_{-1}^1 d\cos\theta \,d^J_{M\lambda }(\theta)
\nonumber\\&&\qquad\qquad \times
\langle p'\theta' M \lambda_1'\lambda_2' | V
|p\theta M \lambda_1\lambda_2\rangle
\label{tm2jm}.
\end{eqnarray}

\section{Summary of Transformations} \label{app:rot:summ}

\doquote
{Oh, I have now a mania for shortness. Whatever I read - my own or other
people's works - it all seems to me not short enough.}
{Anton Chekhov} 

First of all, we would like to be able to transform to $|JMLS\rangle$.
In particular, we would like
\begin{eqnarray}
\langle JMS\lambda | JMLS \rangle
&=& \int d\Omega \,
\langle JMS\lambda | \theta \phi S\lambda \rangle
\langle \theta \phi S\lambda | JMLS \rangle.
\end{eqnarray}
Looking at the first term,
\begin{eqnarray}
\langle JMS\lambda | \theta \phi S\lambda \rangle
&=&
\sqrt{\frac{2J+1}{4\pi}} {\mathcal D}^J_{M\lambda}(\phi,\theta,0).
\end{eqnarray}
The second term takes a little more work.
\begin{eqnarray}
\langle \theta \phi S\lambda | JMLS \rangle
&=&
\langle 0 0 S\lambda | R_{\phi\theta0}^\dagger | JMLS \rangle \nonumber \\
&=&
\sum_{M'}
\langle 0 0 S\lambda | JM'LS \rangle
\langle JM'SL | R_{\phi\theta0}^\dagger | JMLS \rangle \nonumber \\
&=& \sum_{M_L,M_S,M'}
\langle 0 0 S\lambda | L M_L S M_S \rangle
\langle L M_L S M_S | JM'LS \rangle
\nonumber\\&&\qquad\qquad\times
\left( \langle JMSL | R_{\phi\theta0} | JM'LS \rangle \right)^*
\nonumber \\
&=& \sum_{M_L,M_S,M'}
Y_{LM_L}(0,0) \delta_{\lambda,M_S}
\langle L M_L S M_S | JM' \rangle
{{\mathcal D}^J_{MM'}}^*(\phi,\theta,0) \nonumber \\
&=& \sum_{M_L,M_S,M'}
\sqrt{\frac{2L+1}{4\pi}} \delta_{M_L,0} \delta_{\lambda,M_S}
\langle L M_L S M_S | JM' \rangle
{{\mathcal D}^J_{MM'}}^*(\phi,\theta,0) \nonumber\\
&=&
\sqrt{\frac{2L+1}{4\pi}} \langle L 0 S \lambda | J \lambda \rangle
{{\mathcal D}^J_{M\lambda}}^*(\phi,\theta,0)
\end{eqnarray}
Combining these, we get, by virtue of the orthonormality of the
${\mathcal D}$'s, 
\begin{eqnarray}
\langle JMS\lambda | JMLS \rangle 
&=&
\sqrt{\frac{2L+1}{2J+1}} \langle L 0 S \lambda | J \lambda \rangle
\left[ \frac{2J+1}{4\pi} \int d\Omega \,
 {\mathcal D}^J_{M\lambda}   (\phi,\theta,0)
{{\mathcal D}^J_{M\lambda}}^*(\phi,\theta,0) \right] \nonumber \\
&=&
\sqrt{\frac{2L+1}{2J+1}} \langle L 0 S \lambda | J \lambda \rangle.
\end{eqnarray}
Using this, we obtain Machleidt's equation Eq.~(C18) in
Ref.~\cite{Machleidt:1987hj},
\begin{eqnarray}
|JMLS\rangle &=& \sum_\lambda
\sqrt{\frac{2L+1}{2J+1}} \langle L 0;S\lambda | J\lambda \rangle 
|JMS\lambda\rangle.
\end{eqnarray}

For a one particle state, we start with $|\theta\phi S\lambda\rangle$.
From that,
\begin{eqnarray}
|\theta MS\lambda\rangle &=&
\frac{1}{\sqrt{2\pi}} \int_0^{2\pi} d\phi\, e^{iM\phi}
\, |\theta\phi S\lambda\rangle, \\
|JMS\lambda\rangle &=&
\sqrt{\frac{2J+1}{2}} \int_{-1}^1 d\cos\theta\, d^J_{M\lambda}(\theta)
\, |\theta MS\lambda\rangle, \\
|JMLS\rangle &=&
\sqrt{\frac{2L+1}{2J+1}} \langle L 0 S\lambda | J\lambda \rangle
\, |JMS\lambda\rangle.
\end{eqnarray}
Furthermore, for two particle states, we have
$| \theta\phi S_1 \lambda_1 S_2 \lambda_2 \rangle$. However, that is
rewritten as
\begin{eqnarray}
|\theta\phi S\lambda S_1 S_2 \rangle
&=& \sum_{\lambda_1\lambda_2}
\langle \theta\phi S_1 \lambda_1 S_2 \lambda_2
|\theta\phi S\lambda S_1 S_2 \rangle
| \theta\phi S_1 \lambda_1 S_2 \lambda_2 \rangle \\
&=&\sum_{\lambda_1\lambda_2}
\langle 00 S_1 \lambda_1 S_2 \lambda_2 |
R_{\phi\theta0}^\dagger R_{\phi\theta0}
| 00 S\lambda S_1 S_2 \rangle
| \theta\phi S_1 \lambda_1 S_2 \lambda_2 \rangle \\
&=&\sum_{\lambda_1\lambda_2}
\langle 00 S_1 \lambda_1 S_2 \lambda_2 | 00 S\lambda S_1 S_2 \rangle
| \theta\phi S_1 \lambda_1 S_2 \lambda_2 \rangle \\
&=&\sum_{\lambda_1\lambda_2}
\langle S_1 M_1=\lambda_1 S_2 M_2=-\lambda_2 | S M=\lambda S_1 S_2 \rangle
| \theta\phi S_1 \lambda_1 S_2 \lambda_2 \rangle \\
&=&\sum_{\lambda_1\lambda_2}
\langle S_1 \lambda_1 S_2 -\lambda_2 | S \lambda \rangle
| \theta\phi S_1 \lambda_1 S_2 \lambda_2 \rangle.
\end{eqnarray}
Now, the $|\theta\phi S\lambda S_1 S_2 \rangle$ can be used to get
\begin{eqnarray}
|\theta MS\lambda S_1S_2\rangle &=&
\frac{1}{\sqrt{2\pi}} \int_0^{2\pi} d\phi\, e^{iM\phi}
\, |\theta\phi S\lambda S_1S_2\rangle, \\
|JMS\lambda S_1 S_2\rangle &=&
\sqrt{\frac{2J+1}{2}} \int_{-1}^1 d\cos\theta\, d^J_{M\lambda}(\theta)
\, |\theta MS\lambda S_1 S_2\rangle, \\
|JMLS S_1 S_2\rangle &=&
\sqrt{\frac{2L+1}{2J+1}} \langle L 0 S\lambda | J\lambda \rangle
\, |JMS\lambda S_1 S_2\rangle.
\end{eqnarray}

How does this relate to normal spins?  Well, for those,
\begin{eqnarray}
|\theta\phi S M_S S_1 S_2 \rangle
&=&\sum_{M_1M_2}
\langle S_1 M_1 S_2 M_2 | S M_S \rangle | \theta\phi S_1 M_1 S_2 M_2
\rangle. \label{eq:rot:spinhel}
\end{eqnarray}
But the relation of these to helicity is complicated, since 
in general every $M_1$ is connected to every $\lambda_1$.  In
particular,
\begin{eqnarray}
\langle \theta \phi S_1 \lambda_1 &S_2& \lambda_2
| \theta\phi S_1 M_1 S_2 M_2 \rangle \nonumber \\
&=&\langle 00 S_1 \lambda_1 S_2 \lambda_2 |
R_{\phi\theta0}^\dagger R_{\phi\theta0}^{(L)}
| 00 S_1 M_1 S_2 M_2 \rangle \\
&=&
\langle 00 S_1 \lambda_1 S_2 \lambda_2 |
e^{i \theta J_y} e^{i \phi J_z} e^{-i\phi L_z} e^{-i\theta L_y}
| 00 S_1 M_1 S_2 M_2 \rangle \\
&=&
\langle 00 S_1 \lambda_1 S_2 \lambda_2 |
e^{i \theta J_y}
e^{-i\theta L_y}
e^{i\phi S_z}
| 00 S_1 M_1 S_2 M_2 \rangle \\
&=&
\langle 00 S_1 \lambda_1 S_2 \lambda_2 |
e^{i\theta S_y} e^{i\phi S_z} 
| 00 S_1 M_1 S_2 M_2 \rangle \\
&=&
\langle 00 | 00 \rangle
\langle S_1 \lambda_1 | e^{-i\phi S_z} e^{-i\theta S_y} 
| S_1 M_1 \rangle
\langle S_2 -\lambda_2 |
e^{i\theta S_y} e^{i\phi S_z} 
| S_2 M_2 \rangle \\ 
&=&
\langle S_1  \lambda_1 | e^{i\theta S_y} | S_1 M_1 \rangle
\langle S_2 -\lambda_2 | e^{i\theta S_y} | S_2 M_2 \rangle
e^{i \phi (M_1 + M_2)} \\ 
&=&
e^{i \phi (M_1 + M_2)}
d^{S_1}_{M_1, \lambda_1}(\theta)
d^{S_2}_{M_2,-\lambda_2}(\theta) .
\end{eqnarray}
This is easily generalizable, to get
\begin{eqnarray}
\langle \theta \phi \{ S_i \lambda_i \} | \theta\phi \{ S_i M_i \} \rangle
&=& \prod_{i} e^{i \phi M_i} d^{S_i}_{M_i, \lambda_i}(\theta).
\end{eqnarray}
Well, at least this makes sense for 1 and 2 particle states.

This has an implication for parity, $\Pi$.  We get
\begin{eqnarray}
\langle \theta \phi \{ S_i \lambda_i \} | \theta\phi \{ S_i M_i \} \rangle
&=&
\langle \theta \phi \{ S_i \lambda_i \} |
\Pi \Pi | \theta\phi \{ S_i M_i \} \rangle \\
&=& \delta
\langle \pi-\theta \pi+\phi \{ S_i-\lambda_i \}
| \pi-\theta \pi+\phi \{ S_i M_i \} \rangle \\
&=& \delta
 \prod_{i} e^{i (\pi+\phi) M_i} d^{S_i}_{M_i,-\lambda_i}(\pi-\theta) \\
&=& \delta
 \prod_{i} e^{i \phi M_i} 
(-1)^{M_i} (-1)^{S_i-M_i}
d^{S_i}_{M_i, \lambda_i}(\theta) \\
&=& \left[ \delta \prod_i (-1)^{S_i} \right]
 \prod_{i} e^{i \phi M_i} d^{S_i}_{M_i, \lambda_i}(\theta) \\
&=&
 \prod_{i} e^{i \phi M_i} d^{S_i}_{M_i, \lambda_i}(\theta).
\end{eqnarray}
So we see that $\delta = \prod_i (-1)^{S_i}$, so
\begin{eqnarray}
\Pi | \theta \phi S \lambda \rangle
&=& (-1)^S
| \pi-\theta \pi+\phi S -\lambda \rangle \\
\Pi | \theta \phi S_1 \lambda_1 S_2 \lambda_2 \rangle
&=& (-1)^{S_1+S_2}
| \pi-\theta \pi+\phi S_1 -\lambda_1 S_2 -\lambda_2 \rangle.
\end{eqnarray}

Continuing on, we can derive other kets from the normal spin kets,
\begin{eqnarray}
|\theta M_L S M_S S_1 S_2 \rangle
&=& \frac{1}{\sqrt{2\pi}} \int_0^{2\pi} d\phi \,
e^{i M \phi}
|\theta\phi S M_S S_1 S_2 \rangle, \\
|L M_L S M_S S_1 S_2 \rangle
&=&
\int_{-1}^1 d\cos\theta \, Y_{LM_L}(\theta,0)
|\theta M_L S M_S S_1 S_2 \rangle, \\
|J M L S S_1 S_2 \rangle
&=& \sum_{M_L,M_S}
\langle L M_L S M_S | J M \rangle
|L M_L S M_S S_1 S_2 \rangle.
\end{eqnarray}

\vita{
\begin{quotation}
``The life of every individual, if we survey it as a whole and in
general, and only lay stress upon its most significant features, is
really always a tragedy, but gone through in detail, it has the
character of a comedy.  For the deeds and vexations of the day, the
restless irritation of the moment, the desires and fears of the week,
the mishaps of every hour, are all through chance, which is ever bent
upon some jest, scenes of a comedy.

``But the never-satisfied wishes, the frustrated efforts, the hopes
unmercifully crushed by fate, the unfortunate errors of the whole life,
with increasing suffering and death at the end, are always a
tragedy. Thus, as if fate would add derision to the misery of our
existence, our life must contain all the woes of tragedy, and yet we
cannot even assert the dignity of tragic characters, but in the broad
detail of life must inevitably be the foolish characters of a comedy.'' 
\\ ---\textit{Arthur Schopenhauer}
\end{quotation}

Jason Randolph Cooke was born in Binghamton, New York state in
January 26, 1974.  After graduating from Owego Free Academy in Owego,
New York in 1992, he enrolled at the State University of New York at
Stony Brook. He earned a Bachelor of Science degree with honors in
both Physics and Mathematics in May, 1996, and subsequently entered
graduate school at the University of Washington in Seattle that fall.
He earned a Master of Science in Physics and was admitted into
candidacy in the doctoral program in December, 1997. In December, 2001,
he earned a Doctor of Philosophy in Physics at the University of
Washington.}

\end{document}